\documentclass[pdflatex,sn-mathphys-num]{sn-jnl}

\usepackage{graphicx}
\usepackage{multirow}
\usepackage{amsmath,amssymb,amsfonts}
\usepackage{amsthm}
\usepackage{mathrsfs}
\usepackage[title]{appendix}
\usepackage{xcolor}
\usepackage{textcomp}
\usepackage{manyfoot}
\usepackage{booktabs}
\usepackage{algorithm}
\usepackage{algorithmicx}
\usepackage{algpseudocode}
\usepackage{listings}
\usepackage{caption}
\usepackage{seqsplit}
\usepackage{lineno}
\usepackage{cleveref}
\usepackage{longtable}
\usepackage{pdflscape}
\usepackage{array}
\usepackage{makecell}
\usepackage{ltablex}
\usepackage{geometry}
\usepackage{float}

\theoremstyle{thmstyleone}

\theoremstyle{thmstyletwo}

\theoremstyle{thmstylethree}

\begin{document}

\title[Article Title]{Bridging the Genotype-Phenotype Gap with Generative Artificial Intelligence}

\author[1,2]{\fnm{Yangfan} \sur{Liu}}
\equalcont{These authors contributed equally to this work.}

\author*[1,3]{\fnm{Xiong} \sur{Xiong}}\email{xiongxiong@mail.hzau.edu.cn}
\equalcont{These authors contributed equally to this work.}

\author[1,3]{\fnm{Yong} \sur{Liao}}

\author[1,3]{\fnm{Mingli} \sur{Qin}}

\author[6]{\fnm{Zhen} \sur{Huang}}

\author[1,3]{\fnm{Shilin} \sur{Zhu}}

\author[1,3]{\fnm{Lilin} \sur{Yin}}

\author[1,3]{\fnm{Yuhua} \sur{Fu}}

\author[7]{\fnm{Haohao} \sur{Zhang}}

\author[1,3]{\fnm{Jingya} \sur{Xu}}

\author[1,3]{\fnm{Dong} \sur{Yin}}

\author[1,3]{\fnm{Xin} \sur{Huang}}

\author[8]{\fnm{Yuan} \sur{Quan}}

\author[9]{\fnm{Xuan} \sur{Li}}

\author[1,3]{\fnm{Tengfei} \sur{Jiang}}

\author[3,10]{\fnm{Wanneng} \sur{Yang}}

\author[2,11]{\fnm{Xiaohui} \sur{Yuan}}

\author*[4,5]{\fnm{Laurent} \sur{Frantz}}\email{laurent.frantz@lmu.de}
\author*[1,3]{\fnm{Xinyun} \sur{Li}}\email{xyli@mail.hzau.edu.cn}
\author*[1,3]{\fnm{Xiaolei} \sur{Liu}}\email{xiaoleiliu@mail.hzau.edu.cn}
\author*[1,2,3]{\fnm{Shuhong} \sur{Zhao}}\email{zhaoshuhong@yzwlab.cn}

\affil*[1]{\orgdiv{Key Laboratory of Agricultural Animal Genetics, Breeding and Reproduction, Ministry of Education \& College of Animal Science and Technology, Frontiers Science Center for Animal Breeding and Sustainable Production}, \orgname{Huazhong Agricultural University}, \orgaddress{\city{Wuhan}, \state{Hubei}, \country{China}}}

\affil*[2]{\orgname{Yazhouwan National Laboratory}, \orgaddress{\city{Sanya}, \state{Hainan}, \country{China}}}

\affil*[3]{\orgname{Hubei Hongshan Laboratory}, \orgaddress{\city{Wuhan}, \state{Hubei}, \country{China}}}

\affil*[4]{\orgdiv{Palaeogenomics Group, Institute of Palaeoanatomy, Domestication Research and the History of Veterinary Medicine}, \orgname{Ludwig Maximilian University of Munich}, \orgaddress{\city{Munich},  \state{Bavaria}, \country{Germany}}}

\affil*[5]{\orgdiv{School of Biological and Behavioural Sciences}, \orgname{Queen Mary University of London}, \orgaddress{\city{London}, \country{UK}}}

\affil[6]{\orgdiv{Fujian-Hong Kong-Macau-Taiwan Collaborative Laboratory for the Inheritance and Innovation of Traditional Chinese Medicine}, \orgname{Fujian University of Traditional Chinese Medicine}, \orgaddress{\city{Fuzhou}, \state{Fujian}, \country{China}}}

\affil[7]{\orgdiv{School of Computer Science and Technology}, \orgname{Wuhan University of Technology}, \orgaddress{\city{Wuhan}, \state{Hubei}, \country{China}}}

\affil[8]{\orgdiv{College of Informatics}, \orgname{Huazhong Agricultural University}, \orgaddress{\city{Wuhan}, \state{Hubei}, \country{China}}}

\affil[9]{\orgdiv{College of Engineering}, \orgname{Huazhong Agricultural University}, \orgaddress{\city{Wuhan}, \state{Hubei}, \country{China}}}

\affil[10]{\orgdiv{National Key Laboratory of Crop Genetic Improvement, National Center of Plant Gene Research}, \orgname{Huazhong Agricultural University}, \orgaddress{\city{Wuhan}, \state{Hubei}, \country{China}}}

\affil[11]{\orgdiv{College of Plant Protection, College of Mycology}, \orgname{Jilin Agricultural University}, \orgaddress{\city{Changchun}, \state{Jilin}, \country{China}}}

\abstract{The genotype-phenotype gap is a persistent barrier to complex trait genetic dissection, worsened by the explosive growth of genomic data (1.5 billion variants identified in the UK Biobank WGS study) alongside persistently scarce and subjective human-defined phenotypes. Digital phenotyping offers a potential solution, yet existing tools fail to balance scalable non-manual phenotype generation and biological interpretability of these quantitative traits. Here we report AIPheno, the first generative AI-driven "phenotype sequencer" that bridges this gap. It enables high-throughput, unsupervised extraction of digital phenotypes from imaging data and unlocks their biological meaning via generative network analysis. AIPheno transforms imaging modalities into a rich source of quantitative traits, dramatically enhancing cross-species genetic discovery, including novel loci such as \textit{CCBE1} (humans), \textit{KITLG-TMTC3} (domestic pigeons), and \textit{SOD2-IGF2R} (swine). Critically, its generative module decodes AI-derived phenotypes by synthesizing variant-specific images to yield actionable biological insights. For example, it clarifies how the \textit{OCA2-HERC2} locus pleiotropically links pigmentation to retinal vascular traits via vascular visibility modulation. Integrating scalable non-manual phenotyping, enhanced genetic discovery power, and generative mechanistic decoding, AIPheno establishes a transformative closed-loop paradigm. This work addresses the longstanding genotype-phenotype imbalance, redefines digital phenotype utility, and accelerates translation of genetic associations into actionable understanding with profound implications for human health and agriculture.
}

\maketitle

\section{Introduction}\label{sec1}

The genotype-phenotype gap remains one of the most persistent bottlenecks in complex trait genetic dissection, the root cause of which lies in the stark mismatch between genomic data and phenotypic data. Genomic data has undergone explosive growth, for instance, the UK Biobank Whole-Genome Sequencing (WGS) study of 490,640 participants identified 1.5 billion genetic variants\cite{uk2025whole}. By sharp contrast, phenotypic research remains severely lagging. Furthermore, traditional Genome-Wide Association Studies (GWAS) rely entirely on human-defined phenotypes (HDPs). These phenotypes are not only sparse in number and limited in coverage but also burdened with strong subjective classification biases. Such coarse-grained phenotyping capabilities neither match the massive scale of genomic data nor capture subtle true biological differences. Ultimately, this imbalance, which is defined as excess genomic data versus insufficient phenotypic capacity, directly introduces systematic biases into genetic research and severely undermines the power of genetic discovery.

The emergence of advanced digital phenotyping (e.g., imaging, wearable signals) and data-driven representation learning\cite{krizhevsky2012imagenet,bengio2013representation,chen2020improved,chen2020simple,he2022masked,klibaite2025mapping,saunders2025perturb} has partially addressed this by enabling automated phenotyping of digital phenotypic variation, which consistently outperform human-defined phenotypes (HDPs) in enhancing genetic discovery power and identifying novel loci\cite{yun2024unsupervised,kirchler2022transfergwas,xie2024igwas,bonazzola2024unsupervised,patel2024unsupervised,liu2025digital}. For example, Yun et al. \cite{yun2024unsupervised} demonstrated that low-dimensional embeddings generated via Representation Learning for Genetic Discovery (REGLE) encode clinically relevant information in one-dimensional clinical data, specifically the information that conventional expert-defined features do not fully capture. This ability of REGLE-derived embeddings, in turn, enhances the performance of both genetic discovery and disease prediction. Kirchler et al.\cite{kirchler2022transfergwas} and Xie et al.\cite{xie2024igwas} applied transfer learning and self-supervised learning, respectively, to 2D human retinal fundus images, leading to the identification of additional candidate loci associated with eye-related traits and diseases. Bonazzola et al.\cite{bonazzola2024unsupervised} and Patel et al.\cite{patel2024unsupervised} respectively applied unsupervised learning methods to 3D left ventricle mesh data and brain Magnetic Resonance Imaging (MRI) data, with the former identifying 49 significant loci and 25 suggestive loci related to left ventricular morphology, and the latter identifying 97 independent genetic loci (60 replicated) linked to brain structure. Liu et al.\cite{liu2025digital} used wearable-derived features as digital phenotypes to link with underlying genetics, achieving more accurate and objective classification of adolescents with psychiatric disorders than previously possible.

However, a critical gap remains: while these methods excel at detecting genetic associations, they offer little insight into what the learned phenotypes biologically represent or how associated genes exert their effects, mainly because the acquired phenotypes are mutually coupled and lack interpretable modules. Without interpretability, gene-associated digital phenotypes remain abstract constructs that are statistically powerful but biologically opaque. This disconnect impedes the translation of genetic hits into testable hypotheses and creates a troubling paradox: although phenotypic dimension and discovery power continue to grow, biological understanding does not necessarily keep pace.

Existing studies have attempted to address this limitation from three perspectives, though with certain constraints: First, they rely solely on pre-existing indirect evidence, such as annotating identified genetic variants using curated databases, conducting Phenome-Wide Association Studies (PheWAS), or searching for meaningful human-defined phenotypes with high (or genetic) correlation with the trait. These approaches, however, remain constrained by the scope of prior knowledge: they cannot account for phenotypic variations that are either unrecorded in existing gene annotation databases or too complex/abstract to be captured by human-defined phenotypic descriptions. As a result, they fail to uncover novel biological insights that extend beyond the boundaries of currently known information\cite{xie2024igwas,kirchler2022transfergwas}. 
Second, when using heatmaps for interpretability analysis of the image-variation phenotypes (IVPs), these tools only identify static regions of association. One can only infer that genes significantly associated with the IVP affect a specific region in the image by relying on the approximate "regional" correspondence between image variation and the image itself, with brighter areas on the heatmap indicating stronger associations. Notably, heatmaps fail to capture dynamic changes in these associated regions (e.g., how the region's phenotypic features evolve or vary across contexts). For instance, in human brain MRI, while such analyses may associate genes with anatomical regions like the putamen, pallidum, or thalamus, they lack the resolution to specify how these genes modulate the size, structural organization, or other phenotypic variation trends of these regions\cite{patel2024unsupervised,liu2025digital}.
Third, other methods first stratify individuals into subgroups based on a gene-associated phenotypic trait (e.g., high vs. low values of pupil size in domestic pigeon iris images). Then, researchers identify shared image patterns between subgroups images in two ways: one is manual inspection of subgroup images, and the other is computing group-level average images. While this workflow is intended to reveal trait-related anatomical trends, it cannot isolate the target phenotypic variation from confounding factors. These factors include variations in pupil morphology and orientation, all of which also shape subgroup images. As a result, observed image differences cannot be definitively attributed to the target phenotype, precluding precise insights into how the phenotype drives specific anatomical or structural changes in images\cite{kirchler2022transfergwas,bonazzola2024unsupervised}.
Collectively, these methods fall short of bridging the gap between gene discovery and biological insights. They generate lists of genes associated with phenotypes but fail to close the research loop. Without resolving the mechanistic links that connect specific genotypes to distinct phenotypic traits, these genetic hits remain abstract associations. They provide little actionable insight into the biological processes underlying genotype-phenotype relationships.

To bridge this gap, we introduce \textbf{AIPheno} (\textbf{A}utomated Phenotyping and \textbf{I}nterpretation of Image-variation \textbf{Pheno}types), a generative AI framework that functions as a "phenotype sequencer". It links digital phenotypes to genetic associations and further enables the visualization of actual phenotypic differences explained by specific genes, which in turn enables human interpretation of the genotype-phenotype link, offering new biological insights. Built on an Encoder-Generator architecture, AIPheno automatically extracts scalable image-variation phenotypes with high genetic discovery power, while its Generator module synthesizes visual representations of the phenotypic trends captured by each IVP, enabling intuitive, actionable interpretation of how associated genes modulate traits. We validated AIPheno across simulated data and real-world “Image-Genotype” datasets from four species: \textit{Homo sapiens} (human), \textit{Columba livia} (domestic pigeon), \textit{Oryza sativa} (rice), and \textit{Sus scrofa} (swine). AIPheno consistently outperforms both human-defined and deep learning-based phenotyping methods, identifying novel loci such as \textit{CCBE1} in humans, the \textit{KITLG-TMTC3} locus in domestic pigeons, and the \textit{SOD2-IGF2R} locus in swine. Crucially, its interpretable design not only validates reported gene-phenotype associations with visual evidence, but also refines their mechanistic context and reveals biological insights for newly identified loci. By linking digital phenotypes to genes and genes to biological insights, AIPheno establishes a closed-loop paradigm for the genetic dissection of complex traits. 

\section{Results}\label{sec2}
\subsection{Overview of the AIPheno framework}\label{subsec2.1}
\begin{figure}[h]
\centering
\includegraphics[width=1\textwidth]{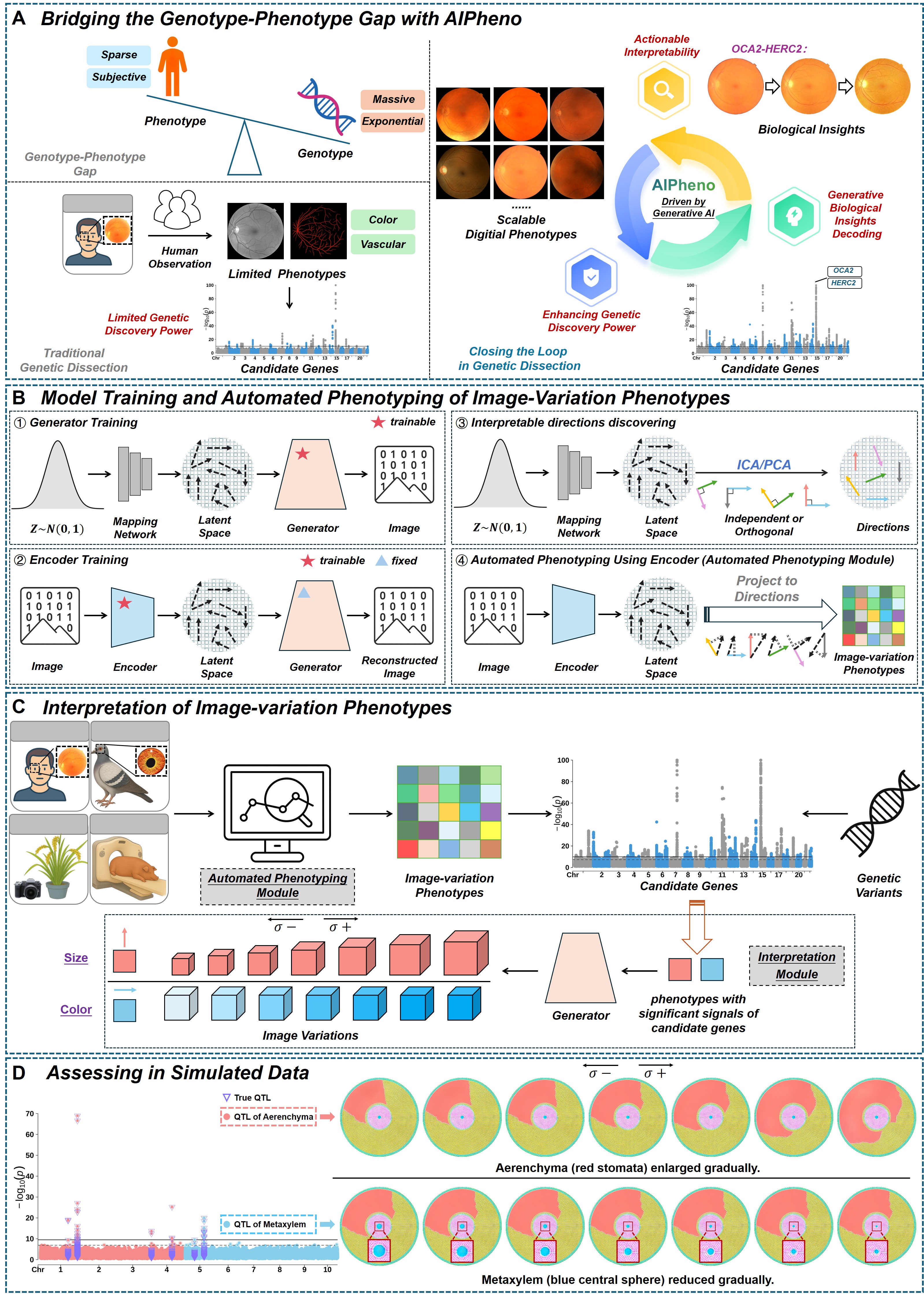}
\caption{}\label{fig1}
\end{figure}

A persistent challenge in genetic dissection is the stark imbalance between the exponential growth of genomic data and the limited, often subjective nature of available phenotypic data. The traditional definition of phenotypes is established through manual observation. This approach can only yield limited phenotypes and fails to capture the variations of high-dimensional image data adequately. Representation learning-based genetic dissection helps alleviate this imbalance by generating a multitude of data-driven digital phenotypes, which significantly boosts genetic discovery power. However, its lack of interpretability gives rise to a troubling paradox: our discovery power continues to grow, yet our biological understanding does not keep pace. This is because the uninterpretable digital phenotypes remain biologically opaque, obscuring the precise image variations associated with genetic variants. Consequently, the crucial step of translating statistically significant associations into mechanistically testable hypotheses is impeded, limiting the biological insights gained from the analysis. Therefore, we propose a novel and efficient data-driven framework for genetic dissection, effectively a "phenotype sequencer". Independent of prior knowledge, this generative AI-driven framework enables the scalable \textbf{A}\-utomated Phenotyping and \textbf{I}\-nterpretation of image-variation \textbf{Pheno}\-types (AIPheno). This interpretable genetic dissection framework, which consists of two core modules for automated phenotyping and interpretation, closes the loop from image-variation phenotypes to genes, which enhances genetic discovery power, and subsequently to biological insight by providing generative biological insights decoding and actionable interpretability (Fig.\ref{fig1}A, Data S1).

Model training and automated phenotyping of IVPs can be divided into four steps (Fig.\ref{fig1}B). First, an unconditional Generator was constructed to approximate the spatial distribution of input image data, which represents an unsupervised learning approach. Second, an additional Encoder was trained (via joint training with the Generator) for rapid image representation acquisition: the Encoder mapped original images to a latent representation, which was input to the Generator for reconstruction; the reconstruction loss (calculated as the difference between original and reconstructed images) was then used to optimize the Encoder. Third, within the latent space of the Generator (which approximates the original image distribution), several interpretable directions existed, which signified a specific variational tendency in the image. To obtain these, data-driven methods such as Independent Component Analysis\cite{lee1998independent} (ICA, for mutually independent directions) and Principal Component Analysis\cite{greenacre2022principal} (PCA, for orthogonal directions) were utilized. Finally, the degree of image variation was measured by projecting the image representation onto the interpretable directions obtained in the third step. This final step concluded the process within the automated phenotyping module, and its output constituted the final IVPs. The automated phenotyping module aims to extract disentangled IVPs from each input image that can effectively generalize image variation phenotypes. We referenced the same notation as GANSpace to denote IVP identifiers\cite{harkonen2020ganspace}. For instance, IVP$(\mathbf{v}_{i}, j)$ was defined as the projection of the representation from the $j$-th layer in the $W+$ space along the $i$-th direction.

Processing of images by the automated phenotyping module yields a large number of IVPs. Each IVP was subsequently subjected to association analysis with genetic variants using a univariate Mixed Linear Model (MLM)\cite{zhang2010mixed}. The significance levels in the association analysis were subjected to Bonferroni correction\cite{haynes2013bonferroni}, which is the most stringent method to control for false positives, based on the number of genetic variants multiplied by the number of IVPs. IVPs that were associated significantly with genetic variants were used for interpretability analysis. In the interpretation module, we conducted direction-wise analysis on the directions that exhibited significant signals in the association analysis by synthesizing high-quality images for visualization (Fig.\ref{fig1}C). We traversed the latent space representation of images along each significant direction (in units of standard deviation) and used the Generator to generate images, thereby observing directional changes. This approach constrains image variation to phenotypic axes alone while preserving other details, enabling actionable visualization of trends (e.g., size and color shifts) to clarify gene function.

To validate AIPheno's genetic discovery and interpretability, we generated a simulated maize root anatomy dataset encompassing two heritable traits: aerenchyma proportion and metaxylem size. To mimic real-world environmental perturbations (e.g., lighting variability), we introduced random brightness and contrast adjustments to each image, independent of genetically driven phenotypic variations. Genomic variation data were derived from real maize datasets. Quantitative Trait Nucleotides (QTNs) linked to aerenchyma proportion were mapped to the first half of all variants (red markers), whereas those associated with metaxylem size were mapped to the second half (blue markers). Details are provided in Methods.

Despite environmental perturbations, AIPheno accurately identified the true loci underlying the two genetically driven image traits, and these results were consistent with GWAS findings. As shown in Fig.\ref{fig1}D, this consistency is reflected in the overlap between red circles (representing aerenchyma QTL), blue circles (representing metaxylem QTL), and purple inverted triangles (representing true QTL). Furthermore, visualizations from the interpretation module aligned with these loci: aerenchyma-associated loci correlated with more pronounced changes in aerenchyma proportion (marked in the red stomatal region), while metaxylem-linked loci corresponded to subtler differences in metaxylem size (marked by the dimensions of the blue central sphere).

Collectively, the GWAS and interpretability analyses of simulated data validated the reliability of both the automated phenotyping module and interpretation module in the AIPheno framework.

\clearpage

\noindent
\textbf{Fig.1 Overview of the AIPheno framework.}
\textbf{A,} Genetic dissection faces a stark imbalance between the massive, exponential growth of genomic data and the sparse, subjective nature of phenotypic data, creating a persistent genotype-phenotype gap. The traditional definition of phenotypes is established through manual observation of data space. This approach only generates limited phenotypes and fails to capture the variations of high-dimensional image data adequately. While representation learning enhances genetic discovery, its lack of interpretability prevents linking variants to specific image variations, thereby impeding biological insights from the association analysis. Driven by generative AI , AIPheno closes the loop from scalable digital phenotypes to genes—a process that involves enhancing genetic discovery power—and subsequently from genes to biological insights, which is achieved through generative biological insights decoding and actionable interpretability. A dynamic video included in the Data S1 provides a more intuitive and comprehensive illustration of the phenomenon presented in this figure, and is recommended for optimal understanding. \textbf{B,} First, a Generator was trained using a latent space as input, which was mapped from a standard normal distribution via a mapping network. With the Generator fixed, an Encoder was then trained to encode images into the latent space, from which the Generator reconstructed the original images. Next, mutually independent or orthogonal directions in the latent space were identified using either Independent Component Analysis (ICA) or Principal Component Analysis (PCA). Finally, images were encoded by the Encoder into the latent space and projected onto these previously identified directions to yield image-variation phenotypes. \textbf{C,} In the AIPheno framework, a large number of image-variation phenotypes were automatically detected in a data-driven manner by the automated phenotyping module. Each image-variation phenotype was subsequently subjected to association analysis with genetic variants using a univariate Mixed Linear Model (MLM). The significance levels from this analysis were adjusted using a Bonferroni correction, based on the number of genetic variants multiplied by the number of image-variation phenotypes, to control for false positives. Image-variation phenotypes that were significantly associated with genetic variants were then used for interpretability analysis. In the interpretation module, a direction-wise analysis was conducted on the significant directions by synthesizing high-quality images for visualization. The image's representation in the latent space was traversed along each significant direction in units of standard deviation, and the Generator created images to show the changes along that direction. Images can be controlled to change only along phenotypic variation directions while preserving other details, enabling the actionable observation of trends (e.g., size and color variations) to facilitate the understanding of gene function. \textbf{D,} To validate the genetic discovery and interpretability of the AIPheno framework, we generated a simulated dataset of maize root anatomy. This dataset included two types of heritable variations: aerenchyma proportion and metaxylem size. To mimic real-world environmental perturbations such as lighting variability, we introduced random brightness and contrast adjustments to each image, independent of the genetically driven phenotypic variations. Simulation results show that true loci were identified by associating genetic variants with the image-variation phenotypes detected by the AIPheno framework's automated phenotyping module. The visualizations from the interpretation module were also consistent with the corresponding loci: aerenchyma loci correspond to more obvious image variations in aerenchyma proportion (marked in the red stomatal region. Also see a dynamic video included in the Data S2), whereas metaxylem loci correspond to more subtle variations in metaxylem size (marked by blue central sphere dimensions. Also see a dynamic video included in the Data S3).
In the plot, red circles denote aerenchyma QTL, blue circles represent metaxylem QTL, and purple inverted triangles indicate the true QTL. All image-variation phenotypes were aggregated to generate the Manhattan plot. 

\subsection{AIPheno enhances the power of genetic discovery with scalable
digitial phenotypes
}\label{subsec2.2}

To validate the practical utility of the AIPheno framework, we constructed "Image-Genotype" datasets across four species: \textit{Homo sapiens} (human), \textit{Columba livia} (domestic pigeon), \textit{Oryza sativa} (rice), and \textit{Sus scrofa} (swine). For human data, we utilized 253,463 public retinal fundus images (from UK Biobank and the Kaggle EyePACS dataset, Fig.\ref{fig_S1}) primarily for model training. Genetic analyses were performed on 76,829 individuals from the UK Biobank cohort, for whom 13,874,430 genetic markers were available. We additionally generated and open-sourced three novel datasets: a domestic pigeon iris RGB image dataset (Fig.\ref{fig_S2}; with genome assembly; 28,903 images and 12,113,557 genetic markers, with 641 individuals used for genetic analysis), a potted rice multi-view RGB image dataset (Fig.\ref{fig_S3}; 327,734 images and 4,321,306 genetic markers, with 529 individuals included in genetic analysis), and a swine Computed Tomography (CT) dataset (Fig.\ref{fig_S4}; 3,803 images and 17,810,683 genetic markers, with 795 individuals used for genetic analysis). Details are provided in Methods.

\begin{figure}[h]
    \centering
    \includegraphics[width=1\textwidth]{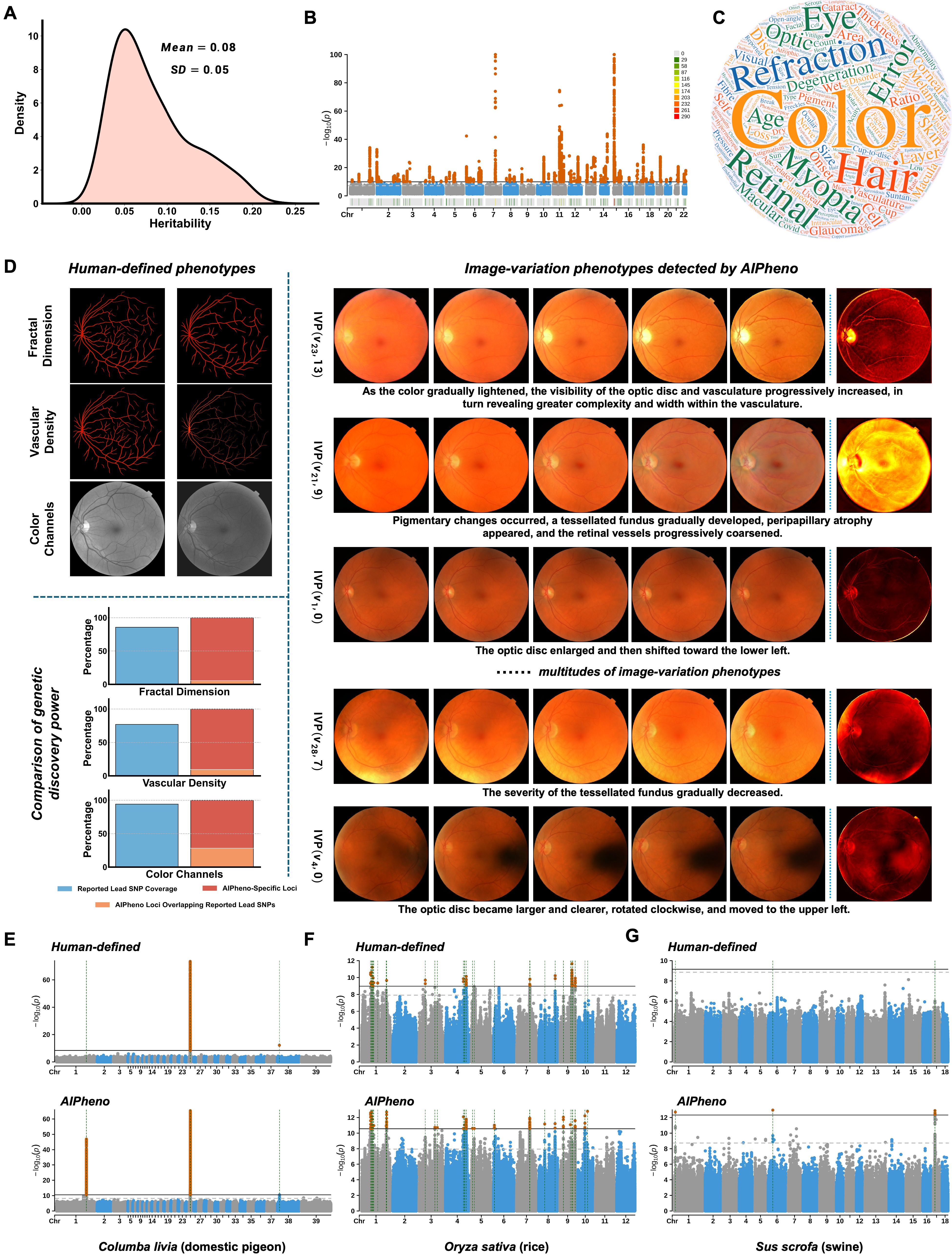}
    \caption{}\label{fig2}
    \end{figure}

For human retinal fundus data, AIPheno's automated phenotyping module processed left-eye images to identify 420 IVPs. Association analyses and heritability estimation (using a mixed linear model with covariate control) yielded heritability estimates ranging from 0.009 to 0.221 (mean=0.08, s.d.=0.05; Fig.\ref{fig2}A), with 6,037 significant SNPs (168 lead SNPs forming 106 loci). These loci were distributed across all chromosomes (Fig.\ref{fig2}B). Gene annotation revealed 84 loci overlapping with GWAS Catalog entries linked to eye traits or pigmentation, while 22 loci were novel. A word cloud of associated traits highlighted "Color" as the most frequent term, followed by eye disease-related terms ("Myopia," "Refraction," and "Error") and eye/pigmentation tissues ("Retinal," "Hair," "Eye," and "Macular") (Fig.\ref{fig2}C). FUMA enrichment analysis (Fig.\ref{fig_S5}) using top GWAS Catalog entries (ranked by \textit{p} value) confirmed associations with eye diseases (e.g., Refractive Error), eye physiology (e.g., Optic Disc Size), and pigmentation (e.g., Hair Color). Furthermore, pigmentation-related terms also topped the GO, KEGG, Reactome, and WikiPathways enrichment results (Fig.\ref{fig_S6}A-D).

To benchmark the power of genetic discovery, we compared AIPheno with HDPs and state-of-the-art deep learning-based phenotyping methods using human data. While HDPs were limited to a few metrics, including vascular branching complexity (fractal dimension), density, and color channels, AIPheno detected a multitude of IVPs, such as pigmentation, retinal lesions, and optic disc/vasculature morphology, thereby capturing more biologically meaningful variation (Fig.\ref{fig2}D). For specific metrics, the vascular branching complexity (fractal dimension) and vascular density metrics from Zekavat et al.\cite{zekavat2022deep} showed 6/7 and 10/13 lead SNPs overlapping AIPheno's loci, respectively. However, they missed 94.3\% (100/106) and 90.6\% (96/106) of AIPheno's unique loci, respectively. Additionally, color channel HDPs (mean R/G/B values) overlapped with 49/53 lead SNPs but missed 71.7\% (76/106) of AIPheno's loci. For deep learning-based approaches, iGWAS (Xie et al.\cite{xie2024igwas}) showed 13/14 lead SNPs overlapping with AIPheno's loci but missed 88.7\% (94/106) of AIPheno's loci. Zhao et al.\cite{zhao2024eye} applied transferGWAS\cite{kirchler2022transfergwas}, using 11 ImageNet pre-trained models and taking the top 10 principal components (PCs) from each model's final layer output as traits for downstream GWAS. Their ensemble results across 11 models showed 70.0\% (432/617) of lead SNPs overlapping with AIPheno's loci, yet 31.1\% (33/106) of AIPheno's loci remained uncovered (Fig.\ref{fig_S7}).

AIPheno also detected numerous IVPs in the domestic pigeon, rice, and swine datasets, exhibiting higher genetic discovery power than HDPs.
For domestic pigeon, AIPheno's automated phenotyping module detected 140 IVPs per image, which encompassed a rich set of image variations: the orientation of the eyeball and pupil, pupil size, and iris color and texture (Fig.\ref{fig_S9}A). These 140 IVPs were significantly associated with two loci. The locus on chromosome 25 was consistent with a traditional case-control phenotyping approach that classifies domestic pigeon irises as either "pearl" or "gravel" (Fig.\ref{fig_S8}), while the other locus, on chromosome 1, was unique to AIPheno (Fig.\ref{fig2}E). 

For rice, AIPheno's automated phenotyping module detected 420 IVPs per image. These IVPs encompassed not only environmental variations, such as plant illumination, but also biologically meaningful image variations, including plant architecture, stem and leaf morphology, color, and texture (Fig.\ref{fig_S9}B). For comparison, we defined 11 HDPs with clear biological significance (Details in Methods). We performed association analyses and compared genetic discovery power for both phenotype types across 22 time points spanning the seedling to mature stages, revealing that AIPheno identified more genetic loci at most time points (Fig.\ref{fig_S10}). Given the highest overlap between the two at T13, we focused subsequent analyses on this time point. GWAS results for all IVPs and HDPs at T13 are shown in Fig.\ref{fig2}F: 10 of the 15 lead SNPs from HDPs overlapped with AIPheno's loci, but missed 54.5\% (12/22) of AIPheno's unique loci.

For swine, AIPheno detected 5,880 IVPs per image. These image variations, which encompass not only changes in fat content and distribution but also alterations in body length and body shape (Fig.\ref{fig_S9}C), were significantly associated with three loci. In contrast, four HDPs (total fat content, average fat content, and image-based body length and body height) were extracted for comparison, none of which yielded significant associations. The results showed that HDPs failed to capture the image variation of fat distribution, while the IVPs detected by AIPheno were able to successfully uncover the genetic basis behind swine fat distribution (Fig.\ref{fig2}G). 

    \clearpage

    \noindent
    \textbf{Fig.2 AIPheno enhances the power of genetic discovery with scalable digitial phenotypes across multi-species datasets.}
    \textbf{A,} Heritability distribution of the 420 image-variation phenotypes of human retinal fundus images detected by AIPheno (mean $=$ 0.08, SD $=$ 0.05, min $=$ 0.009, max $=$ 0.221). \textbf{B,} GWAS Manhattan plot for the Human Retinal Fundus dataset (\textit{p} values were truncated at $1\times10^{-100}$). In this plot, the \textit{p} value for each SNP was the minimum \textit{p} value of that SNP across all image-variation phenotypes GWAS results. The significance threshold was adjusted using a Bonferroni correction for the number of image-variation phenotypes ($p_{\text{original}} = 5\times10^{-8}$, $p_{\text{correction}} = 1.19\times10^{-10}$). The density plot below indicated the number of image-variation phenotypes with significant signals per 2 MB interval. Information for all lead SNPs, genetic loci, and candidate genes is provided in Table S1. \textbf{C,} Word cloud of reported phenotypes from the GWAS Catalog associated with the identified candidate genes. The word cloud results indicate that the most frequent word is Color, followed by terms associated with eye diseases ("Myopia," "Refraction," and "Error"), and then by tissues related to the eye or pigmentation ("Retinal," "Hair," "Eye," and "Macular"). \textbf{D,} Comparison of genetic discovery power between AIPheno and human-defined methods in \textit{Homo sapiens} (human). For AIPheno, a large number of image-variation phenotypes were detected, such as pigmentation, retinal lesions, and the morphology of the optic disc and vasculature (Also see dynamic videos included in the Data S4-S8). For human-defined phenotypes, Fractal Dimension quantifies vascular branching complexity, Vascular Density is defined as the total number of segmented pixels given a consistent field of view and fixed pixel dimensions across all individuals, and Color Channels refer to the mean values of the R, G, and B image channels. Left bar: Percentage of lead SNPs reported by the comparison methods overlapping with AIPheno loci. Bottom right bar: Percentage of AIPheno loci that overlapped with reported lead SNPs. Top right bar: Percentage of AIPheno-specific loci. \textbf{E,} Comparison of genetic discovery power between AIPheno and human-defined methods in \textit{Columba livia} (domestic pigeon). GWAS results for all image-variation phenotypes of AIPheno with significant signals were aggregated to plot the Manhattan plot. The green dashed line highlights significant signals associated with the two types of phenotypes. The human-defined phenotypes are the "pearl" and "gravel" iris categories. The significance threshold of human-defined methods was $4.13\times10^{-9}$. The significance threshold of AIPheno was adjusted using a Bonferroni correction for the number of image-variation phenotypes ($p_{\text{original}}=4.13\times10^{-9}$, $p_{\text{correction}}=2.95\times10^{-11}$). \textbf{F,} Comparison of genetic discovery power between AIPheno and human-defined methods in \textit{Oryza sativa} (rice). The GWAS results with significant signals for image-variation phenotypes of AIPheno and human-defined phenotypes were aggregated separately to plot the Manhattan plots. The green dashed line highlights significant signals associated with the two types of phenotypes. The dashed line indicates the original significance threshold ($p_{\text{original}}=1.16\times10^{-8}$). The effective thresholds were then determined by Bonferroni correction (dividing the original threshold by the number of phenotypes), resulting in $p_{\text{correction}}=1.05\times10^{-9}$ for human-defined methods and $p_{\text{correction}}=2.75\times10^{-11}$ for AIPheno. The human-defined phenotypes are histogram-based mean, histogram-based variance, total projected area, height of the bounding rectangle, width of the bounding rectangle, ratio of total projected area to circumscribed box area, circumscribed box area, ratio of total projected area to hull area, ratio of perimeter to total projected area, perimeter, and green projected area. Information for all lead SNPs, genetic loci, and candidate genes of T13, identified using image-variation phenotypes, is provided in Table \ref{table_s2} to Table \ref{table_s25}. \textbf{G,} Comparison of genetic discovery power between AIPheno and human-defined methods in \textit{Sus scrofa} (swine). GWAS results for all human-defined phenotypes were aggregated to plot the Manhattan plot. GWAS results for all image-variation phenotypes of AIPheno with significant signals were aggregated to plot the Manhattan plot. The green dashed line highlights significant signals associated with the two types of phenotypes. In the Manhattan plot, the dashed threshold line was set at $p_{\text{original}} = 2.81\times10^{-9}$, with $p_{\text{correction}} = 7.02\times10^{-10}$ for human-defined methods and $p_{\text{correction}} = 4.77\times10^{-13}$ for AIPheno. The human-defined phenotypes are total fat content, average fat content, and image-based body length and body height. Table \ref{table_s26} presents information on all lead SNPs, genetic loci, and candidate genes identified using image-variation phenotypes.

\subsection{AIPheno's actionable interpretability validates reported gene-phenotype associations}\label{subsec2.3}
By generating images for visualization, AIPheno validated the consistency of reported gene-phenotype associations using its interpretation module. Next, we will demonstrate this through detailed analyses in two species: domestic pigeon and rice.

For domestic pigeons, three independent studies defined two case-control categories (gravel versus pearl iris, based on the presence or absence of white coloration in the iris) to identify \textit{SLC2A11B}, which carries a protein-truncating nonsense mutation\cite{andrade2021molecular,maclary2021two,si2021genetics}. Loss of function in this gene impairs differentiation of xanthophore-like stromal pigment cells and pteridine biosynthesis, reducing yellow pigment production and resulting in the pearl iris phenotype. Consistent with these findings, AIPheno identified corresponding image variations associated with the same genetic mutation: IVP$(\mathbf{v}_{1}, 11)$ captured the G-to-A nonsense mutation on chromosome 25 (Fig.\ref{fig3}A). Interpretability analyses revealed progressive image changes correlated with genetic variants: the white sector of the iris progressively diminished in area, transitioning into neighboring pigmented regions until color saturation enveloped the entire iris structure, manifesting a gradual phenotypic continuum from pearl iris to gravel iris. As shown in the standard deviation (std) image, image variation was focused on the region outside the pupil and light spot (Fig.\ref{fig3}A).

For rice, the IVP$(\mathbf{v}_{1}, 1)$ mapped to \textit{Os01g0866400} (CGSNL: \textit{MOC2}) (Fig.\ref{fig3}B), a locus repeatedly linked to plant height and growth rate in previous studies\cite{li2020analysis,sandhu2019deciphering,lv2021loci}. Experimental evidence shows that \textit{MOC2} mutations reduce tiller number, cause pale-green leaves, slow growth, and result in dwarfism\cite{koumoto2013rice}, which aligns with AIPheno's interpretability analysis of IVP$(\mathbf{v}_{1}, 1)$. Not only did plant height decrease, but it was also more intuitively observable that the plants exhibited a dwarf phenotype due to a delayed growth rate (Fig.\ref{fig3}B). As shown in the std image, image variation was concentrated at the top and bottom of the plant.

Collectively, these findings from diverse species demonstrate that AIPheno not only validates reported gene-phenotype associations but also offers interpretable visual evidence of how genetic variations manifest as specific phenotypic changes. 

\subsection{AIPheno's actionable interpretability refines insights into established gene-phenotype associations}\label{subsec2.4}
AIPheno's powerful interpretability extends beyond merely validating established gene-phenotype associations, refining these insights by underlying mechanistic links and revealing the subtle phenotypic variations. Below, we will demonstrate this ability in humans and swine.

\begin{figure}[h]
    \centering
    \includegraphics[width=1\textwidth]{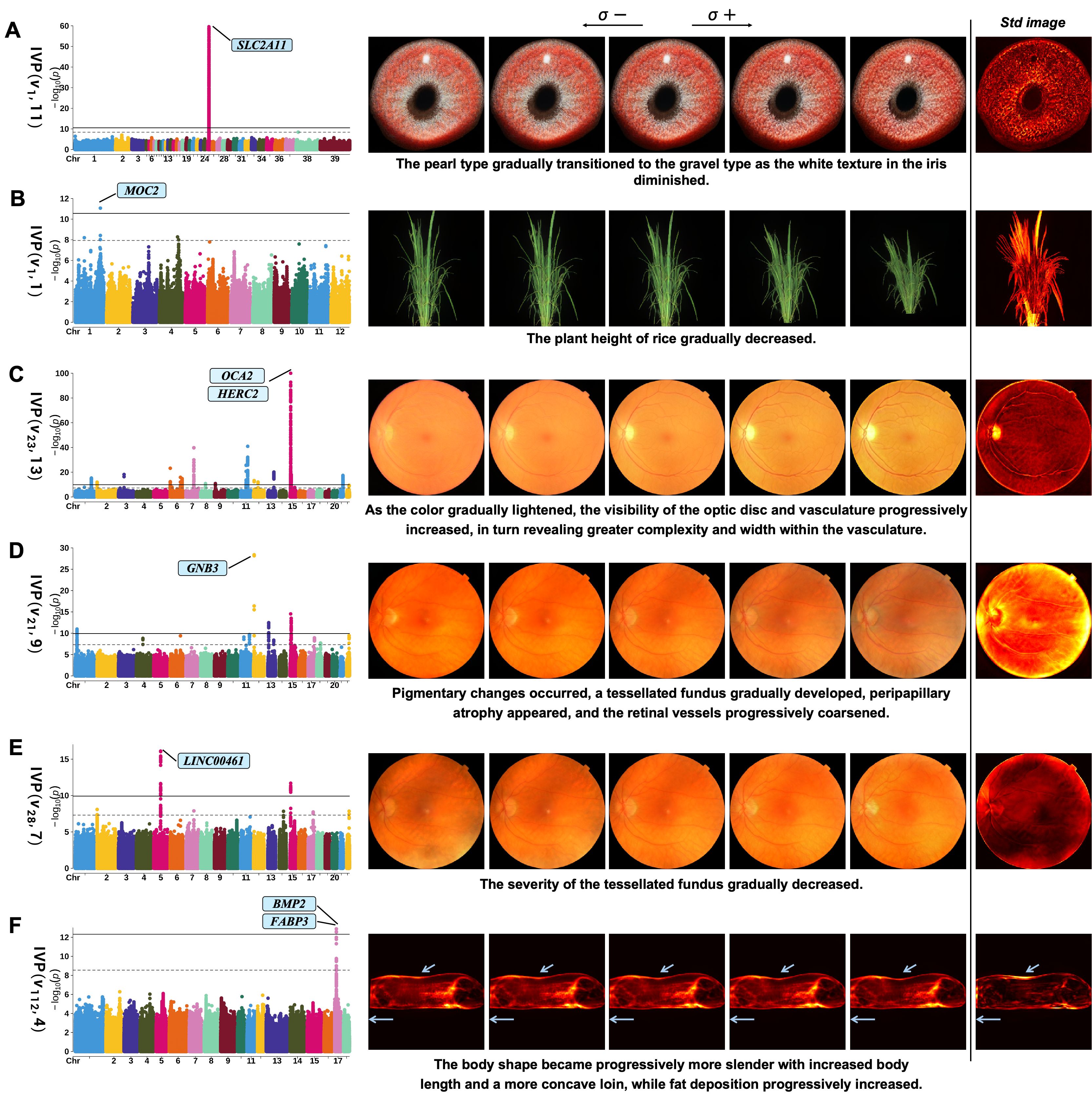}
    \caption{}\label{fig3}
    \end{figure}

In humans, multiple studies have identified the \textit{OCA2}-\textit{HERC2} locus as the most statistically significant locus: For deep learning-based approaches, this included iGWAS\cite{xie2024igwas}, the study by Zhao et al.\cite{zhao2024eye} (excluding the vgg19 model), and AIPheno. This locus's high significance was also observed using HDP-based methods; specifically, studies employing fractal dimension (an indicator quantifying vascular branching complexity), such as those by Zekavat et al.\cite{zekavat2022deep} and Villaplana-Velasco et al.\cite{villaplana2023fine}, similarly highlighted its strong association. In addition, this locus was associated with retinal vessel diameter\cite{ortin2024phenotypic}. Notably, previous studies have further linked this locus to pigmentation traits—including hair color, eye color, and skin pigmentation\cite{liu2010digital,sulem2007genetic,adhikari2019gwas}. This indicates a pleiotropic effect between pigmentation and retinal vascular morphology. However, due to the limitations of human-defined phenotyping and non-interpretable deep learning-based approaches, the reason why these two distinct phenotypes share a common genetic basis remains unclear. From the interpretability analysis of IVP$(\mathbf{v}_{23}, 13)$, we observed that the overall color of the retinal image gradually lightened from orange-red to orange-yellow, a change likely attributable to variations in retinal pigmentation. As the color lightened, the brightness and contrast of the optic disc and vasculature progressively increased, thereby enhancing the visibility of these two tissues. Concurrently, with improved vasculature visibility, a gradual increase in the vasculature's complexity and width was visually observed. As can also be clearly seen from the std image, image variation was not only diffused across the entire fundus region, but it was also concentrated in the optic disc and vascular areas, which further corroborated the aforementioned observation (Fig.\ref{fig3}C). The actionable interpretability analysis of AIPheno revealed coordinated variations between the two phenotypes: pigmentation and retinal vascular morphology.

IVP$(\mathbf{v}_{21}, 9)$ identified a lead SNP, rs5442, mapped to the \textit{GNB3} gene (Fig.\ref{fig3}D). This variant is a highly conserved and deleterious missense mutation (p.Gly272Ser) located within an exon\cite{tedja2018genome}. In the retina, \textit{GNB3} is integral to the phototransduction cascade, where it functions as part of a heterotrimeric G-protein that enhances visual signaling\cite{ritchey2012vision}. Consistent with its critical role, rs5442 has been previously associated with a range of retinal anatomical parameters, including arteriolar width\cite{jiang2023gwas}, macular thickness\cite{gao2019genome}, ganglion cell inner plexiform layer thickness\cite{currant2021genetic}, retinal microvascular diameter\cite{jensen2016novel}, and vascular density\cite{zekavat2022deep}, as well as with ocular diseases such as refractive error\cite{tedja2018genome,hysi2020meta}, myopia\cite{xue2022genome}, and open-angle glaucoma\cite{han2023large}.

Our interpretability analysis of AIPheno elucidated the specific fundus changes attributable to this variant. The primary finding was a progressive coarsening of the vasculature and a decrease in pigmentation across the entire fundus, corresponding to the development of a tessellated fundus. Furthermore, as the severity of the fundus tessellation increased, peripapillary atrophy-like lesions progressively emerged. The std image confirmed that these variations were distributed across the fundus but were particularly concentrated in the tessellated lesions in the periphery and center, peripapillary lesions, and the vasculature (Fig.\ref{fig3}D), highlighting these as key areas of impact for the variant.

These AI-driven observations are consistent with established clinical knowledge. Fundus tessellation and peripapillary atrophy are often observed concurrently in myopic eyes. Their shared pathophysiological basis is the mechanical stretching and thinning of posterior pole tissues caused by excessive axial elongation\cite{cho2016complications,tan2018associations}. While fundus tessellation represents a general, diffuse thinning of the choroid and retinal pigment epithelium, peripapillary atrophy is a more localized and concentrated result of this same tensile stress at the optic disc margin\cite{tan2018associations}.

The strong association between these two clinical signs is well-supported by external evidence. A meta-analysis of population studies, for instance, confirmed that the severity of fundus tessellation is significantly associated with a larger area of peripapillary atrophy\cite{chen2023clinical}. This link is also corroborated in non-human primate models, where highly myopic macaques showed a higher incidence of fundus tessellation and peripapillary atrophy\cite{tian2024fundus}. The consistency between our AIPheno analysis and these clinical and research findings suggests our approach successfully identified the key pathological features. Therefore, our work refines the understanding of the specific retinal fundus image changes caused by the rs5442 variant.

IVP$(\mathbf{v}_{28}, 7)$ identified \textit{LINC00461}, a long non-coding RNA that is the principal target of a disease-associated enhancer at 5q14.3 and serves as the primary transcript for miR-9-2 (Fig.\ref{fig3}E). In retinal progenitor cells and M\"{u}ller glia, regulation of \textit{LINC00461} by this enhancer is critical for the normal timing of retinal neurogenesis, and its disruption alters cell class specification and leads to a reduction in rod photoreceptors\cite{thomas2022cell}. Studies have reported that \textit{LINC00461} is associated not only with retinal anatomical parameters such as macular thickness\cite{gao2019genome} and retinal thickness\cite{jackson2025multi}, representing one of the most significant signals in association analyses, but also with ocular diseases such as age-related macular degeneration\cite{han2020genome} and macular telangiectasia type 2\cite{scerri2017genome}. The interpretability results from AIPheno show that the severity of the tessellated fundus gradually decreased. The std image also revealed that these lesions were primarily concentrated in the fundus periphery, near the optic disc margin and the macula (Fig.\ref{fig3}E). Tessellated fundus is considered the initial stage of myopic maculopathy\cite{ohno2015international}, and studies have also shown that its severity is associated with macular choroidal thickness\cite{yan2015fundus,lyu2021characteristics}. This further corroborates the findings of AIPheno. Therefore, our findings refine the changes in retinal fundus images caused by \textit{LINC00461} variants.

For swine, a peak identified by AIPheno was associated significantly with IVP$(\mathbf{v}_{112}, 4)$ (Fig.\ref{fig3}F). Two candidate genes were identified via positional mapping and literature review: \textit{FABP3} and \textit{BMP2}. In swine, genetic variants of \textit{FABP3} were associated not only with backfat thickness, but also with intramuscular fat content\cite{cho2011association,wang2019association}. Multiple studies have shown that \textit{BMP2} was associated with body shape, body size, and carcass morphology. Fan et al. reported that \textit{BMP2} was associated with increased body length, decreased body depth, and reduced body width\cite{fan2011genome}. \textit{BMP2} was reported by Li et al. to be associated with carcass straight length\cite{li2021further}. Similar observations were made by Zhang et al., who identified associations between \textit{BMP2} and carcass length, body length, body height, and $\text{BMI}_{\text{BL}}$ traits, which suggested that \textit{BMP2} was a strong candidate gene for body size due to its involvement in growth and bone development\cite{zhang2021genome}. The interpretability results of AIPheno demonstrated consistency with previous studies, indicating changes in fat deposition and body length in swine. Interestingly, we discovered a variation in body shape not reported in previous research: the loin of the swine gradually became concave, which made the body shape progressively more slender. As shown in the std image, the image variation was diffused across the entire body region of the swine and is focused on the front and loin areas (Fig.\ref{fig3}F). This visual discovery aligns remarkably with established findings, which identified \textit{BMP2} as a key candidate gene regulating Loin Muscle Depth (LMD)\cite{miao2023integrated}. The depth of the loin muscle directly determines the contour of the pig's back and loin; therefore, when a genetic variant of \textit{BMP2} leads to a shallower muscle, the external morphology naturally presents as a "concave" appearance. Thus, the visualization from AIPheno provides an actionable morphological meaning to these previously established quantitative traits.

    \clearpage

\noindent
\textbf{Fig.3 AIPheno's actionable interpretability validates and refines insights into established gene-phenotype associations.}
Identified genes for which AIPheno validated or refined established gene-phenotype associations are highlighted in blue. In std images, which are visualizations of the standard deviation for each pixel across the generated images, a brighter pixel indicates greater change at that point during image traversal. \textbf{A,} Interpretability analysis of IVP$(\mathbf{v}_{1}, 11)$ in domestic pigeon. The significance threshold was adjusted using a Bonferroni correction for the number of image-variation phenotypes ($p_{\text{original}}=4.13\times10^{-9}$, $p_{\text{correction}}=2.95\times10^{-11}$). A transition from the Pearl-type to the Gravel-type was observed; the white texture in the iris diminished gradually. As shown in the std image, image variation was focused on the region outside the pupil and light spot. Also see a dynamic video included in the Data S9. \textbf{B,} Interpretability analysis of IVP$(\mathbf{v}_{1}, 1)$ in rice. The significance threshold was adjusted using a Bonferroni correction for the number of image-variation phenotypes ($p_{\text{original}}=1.16\times10^{-8}$, $p_{\text{correction}}=2.75\times10^{-11}$). The plant height of rice decreased gradually. As shown in the std image, image variation was concentrated at the top and bottom of the plant. Also see a dynamic video included in the Data S10. \textbf{C,} Interpretability analysis of IVP$(\mathbf{v}_{23}, 13)$ in human. The significance threshold was adjusted using a Bonferroni correction for the number of image-variation phenotypes ($p_{\text{original}}=5\times10^{-8}$, $p_{\text{correction}}=1.19\times10^{-10}$, \textit{p} values were truncated at $1\times10^{-100}$). As the color gradually lightened, the visibility of the optic disc and vasculature progressively increased, in turn revealing greater complexity and width within the vasculature. As shown in the std image, image variation was not only diffused across the entire fundus region, but it was also concentrated in the optic disc and vascular areas. Also see a dynamic video included in the Data S11. \textbf{D,} Interpretability analysis of IVP$(\mathbf{v}_{21}, 9)$ in human. The significance threshold was adjusted using a Bonferroni correction for the number of image-variation phenotypes ($p_{\text{original}}=5\times10^{-8}$, $p_{\text{correction}}=1.19\times10^{-10}$). Pigmentary changes occurred, a tessellated fundus gradually developed, peripapillary atrophy appeared, and the retinal vessels progressively coarsened. The std image also showed that the image variation was distributed across the entire fundus, particularly concentrating on several areas: the tessellated lesions in the periphery and center, peripapillary lesions, and the vasculature. Also see a dynamic video included in the Data S12. \textbf{E,} Interpretability analysis of IVP$(\mathbf{v}_{28}, 7)$ in human. The significance threshold was adjusted using a Bonferroni correction for the number of image-variation phenotypes ($p_{\text{original}}=5\times10^{-8}$, $p_{\text{correction}}=1.19\times10^{-10}$). The severity of the tessellated fundus gradually decreased. The std image also revealed that these lesions were primarily concentrated in the fundus periphery, near the optic disc margin and the macula. Also see a dynamic video included in the Data S13. \textbf{F,} Interpretability analysis of IVP$(\mathbf{v}_{112}, 4)$ in swine. The significance threshold was adjusted using a Bonferroni correction for the number of image-variation phenotypes ($p_{\text{original}}=2.81\times10^{-9}$, $p_{\text{correction}}=4.77\times10^{-13}$). The body shape became progressively more slender with increased body length and a more concave loin, while fat deposition progressively increased. As shown in the std image, the image variation was diffused across the entire body region of the pig and was focused on the front and loin areas. Also see a dynamic video included in the Data S14.

\subsection{AIPheno's actionable interpretability uncovers biological insights into newly identified loci}\label{subsec2.5}
AIPheno not only refines insights into established gene-phenotype associations but, more importantly, also identifies novel loci and reveals their biological insights through actionable interpretability analysis. Next, we demonstrate this capability in three species: domestic pigeon, swine and human.

\begin{figure}[h]
    \centering
    \includegraphics[width=1\textwidth]{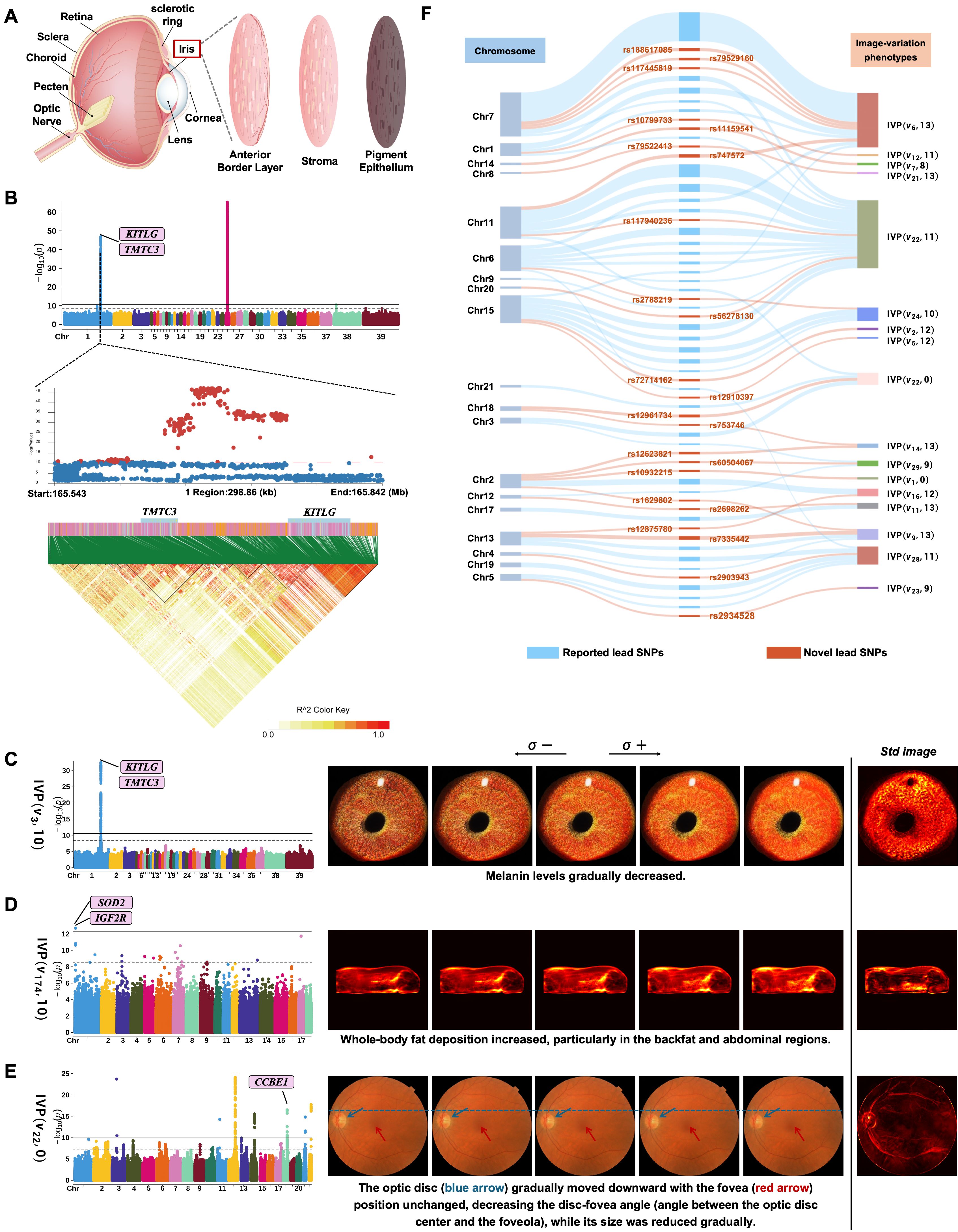}
    \caption{}\label{fig4}
    \end{figure}

The domestic pigeon displays three primary iris color types: yellow to orange "gravel" (wild type), white "pearl", and black "bull" eyes\cite{oliphant1987observations} (Fig.\ref{fig_S8}). Anatomically, the avian iris comprises three layers: the posterior pigment epithelium, which is densely packed with melanin; the stroma, which is a thick middle layer that contains muscle, connective tissue, nerves, and pigment cells; and the anterior border layer, which may include pigment cells or blood vessels in certain species\cite{oliphant1981crystalline,oliphant1987observations,oliphant1987pteridines,hudon1995reflective,sweijd1991histological} (Fig.\ref{fig4}A). In "bull-eyed" pigeons, there is a lack of pigment in both the stroma and the anterior border layer, with the black eye coloration observed in these breeds due primarily to the melanin present in the pigment epithelium\cite{corbett2024mechanistic}. The irises of both gravel and pearl pigeons exhibit anterior stromal pigment cells that contain birefringent crystals, with guanine identified as their primary pigment. The white color of the pearl eye arises because pigment cells deposit only guanine crystals and lack pteridine synthesis\cite{oliphant1987observations}. 

In domestic pigeon, a highly significant signal was identified on chromosome 1. Following genome annotation, positional mapping, and LD block analysis, two candidate genes, \textit{KITLG} and \textit{TMTC3}, were prioritized (Fig.\ref{fig4}B). The \textit{KIT ligand} (\textit{KITLG}) gene, which encodes the ligand for the \textit{KIT}-encoded tyrosine kinase receptor, plays a crucial role in cell development, migration, and melanogenesis regulation\cite{2011KITLG,2011The,wehrle2003role}. Supported by extensive cross-species evidence (e.g., humans, mice, swine, geese, chickens, goats, and threespine stickleback fish), \textit{KITLG} was associated functionally with pigmentation phenotypes in skin, hair, and feathers\cite{guenther2014molecular,cuell2015familial,wang2021identification,wang2009gain,kim2024mapping,ren2021pooled,shen2022genome,talenti2018genomic,miller2007cis}. Furthermore, \textit{KITLG} exhibits a direct association with iris pigmentation. Zazo et al. reported a Waardenburg syndrome type 2 (WS2)-affected family exhibiting heterochromia iridis, which was characterized by abnormal iris pigmentation concurrent with abnormal skin pigmentation, and they demonstrated that these abnormalities were associated with a leucine-to-valine substitution at position 104 (p.Leu104Val) in KITLG\cite{2015Allelic}. Similarly, identical findings were observed in swine. Moscatelli et al. demonstrated in a Large White swine population that both heterochromia iridis and heterochromia iridum were associated with the \textit{KITLG} gene\cite{2020Genome}. In addition to \textit{KITLG}, the \textit{TMTC3} gene was also associated with pigmentation in both humans and chickens\cite{kim2024mapping,shen2022genome,morgan2018genome}. However, such an association has not been reported in the domestic pigeon.

AIPheno's interpretability analysis provided visual evidence consistent with the potential roles of the candidate genes identified from previous studies. A statistically significant peak on chromosome 1 that was identified by IVP$(\mathbf{v}_{3}, 10)$ drove consistent progressive image variations. Initially, pigmented iris regions exhibited interspersed fine black patches, which resulted in an overall darker appearance. Along the positive axis, these black patches diminished progressively, which led to enhanced vibrancy and brightness of the iris (Fig.\ref{fig4}C). Based on this cross-species evidence and interpretability analyses, we propose the following mechanism. Genetic variations in \textit{KITLG} or \textit{TMTC3} may induce differential melanization within the posterior pigment epithelium of the pigeon iris. The std image showed that variation in melanin levels was present throughout the iris, except for the pupil and the light spot. Notably, the mean value of the R channel (quantifying the degree of redness in the image) and the mean value of the V channel (quantifying the brightness of the image), these two HDPs, also identified this locus (Fig.\ref{fig_S11}). However, the definitions of these phenotypes are based on fundamental photometric metrics, which inherently lack biological context, thus making their connection to a specific biological process like melanin accumulation unintuitive.

In swine, AIPheno identified a peak on Chromosome 1, which was associated significantly with IVP$(\mathbf{v}_{174}, 10)$ (Fig.\ref{fig4}D). Through the interpretability analysis, the porcine body parts in the image became increasingly brighter overall, which indicated an increase in the degree of whole-body fat deposition (Fig.\ref{fig4}D). This increase was particularly evident in the backfat and abdominal regions in std image. Based on these image variations, and through positional mapping within the significant locus, two plausible functional candidate genes, \textit{SOD2} and \textit{IGF2R}, were identified. In swine, no association between \textit{SOD2} and fat deposition has been reported in previous research. Post-developmental deletion of skeletal muscle-specific \textit{SOD2} induced systemic physiological effects through escalating mitochondrial deficits and subsequent impairment of energy metabolism in murine models. Additionally, \textit{SOD2} has been implicated in the modulation of lipid pathways, which demonstrated its direct association with both energy homeostasis and lipid metabolic regulation\cite{zhuang2021sod2}. In humans, genetic variants of \textit{SOD2} were related to glucose metabolism\cite{wu2022fasting}. More direct evidence has been demonstrated that genetic polymorphisms in \textit{SOD2} were associated with increased percentage of body fat and accumulation of visceral fat in obese populations, a finding that is directly aligned with our interpretability analysis\cite{hernandez2016genetic}. For \textit{IGF2R}, studies have also reported that its function is related to fat deposition. The \textit{IGF2} locus was associated with muscle mass and fat deposition in swine\cite{nezer1999imprinted}, and its receptor, \textit{IGF2R}, may also be implicated in the variation of these traits. The exonic polymorphism of \textit{IGF2R} was associated with obesity in humans\cite{yang2015association}. Based on interpretability analysis and cross-species evidence, we propose that the \textit{SOD2}-\textit{IGF2R} locus is associated with whole-body fat deposition in swine.

In human retinal fundus images, the disc-fovea angle, which is defined as the angle between the optic disc center and the foveola, serves as a key parameter for the posterior fundus by defining the positional relationship of the optic nerve head relative to the foveola\cite{rohrschneider2004determination}. IVP$(\mathbf{v}_{22}, 0)$ was associated with the disc-fovea angle. The optic disc was moved downward gradually, while the position of the fovea remained almost unchanged, which indicated that a downward shift of the optic disc relative to the fovea had occurred, thus, the disc-fovea angle was decreasing gradually (Fig.\ref{fig4}E). The std image showed that image variation was primarily concentrated in the optic disc and the connecting vasculature, while the fovea region showed almost no variation. Jonas et al. indicated that a larger optic disc-fovea angle was associated with a higher risk of retinal vein occlusion\cite{jonas2015optic}. Compared with participants without the disease, the optic disc-fovea angle was significantly larger in individuals with retinal vein occlusion. Choi et al. demonstrated that the optic disc-fovea angle was associated with the thickness of the Retinal Nerve Fiber Layer (RNFL)\cite{choi2014foveal}. They found that an increase in the optic disc-fovea angle was associated with corresponding changes in RNFL thickness, which potentially led to a greater susceptibility to glaucomatous damage in specific regions of the optic disc. Moreover, Amini et al. demonstrated that the disc-fovea angle itself can be considered an anatomical feature or risk marker associated with glaucoma\cite{amini2014influence}. Apart from the optic disc-fovea angle, IVP$(\mathbf{v}_{22}, 0)$ was also associated with optic disc size because optic disc size was reduced gradually (Fig.\ref{fig4}E). Myopic eyes have larger optic disc sizes\cite{haarman2020complications}. Estimation of the disc size is also important in diagnosing diseases, such as glaucoma, optic disc drusen, pseudopapilledema, optic nerve hypoplasia, and anterior ischemic optic neuropathy\cite{hoffmann2007optic}. Consistent with the aforementioned imaging variations, four lead SNPs were identified within the significant association signals of IVP$(\mathbf{v}_{22}, 0)$, all of which showed the highest significance across all IVPs. Among these, three were mapped to genes associated with optic disc measurement (\textit{DCDC1}) or abnormality of refraction (\textit{RUNX1} and \textit{ADAMTS9}). Notably, the remaining SNP was mapped to \textit{CCBE1}, which is a gene not reported previously to be linked to ocular phenotypes.

In human, given that lead SNPs may exhibit statistical significance across multiple IVPs, a systematic min-p analysis was conducted to identify the IVP that exhibited the strongest statistical association (smallest \textit{p} value) for each lead SNP and a Sankey diagram was constructed to visualize their interrelationships (Fig.\ref{fig4}F). The IVPs showing the strongest statistical association with lead SNPs—those not previously reported in the GWAS Catalog database for pigmentation or eye-related traits—were selected, where blue and red lead SNPs denote previously reported and novel associations, respectively. The thickness of the lines connecting the columns is proportional to the $-\log_{10}(p)$ value of the lead SNP. The image variations captured in IVP$(\mathbf{v}_{6}, 13)$ corresponded to pigmentation (Fig.\ref{fig_S12}). Of the lead SNPs identified for this IVP, four were novel: \textit{TRIM56} (rs79529160, $p = 6.50 \times 10^{-15}$), \textit{TRRAP} and \textit{SMURF1}(rs188617085, $p = 3.59 \times 10^{-19}$), \textit{CYP3A5} (rs117445819, $p = 1.76 \times 10^{-13}$) and \textit{RAB38} (rs747572, $p = 5.56 \times 10^{-20}$). Details for the remaining IVPs are provided in Fig.\ref{fig_S12}.

    \clearpage

\noindent
\textbf{Fig.4 AIPheno's actionable interpretability uncovers biological insights into newly identified genes.} \textbf{A,} Schematic diagram of domestic pigeon eye and iris anatomy. The different eye colors of the domestic pigeon are determined by the iris, which is composed of three layers. The outermost Anterior Border Layer and the middle Stroma contain pigment cells, within which yellow pteridines and reflective purines are found. The innermost Pigment Epithelium is rich in melanin. \textbf{B,} Manhattan plot and LD analysis of GWAS results. Results for all image-variation phenotypes were integrated for plotting, with two threshold lines indicating the significance levels before ($p=4.13\times10^{-9}$) and after ($p=2.95\times10^{-11}$) Bonferroni correction that was based on the number of image-variation phenotypes. Newly identified genes are highlighted in pink. LD analysis was performed within a 300 kb region surrounding the lead SNP on chromosome 1. Significant signals clustered between the \textit{TMTC3} and \textit{KITLG} genes. \textbf{C,} Interpretability analysis of IVP$(\mathbf{v}_{3}, 10)$ in domestic pigeon. The accumulation level of melanin was reduced gradually. The significance threshold was adjusted using a Bonferroni correction for the number of image-variation phenotypes ($p_{\text{original}}=4.13\times10^{-9}$, $p_{\text{correction}}=2.95\times10^{-11}$). The std image showed that variation in melanin levels was present throughout the iris, except for the pupil and the light spot. Also see a dynamic video included in the Data S15. \textbf{D,} Interpretability analysis of IVP$(\mathbf{v}_{174}, 10)$ in swine. The significance threshold was adjusted using a Bonferroni correction for the number of image-variation phenotypes ($p_{\text{original}}=2.81\times10^{-9}$, $p_{\text{correction}}=4.77\times10^{-13}$). Whole-body fat deposition increased, especially in the backfat and abdominal regions. The std image also showed that image variation was primarily concentrated in the backfat and abdominal regions. Also see a dynamic video included in the Data S16. \textbf{E,} Interpretability analysis of IVP$(\mathbf{v}_{22}, 0)$ in human. The significance threshold was adjusted using a Bonferroni correction for the number of image-variation phenotypes ($p_{\text{original}}=5\times10^{-8}$, $p_{\text{correction}}=1.19\times10^{-10}$). The optic disc (blue arrow) gradually moved downward with the fovea (red arrow) position unchanged, decreasing the disc-fovea angle (angle between the optic disc center and the foveola), while its size was reduced gradually. Such variations have been reported to be associated with ocular diseases. The std image showed that image variation was primarily concentrated in the optic disc and the connecting vasculature, while the fovea region showed almost no variation. Also see a dynamic video included in the Data S17. \textbf{F,} A single lead SNP can be statistically significant for multiple image-variation phenotypes. To resolve this, we first conducted a systematic min-p analysis to assign each lead SNP to the single image-variation phenotype with which it had the strongest association (smallest \textit{p} value). Next, we constructed a Sankey diagram to visualize these relationships. Finally, to highlight novel discoveries, we selected associations involving lead SNPs that had no previously reported links to pigmentation or eye-related traits in the GWAS Catalog. In the diagram, chromosomes (left column) are linked to lead SNPs (center column) and their corresponding smallest \textit{p} value image-variation phenotypes (right column). The lead SNPs in the central column are colored red if unreported previously in an eye or pigmentation context and blue if reported. The thickness of the lines connecting the columns is proportional to the $-\log_{10}(p)$ value of the lead SNP.

\section{Discussion}\label{sec3}
Images, which are non-invasive, high-dimensional data acquired in a high-throughput manner, have been used increasingly for phenotyping to bridge the gap in throughput and information complexity between phenomics and genomics. Although numerous studies have demonstrated that data-driven representations of image data capture more effectively data variability than traditional HDPs, few have provided clear interpretations of the image variations that correspond to these representations. This study proposes a framework for scalable automated phenotyping and interpretation of image-variation phenotypes, acting as a "phenotype sequencer". The efficacy of the framework was validated across multispecies datasets, which included human retinal fundus, domestic pigeon iris, potted rice, and swine CT datasets. The results show that AIPheno's automated phenotyping enhances the power of genetic discovery compared with human-defined and other deep learning-based phenotyping methods. AIPheno's interpretability not only validates and refines reported gene-phenotype associations but also uncovers biological insights into newly identified loci, thereby closing the loop in genetic dissection.

We attribute the greater genetic discovery power of AIPheno to three primary reasons. 

First, AIPheno excels at capturing fine-grained image variations, overcoming the limitations of coarse, discrete phenotyping. A key advantage of this approach is its ability to detect subtle, genetically driven variation within phenotypic groups traditionally considered uniform. For instance, domestic pigeon irises are typically categorized as "gravel," "pearl," or "bull," assuming no variation within each group. AIPheno, however, identifies substantial intra-group variation. Even when GWAS was restricted to only the "pearl" or "gravel" group, AIPheno still detected the \textit{KITLG-TMTC3} locus via IVP$(\mathbf{v}_{3}, 10)$, confirming the existence of genetically driven variation within these predefined categories (Fig.\ref{sup_pigeon_two_class}A-B). The interpretability module further visualized these subtle differences as varying melanin levels within both groups (Fig.\ref{sup_pigeon_two_class}C). Moreover, AIPheno identified IVPs for subtle features like eyeball orientation, pupil size, and pupil angle (Fig.\ref{sup_pigeon_corr}A), which notably showed low correlation with HDPs that primarily capture global color variation (Fig.\ref{sup_pigeon_corr}B). This highlights its capacity to uncover novel biological traits missed by conventional, human-defined phenotypes.

Furthermore, this fine-grained approach enables the quantification of complex image variations that are difficult to define or describe manually. For example, in humans, AIPheno successfully quantified the severity of tessellated fundus through IVPs (IVP$(\mathbf{v}_{21}, 9)$ and IVP$(\mathbf{v}_{28}, 7)$). Similarly, in rice, it captured complex traits like drought resistance, which manifest as coordinated color shifts across different plant tissues (Fig.\ref{rice_discussion}A), and developmental rates, reflected in both biomass and subtle color changes (Fig.\ref{rice_discussion}B). These IVPs refined the understanding of image variations associated with validated genes like \textit{GLR1.1} and \textit{SCR8}. Such intricate and multi-faceted phenotypes are challenging to capture and quantify using traditional human-defined methods.

Second, AIPheno effectively captures disentangled IVPs, allowing for the separation of distinct biological features. This is achieved through two complementary mechanisms. First, we employ methods like ICA or PCA to identify statistically independent or orthogonal directions within the latent space of the generator. Second, we leverage the hierarchical structure of the StyleGAN generator, which synthesizes images progressively from coarse to fine resolutions. This layered architecture naturally disentangles different scales of image features, separating high-level morphology from fine-grained color and texture details.

Our multi-species analyses confirm that different layers of AIPheno's generator capture distinct types of biological variation. The initial layers consistently model morphological traits. For example, IVPs from these layers corresponded to the optic disc-fovea angle in humans (IVP$(\mathbf{v}_{22}, 0)$) and overall plant height in rice (IVP$(\mathbf{v}_{1}, 1)$). In contrast, the final layers capture semantic features like color and texture. These layers identified variations in retinal color in humans (IVP$(\mathbf{v}_{23}, 13)$) and iris texture in domestic pigeons (IVP$(\mathbf{v}_{1}, 11)$, IVP$(\mathbf{v}_{3}, 10)$), and changes in whole-body grayscale intensity in swine (IVP$(\mathbf{v}_{174}, 10)$), which directly corresponds to fat deposition, that occurred without altering body shape. Intermediate layers captured a blend of subtle morphology and texture, such as changes in both body shape and fat deposition in swine CT scans (IVP$(\mathbf{v}_{112}, 4)$). This hierarchical disentanglement ensures that individual IVPs represent more specific and biologically interpretable features.

Third, models trained on a large-scale, highly diverse, and comprehensive dataset can uncover richer IVPs. In rice, if the images at the T13 time point are examined alone, no drought-induced plant wilting is observed, whereas such image features appear at other time points. Notably, models trained solely on image data from the T13 time point failed to detect relevant genetic variats. After integrating image data from all time points for model training, a broader spectrum of variations from green to withered phenotypes across the dataset was captured. This integration facilitated the detection of subtle drought stress responses in plants at early stages.

AIPheno achieves actionable and precise interpretability analysis of IVPs used for association studies. First, AIPheno does not require pre-existing indirect evidence to speculate on phenotypic variations; instead, it directly generates high-quality images through the Generator to visually present IVPs, providing novel biological insights for genetic research. Second, AIPheno can intuitively display regions with significant changes via std images and present specific image variations through image traversal. Third, during interpretability analysis, AIPheno controls specific IVP for association studies while holding others constant, ensuring such visualized variations are also precise.

In terms of its framework design and application, two further key advantages were also exhibited by AIPheno. The first advantage is flexibility. The Generator, Encoder, and Direction Discovery components were trained independently, thus the framework was configured flexibly based on the specific data characteristics of the research. For the Encoder, a CNN- or Transformer-based architecture can be chosen and trained either from scratch or by initializing it with a pre-trained model. As for Direction Discovery, methods such as ICA, PCA, or even some unsupervised learning approaches can be implemented\cite{voynov2020unsupervised,ren2021learning}. The second advantage is usability. For researchers whose data align with our study species and acquisition protocols, but which lack sufficient scale for training high-performance models, the pre-trained Encoder, Generator, and Directions based on our large-scale high-quality dataset can be utilized directly to extract IVPs for downstream association and interpretability analyses, provided that consistent image data preprocessing procedures are maintained. For cases where researchers' data differ in study species or acquisition approaches, no time-consuming and labor-intensive manual annotation is required because all model training approaches are based on unsupervised learning.

Currently, the AIPheno framework presents several areas for further optimization and exploration in future work. First, the current AIPheno only processes 2D images. For CT image analysis, 3D CT scans were converted into 2D projections. However, contemporary 3D generators have demonstrated superior performance in generating realistic outputs across domains that included natural objects, humans, and medical imaging\cite{wang2023rodin,anciukevivcius2023renderdiffusion,jung2023conditional,friedrich2024deep}. Consequently, future improvements will be made to AIPheno architectures to enable direct processing of 3D medical data modalities such as CT, MRI, point clouds, and mesh structures. Second, the integration of multimodal data for joint training has not been achieved. Although AIPheno can be applied to different modalities, only single-modal data input is supported in each training session, which restricts the model's ability to extract complementary and richer information from diverse modalities. Additionally, the training of AIPheno remains independent from downstream genetic analysis, which result in the captured image variations that originate potentially from both genetic and non-genetic factors. In future developments, AIPheno should not only leverage phenotypic data from multiple modalities but also implement joint training with genetic data, thereby facilitating the model's learning of rich information from genetically based multi-modal sources. Third, current methods for direction discovery, specifically ICA and PCA, are still based on linear transformations. When the variance of components is very small, the results of interpretability analysis barely reveal image variation. Future trainable and disentangled direction discovery approaches are suggested to be integrated seamlessly into the AIPheno framework in a plug-and-play manner, which will enable enhanced adaptability and interpretability in complex data representations. In the future, we will enable AIPheno to process 3D images, integrate genetic information and other modal data during model training, and develop a learnable Direction Discovery component. These optimizations will enhance AIPheno's capability for complex trait dissection.

Overall, in the genetic dissection of important traits in humans, animals, and plants, AIPheno not only validates and refines established gene-phenotype associations but also uncovers biological insights into newly identified genes, providing an effective framework for human disease risk prediction and precision breeding in animals and plants.

\section{Methods}\label{sec4}
\subsection{Ethical Approval Declarations}
UK Biobank data was accessed through approved project 97563.

Ethics Statement: All experimental procedures adhered to the principles of the Declaration of Helsinki. The experiments that involved swine were approved by the Ethics Committee of Huazhong Agricultural University (Approval No. HZAUSW-2023-0022), and the experiments for domestic pigeons were approved by the Institutional Animal Care and Use Committee (IACUC) of Fujian University of Traditional Chinese Medicine (Approval No. FJTCM IACUC 2024331).

\subsection{Data Collection and Preprocessing}
Four datasets were released in this study for the first time: the simulated maize root dataset, the domestic pigeon iris dataset, the potted rice dataset, and the swine CT dataset. The acquisition of multi-species biological images and genotypic datasets was documented, with data derived from both public repositories and in-house generated sources. Low-quality variants were comprehensively and systematically filtered using standardized criteria to retain high-confidence genetic variation data. In parallel, a series of targeted preprocessing operations were initially performed on the input multi-species bio-images. The main purposes of such preprocessing were to suppress the influence of background noise, to remove low-quality images, and to extract relevant image information through format transformation. Subsequently, the preprocessing pipelines of different species were introduced respectively.

\subsubsection{Simulation Maize Root Dataset}
To validate the effectiveness of the AIPheno framework, a simulated dataset was produced. Maize root anatomical images were generated by the R package GRANAR\cite{heymans2020granar}, and a real maize genotypic dataset was used. The specific simulation process was as follows:

\textbf{Image Data Generation.} Two types of genetically relevant image variations were simulated: one with more obvious variation of aerenchyma proportion, and the other with more subtle variation of metaxylem size. The parameter for \textit{aerenchyma proportion} was set to range from 0.150 to 0.650 with a step size of 0.001, which resulted in 501 values. The parameter for \textit{metaxylem size} was set to range from 0.02 to 0.07, also with a step size of 0.001, which was 51 values. From all combinations of these two parameters, a total of $501\times51=25,551$ images were generated. In addition to these two genetically related image variations, to simulate variations in real data caused by environmental effects, such as lighting, random brightness, and contrast, adjustments were applied to each image using the albumentations package\cite{buslaev2020albumentations}. The adjustment range for both the brightness and contrast factors was set to 0.2.

\textbf{Simulation design.} First, the heritabilities of both aerenchyma proportion and metaxylem size were set to 0.5. Second, SNPs were quality-controlled for a Minor Allele Frequency (MAF) $\geq$ 0.05, which ultimately produced 397,323 SNPs for subsequent analysis. QTNs were divided into three groups, and their contributions to the phenotype all followed normal distributions: the first group consisted of five QTNs that followed a normal distribution $\mathcal{N}(0, 1)$; the second group consisted of five QTNs that followed a normal distribution $\mathcal{N}(0, 0.01)$; the third group consisted of 10 QTNs that followed a normal distribution $\mathcal{N}(0, 0.0001)$. QTNs for both parameters were selected randomly from the first half and the second half of the genetic variants. The simulated phenotypic values were scaled linearly to within the value range of each parameter. Finally, images and genotypes of 2,279 individuals with parameter values that matched the simulated phenotypic values were selected for downstream association analysis.

\subsubsection{Human Retinal Fundus Dataset}
The image data of this dataset were collected from the UK Biobank (UKB) and Kaggle, with a total of 253,463 images that were utilized for model training and evaluation. Genotypic data were available for 76,829 individuals, among which 13,874,430 SNPs were identified. Only images from the left eye were utilized for downstream genetic analysis. 

\textbf{Data Collection.} Access to the UK Biobank data was granted under an approved research protocol, in compliance with the UK Biobank Data Access Agreement. The imaging data were derived from fields 21015 and 21016 of the UK Biobank, which corresponded to the left and right eyes, respectively. Genotypic data were obtained from field 22828, where the imputed genotypes were aligned to the $+$ strand of the reference genome, with their positions reported in GRCh37 coordinates. To enhance data diversity, the EyePACS dataset from Kaggle 
(\url{https://www.kaggle.com/c/diabetic-retinopathy-detection/data}) was utilized, which included both training and test sets that comprised fundus images obtained from healthy individuals and those exhibiting varying severity levels of diabetic retinopathy.    

\textbf{Imaging data.} The image quality control method provided by ContIG was adopted\cite{taleb2022contig}. Initially, images were selected based on successful circle detection using the Hough Transform algorithm, followed by central region cropping. Subsequently, the top and bottom 0.5\% of remaining images that exhibited extreme brightness values were excluded.

\textbf{Genetic data.} The genotypes of individuals with imaging data that passed quality control were extracted, and only SNPs were selected. Filtering was performed based on minor allele frequency (MAF $\geq$ 0.1\%), Hardy-Weinberg equilibrium (HWE significance threshold: $10^{-12}$), imputation quality (information score $\geq$ 0.4), and maximum missingness rates of 10\% for both SNPs and samples. Missing genotypes were subsequently imputed using Beagle software (v5.4)\cite{browning2018one}, followed by a second round of quality control that applied the same criteria. Ultimately, 13,874,430 SNPs were retained for analysis.

\subsubsection{Domestic Pigeon Iris Dataset}
The image data in this dataset were collected primarily through web crawling, with a total of 28,903 images used for model training and evaluation. To facilitate improved genetic analyses, a new, high-quality telomere-to-telomere (T2T) domestic pigeon genome was assembled. Among these, genotypic data were obtained from 641 individuals that contained 12,113,557 SNPs.

\textbf{Data Collection.} The image data were crawled primarily from the Youth Pigeon Information Website (\url{https://qingnianxinge.com/index.html}), with public data from 51 pigeon lofts being collected through this platform. Additionally, a portion of the data was derived from publicly accessible records of pigeon racing competitions. The images of individuals utilized for genetic analysis were captured through our own photography. Genotypic data were derived from whole-genome resequencing of feather samples obtained from domestic pigeons.

\textbf{Imaging data.} A total of 1,936 images were annotated initially to train a U-Net\cite{ronneberger2015u} semantic segmentation model for automatic iris segmentation from the images. After each iris region was segmented from individual images, the iris region was cropped and padded with zero pixels to form a square shape, which ensured that the iris remained centered within the image. Finally, all processed images were resized to 512 $\times$ 512 pixels. This image processing pipeline prevented effectively geometric distortion during square resizing while maintaining consistent dimensions across all images, eliminated positional variations caused by Regions of Interest (ROIs) located at different regions, and further addressed size inconsistencies in ROIs thatresulted from non-uniform distances between the camera and subjects.

\textbf{Sample preparation.} An adult female domestic pigeon was selected for genome assembly. The individual was maintained under standard husbandry conditions at Xinglian Pigeon Industry Co., Ltd.. Ten tissues, which included muscle, skin, heart, liver, gizzard, intestine, ovary, lung, eye, and brain, were dissected, flash-frozen in liquid nitrogen, and stored at -80°C until further processing.

\textbf{Genomic DNA preparation and sequencing.} Total genomic DNA (gDNA) was isolated from muscle using the Sodium Dodecyl Sulfate (SDS) lysis protocol. DNA integrity was evaluated using the Agilent Bioanalyzer 2100 system (Agilent Technologies, CA, USA), and quantification was performed using a Qubit 3.0 Fluorometer (Invitrogen, Waltham, MA, USA). Three sequencing libraries were constructed: (1) a short-read library compatible with the DNBSEQ-T7 sequencing platform (MGI Tech, Shenzhen, China), (2) a high-fidelity (HiFi) library for PacBio SMRT sequencing (Pacific Biosciences, CA, USA), and (3) an ultra-long-read library for the Nanopore PromethION platform (Oxford Nanopore Technologies, Oxford Science Park, UK).

\textbf{Chromosome conformation capture (Hi-C) experiment.} Hi-C technology was used for chromosomal scaffolding. Biotinylated DNA fragments were sheared mechanically to 300-500 bp using a Covaris S220 sonicator, followed by streptavidin bead enrichment (Dynabeads M-280, Thermo Fisher Scientific) to generate paired-end Hi-C libraries. Sequencing was performed on the DNBSEQ-T7 system (MGI Tech).

\textbf{RNA extraction and sequencing.} Total RNA was extracted from all 10 tissues using the TRIzol reagent (Thermo Fisher Scientific, MA, USA). Pooled RNA-seq libraries were prepared with the TruSeq RNA Library Preparation Kit v2 (Illumina, CA, USA) and sequenced on the DNBSEQ-T7 platform. Raw reads were processed using fastp ($\texttt{---l}=50$, v0.19.3)\cite{chen2018fastp} to remove adapter sequences and low-quality bases (Phred score $<$ 20), which yielded clean reads for downstream analysis.

\textbf{Telomere-to-telomere genome assembly.} HiFi reads, Oxford Nanopore Technologies (ONT) ultra-long reads, and Hi-C data were assembled into contigs using Hifiasm (v0.17.4-r455)\cite{cheng2021haplotype} with default parameters (GitHub: \url{https://github.com/chhylp123/hifiasm}). Chromosome-level scaffolding was performed by integrating Hi-C data with Juicer (v1.6)\cite{durand2016juicer} and HapHiC (v1.0.1)\cite{zeng2024chromosome}. Manual refinement and chromosome ordering were conducted using Juicebox (v1.11.08)\cite{durand2016juicebox}.

To resolve telomeric regions, ONT ultra-long reads were aligned to the assembly with Minimap2 (v2.24-r1122, parameters: \texttt{---ax map-ont})\cite{li2018minimap2}. Reads mapped within 100 bp of chromosome ends were analyzed for telomeric repeat motifs ("CCCTAA"/"TTAGGG")\cite{vega2003getting}. Reads with the highest motif frequency were designated as reference (REF), and others were classified as query (QUERY). Consensus telomeric sequences were generated using medaka\_consensus (v1.2.1, model: r941\_prom\_fast\_g507, \url{https://github.com/nanoporetech/medaka}) for REF and QUERY reads independently. These sequences were then aligned to chromosome termini using Nucmer (v4.0.0rc1)\cite{kurtz2004versatile}, and regions with $\geq90$\% alignment coverage were integrated to replace original telomeric sequences. 

Scaffold gaps were resolved using TGS-GapCloser (v1.2.0, parameters: \texttt{---min\_nread 10})\cite{xu2020tgs}, 
which leveraged ONT read-to-contig coverage relationships. Final genome polishing was performed with Pilon 
(v1.23, parameters: \texttt{---fix snps}, \texttt{indels ---changes})\cite{walker2014pilon} using Illumina short reads.

\textbf{Repeat and gene annotation.} Repetitive elements in the genome were annotated using a combined strategy of homology-based and \textit{de novo} prediction methods. Homology-based identification was performed with RepeatMasker and RepeatProteinMask (RepBase library) to detect known repeats, and \textit{de novo} prediction utilized RepeatModeler and LTR-FINDER to identify novel repetitive sequences based on structural features. Tandem repeats were predicted using TRF (v4.09) with default parameters. 

For protein-coding gene annotation, we used a multi-evidence integration approach was employed. Homologous genes were identified by aligning known protein sequences from \textit{Columba livia} (GCF\_036013475.1, GCA\_032206205.1, GCA\_000337935.2) and \textit{Columba guinea} (GCA\_032206185.1) to the genome using Miniprot (v0.11-r234)\cite{li2023protein} with the parameter \texttt{-j 1}. \textit{De novo} gene prediction was conducted using Augustus\cite{stanke2006augustus} and Genscan\cite{burge1997prediction} to model potential coding regions. Transcriptomic evidence was incorporated by aligning RNA-seq reads to the genome with HISAT2\cite{kim2019graph}, followed by transcript assembly using StringTie\cite{pertea2015stringtie} and coding sequence prediction using TransDecoder. These heterogeneous datasets (i.e., homology alignments, \textit{de novo} predictions, and transcript-derived evidence) were integrated into a non-redundant gene set using MAKER\cite{holt2011maker2}. With priority given to transcript-supported and homology-based annotations, the final gene set was refined using HiFAP\cite{sun2024chromosomal} to resolve conflicts and to ensure consistency. Genome annotation was assessed with BUSCO (v5.5.0) against the aves\_odb10 lineage, which achieved 99.80\% completeness. Non-coding RNAs were annotated as follows: tRNAscan-SE\cite{lowe1997trnascan} for tRNA identification, BLASTN\cite{altschul1990basic} against homologous rRNA references for rRNA detection, and INFERNAL\cite{griffiths2005rfam} with Rfam covariance models for miRNA and snRNA prediction.

Functional annotation of protein-coding genes included DIAMOND (v2.0.14) alignment (blastp, $e-value\leq1e-5$) to TrEMBL, SwissProt, NR, KOG/COG, and KEGG databases, KOBAS (v3.0)\cite{bu2021kobas} for KEGG pathway mapping, InterProScan (v5.61-93.0)\cite{jones2014interproscan} for domain/motif analysis, and HMMER3 (v3.3.1)\cite{mistry2013challenges} for Pfam and annotation transcription factors.

\textbf{Genetic data.} All collected WGS data were processed using standard bioinformatics pipelines. By using fastp (v0.12.4)\cite{chen2018fastp}, raw data were first trimmed by removing adapters and low-quality (\texttt{-W 4 -M 20 -q 20 -u 40 -n 5 -l 15}) bases. The remaining high-quality reads were aligned against the reference sequence using BWA (v0.7.17)\cite{li2009fast}. Uniquely mapped reads were used for detection of short variants with Sentieon (v2023.08)\cite{freed2017sentieon}. To obtain highly confident short variants, samples with sequencing depth $<$ 3 and coverage $<$ 70\% were removed. GATK (v4.0.3.0)\cite{DePristo_2011} was next used the parameter \texttt{QUAL < 30.0 | | QD < 2.0 | | FS > 60.0 | | MQ < 40.0 | | SOR > 3.0 | | ReadPosRankSum < -8.0 / QUAL < 30.0 | | QD < 2.0 | | FS > 200.0 | | SOR > 10.0 | | ReadPosRankSum < -20.0 | | MQ < 40.0 | | MQRankSum < -12.5} to retain high-quality. The SNP data were filtered through criteria that included a MAF threshold $\geq$ 0.05, maximum missing genotype rates per variant of 0.1, and maximum missing genotype rates per sample of 0.1. The missing genotypes were imputed using Beagle software (v5.4), followed by quality control with the same filtering criteria, which resulted in a final set of 12,113,557 SNPs.

\subsubsection{Potted Rice Dataset}
This dataset was compried of side-view images captured from 529 accessions across 22 developmental timepoints from seedling to maturity stages. A total of 327,734 images were utilized for model training and evaluation. The genotypic data of the 529 individuals were characterized by 4,321,306 SNPs. 

\textbf{Data Collection.} A previous study provided detailed information on the rice accessions and their genotyping\cite{yang2014combining}. Two biological replicates were established for each accession. Each replicate was imaged from 15 distinct side-view angles to ensure complete $360^\circ$ circumferential coverage of the plant. Twenty-two time points were designated from the seeding to heading stages. Imaging was conducted at each time point using the aforementioned protocol.

\textbf{Imaging data.} The foreground region of each image was cropped and pasted onto the center of a blank square image of uniform size. The composite images were then resized uniformly to 256 $\times$ 256. This approach  eliminated variations caused by inconsistent positioning of plants within images and prevented deformation induced by direct resizing of all images to a uniform dimension.

\textbf{Genetic data.} The SNP data were processed using filtering criteria with a MAF threshold $\geq$ 0.05, with maximum missing genotypic rates per variant and per sample of 0.1. Missing genotypes were filled using Beagle software (v5.4), and quality control was reapplied with identical thresholds, which yielded a final dataset of 4,321,306 SNPs.

\subsubsection{Swine CT Dataset}
The dataset was comprised of two-dimensional CT fat projection images of 3,803 boars from four breeds (i.e., Landrace, Yorkshire, Pietrain, and Duroc), which were utilized for model training and evaluation. Genotypic data were available for 795 individuals, which contained 17,810,683 SNPs.

\textbf{Data Collection.} 
After anesthesia, the pigs were positioned prone and scanned using a Siemens AS plus CT scanner with the following parameters: 110 kV/160 mA, 512 $\times$ 512 matrix, axial orientation, and 5 mm slice thickness. The remaining body parts of the boar, excluding the head, were scanned. The genotypic data were obtained through whole-genome resequencing of ear tissue samples.

\textbf{Imaging data.} A dataset of 80 annotated samples was utilized to train a 3D semantic segmentation model for CT images using nnU-Net\cite{isensee2021nnu}, with the purpose of removing extraneous targets, such as CT tables from boar body scans. After nnU-Net segmentation, a threshold segmentation (-200 $<$ HU $<$ 0) was performed to isolate whole-body fat voxels\cite{gjerlaug2012genetic}. Subsequently, the entire CT image was interpolated linearly to achieve isotropic voxels with a spacing of 1 $\times$ 1 $\times$ 1, which ensured alignment with real-world physical dimensions. Because higher HU values indicated greater voxel density, the HU values of all voxels were increased uniformly by 200 to guarantee 
non-negative pixel values post-projection. Then, a sagittal projection of the 3D volume was generated by summing all voxel values along the same directional axis, which resulted in a 2D projection image. After projection, the foreground within the image was cropped out, then pasted into the center of a uniformly sized square blank image. The maximum value across all images was identified, and the image was normalized accordingly. Finally, the image was scaled to the 0-255 range and saved as a PNG image in uint8 format.

After preprocessing with the aforementioned methods, image variations caused by positional differences among individuals were eliminated. Additionally, a unified normalization procedure was applied, which enabled the comparability of pixel values across different individuals. A higher pixel value was indicative of a greater fat content in the corresponding projection direction.

\textbf{Genetic data.} A population of 795 individuals from four breeds (i.e., Duroc, Landrace, Pietrain, and Large White) was subjected to next-generation short-read sequencing.  Initially, quality control of the raw FASTQ sequencing data was performed using fastp (v0.12.4), whereby adapters and low-quality bases were removed. Subsequently, the high-quality data that remained after quality control were aligned to the swine reference genome (\textit{Sus Scrofa} 11.1) using BWA (v0.7.17) software. Uniquely mapped reads were used for detection of short variants with Sentieon (v2023.08). If multiple sequences were mapped to the same genomic location, only the unique alignment that possessed the highest mapping quality was retained. The SNP dataset was filtered based on criteria that included a MAF threshold $\geq$ 0.05 and maximum missing genotypic rates of 0.1 for both variants and samples. Missing genotypes were imputed using Beagle software (v5.4), followed by reapplying the same quality control thresholds, which ultimately yielded a final dataset of 17,810,683 SNPs. 

\subsection{The Encoder-Generator Framework of AIPheno}\label{The Encoder-Generator Framework of AIPheno}
The primary and core objective of AIPheno is to acquire high-dimensional phenotypic variations from the input images and to perform interpretable analyses on those phenotypes that exhibit significant signals in GWAS. In pursuit of this goal, we have devised a deep learning framework predicated on an encoder-generator architecture (Fig.\ref{fig_S16}). Subsequently, the specific structures of the Generator and the Encoder will be introduced.

\textbf{Generator.} In consideration of the limited volume of bio-image data, an unconditional StyleGAN\cite{karras2019style,karras2020training} network was selected to construct the generator. The fundamental structure of a StyleGAN consists of a standard gaussian probability distribution $p(z)$, from which a latent vector $z$ is sampled, and a Mapping network $M(z)$ maps the latent vector $z$ to the $w$ space, and then a synthesis network $S(z)$ is utilized to generate an output image $I$. This can be represented mathematically as $z\sim p(z)$, $w=M(z)$ and $I=S(w)$. The neural network $S(\cdot)$ can be disassembled further into a sequence of $L$ intermediate layers, denoted as $S_{1},S_{2},...,S_{L}$. The initial layer takes the latent vector $w$ as the input and yields a feature tensor $Y_{1}=S_{1}(w)$. For the subsequent layers, each layer generates features as a function of the output of the preceding layer, that is, $Y_{i}=S_{i}(Y_{i-1})$. The final output of the last layer, $I=S_{L}(Y_{L-1})$, is an RGB image. In this study, we used the Generator for unsupervised training to fit the distribution of the original input bio-images. For all the preprocessed bio-image data, we first resized the image to 256 $\times$ 256. The value of $L$ was set to 14, and the dimension of $w$ was 512. To enhance the ability to restore image details, we mapped the $w\in \mathbb{R} ^{1\times 512}$ space into the $w+\in \mathbb{R} ^{14\times 512}$ space\cite{kim2021exploiting} (Fig.\ref{fig_S18}A). Therefore, the mathematical representation of generating images using the synthesis network was transformed into $I=S(w+)$. Notice that each vector in $w+$ was first projected by the affine layer $A_{i}$, then it affected the corresponding layers $S_{i}$ by modulating on the convolution kernels, where $i=1,2,...,L$. Fig.\ref{sup_method_un_human}, Fig.\ref{sup_method_un_pigeon}, Fig.\ref{sup_method_un_rice} and Fig.\ref{sup_method_un_swine} respectively present the images generated unconditionally from retinal fundus of human, irises of domestic pigeon, potted rice, and swine CT, which demonstrated that the generator possessed the capacity to fit the spatial distribution of the original input data.

\textbf{Encoder.} The objective of the Encoder is to achieve precise image reconstruction in combination with our proposed Generator, thereby obtaining the latent vector of the input image within the $w+$ space. This was represented mathematically as given a bio-image $I\in \mathbb{R} ^{H\times W\times C}$; the Encoder was able to specify $L$ different style vectors denoted by $w_{i}\in \mathbb{R} ^{1\times C}$, where $i=1,2,...,L$ was the index of the vector injected into the different stages of the Generator. The overall structure of the Encoder was divided into two components (Fig.\ref{fig_S16}). One component consisted of the global and local image representation modules, and the other component was composed of $L$ learnable tokens. Within the first component, the input image $I$ was encoded by Swin\cite{liu2021swin} Backbone, which generated a series of local image features $F_{local}^{0}$ to $F_{local}^{3}$ at multi-resolutions and a global image feature $F_{global}$. The second component utilized a Cross-Modality Multi-Head Attention (X-MHA)\cite{li2022grounded} to interact with the local information of the image and then fused it with the global image representation to obtain the final representation in the $w+$ space. Four layers of X-MHA were incorporated with the intention of fusing the local features of images at various resolutions. The whole process was written as according to Eq.\ref{eq1}, where $X_{f2t}$ represents the fusion pathway from local image feature to learnable tokens $X$, and $\widehat{F}_{t2f}$ represents the opposite fusion pathway. Notice that when $i=0$, the feature fusion has not been performed yet, and the corresponding formula is shown as Formula Eq.\ref{eq2}. Subsequent to passing through X-MHA, the features of learnable tokens will proceed directly into the X-MHA of the next layer. Concurrently, the local image feature $F_{t2f}^{i+1}$ was subjected to up-sampling processing and was integrated with the local image feature $F_{local}^{i+1}$ of the next layer prior to entering the X-MHA (Eq.\ref{eq3}).

\begin{equation}
    X_{f2t}^{i+1}, F_{t2f}^{i+1}=\emph{X-MHA}(X_{f2t}^{i}, \widehat{F}_{t2f}^{i}), i=0,1,2,3\label{eq1}
\end{equation}

\begin{equation}
    X_{f2t}^{0}=X, \widehat{F}_{t2f}^{0}=F_{local}^{0}\label{eq2}
\end{equation}

\begin{equation}
    \widehat{F}_{t2f}^{i+1}=UP(F_{t2f}^{i+1})+F_{local}^{i+1}, i=0,1,2\label{eq3}
\end{equation}

For simplicity, the preprocessed input bio-images were resized uniformly to a specific dimension of 256 $\times$ 256. In this context, the parameters $L$ and $C$ were set as 14 and 512, respectively. Moreover, we selected the network size of Swin-Large and employed the weights pre-trained on the basis of the ImageNet-22K\cite{russakovsky2015imagenet} data to initialize the Backbone network. Fig.\ref{fig_S17} presents an overview of the proposed X-MHA. More precisely, $X\in \mathbb{R} ^{14\times 512}$ learnable tokens were initialized randomly. Subsequently, within the X-MHA, the learnable tokens $X$ and the flattened local feature after fusion $\widehat{F}_{t2f}^{i}\in \mathbb{R} ^{HW\times 512}$ of the image functioned as Query and Key-Value, respectively, to participate in the cross-fusion process; each head computed the context vectors of one modality by attending to the other modality. With respect to the $X_{f2t}$, the query $Q$, key $K$, and value $V$ were all projected in accordance with Eq.\ref{eq4}. Conversely, the $F_{t2f}$ was as depicted in Eq.\ref{eq5}. It is noted that $W_{Q}^{\emph{X-MHA}}$, $W_{K}^{\emph{X-MHA}}$, and $W_{V}^{\emph{X-MHA}}\in \mathbb{R} ^{512\times 512}$ were learnable projection heads in the X-MHA module. The calculation formula of \emph{Attention} is presented in Eq.\ref{eq6}, where the feature dimension $d$ was 512. The final update results from the Multi-Head Attention are given in Eq.\ref{eq7}, where $W_{O}^{\emph{X-MHA}}\in \mathbb{R} ^{512\times 512}$ was also learnable and was responsible for fusing the results $Attn$ from different heads $h$. In this case, $H$ was set to 4.

\begin{equation}
    Q_{i}=X_{f2t}^{i}W_{Q}^{\emph{X-MHA}}, K_{i}=\widehat{F}_{t2f}^{i}W_{K}^{\emph{X-MHA}}, V_{i}=\widehat{F}_{t2f}^{i}W_{V}^{\emph{X-MHA}}\label{eq4}
\end{equation}

\begin{equation}
    Q_{i}=\widehat{F}_{t2f}^{i}W_{Q}^{\emph{X-MHA}}, K_{i}=X_{f2t}^{i}W_{K}^{\emph{X-MHA}}, V_{i}=X_{f2t}^{i}W_{V}^{\emph{X-MHA}}\label{eq5}
\end{equation}

\begin{equation}
    Attn(Q_{i},K_{i},V_{i})=Softmax(\frac{Q_{i}K_{i}^{T}}{\sqrt{d} })V_{i},i=0,1,2,3\label{eq6}
\end{equation}

\begin{equation}
    MHA(Q_{i},K_{i},V_{i})=[Attn(Q_{i},K_{i},V_{i})]_{h=1:H}W_{O}^{\emph{X-MHA}}\label{eq7}
\end{equation}

\subsection{Model Training}\label{Model Training}
The entire Encoder-Generator framework adopted a two-step training approach. First, a Generator was trained independently to adapt to the distribution of the input bio-images. Subsequently, the parameters of the trained Generator were held constant, and the Encoder and Generator were trained jointly with the aim of obtaining image representation. Because the entire training process was unsupervised, each dataset was split randomly into a training set and a validation set; the training set accounted for 95\% and the validation set for 5\% (with the exception of the swine CT dataset, for which the GWAS individuals were used as the validation set, and the remainder as the training set). The following subsequent section provides a detailed description of the two steps.

\textbf{Generator Training.} In the initial stage, the processed training, multi-species bio-images were resized to a dimension of 256 $\times$ 256 and converted into the Lightning Memory-Mapped Database (LMDB) format for storage purposes. These processed images were then used for the subsequent training of the Generator. Subsequently, a normalization procedure was conducted out on the three RGB channels of the images. Specifically, the means of the channels were set to [0.5, 0.5, 0.5], and the variances were set to [0.5, 0.5, 0.5]. The initial learning rate, denoted as lr, was established as 0.002, and the training batch size was configured to be 24. The remaining training parameters were assigned the default values as described in StyleGAN2\cite{karras2020training}. The entire training process was implemented using a single 40G-A100 GPU, and we adopted the Adam optimization strategy. The total number of training iterations was set to 250,000. Upon the attainment of convergence in the training of the Generator, we used an early stopping strategy to obtain the final model, thereby achieving a reduction in the training time.

\textbf{Encoder Training.} We adopted the strategy of training the Encoder and the Generator jointly to obtain the Encoder model. Similar to the training of the Generator, the preprocessed, multi-species bio-images were also resized to 256 $\times$ 256. During training, a random horizontal flip was applied to the input bio-images with a probability of 0.5 for data augmentation. A normalization operation was then performed on the three RGB channels of the images, with the means set to [0.5, 0.5, 0.5] and the variances set to [0.5, 0.5, 0.5]. The training batch size was configured as 16, and the entire training process was carried out using a single 40G-A100 GPU, with the Adam optimization strategy. The total number of training iterations was 600,000. Once the Encoder model converged, we used an early stopping strategy to obtain the final model. Notice that, during training, the synthesis backbone of Generator that included the affine layer $A$ was strictly fixed. Furthermore, the pixel-wise Mean Squared Error (MSE) loss and Learned Perceptual Image Patch Similarity (LPIPS\cite{zhang2018unreasonable}) were utilized to measure the pixel-level and perceptual-level similarities between the input image $I$ and the reconstructed image $I^{'}$. Simultaneously, LPIPS, which was calculated based on the features within an Inception\cite{simonyan2014very} net $VGG(\cdot)$, was also used. It formulated the objective as Eq.\ref{eq8}.

\begin{equation}
    L_{LPIPS}=||VGG(I)-VGG(I^{'})||_{2}\label{eq8}
\end{equation}

\begin{equation}
    L_{total}=\lambda _{LPIPS}L_{LPIPS}+\lambda _{MSE}L_{MSE}\label{eq9}
\end{equation}

As shown in Eq.\ref{eq9}, the overall loss function was composed of two parts. $\lambda _{LPIPS}$ and $\lambda _{MSE}$ acted as weights for balancing each loss. During the training process, for Human Retinal Fundus Dataset, Domestic Pigeon Iris Dataset and Swine CT Dataset, the initial learning rate was configured as lr$=$0.0001, $\lambda _{LPIPS}$ was set to 4, and $\lambda _{MSE}$ was set to 1. In contrast, for Potted Rice Dataset, the initial learning rate was set as lr$=$0.001, $\lambda _{LPIPS}$ was set to 0.8, and $\lambda _{MSE}$ was set to 1.

Fig.\ref{sup_framework_generation} showed the reconstructed input images of four datasets. Results demonstrated that the Encoder exhibited robust capability to restore the overall image information, with consistent preservation of fine details.

\subsection{Phenotype Extraction}
The high-dimensional features extracted by the Encoder we proposed have reflected satisfactorily the variation information of the original bio-images. Nevertheless, GWAS processes single-dimensional values separately, and the deep features were mutually disentangled in different dimensions, which was not conducive to the subsequent interpretable analysis of phenotypes. Therefore, the objective of this section was to utilize the pre-trained Generator to discover the mutually independent and interpretable directions hidden in the $W$ space, and output the projection values of the input bio-images onto these directions as the phenotypes that described the image variation. 

\textbf{Discovering Interpretable Directions.} A data-driven Independent Component Analysis (ICA)\cite{hyvarinen2000independent} approach was used to discover the interpretable directions. For a fixed and trained synthesis network, a randomly sampled $z$ from the standard normal distribution $p(z)$, which was mapped to $w$ using the mapping network, corresponded to a generated image $I$ (Fig.\ref{fig_S18}A). This process was denoted as $\emph{z-w-I}$, that is, each $z$ and $w$ corresponded to a generated image. 

Generally, the quantity of original bio-images is rather limited. Direct exploration of directions is prone to introducing biases. Consequently, we sampled 100,000 times from the standard normal distribution $p(z)$ to simulate and to generate a set of 100,000 images that conformed to the distribution of the original bio-images, and this generated image set was then utilized for the discovery of meaningful directions (Fig.\ref{fig_S18}B). In the $W$ space, each image was  represented by a vector $w$ with $C=512$ dimensions. Meanwhile, a direction in the $W$ space was denoted by a vector with $C=512$ dimensions. The core objective of ICA was to decompose the observed mixed vectors into several mutually independent source vectors. Suppose there existed multiple source vectors that were mixed through an unknown linear mixing matrix. ICA aimed to estimate the mixing matrix and the source vectors through statistical methods, thereby achieving vector separation. In this study, we used the FastICA \cite{langlois2010introduction} algorithm in the \emph{Scikit-Learn} python package\cite{hao2019machine} to solve for the independent components of ICA; the number of independent components of ICA after solution was $K$, which produced $K\times 512$ interpretable directions.

\textbf{Project and Phenotype Extraction.} For any image within the population to be analyzed genetically, a trained Encoder network was adopted to obtain the representation in the $W+$ space. Subsequently, the $L$-layer features in the $W+$ space were respectively dot-multiplied with the $K$ directions to acquire the projection values (Fig.\ref{fig_S18}C). Notice that the dimension of the image representation in the $W+$ space was $L\times C$, and the dimension of the directions was $K \times C$. Therefore, $L \times K$ projection values were obtained finally, which constituted the final phenotype of the input bio-image. Specifically, we set $L=14$ and $C=512$ in this study. For different image data, the value of $K$ varied. However, to reduce the computational cost of GWAS, it was not recommended that the value of $K$ exceed 30.

\subsection{Interpretation Analysis}

\textbf{Direction Traversing.} For the phenotypic dimensions that exhibited significant signals within the GWAS analysis, we performed a traversal operation along their corresponding directions to acquire the image phenotypes subsequent to the traversal. Thereafter, the well-trained Generator was utilized to convert traversaled phenotypes into corresponding images $\hat{I}$, thereby providing an actionable visualization of the specific implications of the phenotypic dimension. In essence, this process constituted an interpretability analysis. More concretely, prior section\ref{Model Training} proposed to use a certain direction $D\in \mathbb{R} ^{512}$ in the $W$ space to represent a semantic concept. After identifying a semantically meaningful direction, the traversing was achieved using the following Eq.\ref{eq10}, which was applied in image editing\cite{harkonen2020ganspace,shen2021closed,song2023householder}. 

\begin{equation}
    \hat{I}=S(\hat{w})=S(w+\alpha D)\label{eq10}
\end{equation}

In other words, the target semantic was modified by displacing the latent code $w\in \mathbb{R} ^{512}$ (in $W$ space) linearly along the identified direction $D$. The parameter $\alpha $ represented the manipulation intensity.

\textbf{Layer-Wise Interpretation.} As introduced in Section\ref{The Encoder-Generator Framework of AIPheno}, it is common practice for us to utilize the $W+$ space as a substitute for the $W$ space space to represent the input image features. Moreover, the $L$ layers within the $W+$ space acted respectively on the corresponding affine layer $A_{i}$ in the Generator, thereby generating images iteratively. Consequently, we also extended the traversal of directions to be carried out within the $W+$ space. Suppose that the phenotype of the $i$-th layer in the $W+$ space of an image yielded a significant locus. Then, only the phenotype of this layer was traversed along direction $D$, but the phenotypes of the other layers  remained unchanged. The traversed phenotype $\hat{w_{i}}$ was then fed into the trained Generator together with the phenotypes of the other layers to generate an image $\hat{I}$. The specific calculation formulas are shown in Eq.\ref{eq11} and Eq.\ref{eq12}.

\begin{equation}
    \hat{w_{i}}=w_{i}+\alpha D\label{eq11}
\end{equation}

\begin{equation}
    \hat{I}=S((w_{1}, w_{2},...,\hat{w_{i}}, w_{L}))\label{eq12}
\end{equation}

By virtue of the layer-by-layer iterative generation approach adopted during image generation, in the $L$ layers, the shallow-layer $w_{i}$($i=0,1,...,6$) represented information, such as image texture and contour, and the deep-layer $w_{i}$($i=7,8,...,L$) represented image variations that were more meaningful semantically, such as pigment deposition in human retinal images, and drought resistance or developmental rate in rice. In this way, phenotypic extraction in the $W+$ space achieved layer-wise interpretability of bio-image variation.

\subsection{Genetic Analysis}

\subsubsection{Simulation Study}
\textbf{Phenotypes.} Because the $W+$ space was decomposed into 30 directions using ICA, 420 IVPs were included in the GWAS analysis.

\textbf{GWAS.} GWAS analysis was performed using the MLM method in rMVP\cite{yin2021rmvp} with the top three genetic principal components were included as covariates. To maintain the false discovery rate $<$ 5\%, Bonferroni correction was applied by dividing the significance level by both the number of markers and the number of phenotypes. The significance threshold for the IVPs was calculated as $\frac{0.05}{397,323 \times 420} = 3.00 \times 10^{-10}$.

\textbf{Identification of true loci.} Genetic variants within 250 kb upstream and downstream of each QTN were defined as true loci.

\subsubsection{Human}
\textbf{Phenotypes.} Only the left-eye images were utilized to extract phenotypes for downstream analysis. As the $W+$ space was decomposed into 30 directions using ICA, the number of IVPs employed in the GWAS analysis amounted to 420. For comparison, a $400 \times 400$ pixel patch at coordinates [600:1000, 800:1200] was cropped, and the average intensities of its red, green, and blue channels were taken as the quantitative traits\cite{xie2024igwas}.

\textbf{GWAS.} Given the substantial size of the UKB dataset, traditional GWAS methods become computationally prohibitive when analyzing numerous phenotypes. To address this, we implemented a three-step approach for UKB GWAS analysis. First, the kinship matrix was constructed using HIBLUP\cite{yin2023hiblup}, followed by calculation of the top 10 principal components (PCs) from its principal component analysis (PCA). Subsequently, age, sex, and the top 10 genetic PCs were incorporated as covariates, while the kinship matrix and all phenotypes were input into rMVP. We used restricted maximum likelihood (REML) methodology to estimate variance components and heritability for all phenotypes, with eigenvalues and eigenvectors of each phenotype  computed simultaneously. Finally, all phenotypes, their eigenvalues, eigenvectors, variance components, and heritability were input into rMVP. Analysis of all traits was implemented in a single step based on the MLM. To maintain the false discovery rate $<$ 5\%, we used a stringent threshold: Bonferroni correction was applied by dividing the conventional threshold of $5 \times 10^{-8}$ by the number of tested phenotypes, which yielded an adjusted threshold of $\frac{5 \times 10^{-8}}{420} = 1.19 \times 10^{-10}$. For HDPs, the threshold was $\frac{5 \times 10^{-8}}{3} = 1.67 \times 10^{-8}$.

\textbf{Identification of lead SNPs and genomic loci.} The SNP2GENE function of FUMA was utilized to identify lead SNPs and genomic loci. Initially, GWAS results from all phenotypes were integrated into a single summary statistics file (hereafter referred to as the "minP file") by identifying the minimum \textit{p} value for each SNP across all phenotypes. We used a two-step clumping procedure. First, significant SNPs (i.e., the \textit{p} value was $p < 1.19 \times 10^{-10}$ for IVPs and $p < 1.67 \times 10^{-8}$ for HDPs.) were clumped using an LD threshold of $r^{2} = 0.6$ to derive independent significant SNPs. Thereafter, the independent significant SNPs were reclumped at an LD threshold of $r^{2} = 0.1$ to determine independent lead SNPs. A genomic locus was the minimal continuous segment that encompassed all SNPs (i.e., both GWAS markers and 1000 Genomes reference panel markers that surpassed the MAF cutoff) showing an $r^{2} > 0.1$ with the lead SNPs. Adjacent regions separated by $\leq$ 250 kb were combined into a single locus.

\textbf{Identification of candidate genes.} We performed gene mapping using the positional mapping function of the FUMA SNP2GENE module, with the maximum distance parameter maintained at the default setting of 10 kb. In instances where a lead SNP failed to be mapped to any gene, the GWAS Catalog database (\url{https://www.ebi.ac.uk/gwas/home}) or the LDtrait Tool developed by the National Institutes of Health (\url{https://ldlink.nih.gov/?tab=ldtrait}) was examined systematically to determine the presence of identical or highly linked SNPs with previously documented gene associations.

\textbf{Querying GWAS catalog.} All mapped genes were inspected manually in the GWAS Catalog database (The query date was November 2024). All traits that were reported previously to be associated with these genes were collected, with a specific focus on identifying documented ocular-related or pigmentation-related traits.

\textbf{Enrichment analysis.} The enrichment analyses for GO Biological Process, KEGG, Reactome, WikiPathways, and GWAS Catalog were all implemented using the GENE2FUNC module of FUMA.

\subsubsection{Domestic pigeon}
\textbf{Phenotypes.} Because the $W+$ space was decomposed into 10 directions using ICA, 140 IVPs were utilized for GWAS analysis. For comparison, eight HDPs were extracted: gravel-type iris and pearl-type iris that served as case-control traits, the mean values of the three RGB channels in the foreground region, the mean values of the three HSV channels in the foreground region, and the mean value of the grayscale image in the foreground region.

\textbf{GWAS.} Because the images of sequenced individuals were obtained from two separate captures, the GWAS analysis was performed using the MLM in the rMVP package, with batch effects and the top three genetic PCs included as covariates. To maintain the false discovery rate $<$ 5\%, the Bonferroni correction was performed by dividing the significance level by both the number of markers and the number of phenotypes. Due to the different numbers of phenotypes between IVPs ($n=140$) and HDPs ($n=8$), distinct thresholds were applied: for IVPs, the adjusted threshold was calculated as $\frac{0.05}{12,113,557 \times 140} = 2.95 \times 10^{-11}$, and for HDPs as $\frac{0.05}{12,113,557 \times 8} = 5.16 \times 10^{-10}$.

\textbf{Identification of candidate genes.} LDBlock analysis was performed using LDBlockShow\cite{dong2021ldblockshow}. The input GWAS results file that was with the \texttt{-InGWAS} parameter contained the minimum \textit{p} value (minP) for each SNP. The input annotation file that was specified with the \texttt{-InGFF} parameter provided functional annotations for protein-coding genes of the optimal transcript isoforms. The analyzed region spanned 150 kb upstream and downstream of the top SNP.

\subsubsection{Rice}
\textbf{Phenotypes.} Because the $W+$ space was decomposed into 30 directions using ICA, 420 IVPs were included in the GWAS analysis. For comparison, 11 HDPs with relatively clear biological significance were extracted, specifically: histogram-based mean (M\_TEX), histogram-based variance (SE\_TEX), total projected area (TPA), height of the bounding rectangle (H), width of the bounding rectangle (W), ratio of total projected area to circumscribed box area (TBR), circumscribed box area (CBA), ratio of total projected area to hull area (THR), ratio of perimeter to total projected area (PAR), perimeter (P), and green projected area (GPA).

\textbf{GWAS.} GWAS analysis was performed using the MLM method in rMVP with the top three genetic principal components included as covariates. To maintain the false discovery rate $<$ 5\%, the Bonferroni correction was applied by dividing the significance level by both the number of markers and the number of phenotypes. Due to differences in phenotype counts between the IVPs ($n=420$) and HDPs ($n=11$), distinct thresholds were adopted: for IVPs, the adjusted threshold was calculated as $\frac{0.05}{4,321,306 \times 420} = 2.75 \times 10^{-11}$, and for HDPs as $\frac{0.05}{4,321,306 \times 11} = 1.05 \times 10^{-9}$.

\textbf{Identification of lead SNPs and genomic loci.} Lead SNPs were determined by implementing the clumping function of PLINK2\cite{purcell2007plink} to cluster SNPs that were associated significantly. For the IVPs, the parameters were configured as follows: $\texttt{--clump-p1}=2.75\times 10^{-11}$, $\texttt{--clump-p2}=1.16\times 10^{-8}$, $\texttt{--clump-kb}=300$, and $\texttt{--r2} = 0.25$. In parallel, the HDP parameters were established with $\texttt{--clump-p1}=1.05\times10^{-9}$,  $\texttt{--clump-p2}=1.16\times 10^{-8}$, $\texttt{--clump-kb}=300$, and $\texttt{--r2}=0.25$. The minP file was utilized as the input file for both analytical approaches. The region spanning 300 kb on each side of the lead SNP was defined as a locus. The parameter configurations were derived from prior research studies\cite{yang2014combining,yang2015genome,guo2018genome}.

\textbf{Identification of candidate genes.} The coordinates of the lead SNPs were converted from MSU6.1 to MSU7 using the convtool provided in RiceVarMap v2.0 (\url{http://ricevarmap.ncpgr.cn/})\cite{zhao2021inferred}. All potential candidate genes located within 100-kb flanking regions of the lead SNP were identified systematically. The querying procedure was performed using PyRice, a unified programming API that enabled simultaneous access to all supported databases with standardized output formats\cite{do2021pyrice}. Subsequently, manual validation was conducted using the Rice Annotation Project Database (\url{https://rapdb.dna.affrc.go.jp/index.html}) for quality assurance.

\subsubsection{Swine}
\textbf{Phenotypes.} Due to the limited dataset size, the encoder for this species was trained using transfer learning with Swin Transformer that was pre-trained on the ImageNet dataset (\seqsplit{swin\_large\_patch4\_window12\_384\_22k})\cite{liu2021swin}. The $W+$ space was decomposed into 5,880 principal components through PCA. Consequently, 5,880 IVPs were included in the GWAS analysis. For comparison, four HDPs were extracted, specifically: total fat content, average fat content, body length based on image and body height based on image.

\textbf{GWAS.} GWAS analysis was performed using the MLM method in rMVP with the top three genetic principal components were included as covariates. To maintain the false discovery rate $<$ 5\%, Bonferroni correction was applied by dividing the significance level by both the number of markers and the number of phenotypes. The significance threshold for the IVPs was calculated as $\frac{0.05}{17,810,683 \times 5,880} = 4.77 \times 10^{-13}$ and for HDPs as $\frac{0.05}{17,810,683 \times 4} = 7.02 \times 10^{-10}$.

\textbf{Identification of lead SNPs and genomic loci.} The clump function of PLINK2 was utilized to identify lead SNP by applying the specific parameters ($\texttt{--clump-p1}=4.77\times 10^{-13}$, $\texttt{--clump-p2}=2.81\times 10^{-9}$, $\texttt{--clump-kb}=500$, and $\texttt{--r2} = 0.1$), and the minP file served as the input for analysis. A genomic locus was defined as a 500 kb region spanning both sides of the lead SNP.

\textbf{Identification of candidate genes.} Genes located within 500 kb upstream and downstream of the lead SNP were considered potential candidate genes. The entire query process was conducted on the IAnimal database (\url{https://ianimal.pro/index})\cite{fu2023ianimal}.
\backmatter

\bmhead{Correspondence and requests for materials}
Correspondence and requests for materials should be addressed to Shuhong Zhao.

\section*{Declarations}

\subsection*{Funding}
This work was supported by the National Natural Science Foundation of China [32221005, 32494801], the National Key 
Research and Development Program of China [2021YFD1300800], the Fundamental Research Funds for the Central Universities [2662023PY008], the earmarked fund for CARS [CARS-35], and Agricultural Science and Technology Major Project.

\subsection*{Conflict of interest}
The authors declare no competing interests.

\subsection*{Consent for publication}
Not applicable

\subsection*{Data availability}
The data will be released publicly upon the article's publication.

\subsection*{Materials availability}
Not applicable

\subsection*{Code availability}
The code will be released publicly upon the article's publication.

\subsection*{Author contribution}
S.H.Z., X.L.L., X.Y.L., L.F., and X.X. conceived the study; X.L.L. and X.X. designed the experiments; Y.F.L. performed the data collection and analyses under the assistance and guidance from Y.L., M.L.Q., Z.H., S.L.Z., L.L.Y, Y.H.F., J.Y.X., D.Y., X.H., Y.Q., T.F.J., W.N.Y., X.L.L., and S.H.Z.; X.X. developed and evaluated the model under the assistance and guidance from H.H.Z., X.L. and X.H.Y., and X.Y.L.; Y.F.L. and X.X. drafted the paper, X.L.L., L.F., X.Y.L., and S.H.Z. modified the paper; all authors reviewed and approved the final manuscript.

\bibliography{aipheno}


\begin{thebibliography}{153}
\ifx \bisbn   \undefined \def \bisbn  #1{ISBN #1}\fi
\ifx \binits  \undefined \def \binits#1{#1}\fi
\ifx \bauthor  \undefined \def \bauthor#1{#1}\fi
\ifx \batitle  \undefined \def \batitle#1{#1}\fi
\ifx \bjtitle  \undefined \def \bjtitle#1{#1}\fi
\ifx \bvolume  \undefined \def \bvolume#1{\textbf{#1}}\fi
\ifx \byear  \undefined \def \byear#1{#1}\fi
\ifx \bissue  \undefined \def \bissue#1{#1}\fi
\ifx \bfpage  \undefined \def \bfpage#1{#1}\fi
\ifx \blpage  \undefined \def \blpage #1{#1}\fi
\ifx \burl  \undefined \def \burl#1{\textsf{#1}}\fi
\ifx \doiurl  \undefined \def \doiurl#1{\url{https://doi.org/#1}}\fi
\ifx \betal  \undefined \def \betal{\textit{et al.}}\fi
\ifx \binstitute  \undefined \def \binstitute#1{#1}\fi
\ifx \binstitutionaled  \undefined \def \binstitutionaled#1{#1}\fi
\ifx \bctitle  \undefined \def \bctitle#1{#1}\fi
\ifx \beditor  \undefined \def \beditor#1{#1}\fi
\ifx \bpublisher  \undefined \def \bpublisher#1{#1}\fi
\ifx \bbtitle  \undefined \def \bbtitle#1{#1}\fi
\ifx \bedition  \undefined \def \bedition#1{#1}\fi
\ifx \bseriesno  \undefined \def \bseriesno#1{#1}\fi
\ifx \blocation  \undefined \def \blocation#1{#1}\fi
\ifx \bsertitle  \undefined \def \bsertitle#1{#1}\fi
\ifx \bsnm \undefined \def \bsnm#1{#1}\fi
\ifx \bsuffix \undefined \def \bsuffix#1{#1}\fi
\ifx \bparticle \undefined \def \bparticle#1{#1}\fi
\ifx \barticle \undefined \def \barticle#1{#1}\fi
\bibcommenthead
\ifx \bconfdate \undefined \def \bconfdate #1{#1}\fi
\ifx \botherref \undefined \def \botherref #1{#1}\fi
\ifx \url \undefined \def \url#1{\textsf{#1}}\fi
\ifx \bchapter \undefined \def \bchapter#1{#1}\fi
\ifx \bbook \undefined \def \bbook#1{#1}\fi
\ifx \bcomment \undefined \def \bcomment#1{#1}\fi
\ifx \oauthor \undefined \def \oauthor#1{#1}\fi
\ifx \citeauthoryear \undefined \def \citeauthoryear#1{#1}\fi
\ifx \endbibitem  \undefined \def \endbibitem {}\fi
\ifx \bconflocation  \undefined \def \bconflocation#1{#1}\fi
\ifx \arxivurl  \undefined \def \arxivurl#1{\textsf{#1}}\fi
\csname PreBibitemsHook\endcsname

\bibitem[\protect\citeauthoryear{Consortium et~al.}{2025}]{uk2025whole}
\begin{barticle}
\bauthor{\bsnm{Consortium}, \binits{U.B.W.-G.S.}}, \betal:
\batitle{Whole-genome sequencing of 490,640 uk biobank participants}.
\bjtitle{Nature}
\bvolume{645}(\bissue{8081}),
\bfpage{692}
(\byear{2025})
\doiurl{10.1038/s41586-025-09272-9}
\end{barticle}
\endbibitem

\bibitem[\protect\citeauthoryear{Krizhevsky et~al.}{2012}]{krizhevsky2012imagenet}
\begin{botherref}
\oauthor{\bsnm{Krizhevsky}, \binits{A.}},
\oauthor{\bsnm{Sutskever}, \binits{I.}},
\oauthor{\bsnm{Hinton}, \binits{G.E.}}:
Imagenet classification with deep convolutional neural networks.
Adv. Neural Inf. Process. Syst.
\textbf{25}
(2012)
\doiurl{10.1145/3065386}
\end{botherref}
\endbibitem

\bibitem[\protect\citeauthoryear{Bengio et~al.}{2013}]{bengio2013representation}
\begin{barticle}
\bauthor{\bsnm{Bengio}, \binits{Y.}},
\bauthor{\bsnm{Courville}, \binits{A.}},
\bauthor{\bsnm{Vincent}, \binits{P.}}:
\batitle{Representation learning: a review and new perspectives}.
\bjtitle{IEEE Trans. Pattern Anal. Mach. Intell.}
\bvolume{35}(\bissue{8}),
\bfpage{1798}--\blpage{1828}
(\byear{2013})
\doiurl{10.1109/tpami.2013.50}
\end{barticle}
\endbibitem

\bibitem[\protect\citeauthoryear{Chen et~al.}{2020a}]{chen2020improved}
\begin{barticle}
\bauthor{\bsnm{Chen}, \binits{X.}},
\bauthor{\bsnm{Fan}, \binits{H.}},
\bauthor{\bsnm{Girshick}, \binits{R.}},
\bauthor{\bsnm{He}, \binits{K.}}:
\batitle{Improved baselines with momentum contrastive learning}.
\bjtitle{Preprint at arXiv}
(\byear{2020})
\doiurl{10.48550/arXiv.2003.04297}
\end{barticle}
\endbibitem

\bibitem[\protect\citeauthoryear{Chen et~al.}{2020b}]{chen2020simple}
\begin{bchapter}
\bauthor{\bsnm{Chen}, \binits{T.}},
\bauthor{\bsnm{Kornblith}, \binits{S.}},
\bauthor{\bsnm{Norouzi}, \binits{M.}},
\bauthor{\bsnm{Hinton}, \binits{G.}}:
\bctitle{A simple framework for contrastive learning of visual representations}.
In: \bbtitle{Proceedings of the 37th International Conference on Machine Learning (ICML)}.
\bsertitle{Proceedings of Machine Learning Research},
vol. \bseriesno{119},
pp. \bfpage{1597}--\blpage{1607}.
\bpublisher{PMLR},
\blocation{Virtual}
(\byear{2020}).
\doiurl{10.5555/3524938.3525087} .
\bcomment{International Machine Learning Society}
\end{bchapter}
\endbibitem

\bibitem[\protect\citeauthoryear{He et~al.}{2022}]{he2022masked}
\begin{bchapter}
\bauthor{\bsnm{He}, \binits{K.}},
\bauthor{\bsnm{Chen}, \binits{X.}},
\bauthor{\bsnm{Xie}, \binits{S.}},
\bauthor{\bsnm{Li}, \binits{Y.}},
\bauthor{\bsnm{Doll{\'a}r}, \binits{P.}},
\bauthor{\bsnm{Girshick}, \binits{R.}}:
\bctitle{Masked autoencoders are scalable vision learners}.
In: \bbtitle{Proceedings of the IEEE/CVF Conference on Computer Vision and Pattern Recognition (CVPR)},
pp. \bfpage{16000}--\blpage{16009}.
\bpublisher{IEEE Computer Society},
\blocation{New Orleans, LA, USA}
(\byear{2022}).
\doiurl{10.1109/CVPR52688.2022.01553} .
\bcomment{IEEE/CVF}
\end{bchapter}
\endbibitem

\bibitem[\protect\citeauthoryear{Klibaite et~al.}{2025}]{klibaite2025mapping}
\begin{barticle}
\bauthor{\bsnm{Klibaite}, \binits{U.}},
\bauthor{\bsnm{Li}, \binits{T.}},
\bauthor{\bsnm{Aldarondo}, \binits{D.}},
\bauthor{\bsnm{Akoad}, \binits{J.F.}},
\bauthor{\bsnm{{\"O}lveczky}, \binits{B.P.}},
\bauthor{\bsnm{Dunn}, \binits{T.W.}}:
\batitle{Mapping the landscape of social behavior}.
\bjtitle{Cell}
\bvolume{188}(\bissue{8}),
\bfpage{2249}--\blpage{2266}
(\byear{2025})
\doiurl{10.1016/j.cell.2025.01.044}
\end{barticle}
\endbibitem

\bibitem[\protect\citeauthoryear{Saunders et~al.}{2025}]{saunders2025perturb}
\begin{barticle}
\bauthor{\bsnm{Saunders}, \binits{R.A.}},
\bauthor{\bsnm{Allen}, \binits{W.E.}},
\bauthor{\bsnm{Pan}, \binits{X.}},
\bauthor{\bsnm{Sandhu}, \binits{J.}},
\bauthor{\bsnm{Lu}, \binits{J.}},
\bauthor{\bsnm{Lau}, \binits{T.K.}},
\bauthor{\bsnm{Smolyar}, \binits{K.}},
\bauthor{\bsnm{Sullivan}, \binits{Z.A.}},
\bauthor{\bsnm{Dulac}, \binits{C.}},
\bauthor{\bsnm{Weissman}, \binits{J.S.}}, \betal:
\batitle{Perturb-multimodal: A platform for pooled genetic screens with imaging and sequencing in intact mammalian tissue}.
\bjtitle{Cell}
(\byear{2025})
\doiurl{10.1016/j.cell.2025.05.022}
\end{barticle}
\endbibitem

\bibitem[\protect\citeauthoryear{Yun et~al.}{2024}]{yun2024unsupervised}
\begin{barticle}
\bauthor{\bsnm{Yun}, \binits{T.}},
\bauthor{\bsnm{Cosentino}, \binits{J.}},
\bauthor{\bsnm{Behsaz}, \binits{B.}},
\bauthor{\bsnm{McCaw}, \binits{Z.R.}},
\bauthor{\bsnm{Hill}, \binits{D.}},
\bauthor{\bsnm{Luben}, \binits{R.}},
\bauthor{\bsnm{Lai}, \binits{D.}},
\bauthor{\bsnm{Bates}, \binits{J.}},
\bauthor{\bsnm{Yang}, \binits{H.}},
\bauthor{\bsnm{Schwantes-An}, \binits{T.-H.}}, \betal:
\batitle{Unsupervised representation learning on high-dimensional clinical data improves genomic discovery and prediction}.
\bjtitle{Nat. Genet.}
\bvolume{56}(\bissue{8}),
\bfpage{1604}--\blpage{1613}
(\byear{2024})
\doiurl{10.1038/s41588-024-01831-6}
\end{barticle}
\endbibitem

\bibitem[\protect\citeauthoryear{Kirchler et~al.}{2022}]{kirchler2022transfergwas}
\begin{barticle}
\bauthor{\bsnm{Kirchler}, \binits{M.}},
\bauthor{\bsnm{Konigorski}, \binits{S.}},
\bauthor{\bsnm{Norden}, \binits{M.}},
\bauthor{\bsnm{Meltendorf}, \binits{C.}},
\bauthor{\bsnm{Kloft}, \binits{M.}},
\bauthor{\bsnm{Schurmann}, \binits{C.}},
\bauthor{\bsnm{Lippert}, \binits{C.}}:
\batitle{transfergwas: Gwas of images using deep transfer learning}.
\bjtitle{Bioinformatics}
\bvolume{38}(\bissue{14}),
\bfpage{3621}--\blpage{3628}
(\byear{2022})
\doiurl{10.1093/bioinformatics/btac369}
\end{barticle}
\endbibitem

\bibitem[\protect\citeauthoryear{Xie et~al.}{2024}]{xie2024igwas}
\begin{barticle}
\bauthor{\bsnm{Xie}, \binits{Z.}},
\bauthor{\bsnm{Zhang}, \binits{T.}},
\bauthor{\bsnm{Kim}, \binits{S.}},
\bauthor{\bsnm{Lu}, \binits{J.}},
\bauthor{\bsnm{Zhang}, \binits{W.}},
\bauthor{\bsnm{Lin}, \binits{C.-H.}},
\bauthor{\bsnm{Wu}, \binits{M.-R.}},
\bauthor{\bsnm{Davis}, \binits{A.}},
\bauthor{\bsnm{Channa}, \binits{R.}},
\bauthor{\bsnm{Giancardo}, \binits{L.}}, \betal:
\batitle{igwas: Image-based genome-wide association of self-supervised deep phenotyping of retina fundus images}.
\bjtitle{PLoS Genet.}
\bvolume{20}(\bissue{5}),
\bfpage{1011273}
(\byear{2024})
\doiurl{10.1371/journal.pgen.1011273}
\end{barticle}
\endbibitem

\bibitem[\protect\citeauthoryear{Bonazzola et~al.}{2024}]{bonazzola2024unsupervised}
\begin{barticle}
\bauthor{\bsnm{Bonazzola}, \binits{R.}},
\bauthor{\bsnm{Ferrante}, \binits{E.}},
\bauthor{\bsnm{Ravikumar}, \binits{N.}},
\bauthor{\bsnm{Xia}, \binits{Y.}},
\bauthor{\bsnm{Keavney}, \binits{B.}},
\bauthor{\bsnm{Plein}, \binits{S.}},
\bauthor{\bsnm{Syeda-Mahmood}, \binits{T.}},
\bauthor{\bsnm{Frangi}, \binits{A.F.}}:
\batitle{Unsupervised ensemble-based phenotyping enhances discoverability of genes related to left-ventricular morphology}.
\bjtitle{Nat. Mach. Intell.}
\bvolume{6}(\bissue{3}),
\bfpage{291}--\blpage{306}
(\byear{2024})
\doiurl{10.1038/s42256-024-00801-1}
\end{barticle}
\endbibitem

\bibitem[\protect\citeauthoryear{Patel et~al.}{2024}]{patel2024unsupervised}
\begin{barticle}
\bauthor{\bsnm{Patel}, \binits{K.}},
\bauthor{\bsnm{Xie}, \binits{Z.}},
\bauthor{\bsnm{Yuan}, \binits{H.}},
\bauthor{\bsnm{Islam}, \binits{S.M.S.}},
\bauthor{\bsnm{Xie}, \binits{Y.}},
\bauthor{\bsnm{He}, \binits{W.}},
\bauthor{\bsnm{Zhang}, \binits{W.}},
\bauthor{\bsnm{Gottlieb}, \binits{A.}},
\bauthor{\bsnm{Chen}, \binits{H.}},
\bauthor{\bsnm{Giancardo}, \binits{L.}}, \betal:
\batitle{Unsupervised deep representation learning enables phenotype discovery for genetic association studies of brain imaging}.
\bjtitle{Commun. Biol.}
\bvolume{7}(\bissue{1}),
\bfpage{414}
(\byear{2024})
\doiurl{10.1038/s42003-024-06096-7}
\end{barticle}
\endbibitem

\bibitem[\protect\citeauthoryear{Liu et~al.}{2025}]{liu2025digital}
\begin{barticle}
\bauthor{\bsnm{Liu}, \binits{J.J.}},
\bauthor{\bsnm{Borsari}, \binits{B.}},
\bauthor{\bsnm{Li}, \binits{Y.}},
\bauthor{\bsnm{Liu}, \binits{S.X.}},
\bauthor{\bsnm{Gao}, \binits{Y.}},
\bauthor{\bsnm{Xin}, \binits{X.}},
\bauthor{\bsnm{Lou}, \binits{S.}},
\bauthor{\bsnm{Jensen}, \binits{M.}},
\bauthor{\bsnm{Garrido-Martin}, \binits{D.}},
\bauthor{\bsnm{Verplaetse}, \binits{T.L.}}, \betal:
\batitle{Digital phenotyping from wearables using ai characterizes psychiatric disorders and identifies genetic associations}.
\bjtitle{Cell}
\bvolume{188}(\bissue{2}),
\bfpage{515}--\blpage{529}
(\byear{2025})
\doiurl{10.1016/j.cell.2024.11.012}
\end{barticle}
\endbibitem

\bibitem[\protect\citeauthoryear{Lee}{1998}]{lee1998independent}
\begin{bchapter}
\bauthor{\bsnm{Lee}, \binits{T.-W.}}:
\bctitle{Independent component analysis}.
In: \beditor{\bsnm{Hyv{\"a}rinen}, \binits{A.}},
\beditor{\bsnm{Karhunen}, \binits{J.}},
\beditor{\bsnm{Oja}, \binits{E.}} (eds.)
\bbtitle{Independent Component Analysis: Theory and Applications},
pp. \bfpage{27}--\blpage{66}.
\bpublisher{Springer},
\blocation{Boston, MA}
(\byear{1998}).
\doiurl{10.1007/978-1-4757-2851-4_2}
\end{bchapter}
\endbibitem

\bibitem[\protect\citeauthoryear{Greenacre et~al.}{2022}]{greenacre2022principal}
\begin{barticle}
\bauthor{\bsnm{Greenacre}, \binits{M.}},
\bauthor{\bsnm{Groenen}, \binits{P.J.F.}},
\bauthor{\bsnm{Hastie}, \binits{T.}},
\bauthor{\bsnm{d’Enza}, \binits{A.I.}},
\bauthor{\bsnm{Markos}, \binits{A.}},
\bauthor{\bsnm{Tuzhilina}, \binits{E.}}:
\batitle{Principal component analysis}.
\bjtitle{Nat. Rev. Methods Primers}
\bvolume{2}(\bissue{1}),
\bfpage{100}
(\byear{2022})
\doiurl{10.1038/s43586-022-00184-w}
\end{barticle}
\endbibitem

\bibitem[\protect\citeauthoryear{H{\"a}rk{\"o}nen et~al.}{2020}]{harkonen2020ganspace}
\begin{barticle}
\bauthor{\bsnm{H{\"a}rk{\"o}nen}, \binits{E.}},
\bauthor{\bsnm{Hertzmann}, \binits{A.}},
\bauthor{\bsnm{Lehtinen}, \binits{J.}},
\bauthor{\bsnm{Paris}, \binits{S.}}:
\batitle{Ganspace: Discovering interpretable gan controls}.
\bjtitle{Adv. Neural Inf. Process. Syst.}
\bvolume{33},
\bfpage{9841}--\blpage{9850}
(\byear{2020})
\doiurl{10.48550/arXiv.2004.02546}
\end{barticle}
\endbibitem

\bibitem[\protect\citeauthoryear{Zhang et~al.}{2010}]{zhang2010mixed}
\begin{barticle}
\bauthor{\bsnm{Zhang}, \binits{Z.}},
\bauthor{\bsnm{Ersoz}, \binits{E.}},
\bauthor{\bsnm{Lai}, \binits{C.-Q.}},
\bauthor{\bsnm{Todhunter}, \binits{R.J.}},
\bauthor{\bsnm{Tiwari}, \binits{H.K.}},
\bauthor{\bsnm{Gore}, \binits{M.A.}},
\bauthor{\bsnm{Bradbury}, \binits{P.J.}},
\bauthor{\bsnm{Yu}, \binits{J.}},
\bauthor{\bsnm{Arnett}, \binits{D.K.}},
\bauthor{\bsnm{Ordovas}, \binits{J.M.}}, \betal:
\batitle{Mixed linear model approach adapted for genome-wide association studies}.
\bjtitle{Nat. Genet.}
\bvolume{42}(\bissue{4}),
\bfpage{355}--\blpage{360}
(\byear{2010})
\doiurl{10.59350/f0mqf-dem95}
\end{barticle}
\endbibitem

\bibitem[\protect\citeauthoryear{Haynes}{2013}]{haynes2013bonferroni}
\begin{bchapter}
\bauthor{\bsnm{Haynes}, \binits{W.}}:
\bctitle{Bonferroni correction}.
In: \beditor{\bsnm{Dubitzky}, \binits{W.}},
\beditor{\bsnm{Wolkenhauer}, \binits{O.}},
\beditor{\bsnm{Cho}, \binits{K.-H.}},
\beditor{\bsnm{Yokota}, \binits{H.}} (eds.)
\bbtitle{Encyclopedia of Systems Biology},
pp. \bfpage{154}--\blpage{154}.
\bpublisher{Springer},
\blocation{New York, NY}
(\byear{2013}).
\doiurl{10.1007/978-1-4419-9863-7_1213}
\end{bchapter}
\endbibitem

\bibitem[\protect\citeauthoryear{Zekavat et~al.}{2022}]{zekavat2022deep}
\begin{barticle}
\bauthor{\bsnm{Zekavat}, \binits{S.M.}},
\bauthor{\bsnm{Raghu}, \binits{V.K.}},
\bauthor{\bsnm{Trinder}, \binits{M.}},
\bauthor{\bsnm{Ye}, \binits{Y.}},
\bauthor{\bsnm{Koyama}, \binits{S.}},
\bauthor{\bsnm{Honigberg}, \binits{M.C.}},
\bauthor{\bsnm{Yu}, \binits{Z.}},
\bauthor{\bsnm{Pampana}, \binits{A.}},
\bauthor{\bsnm{Urbut}, \binits{S.}},
\bauthor{\bsnm{Haidermota}, \binits{S.}}, \betal:
\batitle{Deep learning of the retina enables phenome-and genome-wide analyses of the microvasculature}.
\bjtitle{Circulation}
\bvolume{145}(\bissue{2}),
\bfpage{134}--\blpage{150}
(\byear{2022})
\doiurl{10.1161/circulationaha.121.057709}
\end{barticle}
\endbibitem

\bibitem[\protect\citeauthoryear{Zhao et~al.}{2024}]{zhao2024eye}
\begin{barticle}
\bauthor{\bsnm{Zhao}, \binits{B.}},
\bauthor{\bsnm{Li}, \binits{Y.}},
\bauthor{\bsnm{Fan}, \binits{Z.}},
\bauthor{\bsnm{Wu}, \binits{Z.}},
\bauthor{\bsnm{Shu}, \binits{J.}},
\bauthor{\bsnm{Yang}, \binits{X.}},
\bauthor{\bsnm{Yang}, \binits{Y.}},
\bauthor{\bsnm{Wang}, \binits{X.}},
\bauthor{\bsnm{Li}, \binits{B.}},
\bauthor{\bsnm{Wang}, \binits{X.}}, \betal:
\batitle{Eye-brain connections revealed by multimodal retinal and brain imaging genetics}.
\bjtitle{Nat. Commun.}
\bvolume{15}(\bissue{1}),
\bfpage{6064}
(\byear{2024})
\doiurl{10.1101/2023.02.16.23286035}
\end{barticle}
\endbibitem

\bibitem[\protect\citeauthoryear{Andrade et~al.}{2021}]{andrade2021molecular}
\begin{barticle}
\bauthor{\bsnm{Andrade}, \binits{P.}},
\bauthor{\bsnm{Gazda}, \binits{M.A.}},
\bauthor{\bsnm{Ara{\'u}jo}, \binits{P.M.}},
\bauthor{\bsnm{Afonso}, \binits{S.}},
\bauthor{\bsnm{Rasmussen}, \binits{J.A.}},
\bauthor{\bsnm{Marques}, \binits{C.I.}},
\bauthor{\bsnm{Lopes}, \binits{R.J.}},
\bauthor{\bsnm{Gilbert}, \binits{M.T.P.}},
\bauthor{\bsnm{Carneiro}, \binits{M.}}:
\batitle{Molecular parallelisms between pigmentation in the avian iris and the integument of ectothermic vertebrates}.
\bjtitle{PLoS Genet.}
\bvolume{17}(\bissue{2}),
\bfpage{1009404}
(\byear{2021})
\doiurl{10.1371/journal.pgen.1009404}
\end{barticle}
\endbibitem

\bibitem[\protect\citeauthoryear{Maclary et~al.}{2021}]{maclary2021two}
\begin{barticle}
\bauthor{\bsnm{Maclary}, \binits{E.T.}},
\bauthor{\bsnm{Phillips}, \binits{B.}},
\bauthor{\bsnm{Wauer}, \binits{R.}},
\bauthor{\bsnm{Boer}, \binits{E.F.}},
\bauthor{\bsnm{Bruders}, \binits{R.}},
\bauthor{\bsnm{Gilvarry}, \binits{T.}},
\bauthor{\bsnm{Holt}, \binits{C.}},
\bauthor{\bsnm{Yandell}, \binits{M.}},
\bauthor{\bsnm{Shapiro}, \binits{M.D.}}:
\batitle{Two genomic loci control three eye colors in the domestic pigeon (columba livia)}.
\bjtitle{Mol. Biol. Evol.}
\bvolume{38}(\bissue{12}),
\bfpage{5376}--\blpage{5390}
(\byear{2021})
\doiurl{10.1093/molbev/msab260}
\end{barticle}
\endbibitem

\bibitem[\protect\citeauthoryear{Si et~al.}{2021}]{si2021genetics}
\begin{barticle}
\bauthor{\bsnm{Si}, \binits{S.}},
\bauthor{\bsnm{Xu}, \binits{X.}},
\bauthor{\bsnm{Zhuang}, \binits{Y.}},
\bauthor{\bsnm{Gao}, \binits{X.}},
\bauthor{\bsnm{Zhang}, \binits{H.}},
\bauthor{\bsnm{Zou}, \binits{Z.}},
\bauthor{\bsnm{Luo}, \binits{S.-J.}}:
\batitle{The genetics and evolution of eye color in domestic pigeons (columba livia)}.
\bjtitle{PLoS Genet.}
\bvolume{17}(\bissue{8}),
\bfpage{1009770}
(\byear{2021})
\doiurl{10.1371/journal.pgen.1009770}
\end{barticle}
\endbibitem

\bibitem[\protect\citeauthoryear{Li et~al.}{2020}]{li2020analysis}
\begin{barticle}
\bauthor{\bsnm{Li}, \binits{X.}},
\bauthor{\bsnm{Chen}, \binits{Z.}},
\bauthor{\bsnm{Zhang}, \binits{G.}},
\bauthor{\bsnm{Lu}, \binits{H.}},
\bauthor{\bsnm{Qin}, \binits{P.}},
\bauthor{\bsnm{Qi}, \binits{M.}},
\bauthor{\bsnm{Yu}, \binits{Y.}},
\bauthor{\bsnm{Jiao}, \binits{B.}},
\bauthor{\bsnm{Zhao}, \binits{X.}},
\bauthor{\bsnm{Gao}, \binits{Q.}}, \betal:
\batitle{Analysis of genetic architecture and favorable allele usage of agronomic traits in a large collection of chinese rice accessions}.
\bjtitle{Sci. China Life Sci.}
\bvolume{63},
\bfpage{1688}--\blpage{1702}
(\byear{2020})
\doiurl{10.1007/s11427-019-1682-6}
\end{barticle}
\endbibitem

\bibitem[\protect\citeauthoryear{Sandhu et~al.}{2019}]{sandhu2019deciphering}
\begin{barticle}
\bauthor{\bsnm{Sandhu}, \binits{N.}},
\bauthor{\bsnm{Subedi}, \binits{S.R.}},
\bauthor{\bsnm{Singh}, \binits{V.K.}},
\bauthor{\bsnm{Sinha}, \binits{P.}},
\bauthor{\bsnm{Kumar}, \binits{S.}},
\bauthor{\bsnm{Singh}, \binits{S.P.}},
\bauthor{\bsnm{Ghimire}, \binits{S.K.}},
\bauthor{\bsnm{Pandey}, \binits{M.}},
\bauthor{\bsnm{Yadaw}, \binits{R.B.}},
\bauthor{\bsnm{Varshney}, \binits{R.K.}}, \betal:
\batitle{Deciphering the genetic basis of root morphology, nutrient uptake, yield, and yield-related traits in rice under dry direct-seeded cultivation systems}.
\bjtitle{Sci. Rep.}
\bvolume{9}(\bissue{1}),
\bfpage{9334}
(\byear{2019})
\doiurl{10.1038/s41598-019-45770-3}
\end{barticle}
\endbibitem

\bibitem[\protect\citeauthoryear{Lv et~al.}{2021}]{lv2021loci}
\begin{barticle}
\bauthor{\bsnm{Lv}, \binits{Y.}},
\bauthor{\bsnm{Ma}, \binits{J.}},
\bauthor{\bsnm{Wang}, \binits{Y.}},
\bauthor{\bsnm{Wang}, \binits{Q.}},
\bauthor{\bsnm{Lu}, \binits{X.}},
\bauthor{\bsnm{Hu}, \binits{H.}},
\bauthor{\bsnm{Qian}, \binits{Q.}},
\bauthor{\bsnm{Guo}, \binits{L.}},
\bauthor{\bsnm{Shang}, \binits{L.}}:
\batitle{Loci and natural alleles for low-nitrogen-induced growth response revealed by the genome-wide association study analysis in rice (oryza sativa l.)}.
\bjtitle{Front. Plant Sci.}
\bvolume{12},
\bfpage{770736}
(\byear{2021})
\doiurl{10.3389/fpls.2021.770736}
\end{barticle}
\endbibitem

\bibitem[\protect\citeauthoryear{Koumoto et~al.}{2013}]{koumoto2013rice}
\begin{barticle}
\bauthor{\bsnm{Koumoto}, \binits{T.}},
\bauthor{\bsnm{Shimada}, \binits{H.}},
\bauthor{\bsnm{Kusano}, \binits{H.}},
\bauthor{\bsnm{She}, \binits{K.-C.}},
\bauthor{\bsnm{Iwamoto}, \binits{M.}},
\bauthor{\bsnm{Takano}, \binits{M.}}:
\batitle{Rice monoculm mutation moc2, which inhibits outgrowth of the second tillers, is ascribed to lack of a fructose-1, 6-bisphosphatase}.
\bjtitle{Plant Biotechnol.}
\bvolume{30}(\bissue{1}),
\bfpage{47}--\blpage{56}
(\byear{2013})
\doiurl{10.5511/plantbiotechnology.12.1210a}
\end{barticle}
\endbibitem

\bibitem[\protect\citeauthoryear{Villaplana-Velasco et~al.}{2023}]{villaplana2023fine}
\begin{barticle}
\bauthor{\bsnm{Villaplana-Velasco}, \binits{A.}},
\bauthor{\bsnm{Pigeyre}, \binits{M.}},
\bauthor{\bsnm{Engelmann}, \binits{J.}},
\bauthor{\bsnm{Rawlik}, \binits{K.}},
\bauthor{\bsnm{Canela-Xandri}, \binits{O.}},
\bauthor{\bsnm{Tochel}, \binits{C.}},
\bauthor{\bsnm{Lona-Durazo}, \binits{F.}},
\bauthor{\bsnm{Mookiah}, \binits{M.R.}},
\bauthor{\bsnm{Doney}, \binits{A.}},
\bauthor{\bsnm{Parra}, \binits{E.J.}}, \betal:
\batitle{Fine-mapping of retinal vascular complexity loci identifies notch regulation as a shared mechanism with myocardial infarction outcomes}.
\bjtitle{Commun. Biol.}
\bvolume{6}(\bissue{1}),
\bfpage{523}
(\byear{2023})
\doiurl{10.1038/s42003-023-04836-9}
\end{barticle}
\endbibitem

\bibitem[\protect\citeauthoryear{Ort{\'\i}n~Vela et~al.}{2024}]{ortin2024phenotypic}
\begin{barticle}
\bauthor{\bsnm{Ort{\'\i}n~Vela}, \binits{S.}},
\bauthor{\bsnm{Beyeler}, \binits{M.J.}},
\bauthor{\bsnm{Trofimova}, \binits{O.}},
\bauthor{\bsnm{Iuliani}, \binits{I.}},
\bauthor{\bsnm{Vargas~Quiros}, \binits{J.D.}},
\bauthor{\bsnm{Vries}, \binits{V.A.}},
\bauthor{\bsnm{Meloni}, \binits{I.}},
\bauthor{\bsnm{Elwakil}, \binits{A.}},
\bauthor{\bsnm{Hoogewoud}, \binits{F.}},
\bauthor{\bsnm{Liefers}, \binits{B.}}, \betal:
\batitle{Phenotypic and genetic characteristics of retinal vascular parameters and their association with diseases}.
\bjtitle{Nat. Commun.}
\bvolume{15}(\bissue{1}),
\bfpage{9593}
(\byear{2024})
\doiurl{10.1038/s41467-024-52334-1}
\end{barticle}
\endbibitem

\bibitem[\protect\citeauthoryear{Liu et~al.}{2010}]{liu2010digital}
\begin{barticle}
\bauthor{\bsnm{Liu}, \binits{F.}},
\bauthor{\bsnm{Wollstein}, \binits{A.}},
\bauthor{\bsnm{Hysi}, \binits{P.G.}},
\bauthor{\bsnm{Ankra-Badu}, \binits{G.A.}},
\bauthor{\bsnm{Spector}, \binits{T.D.}},
\bauthor{\bsnm{Park}, \binits{D.}},
\bauthor{\bsnm{Zhu}, \binits{G.}},
\bauthor{\bsnm{Larsson}, \binits{M.}},
\bauthor{\bsnm{Duffy}, \binits{D.L.}},
\bauthor{\bsnm{Montgomery}, \binits{G.W.}}, \betal:
\batitle{Digital quantification of human eye color highlights genetic association of three new loci}.
\bjtitle{PLoS Genet.}
\bvolume{6}(\bissue{5}),
\bfpage{1000934}
(\byear{2010})
\doiurl{10.1371/journal.pgen.1000934}
\end{barticle}
\endbibitem

\bibitem[\protect\citeauthoryear{Sulem et~al.}{2007}]{sulem2007genetic}
\begin{barticle}
\bauthor{\bsnm{Sulem}, \binits{P.}},
\bauthor{\bsnm{Gudbjartsson}, \binits{D.F.}},
\bauthor{\bsnm{Stacey}, \binits{S.N.}},
\bauthor{\bsnm{Helgason}, \binits{A.}},
\bauthor{\bsnm{Rafnar}, \binits{T.}},
\bauthor{\bsnm{Magnusson}, \binits{K.P.}},
\bauthor{\bsnm{Manolescu}, \binits{A.}},
\bauthor{\bsnm{Karason}, \binits{A.}},
\bauthor{\bsnm{Palsson}, \binits{A.}},
\bauthor{\bsnm{Thorleifsson}, \binits{G.}}, \betal:
\batitle{Genetic determinants of hair, eye and skin pigmentation in europeans}.
\bjtitle{Nat. Genet.}
\bvolume{39}(\bissue{12}),
\bfpage{1443}--\blpage{1452}
(\byear{2007})
\doiurl{10.1038/ng.2007.13}
\end{barticle}
\endbibitem

\bibitem[\protect\citeauthoryear{Adhikari et~al.}{2019}]{adhikari2019gwas}
\begin{barticle}
\bauthor{\bsnm{Adhikari}, \binits{K.}},
\bauthor{\bsnm{Mendoza-Revilla}, \binits{J.}},
\bauthor{\bsnm{Sohail}, \binits{A.}},
\bauthor{\bsnm{Fuentes-Guajardo}, \binits{M.}},
\bauthor{\bsnm{Lampert}, \binits{J.}},
\bauthor{\bsnm{Chac{\'o}n-Duque}, \binits{J.C.}},
\bauthor{\bsnm{Hurtado}, \binits{M.}},
\bauthor{\bsnm{Villegas}, \binits{V.}},
\bauthor{\bsnm{Granja}, \binits{V.}},
\bauthor{\bsnm{Acu{\~n}a-Alonzo}, \binits{V.}}, \betal:
\batitle{A gwas in latin americans highlights the convergent evolution of lighter skin pigmentation in eurasia}.
\bjtitle{Nat. Commun.}
\bvolume{10}(\bissue{1}),
\bfpage{358}
(\byear{2019})
\doiurl{10.1038/s41467-018-08147-0}
\end{barticle}
\endbibitem

\bibitem[\protect\citeauthoryear{Tedja et~al.}{2018}]{tedja2018genome}
\begin{barticle}
\bauthor{\bsnm{Tedja}, \binits{M.S.}},
\bauthor{\bsnm{Wojciechowski}, \binits{R.}},
\bauthor{\bsnm{Hysi}, \binits{P.G.}},
\bauthor{\bsnm{Eriksson}, \binits{N.}},
\bauthor{\bsnm{Furlotte}, \binits{N.A.}},
\bauthor{\bsnm{Verhoeven}, \binits{V.J.M.}},
\bauthor{\bsnm{Iglesias}, \binits{A.I.}},
\bauthor{\bsnm{Meester-Smoor}, \binits{M.A.}},
\bauthor{\bsnm{Tompson}, \binits{S.W.}},
\bauthor{\bsnm{Fan}, \binits{Q.}}, \betal:
\batitle{Genome-wide association meta-analysis highlights light-induced signaling as a driver for refractive error}.
\bjtitle{Nat. Genet.}
\bvolume{50}(\bissue{6}),
\bfpage{834}--\blpage{848}
(\byear{2018})
\doiurl{10.1038/s41588-018-0127-7}
\end{barticle}
\endbibitem

\bibitem[\protect\citeauthoryear{Ritchey et~al.}{2012}]{ritchey2012vision}
\begin{barticle}
\bauthor{\bsnm{Ritchey}, \binits{E.R.}},
\bauthor{\bsnm{Zelinka}, \binits{C.}},
\bauthor{\bsnm{Tang}, \binits{J.}},
\bauthor{\bsnm{Liu}, \binits{J.}},
\bauthor{\bsnm{Code}, \binits{K.A.}},
\bauthor{\bsnm{Petersen-Jones}, \binits{S.}},
\bauthor{\bsnm{Fischer}, \binits{A.J.}}:
\batitle{Vision-guided ocular growth in a mutant chicken model with diminished visual acuity}.
\bjtitle{Exp. Eye Res.}
\bvolume{102},
\bfpage{59}--\blpage{69}
(\byear{2012})
\doiurl{10.1016/j.exer.2012.07.001}
\end{barticle}
\endbibitem

\bibitem[\protect\citeauthoryear{Jiang et~al.}{2023}]{jiang2023gwas}
\begin{barticle}
\bauthor{\bsnm{Jiang}, \binits{X.}},
\bauthor{\bsnm{Hysi}, \binits{P.G.}},
\bauthor{\bsnm{Khawaja}, \binits{A.P.}},
\bauthor{\bsnm{Mahroo}, \binits{O.A.}},
\bauthor{\bsnm{Xu}, \binits{Z.}},
\bauthor{\bsnm{Hammond}, \binits{C.J.}},
\bauthor{\bsnm{Foster}, \binits{P.J.}},
\bauthor{\bsnm{Welikala}, \binits{R.A.}},
\bauthor{\bsnm{Barman}, \binits{S.A.}},
\bauthor{\bsnm{Whincup}, \binits{P.H.}}, \betal:
\batitle{Gwas on retinal vasculometry phenotypes}.
\bjtitle{PLoS Genet.}
\bvolume{19}(\bissue{2}),
\bfpage{1010583}
(\byear{2023})
\doiurl{10.1371/journal.pgen.1010583}
\end{barticle}
\endbibitem

\bibitem[\protect\citeauthoryear{Gao et~al.}{2019}]{gao2019genome}
\begin{barticle}
\bauthor{\bsnm{Gao}, \binits{X.R.}},
\bauthor{\bsnm{Huang}, \binits{H.}},
\bauthor{\bsnm{Kim}, \binits{H.}}:
\batitle{Genome-wide association analyses identify 139 loci associated with macular thickness in the uk biobank cohort}.
\bjtitle{Hum. Mol. Genet.}
\bvolume{28}(\bissue{7}),
\bfpage{1162}--\blpage{1172}
(\byear{2019})
\doiurl{10.1093/hmg/ddy422}
\end{barticle}
\endbibitem

\bibitem[\protect\citeauthoryear{Currant et~al.}{2021}]{currant2021genetic}
\begin{barticle}
\bauthor{\bsnm{Currant}, \binits{H.}},
\bauthor{\bsnm{Hysi}, \binits{P.}},
\bauthor{\bsnm{Fitzgerald}, \binits{T.W.}},
\bauthor{\bsnm{Gharahkhani}, \binits{P.}},
\bauthor{\bsnm{Bonnemaijer}, \binits{P.W.M.}},
\bauthor{\bsnm{Senabouth}, \binits{A.}},
\bauthor{\bsnm{Hewitt}, \binits{A.W.}},
\bauthor{\bsnm{Eye}, \binits{U.B.}},
\bauthor{\bsnm{Consortium}, \binits{V.}},
\bauthor{\bsnm{Consortium}, \binits{I.G.G.}},
\bauthor{\bsnm{Atan}, \binits{D.}}, \betal:
\batitle{Genetic variation affects morphological retinal phenotypes extracted from uk biobank optical coherence tomography images}.
\bjtitle{PLoS Genet.}
\bvolume{17}(\bissue{5}),
\bfpage{1009497}
(\byear{2021})
\doiurl{10.1371/journal.pgen.1009497}
\end{barticle}
\endbibitem

\bibitem[\protect\citeauthoryear{Jensen et~al.}{2016}]{jensen2016novel}
\begin{barticle}
\bauthor{\bsnm{Jensen}, \binits{R.A.}},
\bauthor{\bsnm{Sim}, \binits{X.}},
\bauthor{\bsnm{Smith}, \binits{A.V.}},
\bauthor{\bsnm{Li}, \binits{X.}},
\bauthor{\bsnm{Jakobsd{\'o}ttir}, \binits{J.}},
\bauthor{\bsnm{Cheng}, \binits{C.-Y.}},
\bauthor{\bsnm{Brody}, \binits{J.A.}},
\bauthor{\bsnm{Cotch}, \binits{M.F.}},
\bauthor{\bsnm{Mcknight}, \binits{B.}},
\bauthor{\bsnm{Klein}, \binits{R.}}, \betal:
\batitle{Novel genetic loci associated with retinal microvascular diameter}.
\bjtitle{Circ. Cardiovasc. Genet.}
\bvolume{9}(\bissue{1}),
\bfpage{45}--\blpage{54}
(\byear{2016})
\doiurl{10.1161/CIRCGENETICS.115.001142}
\end{barticle}
\endbibitem

\bibitem[\protect\citeauthoryear{Hysi et~al.}{2020}]{hysi2020meta}
\begin{barticle}
\bauthor{\bsnm{Hysi}, \binits{P.G.}},
\bauthor{\bsnm{Choquet}, \binits{H.}},
\bauthor{\bsnm{Khawaja}, \binits{A.P.}},
\bauthor{\bsnm{Wojciechowski}, \binits{R.}},
\bauthor{\bsnm{Tedja}, \binits{M.S.}},
\bauthor{\bsnm{Yin}, \binits{J.}},
\bauthor{\bsnm{Simcoe}, \binits{M.J.}},
\bauthor{\bsnm{Patasova}, \binits{K.}},
\bauthor{\bsnm{Mahroo}, \binits{O.A.}},
\bauthor{\bsnm{Thai}, \binits{K.K.}}, \betal:
\batitle{Meta-analysis of 542,934 subjects of european ancestry identifies new genes and mechanisms predisposing to refractive error and myopia}.
\bjtitle{Nat. Genet.}
\bvolume{52}(\bissue{4}),
\bfpage{401}--\blpage{407}
(\byear{2020})
\doiurl{10.1038/s41588-020-0599-0}
\end{barticle}
\endbibitem

\bibitem[\protect\citeauthoryear{Xue et~al.}{2022}]{xue2022genome}
\begin{botherref}
\oauthor{\bsnm{Xue}, \binits{Z.}},
\oauthor{\bsnm{Yuan}, \binits{J.}},
\oauthor{\bsnm{Chen}, \binits{F.}},
\oauthor{\bsnm{Yao}, \binits{Y.}},
\oauthor{\bsnm{Xing}, \binits{S.}},
\oauthor{\bsnm{Yu}, \binits{X.}},
\oauthor{\bsnm{Li}, \binits{K.}},
\oauthor{\bsnm{Wang}, \binits{C.}},
\oauthor{\bsnm{Bao}, \binits{J.}},
\oauthor{\bsnm{Qu}, \binits{J.}}, et al.:
Genome-wide association meta-analysis of 88,250 individuals highlights pleiotropic mechanisms of five ocular diseases in uk biobank.
EBioMedicine
\textbf{82}
(2022)
\doiurl{10.1016/j.ebiom.2022.104161}
\end{botherref}
\endbibitem

\bibitem[\protect\citeauthoryear{Han et~al.}{2023}]{han2023large}
\begin{barticle}
\bauthor{\bsnm{Han}, \binits{X.}},
\bauthor{\bsnm{Gharahkhani}, \binits{P.}},
\bauthor{\bsnm{Hamel}, \binits{A.R.}},
\bauthor{\bsnm{Ong}, \binits{J.S.}},
\bauthor{\bsnm{Renter{\'\i}a}, \binits{M.E.}},
\bauthor{\bsnm{Mehta}, \binits{P.}},
\bauthor{\bsnm{Dong}, \binits{X.}},
\bauthor{\bsnm{Pasutto}, \binits{F.}},
\bauthor{\bsnm{Hammond}, \binits{C.}},
\bauthor{\bsnm{Young}, \binits{T.L.}}, \betal:
\batitle{Large-scale multitrait genome-wide association analyses identify hundreds of glaucoma risk loci}.
\bjtitle{Nat. Genet.}
\bvolume{55}(\bissue{7}),
\bfpage{1116}--\blpage{1125}
(\byear{2023})
\doiurl{10.1038/s41588-023-01428-5}
\end{barticle}
\endbibitem

\bibitem[\protect\citeauthoryear{Cho et~al.}{2016}]{cho2016complications}
\begin{barticle}
\bauthor{\bsnm{Cho}, \binits{B.-J.}},
\bauthor{\bsnm{Shin}, \binits{J.Y.}},
\bauthor{\bsnm{Yu}, \binits{H.G.}}:
\batitle{Complications of pathologic myopia}.
\bjtitle{Eye Contact Lens}
\bvolume{42}(\bissue{1}),
\bfpage{9}--\blpage{15}
(\byear{2016})
\doiurl{10.1097/ICL.0000000000000223}
\end{barticle}
\endbibitem

\bibitem[\protect\citeauthoryear{Tan et~al.}{2018}]{tan2018associations}
\begin{barticle}
\bauthor{\bsnm{Tan}, \binits{N.Y.Q.}},
\bauthor{\bsnm{Tham}, \binits{Y.-C.}},
\bauthor{\bsnm{Ding}, \binits{Y.}},
\bauthor{\bsnm{Yasuda}, \binits{M.}},
\bauthor{\bsnm{Sabanayagam}, \binits{C.}},
\bauthor{\bsnm{Saw}, \binits{S.-M.}},
\bauthor{\bsnm{Wang}, \binits{J.J.}},
\bauthor{\bsnm{Mitchell}, \binits{P.}},
\bauthor{\bsnm{Wong}, \binits{T.Y.}},
\bauthor{\bsnm{Cheng}, \binits{C.-Y.}}:
\batitle{Associations of peripapillary atrophy and fundus tessellation with diabetic retinopathy}.
\bjtitle{Ophthalmology Retina}
\bvolume{2}(\bissue{6}),
\bfpage{574}--\blpage{581}
(\byear{2018})
\doiurl{10.1016/j.oret.2017.09.019}
\end{barticle}
\endbibitem

\bibitem[\protect\citeauthoryear{Chen et~al.}{2023}]{chen2023clinical}
\begin{barticle}
\bauthor{\bsnm{Chen}, \binits{X.-Y.}},
\bauthor{\bsnm{He}, \binits{H.-L.}},
\bauthor{\bsnm{Xu}, \binits{J.}},
\bauthor{\bsnm{Liu}, \binits{Y.-X.}},
\bauthor{\bsnm{Jin}, \binits{Z.-B.}}:
\batitle{Clinical features of fundus tessellation and its relationship with myopia: a systematic review and meta-analysis}.
\bjtitle{Ophthalmol. Ther.}
\bvolume{12}(\bissue{6}),
\bfpage{3159}--\blpage{3175}
(\byear{2023})
\doiurl{10.1007/s40123-023-00802-0}
\end{barticle}
\endbibitem

\bibitem[\protect\citeauthoryear{Tian et~al.}{2024}]{tian2024fundus}
\begin{barticle}
\bauthor{\bsnm{Tian}, \binits{J.}},
\bauthor{\bsnm{Wu}, \binits{J.}},
\bauthor{\bsnm{Liu}, \binits{W.}},
\bauthor{\bsnm{Chen}, \binits{K.}},
\bauthor{\bsnm{Zhu}, \binits{S.}},
\bauthor{\bsnm{Lin}, \binits{C.}},
\bauthor{\bsnm{Liu}, \binits{H.}},
\bauthor{\bsnm{Hou}, \binits{S.}},
\bauthor{\bsnm{Huang}, \binits{Z.}},
\bauthor{\bsnm{Zhu}, \binits{Y.}}, \betal:
\batitle{Fundus tessellation and parapapillary atrophy, as ocular characteristics of spontaneously high myopia in macaques: The non-human primates eye study}.
\bjtitle{Transl. Vis. Sci. Technol.}
\bvolume{13}(\bissue{5}),
\bfpage{8}--\blpage{8}
(\byear{2024})
\doiurl{10.1167/tvst.13.5.8}
\end{barticle}
\endbibitem

\bibitem[\protect\citeauthoryear{Thomas et~al.}{2022}]{thomas2022cell}
\begin{barticle}
\bauthor{\bsnm{Thomas}, \binits{E.D.}},
\bauthor{\bsnm{Timms}, \binits{A.E.}},
\bauthor{\bsnm{Giles}, \binits{S.}},
\bauthor{\bsnm{Harkins-Perry}, \binits{S.}},
\bauthor{\bsnm{Lyu}, \binits{P.}},
\bauthor{\bsnm{Hoang}, \binits{T.}},
\bauthor{\bsnm{Qian}, \binits{J.}},
\bauthor{\bsnm{Jackson}, \binits{V.E.}},
\bauthor{\bsnm{Bahlo}, \binits{M.}},
\bauthor{\bsnm{Blackshaw}, \binits{S.}}, \betal:
\batitle{Cell-specific cis-regulatory elements and mechanisms of non-coding genetic disease in human retina and retinal organoids}.
\bjtitle{Dev. Cell}
\bvolume{57}(\bissue{6}),
\bfpage{820}--\blpage{836}
(\byear{2022})
\doiurl{10.1016/j.devcel.2022.02.018}
\end{barticle}
\endbibitem

\bibitem[\protect\citeauthoryear{Jackson et~al.}{2025}]{jackson2025multi}
\begin{barticle}
\bauthor{\bsnm{Jackson}, \binits{V.E.}},
\bauthor{\bsnm{Wu}, \binits{Y.}},
\bauthor{\bsnm{Bonelli}, \binits{R.}},
\bauthor{\bsnm{Owen}, \binits{J.P.}},
\bauthor{\bsnm{Scott}, \binits{L.W.}},
\bauthor{\bsnm{Farashi}, \binits{S.}},
\bauthor{\bsnm{Kihara}, \binits{Y.}},
\bauthor{\bsnm{Gantner}, \binits{M.L.}},
\bauthor{\bsnm{Egan}, \binits{C.}},
\bauthor{\bsnm{Williams}, \binits{K.M.}}, \betal:
\batitle{Multi-omic spatial effects on high-resolution ai-derived retinal thickness}.
\bjtitle{Nat. Commun.}
\bvolume{16}(\bissue{1}),
\bfpage{1317}
(\byear{2025})
\doiurl{10.1038/s41467-024-55635-7}
\end{barticle}
\endbibitem

\bibitem[\protect\citeauthoryear{Han et~al.}{2020}]{han2020genome}
\begin{barticle}
\bauthor{\bsnm{Han}, \binits{X.}},
\bauthor{\bsnm{Gharahkhani}, \binits{P.}},
\bauthor{\bsnm{Mitchell}, \binits{P.}},
\bauthor{\bsnm{Liew}, \binits{G.}},
\bauthor{\bsnm{Hewitt}, \binits{A.W.}},
\bauthor{\bsnm{MacGregor}, \binits{S.}}:
\batitle{Genome-wide meta-analysis identifies novel loci associated with age-related macular degeneration}.
\bjtitle{J. Hum. Genet.}
\bvolume{65}(\bissue{8}),
\bfpage{657}--\blpage{665}
(\byear{2020})
\doiurl{10.1038/s10038-020-0750-x}
\end{barticle}
\endbibitem

\bibitem[\protect\citeauthoryear{Scerri et~al.}{2017}]{scerri2017genome}
\begin{barticle}
\bauthor{\bsnm{Scerri}, \binits{T.S.}},
\bauthor{\bsnm{Quaglieri}, \binits{A.}},
\bauthor{\bsnm{Cai}, \binits{C.}},
\bauthor{\bsnm{Zernant}, \binits{J.}},
\bauthor{\bsnm{Matsunami}, \binits{N.}},
\bauthor{\bsnm{Baird}, \binits{L.}},
\bauthor{\bsnm{Scheppke}, \binits{L.}},
\bauthor{\bsnm{Bonelli}, \binits{R.}},
\bauthor{\bsnm{Yannuzzi}, \binits{L.A.}},
\bauthor{\bsnm{Friedlander}, \binits{M.}}, \betal:
\batitle{Genome-wide analyses identify common variants associated with macular telangiectasia type 2}.
\bjtitle{Nat. Genet.}
\bvolume{49}(\bissue{4}),
\bfpage{559}--\blpage{567}
(\byear{2017})
\doiurl{10.1038/ng.3799}
\end{barticle}
\endbibitem

\bibitem[\protect\citeauthoryear{Ohno-Matsui et~al.}{2015}]{ohno2015international}
\begin{barticle}
\bauthor{\bsnm{Ohno-Matsui}, \binits{K.}},
\bauthor{\bsnm{Kawasaki}, \binits{R.}},
\bauthor{\bsnm{Jonas}, \binits{J.B.}},
\bauthor{\bsnm{Cheung}, \binits{C.M.G.}},
\bauthor{\bsnm{Saw}, \binits{S.-M.}},
\bauthor{\bsnm{Verhoeven}, \binits{V.J.M.}},
\bauthor{\bsnm{Klaver}, \binits{C.C.W.}},
\bauthor{\bsnm{Moriyama}, \binits{M.}},
\bauthor{\bsnm{Shinohara}, \binits{K.}},
\bauthor{\bsnm{Kawasaki}, \binits{Y.}}, \betal:
\batitle{International photographic classification and grading system for myopic maculopathy}.
\bjtitle{Am. J. Ophthalmol.}
\bvolume{159}(\bissue{5}),
\bfpage{877}--\blpage{883}
(\byear{2015})
\doiurl{10.1016/j.ajo.2015.01.022}
\end{barticle}
\endbibitem

\bibitem[\protect\citeauthoryear{Yan et~al.}{2015}]{yan2015fundus}
\begin{barticle}
\bauthor{\bsnm{Yan}, \binits{Y.N.}},
\bauthor{\bsnm{Wang}, \binits{Y.X.}},
\bauthor{\bsnm{Xu}, \binits{L.}},
\bauthor{\bsnm{Xu}, \binits{J.}},
\bauthor{\bsnm{Wei}, \binits{W.B.}},
\bauthor{\bsnm{Jonas}, \binits{J.B.}}:
\batitle{Fundus tessellation: prevalence and associated factors: the beijing eye study 2011}.
\bjtitle{Ophthalmology}
\bvolume{122}(\bissue{9}),
\bfpage{1873}--\blpage{1880}
(\byear{2015})
\doiurl{10.1016/j.ophtha.2015.05.031}
\end{barticle}
\endbibitem

\bibitem[\protect\citeauthoryear{Lyu et~al.}{2021}]{lyu2021characteristics}
\begin{barticle}
\bauthor{\bsnm{Lyu}, \binits{H.}},
\bauthor{\bsnm{Chen}, \binits{Q.}},
\bauthor{\bsnm{Hu}, \binits{G.}},
\bauthor{\bsnm{Shi}, \binits{Y.}},
\bauthor{\bsnm{Ye}, \binits{L.}},
\bauthor{\bsnm{Yin}, \binits{Y.}},
\bauthor{\bsnm{Fan}, \binits{Y.}},
\bauthor{\bsnm{Zou}, \binits{H.}},
\bauthor{\bsnm{He}, \binits{J.}},
\bauthor{\bsnm{Zhu}, \binits{J.}}, \betal:
\batitle{Characteristics of fundal changes in fundus tessellation in young adults}.
\bjtitle{Front. Med.}
\bvolume{8},
\bfpage{616249}
(\byear{2021})
\doiurl{10.3389/fmed.2021.616249}
\end{barticle}
\endbibitem

\bibitem[\protect\citeauthoryear{Cho et~al.}{2011}]{cho2011association}
\begin{barticle}
\bauthor{\bsnm{Cho}, \binits{K.H.}},
\bauthor{\bsnm{Kim}, \binits{M.J.}},
\bauthor{\bsnm{Jeon}, \binits{G.J.}},
\bauthor{\bsnm{Chung}, \binits{H.Y.}}:
\batitle{Association of genetic variants for fabp3 gene with back fat thickness and intramuscular fat content in pig}.
\bjtitle{Mol. Biol. Rep.}
\bvolume{38},
\bfpage{2161}--\blpage{2166}
(\byear{2011})
\doiurl{10.1007/s11033-010-0344-3}
\end{barticle}
\endbibitem

\bibitem[\protect\citeauthoryear{Wang et~al.}{2019}]{wang2019association}
\begin{barticle}
\bauthor{\bsnm{Wang}, \binits{B.}},
\bauthor{\bsnm{Li}, \binits{P.}},
\bauthor{\bsnm{Zhou}, \binits{W.}},
\bauthor{\bsnm{Gao}, \binits{C.}},
\bauthor{\bsnm{Liu}, \binits{H.}},
\bauthor{\bsnm{Li}, \binits{H.}},
\bauthor{\bsnm{Niu}, \binits{P.}},
\bauthor{\bsnm{Zhang}, \binits{Z.}},
\bauthor{\bsnm{Li}, \binits{Q.}},
\bauthor{\bsnm{Zhou}, \binits{J.}}, \betal:
\batitle{Association of twelve candidate gene polymorphisms with the intramuscular fat content and average backfat thickness of chinese suhuai pigs}.
\bjtitle{Animals}
\bvolume{9}(\bissue{11}),
\bfpage{858}
(\byear{2019})
\doiurl{10.3390/ani9110858}
\end{barticle}
\endbibitem

\bibitem[\protect\citeauthoryear{Fan et~al.}{2011}]{fan2011genome}
\begin{barticle}
\bauthor{\bsnm{Fan}, \binits{B.}},
\bauthor{\bsnm{Onteru}, \binits{S.K.}},
\bauthor{\bsnm{Du}, \binits{Z.-Q.}},
\bauthor{\bsnm{Garrick}, \binits{D.J.}},
\bauthor{\bsnm{Stalder}, \binits{K.J.}},
\bauthor{\bsnm{Rothschild}, \binits{M.F.}}:
\batitle{Genome-wide association study identifies loci for body composition and structural soundness traits in pigs}.
\bjtitle{PLoS One}
\bvolume{6}(\bissue{2}),
\bfpage{14726}
(\byear{2011})
\doiurl{10.1371/journal.pone.0014726}
\end{barticle}
\endbibitem

\bibitem[\protect\citeauthoryear{Li et~al.}{2021}]{li2021further}
\begin{barticle}
\bauthor{\bsnm{Li}, \binits{L.-Y.}},
\bauthor{\bsnm{Xiao}, \binits{S.-J.}},
\bauthor{\bsnm{Tu}, \binits{J.-M.}},
\bauthor{\bsnm{Zhang}, \binits{Z.-K.}},
\bauthor{\bsnm{Zheng}, \binits{H.}},
\bauthor{\bsnm{Huang}, \binits{L.-B.}},
\bauthor{\bsnm{Huang}, \binits{Z.-Y.}},
\bauthor{\bsnm{Yan}, \binits{M.}},
\bauthor{\bsnm{Liu}, \binits{X.-D.}},
\bauthor{\bsnm{Guo}, \binits{Y.-M.}}:
\batitle{A further survey of the quantitative trait loci affecting swine body size and carcass traits in five related pig populations}.
\bjtitle{Anim. Genet.}
\bvolume{52}(\bissue{5}),
\bfpage{621}--\blpage{632}
(\byear{2021})
\doiurl{10.1111/age.13112}
\end{barticle}
\endbibitem

\bibitem[\protect\citeauthoryear{Zhang et~al.}{2021}]{zhang2021genome}
\begin{barticle}
\bauthor{\bsnm{Zhang}, \binits{H.}},
\bauthor{\bsnm{Zhuang}, \binits{Z.}},
\bauthor{\bsnm{Yang}, \binits{M.}},
\bauthor{\bsnm{Ding}, \binits{R.}},
\bauthor{\bsnm{Quan}, \binits{J.}},
\bauthor{\bsnm{Zhou}, \binits{S.}},
\bauthor{\bsnm{Gu}, \binits{T.}},
\bauthor{\bsnm{Xu}, \binits{Z.}},
\bauthor{\bsnm{Zheng}, \binits{E.}},
\bauthor{\bsnm{Cai}, \binits{G.}}, \betal:
\batitle{Genome-wide detection of genetic loci and candidate genes for body conformation traits in duroc$\times$ landrace$\times$ yorkshire crossbred pigs}.
\bjtitle{Front. Genet.}
\bvolume{12},
\bfpage{664343}
(\byear{2021})
\doiurl{10.3389/fgene.2021.664343}
\end{barticle}
\endbibitem

\bibitem[\protect\citeauthoryear{Miao et~al.}{2023}]{miao2023integrated}
\begin{barticle}
\bauthor{\bsnm{Miao}, \binits{Y.}},
\bauthor{\bsnm{Zhao}, \binits{Y.}},
\bauthor{\bsnm{Wan}, \binits{S.}},
\bauthor{\bsnm{Mei}, \binits{Q.}},
\bauthor{\bsnm{Wang}, \binits{H.}},
\bauthor{\bsnm{Fu}, \binits{C.}},
\bauthor{\bsnm{Li}, \binits{X.}},
\bauthor{\bsnm{Zhao}, \binits{S.}},
\bauthor{\bsnm{Xu}, \binits{X.}},
\bauthor{\bsnm{Xiang}, \binits{T.}}:
\batitle{Integrated analysis of genome-wide association studies and 3d epigenomic characteristics reveal the bmp2 gene regulating loin muscle depth in yorkshire pigs}.
\bjtitle{PLoS Genet.}
\bvolume{19}(\bissue{6}),
\bfpage{1010820}
(\byear{2023})
\doiurl{10.1371/journal.pgen.1010820}
\end{barticle}
\endbibitem

\bibitem[\protect\citeauthoryear{Oliphant}{1987}]{oliphant1987observations}
\begin{barticle}
\bauthor{\bsnm{Oliphant}, \binits{L.W.}}:
\batitle{Observations on the pigmentation of the pigeon iris}.
\bjtitle{Pigment Cell Res.}
\bvolume{1}(\bissue{3}),
\bfpage{202}--\blpage{208}
(\byear{1987})
\doiurl{10.1111/j.1600-0749.1987.tb00414.x}
\end{barticle}
\endbibitem

\bibitem[\protect\citeauthoryear{Oliphant}{1981}]{oliphant1981crystalline}
\begin{barticle}
\bauthor{\bsnm{Oliphant}, \binits{L.W.}}:
\batitle{Crystalline pteridines in the stromal pigment cells of the iris of the great horned owl}.
\bjtitle{Cell Tissue Res.}
\bvolume{217},
\bfpage{387}--\blpage{395}
(\byear{1981})
\doiurl{10.1007/bf00233588}
\end{barticle}
\endbibitem

\bibitem[\protect\citeauthoryear{Oliphant}{1987}]{oliphant1987pteridines}
\begin{barticle}
\bauthor{\bsnm{Oliphant}, \binits{L.W.}}:
\batitle{Pteridines and purines as major pigments of the avian iris}.
\bjtitle{Pigment Cell Res.}
\bvolume{1}(\bissue{2}),
\bfpage{129}--\blpage{131}
(\byear{1987})
\doiurl{10.1111/j.1600-0749.1987.tb00401.x}
\end{barticle}
\endbibitem

\bibitem[\protect\citeauthoryear{Hudon and Oliphant}{1995}]{hudon1995reflective}
\begin{barticle}
\bauthor{\bsnm{Hudon}, \binits{J.}},
\bauthor{\bsnm{Oliphant}, \binits{L.W.}}:
\batitle{Reflective organelles in the anterior pigment epithelium of the iris of the european starling sturnus vulgaris}.
\bjtitle{Cell Tissue Res.}
\bvolume{280},
\bfpage{383}--\blpage{389}
(\byear{1995})
\doiurl{10.1007/bf00307811}
\end{barticle}
\endbibitem

\bibitem[\protect\citeauthoryear{Sweijd and Craig}{1991}]{sweijd1991histological}
\begin{botherref}
\oauthor{\bsnm{Sweijd}, \binits{N.}},
\oauthor{\bsnm{Craig}, \binits{A.J.F.K.}}:
Histological basis of age-related changes in iris color in the african pied starling (spreo bicolor).
The Auk,
53--59
(1991)
\end{botherref}
\endbibitem

\bibitem[\protect\citeauthoryear{Corbett et~al.}{2024}]{corbett2024mechanistic}
\begin{barticle}
\bauthor{\bsnm{Corbett}, \binits{E.C.}},
\bauthor{\bsnm{Brumfield}, \binits{R.T.}},
\bauthor{\bsnm{Faircloth}, \binits{B.C.}}:
\batitle{The mechanistic, genetic and evolutionary causes of bird eye colour variation}.
\bjtitle{Ibis}
\bvolume{166}(\bissue{2}),
\bfpage{560}--\blpage{589}
(\byear{2024})
\doiurl{10.1111/ibi.13276}
\end{barticle}
\endbibitem

\bibitem[\protect\citeauthoryear{Amyere et~al.}{2011}]{2011KITLG}
\begin{barticle}
\bauthor{\bsnm{Amyere}, \binits{M.}},
\bauthor{\bsnm{Vogt}, \binits{T.}},
\bauthor{\bsnm{Hoo}, \binits{J.}},
\bauthor{\bsnm{Brandrup}, \binits{F.}},
\bauthor{\bsnm{Bygum}, \binits{A.}},
\bauthor{\bsnm{Boon}, \binits{L.}},
\bauthor{\bsnm{Vikkula}, \binits{M.}}:
\batitle{Kitlg mutations cause familial progressive hyper- and hypopigmentation}.
\bjtitle{J. Investig. Dermatol.}
\bvolume{131}(\bissue{6}),
\bfpage{1234}--\blpage{1239}
(\byear{2011})
\doiurl{10.1038/jid.2011.29}
\end{barticle}
\endbibitem

\bibitem[\protect\citeauthoryear{Picardo and Cardinali}{2011}]{2011The}
\begin{barticle}
\bauthor{\bsnm{Picardo}, \binits{M.}},
\bauthor{\bsnm{Cardinali}, \binits{G.}}:
\batitle{The genetic determination of skin pigmentation: Kitlg and the kitlg/c-kit pathway as key players in the onset of human familial pigmentary diseases}.
\bjtitle{J. Investig. Dermatol.}
\bvolume{131}(\bissue{6}),
\bfpage{1182}--\blpage{1185}
(\byear{2011})
\doiurl{10.1038/jid.2011.67}
\end{barticle}
\endbibitem

\bibitem[\protect\citeauthoryear{Wehrle-Haller}{2003}]{wehrle2003role}
\begin{barticle}
\bauthor{\bsnm{Wehrle-Haller}, \binits{B.}}:
\batitle{The role of kit-ligand in melanocyte development and epidermal homeostasis}.
\bjtitle{Pigment Cell Res.}
\bvolume{16}(\bissue{3}),
\bfpage{287}--\blpage{296}
(\byear{2003})
\doiurl{10.1034/j.1600-0749.2003.00055.x}
\end{barticle}
\endbibitem

\bibitem[\protect\citeauthoryear{Guenther et~al.}{2014}]{guenther2014molecular}
\begin{barticle}
\bauthor{\bsnm{Guenther}, \binits{C.A.}},
\bauthor{\bsnm{Tasic}, \binits{B.}},
\bauthor{\bsnm{Luo}, \binits{L.}},
\bauthor{\bsnm{Bedell}, \binits{M.A.}},
\bauthor{\bsnm{Kingsley}, \binits{D.M.}}:
\batitle{A molecular basis for classic blond hair color in europeans}.
\bjtitle{Nat. Genet.}
\bvolume{46}(\bissue{7}),
\bfpage{748}--\blpage{752}
(\byear{2014})
\doiurl{10.1038/ng.2991}
\end{barticle}
\endbibitem

\bibitem[\protect\citeauthoryear{Cuell et~al.}{2015}]{cuell2015familial}
\begin{barticle}
\bauthor{\bsnm{Cuell}, \binits{A.}},
\bauthor{\bsnm{Bansal}, \binits{N.}},
\bauthor{\bsnm{Cole}, \binits{T.}},
\bauthor{\bsnm{Kaur}, \binits{M.R.}},
\bauthor{\bsnm{Lee}, \binits{J.}},
\bauthor{\bsnm{Loffeld}, \binits{A.}},
\bauthor{\bsnm{Moss}, \binits{C.}},
\bauthor{\bsnm{O'donnell}, \binits{M.}},
\bauthor{\bsnm{Takeichi}, \binits{T.}},
\bauthor{\bsnm{Thind}, \binits{C.K.}}, \betal:
\batitle{Familial progressive hyper-and hypopigmentation and malignancy in two families with new mutations in kitlg}.
\bjtitle{Clin. Exp. Dermatol.}
\bvolume{40}(\bissue{8}),
\bfpage{860}--\blpage{864}
(\byear{2015})
\doiurl{10.1111/ced.12702}
\end{barticle}
\endbibitem

\bibitem[\protect\citeauthoryear{Wang et~al.}{2021}]{wang2021identification}
\begin{barticle}
\bauthor{\bsnm{Wang}, \binits{J.}},
\bauthor{\bsnm{Li}, \binits{W.}},
\bauthor{\bsnm{Zhou}, \binits{N.}},
\bauthor{\bsnm{Liu}, \binits{J.}},
\bauthor{\bsnm{Zhang}, \binits{S.}},
\bauthor{\bsnm{Li}, \binits{X.}},
\bauthor{\bsnm{Li}, \binits{Z.}},
\bauthor{\bsnm{Yang}, \binits{Z.}},
\bauthor{\bsnm{Sun}, \binits{M.}},
\bauthor{\bsnm{Li}, \binits{M.}}:
\batitle{Identification of a novel mutation in the kitlg gene in a chinese family with familial progressive hyper-and hypopigmentation}.
\bjtitle{BMC Med. Genomics}
\bvolume{14}(\bissue{1}),
\bfpage{12}
(\byear{2021})
\doiurl{10.21203/rs.3.rs-28033/v1}
\end{barticle}
\endbibitem

\bibitem[\protect\citeauthoryear{Wang et~al.}{2009}]{wang2009gain}
\begin{barticle}
\bauthor{\bsnm{Wang}, \binits{Z.-Q.}},
\bauthor{\bsnm{Si}, \binits{L.}},
\bauthor{\bsnm{Tang}, \binits{Q.}},
\bauthor{\bsnm{Lin}, \binits{D.}},
\bauthor{\bsnm{Fu}, \binits{Z.}},
\bauthor{\bsnm{Zhang}, \binits{J.}},
\bauthor{\bsnm{Cui}, \binits{B.}},
\bauthor{\bsnm{Zhu}, \binits{Y.}},
\bauthor{\bsnm{Kong}, \binits{X.}},
\bauthor{\bsnm{Deng}, \binits{M.}}, \betal:
\batitle{Gain-of-function mutation of kit ligand on melanin synthesis causes familial progressive hyperpigmentation}.
\bjtitle{Am. J. Hum. Genet.}
\bvolume{84}(\bissue{5}),
\bfpage{672}--\blpage{677}
(\byear{2009})
\doiurl{10.1016/j.ajhg.2009.03.019}
\end{barticle}
\endbibitem

\bibitem[\protect\citeauthoryear{Kim et~al.}{2024}]{kim2024mapping}
\begin{barticle}
\bauthor{\bsnm{Kim}, \binits{B.}},
\bauthor{\bsnm{Kim}, \binits{D.S.}},
\bauthor{\bsnm{Shin}, \binits{J.-G.}},
\bauthor{\bsnm{Leem}, \binits{S.}},
\bauthor{\bsnm{Cho}, \binits{M.}},
\bauthor{\bsnm{Kim}, \binits{H.}},
\bauthor{\bsnm{Gu}, \binits{K.-N.}},
\bauthor{\bsnm{Seo}, \binits{J.Y.}},
\bauthor{\bsnm{You}, \binits{S.W.}},
\bauthor{\bsnm{Martin}, \binits{A.R.}}, \betal:
\batitle{Mapping and annotating genomic loci to prioritize genes and implicate distinct polygenic adaptations for skin color}.
\bjtitle{Nat. Commun.}
\bvolume{15}(\bissue{1}),
\bfpage{4874}
(\byear{2024})
\doiurl{10.1038/s41467-024-49031-4}
\end{barticle}
\endbibitem

\bibitem[\protect\citeauthoryear{Ren et~al.}{2021}]{ren2021pooled}
\begin{barticle}
\bauthor{\bsnm{Ren}, \binits{S.}},
\bauthor{\bsnm{Lyu}, \binits{G.}},
\bauthor{\bsnm{Irwin}, \binits{D.M.}},
\bauthor{\bsnm{Liu}, \binits{X.}},
\bauthor{\bsnm{Feng}, \binits{C.}},
\bauthor{\bsnm{Luo}, \binits{R.}},
\bauthor{\bsnm{Zhang}, \binits{J.}},
\bauthor{\bsnm{Sun}, \binits{Y.}},
\bauthor{\bsnm{Shang}, \binits{S.}},
\bauthor{\bsnm{Zhang}, \binits{S.}}, \betal:
\batitle{Pooled sequencing analysis of geese (anser cygnoides) reveals genomic variations associated with feather color}.
\bjtitle{Front. Genet.}
\bvolume{12},
\bfpage{650013}
(\byear{2021})
\doiurl{10.3389/fgene.2021.650013}
\end{barticle}
\endbibitem

\bibitem[\protect\citeauthoryear{Shen et~al.}{2022}]{shen2022genome}
\begin{barticle}
\bauthor{\bsnm{Shen}, \binits{Q.}},
\bauthor{\bsnm{Zhou}, \binits{J.}},
\bauthor{\bsnm{Li}, \binits{J.}},
\bauthor{\bsnm{Zhao}, \binits{X.}},
\bauthor{\bsnm{Zheng}, \binits{L.}},
\bauthor{\bsnm{Bao}, \binits{H.}},
\bauthor{\bsnm{Wu}, \binits{C.}}:
\batitle{Genome-wide association study identifies candidate genes for stripe pattern feather color of rhode island red chicks}.
\bjtitle{Genes}
\bvolume{13}(\bissue{9}),
\bfpage{1511}
(\byear{2022})
\doiurl{10.3390/genes13091511}
\end{barticle}
\endbibitem

\bibitem[\protect\citeauthoryear{Talenti et~al.}{2018}]{talenti2018genomic}
\begin{barticle}
\bauthor{\bsnm{Talenti}, \binits{A.}},
\bauthor{\bsnm{Bertolini}, \binits{F.}},
\bauthor{\bsnm{Williams}, \binits{J.}},
\bauthor{\bsnm{Moaeen-ud-Din}, \binits{M.}},
\bauthor{\bsnm{Frattini}, \binits{S.}},
\bauthor{\bsnm{Coizet}, \binits{B.}},
\bauthor{\bsnm{Pagnacco}, \binits{G.}},
\bauthor{\bsnm{Reecy}, \binits{J.}},
\bauthor{\bsnm{Rothschild}, \binits{M.F.}},
\bauthor{\bsnm{Crepaldi}, \binits{P.}}, \betal:
\batitle{Genomic analysis suggests kitlg is responsible for a roan pattern in two pakistani goat breeds}.
\bjtitle{J. Hered.}
\bvolume{109}(\bissue{3}),
\bfpage{315}--\blpage{319}
(\byear{2018})
\doiurl{10.1093/jhered/esx093}
\end{barticle}
\endbibitem

\bibitem[\protect\citeauthoryear{Miller et~al.}{2007}]{miller2007cis}
\begin{barticle}
\bauthor{\bsnm{Miller}, \binits{C.T.}},
\bauthor{\bsnm{Beleza}, \binits{S.}},
\bauthor{\bsnm{Pollen}, \binits{A.A.}},
\bauthor{\bsnm{Schluter}, \binits{D.}},
\bauthor{\bsnm{Kittles}, \binits{R.A.}},
\bauthor{\bsnm{Shriver}, \binits{M.D.}},
\bauthor{\bsnm{Kingsley}, \binits{D.M.}}:
\batitle{cis-regulatory changes in kit ligand expression and parallel evolution of pigmentation in sticklebacks and humans}.
\bjtitle{Cell}
\bvolume{131}(\bissue{6}),
\bfpage{1179}--\blpage{1189}
(\byear{2007})
\doiurl{10.1016/j.cell.2007.10.055}
\end{barticle}
\endbibitem

\bibitem[\protect\citeauthoryear{Zazoseco et~al.}{2015}]{2015Allelic}
\begin{barticle}
\bauthor{\bsnm{Zazoseco}, \binits{C.}},
\bauthor{\bsnm{Serrodecastro}, \binits{L.}},
\bauthor{\bsnm{Vannierop}, \binits{J.W.}},
\bauthor{\bsnm{Mor{\'\i}n}, \binits{M.}},
\bauthor{\bsnm{Jhangiani}, \binits{S.}},
\bauthor{\bsnm{Verver}, \binits{E.J.J.}},
\bauthor{\bsnm{Schraders}, \binits{M.}},
\bauthor{\bsnm{Maiwald}, \binits{N.}},
\bauthor{\bsnm{Wesdorp}, \binits{M.}},
\bauthor{\bsnm{Venselaar}, \binits{H.}}:
\batitle{Allelic mutations of kitlg, encoding kit ligand, cause asymmetric and unilateral hearing loss and waardenburg syndrome type 2}.
\bjtitle{Am. J. Hum. Genet.}
\bvolume{97}(\bissue{5}),
\bfpage{647}--\blpage{660}
(\byear{2015})
\doiurl{10.3410/f.725901771.793511836}
\end{barticle}
\endbibitem

\bibitem[\protect\citeauthoryear{Moscatelli et~al.}{2020}]{2020Genome}
\begin{barticle}
\bauthor{\bsnm{Moscatelli}, \binits{G.}},
\bauthor{\bsnm{Bovo}, \binits{S.}},
\bauthor{\bsnm{Schiavo}, \binits{G.}},
\bauthor{\bsnm{Mazzoni}, \binits{G.}},
\bauthor{\bsnm{Bertolini}, \binits{F.}},
\bauthor{\bsnm{Dall'Olio}, \binits{S.}},
\bauthor{\bsnm{Fontanesi}, \binits{L.}}:
\batitle{Genome-wide association studies for iris pigmentation and heterochromia patterns in large white pigs}.
\bjtitle{Anim. Genet.}
\bvolume{51}(\bissue{3}),
\bfpage{409}--\blpage{419}
(\byear{2020})
\doiurl{10.1111/age.12930}
\end{barticle}
\endbibitem

\bibitem[\protect\citeauthoryear{Morgan et~al.}{2018}]{morgan2018genome}
\begin{barticle}
\bauthor{\bsnm{Morgan}, \binits{M.D.}},
\bauthor{\bsnm{Pairo-Castineira}, \binits{E.}},
\bauthor{\bsnm{Rawlik}, \binits{K.}},
\bauthor{\bsnm{Canela-Xandri}, \binits{O.}},
\bauthor{\bsnm{Rees}, \binits{J.}},
\bauthor{\bsnm{Sims}, \binits{D.}},
\bauthor{\bsnm{Tenesa}, \binits{A.}},
\bauthor{\bsnm{Jackson}, \binits{I.J.}}:
\batitle{Genome-wide study of hair colour in uk biobank explains most of the snp heritability}.
\bjtitle{Nat. Commun.}
\bvolume{9}(\bissue{1}),
\bfpage{5271}
(\byear{2018})
\doiurl{10.1038/s41467-018-07691-z}
\end{barticle}
\endbibitem

\bibitem[\protect\citeauthoryear{Zhuang et~al.}{2021}]{zhuang2021sod2}
\begin{barticle}
\bauthor{\bsnm{Zhuang}, \binits{A.}},
\bauthor{\bsnm{Yang}, \binits{C.}},
\bauthor{\bsnm{Liu}, \binits{Y.}},
\bauthor{\bsnm{Tan}, \binits{Y.}},
\bauthor{\bsnm{Bond}, \binits{S.T.}},
\bauthor{\bsnm{Walker}, \binits{S.}},
\bauthor{\bsnm{Sikora}, \binits{T.}},
\bauthor{\bsnm{Laskowski}, \binits{A.}},
\bauthor{\bsnm{Sharma}, \binits{A.}},
\bauthor{\bsnm{Haan}, \binits{J.B.}}, \betal:
\batitle{Sod2 in skeletal muscle: New insights from an inducible deletion model}.
\bjtitle{Redox Biol.}
\bvolume{47},
\bfpage{102135}
(\byear{2021})
\doiurl{10.1016/j.redox.2021.102135}
\end{barticle}
\endbibitem

\bibitem[\protect\citeauthoryear{Wu et~al.}{2022}]{wu2022fasting}
\begin{barticle}
\bauthor{\bsnm{Wu}, \binits{N.}},
\bauthor{\bsnm{Zhai}, \binits{X.}},
\bauthor{\bsnm{Yuan}, \binits{F.}},
\bauthor{\bsnm{Li}, \binits{J.}},
\bauthor{\bsnm{Yu}, \binits{N.}},
\bauthor{\bsnm{Zhang}, \binits{F.}},
\bauthor{\bsnm{Li}, \binits{D.}},
\bauthor{\bsnm{Wang}, \binits{J.}},
\bauthor{\bsnm{Zhang}, \binits{L.}},
\bauthor{\bsnm{Shi}, \binits{Y.}}, \betal:
\batitle{Fasting glucose mediates the influence of genetic variants of sod2 gene on lean non-alcoholic fatty liver disease}.
\bjtitle{Front. Genet.}
\bvolume{13},
\bfpage{970854}
(\byear{2022})
\doiurl{10.3389/fgene.2022.970854}
\end{barticle}
\endbibitem

\bibitem[\protect\citeauthoryear{Hern{\'a}ndez-Guerrero et~al.}{2016}]{hernandez2016genetic}
\begin{barticle}
\bauthor{\bsnm{Hern{\'a}ndez-Guerrero}, \binits{C.}},
\bauthor{\bsnm{Hern{\'a}ndez-Ch{\'a}vez}, \binits{P.}},
\bauthor{\bsnm{Romo-Palafox}, \binits{I.}},
\bauthor{\bsnm{Blanco-Melo}, \binits{G.}},
\bauthor{\bsnm{Parra-Carriedo}, \binits{A.}},
\bauthor{\bsnm{P{\'e}rez-Lizaur}, \binits{A.}}:
\batitle{Genetic polymorphisms in sod (rs2070424, rs7880) and cat (rs7943316, rs1001179) enzymes are associated with increased body fat percentage and visceral fat in an obese population from central mexico}.
\bjtitle{Arch. Med. Res.}
\bvolume{47}(\bissue{5}),
\bfpage{331}--\blpage{339}
(\byear{2016})
\doiurl{10.1016/j.arcmed.2016.08.007}
\end{barticle}
\endbibitem

\bibitem[\protect\citeauthoryear{Nezer et~al.}{1999}]{nezer1999imprinted}
\begin{barticle}
\bauthor{\bsnm{Nezer}, \binits{C.}},
\bauthor{\bsnm{Moreau}, \binits{L.}},
\bauthor{\bsnm{Brouwers}, \binits{B.}},
\bauthor{\bsnm{Coppieters}, \binits{W.}},
\bauthor{\bsnm{Detilleux}, \binits{J.}},
\bauthor{\bsnm{Hanset}, \binits{R.}},
\bauthor{\bsnm{Karim}, \binits{L.}},
\bauthor{\bsnm{Kvasz}, \binits{A.}},
\bauthor{\bsnm{Leroy}, \binits{P.}},
\bauthor{\bsnm{Georges}, \binits{M.}}:
\batitle{An imprinted qtl with major effect on muscle mass and fat deposition maps to the igf2 locus in pigs}.
\bjtitle{Nat. Genet.}
\bvolume{21}(\bissue{2}),
\bfpage{155}--\blpage{156}
(\byear{1999})
\doiurl{10.1038/5935}
\end{barticle}
\endbibitem

\bibitem[\protect\citeauthoryear{Yang}{2015}]{yang2015association}
\begin{barticle}
\bauthor{\bsnm{Yang}, \binits{S.-A.}}:
\batitle{Association between exonic polymorphism (rs629849, gly1619arg) of igf2r gene and obesity in korean population}.
\bjtitle{J. Exerc. Rehabil.}
\bvolume{11}(\bissue{5}),
\bfpage{282}
(\byear{2015})
\doiurl{10.12965/jer.150239}
\end{barticle}
\endbibitem

\bibitem[\protect\citeauthoryear{Rohrschneider}{2004}]{rohrschneider2004determination}
\begin{barticle}
\bauthor{\bsnm{Rohrschneider}, \binits{K.}}:
\batitle{Determination of the location of the fovea on the fundus}.
\bjtitle{Invest. Ophthalmol. Vis. Sci.}
\bvolume{45}(\bissue{9}),
\bfpage{3257}--\blpage{3258}
(\byear{2004})
\doiurl{10.1167/iovs.03-1157}
\end{barticle}
\endbibitem

\bibitem[\protect\citeauthoryear{Jonas et~al.}{2015}]{jonas2015optic}
\begin{barticle}
\bauthor{\bsnm{Jonas}, \binits{R.A.}},
\bauthor{\bsnm{Wang}, \binits{Y.X.}},
\bauthor{\bsnm{Yang}, \binits{H.}},
\bauthor{\bsnm{Li}, \binits{J.J.}},
\bauthor{\bsnm{Xu}, \binits{L.}},
\bauthor{\bsnm{Panda-Jonas}, \binits{S.}},
\bauthor{\bsnm{Jonas}, \binits{J.B.}}:
\batitle{Optic disc-fovea angle: the beijing eye study 2011}.
\bjtitle{PLoS One}
\bvolume{10}(\bissue{11}),
\bfpage{0141771}
(\byear{2015})
\doiurl{10.1371/journal.pone.0141771}
\end{barticle}
\endbibitem

\bibitem[\protect\citeauthoryear{Choi et~al.}{2014}]{choi2014foveal}
\begin{barticle}
\bauthor{\bsnm{Choi}, \binits{J.A.}},
\bauthor{\bsnm{Kim}, \binits{J.-S.}},
\bauthor{\bsnm{Park}, \binits{H.-Y.L.}},
\bauthor{\bsnm{Park}, \binits{H.}},
\bauthor{\bsnm{Park}, \binits{C.K.}}:
\batitle{The foveal position relative to the optic disc and the retinal nerve fiber layer thickness profile in myopia}.
\bjtitle{Invest. Ophthalmol. Vis. Sci.}
\bvolume{55}(\bissue{3}),
\bfpage{1419}--\blpage{1426}
(\byear{2014})
\doiurl{10.1167/iovs.13-13604}
\end{barticle}
\endbibitem

\bibitem[\protect\citeauthoryear{Amini et~al.}{2014}]{amini2014influence}
\begin{barticle}
\bauthor{\bsnm{Amini}, \binits{N.}},
\bauthor{\bsnm{Nowroozizadeh}, \binits{S.}},
\bauthor{\bsnm{Cirineo}, \binits{N.}},
\bauthor{\bsnm{Henry}, \binits{S.}},
\bauthor{\bsnm{Chang}, \binits{T.}},
\bauthor{\bsnm{Chou}, \binits{T.}},
\bauthor{\bsnm{Coleman}, \binits{A.L.}},
\bauthor{\bsnm{Caprioli}, \binits{J.}},
\bauthor{\bsnm{Nouri-Mahdavi}, \binits{K.}}:
\batitle{Influence of the disc--fovea angle on limits of rnfl variability and glaucoma discrimination}.
\bjtitle{Invest. Ophthalmol. Vis. Sci.}
\bvolume{55}(\bissue{11}),
\bfpage{7332}--\blpage{7342}
(\byear{2014})
\doiurl{10.1167/iovs.14-14962}
\end{barticle}
\endbibitem

\bibitem[\protect\citeauthoryear{Haarman et~al.}{2020}]{haarman2020complications}
\begin{barticle}
\bauthor{\bsnm{Haarman}, \binits{A.E.G.}},
\bauthor{\bsnm{Enthoven}, \binits{C.A.}},
\bauthor{\bsnm{Tideman}, \binits{J.W.L.}},
\bauthor{\bsnm{Tedja}, \binits{M.S.}},
\bauthor{\bsnm{Verhoeven}, \binits{V.J.M.}},
\bauthor{\bsnm{Klaver}, \binits{C.C.W.}}:
\batitle{The complications of myopia: a review and meta-analysis}.
\bjtitle{Invest. Ophthalmol. Vis. Sci.}
\bvolume{61}(\bissue{4}),
\bfpage{49}--\blpage{49}
(\byear{2020})
\doiurl{10.1167/iovs.61.4.49}
\end{barticle}
\endbibitem

\bibitem[\protect\citeauthoryear{Hoffmann et~al.}{2007}]{hoffmann2007optic}
\begin{barticle}
\bauthor{\bsnm{Hoffmann}, \binits{E.M.}},
\bauthor{\bsnm{Zangwill}, \binits{L.M.}},
\bauthor{\bsnm{Crowston}, \binits{J.G.}},
\bauthor{\bsnm{Weinreb}, \binits{R.N.}}:
\batitle{Optic disk size and glaucoma}.
\bjtitle{Surv. Ophthalmol.}
\bvolume{52}(\bissue{1}),
\bfpage{32}--\blpage{49}
(\byear{2007})
\doiurl{10.5005/jp/books/18037_3}
\end{barticle}
\endbibitem

\bibitem[\protect\citeauthoryear{Voynov and Babenko}{2020}]{voynov2020unsupervised}
\begin{bchapter}
\bauthor{\bsnm{Voynov}, \binits{A.}},
\bauthor{\bsnm{Babenko}, \binits{A.}}:
\bctitle{Unsupervised discovery of interpretable directions in the gan latent space}.
In: \bbtitle{Proceedings of the 37th International Conference on Machine Learning (ICML)}.
\bsertitle{Proceedings of Machine Learning Research},
vol. \bseriesno{119},
pp. \bfpage{9786}--\blpage{9796}.
\bpublisher{PMLR},
\blocation{Virtual}
(\byear{2020}).
\doiurl{10.5555/3524938.3525845} .
\bcomment{International Machine Learning Society}
\end{bchapter}
\endbibitem

\bibitem[\protect\citeauthoryear{Ren et~al.}{2022}]{ren2021learning}
\begin{bchapter}
\bauthor{\bsnm{Ren}, \binits{X.}},
\bauthor{\bsnm{Yang}, \binits{T.}},
\bauthor{\bsnm{Wang}, \binits{Y.}},
\bauthor{\bsnm{Zeng}, \binits{W.}}:
\bctitle{Learning disentangled representation by exploiting pretrained generative models: A contrastive learning view}.
In: \bbtitle{ICLR}
(\byear{2022}).
\doiurl{10.48550/arXiv.2102.10543}
\end{bchapter}
\endbibitem

\bibitem[\protect\citeauthoryear{Wang et~al.}{2023}]{wang2023rodin}
\begin{bchapter}
\bauthor{\bsnm{Wang}, \binits{T.}},
\bauthor{\bsnm{Zhang}, \binits{B.}},
\bauthor{\bsnm{Zhang}, \binits{T.}},
\bauthor{\bsnm{Gu}, \binits{S.}},
\bauthor{\bsnm{Bao}, \binits{J.}},
\bauthor{\bsnm{Baltrusaitis}, \binits{T.}},
\bauthor{\bsnm{Shen}, \binits{J.}},
\bauthor{\bsnm{Chen}, \binits{D.}},
\bauthor{\bsnm{Wen}, \binits{F.}},
\bauthor{\bsnm{Chen}, \binits{Q.}}, \betal:
\bctitle{Rodin: A generative model for sculpting 3d digital avatars using diffusion}.
In: \bbtitle{Proceedings of the IEEE/CVF Conference on Computer Vision and Pattern Recognition (CVPR)},
pp. \bfpage{4563}--\blpage{4573}.
\bpublisher{IEEE Computer Society},
\blocation{Vancouver, BC, Canada}
(\byear{2023}).
\doiurl{10.1109/CVPR52729.2023.00443} .
\bcomment{IEEE/CVF}
\end{bchapter}
\endbibitem

\bibitem[\protect\citeauthoryear{Anciukevi{\v{c}}ius et~al.}{2023}]{anciukevivcius2023renderdiffusion}
\begin{bchapter}
\bauthor{\bsnm{Anciukevi{\v{c}}ius}, \binits{T.}},
\bauthor{\bsnm{Xu}, \binits{Z.}},
\bauthor{\bsnm{Fisher}, \binits{M.}},
\bauthor{\bsnm{Henderson}, \binits{P.}},
\bauthor{\bsnm{Bilen}, \binits{H.}},
\bauthor{\bsnm{Mitra}, \binits{N.J.}},
\bauthor{\bsnm{Guerrero}, \binits{P.}}:
\bctitle{Renderdiffusion: Image diffusion for 3d reconstruction, inpainting and generation}.
In: \bbtitle{Proceedings of the IEEE/CVF Conference on Computer Vision and Pattern Recognition (CVPR)},
pp. \bfpage{12608}--\blpage{12618}.
\bpublisher{IEEE Computer Society},
\blocation{Vancouver, BC, Canada}
(\byear{2023}).
\doiurl{10.1109/CVPR52729.2023.01213} .
\bcomment{IEEE/CVF}
\end{bchapter}
\endbibitem

\bibitem[\protect\citeauthoryear{Jung et~al.}{2023}]{jung2023conditional}
\begin{barticle}
\bauthor{\bsnm{Jung}, \binits{E.}},
\bauthor{\bsnm{Luna}, \binits{M.}},
\bauthor{\bsnm{Park}, \binits{S.H.}}:
\batitle{Conditional gan with 3d discriminator for mri generation of alzheimer’s disease progression}.
\bjtitle{Pattern Recognit.}
\bvolume{133},
\bfpage{109061}
(\byear{2023})
\doiurl{10.1016/j.patcog.2022.109061}
\end{barticle}
\endbibitem

\bibitem[\protect\citeauthoryear{Friedrich et~al.}{2024}]{friedrich2024deep}
\begin{bchapter}
\bauthor{\bsnm{Friedrich}, \binits{P.}},
\bauthor{\bsnm{Frisch}, \binits{Y.}},
\bauthor{\bsnm{Cattin}, \binits{P.C.}}:
\bctitle{Deep generative models for 3d medical image synthesis}.
In: \beditor{\bsnm{Maier-Hein}, \binits{K.}},
\beditor{\bsnm{Ourselin}, \binits{S.}},
\beditor{\bsnm{Joskowicz}, \binits{L.}} (eds.)
\bbtitle{Generative Machine Learning Models in Medical Image Computing},
pp. \bfpage{255}--\blpage{278}.
\bpublisher{Springer},
\blocation{Cham, Switzerland}
(\byear{2024}).
\doiurl{10.1007/978-3-031-80965-1_13}
\end{bchapter}
\endbibitem

\bibitem[\protect\citeauthoryear{Heymans et~al.}{2020}]{heymans2020granar}
\begin{barticle}
\bauthor{\bsnm{Heymans}, \binits{A.}},
\bauthor{\bsnm{Couvreur}, \binits{V.}},
\bauthor{\bsnm{LaRue}, \binits{T.}},
\bauthor{\bsnm{Paez-Garcia}, \binits{A.}},
\bauthor{\bsnm{Lobet}, \binits{G.}}:
\batitle{Granar, a computational tool to better understand the functional importance of monocotyledon root anatomy}.
\bjtitle{Plant Physiol.}
\bvolume{182}(\bissue{2}),
\bfpage{707}--\blpage{720}
(\byear{2020})
\doiurl{10.1104/pp.19.00617}
\end{barticle}
\endbibitem

\bibitem[\protect\citeauthoryear{Buslaev et~al.}{2020}]{buslaev2020albumentations}
\begin{barticle}
\bauthor{\bsnm{Buslaev}, \binits{A.}},
\bauthor{\bsnm{Iglovikov}, \binits{V.I.}},
\bauthor{\bsnm{Khvedchenya}, \binits{E.}},
\bauthor{\bsnm{Parinov}, \binits{A.}},
\bauthor{\bsnm{Druzhinin}, \binits{M.}},
\bauthor{\bsnm{Kalinin}, \binits{A.A.}}:
\batitle{Albumentations: fast and flexible image augmentations}.
\bjtitle{Information}
\bvolume{11}(\bissue{2}),
\bfpage{125}
(\byear{2020})
\doiurl{10.3390/info11020125}
\end{barticle}
\endbibitem

\bibitem[\protect\citeauthoryear{Taleb et~al.}{2022}]{taleb2022contig}
\begin{bchapter}
\bauthor{\bsnm{Taleb}, \binits{A.}},
\bauthor{\bsnm{Kirchler}, \binits{M.}},
\bauthor{\bsnm{Monti}, \binits{R.}},
\bauthor{\bsnm{Lippert}, \binits{C.}}:
\bctitle{Contig: Self-supervised multimodal contrastive learning for medical imaging with genetics}.
In: \bbtitle{Proceedings of the IEEE/CVF Conference on Computer Vision and Pattern Recognition (CVPR)},
pp. \bfpage{20908}--\blpage{20921}.
\bpublisher{IEEE Computer Society},
\blocation{New Orleans, LA, USA}
(\byear{2022}).
\doiurl{10.1109/CVPR52688.2022.02024} .
\bcomment{IEEE/CVF}
\end{bchapter}
\endbibitem

\bibitem[\protect\citeauthoryear{Browning et~al.}{2018}]{browning2018one}
\begin{barticle}
\bauthor{\bsnm{Browning}, \binits{B.L.}},
\bauthor{\bsnm{Zhou}, \binits{Y.}},
\bauthor{\bsnm{Browning}, \binits{S.R.}}:
\batitle{A one-penny imputed genome from next-generation reference panels}.
\bjtitle{Am. J. Hum. Genet.}
\bvolume{103}(\bissue{3}),
\bfpage{338}--\blpage{348}
(\byear{2018})
\doiurl{10.1016/j.ajhg.2018.07.015}
\end{barticle}
\endbibitem

\bibitem[\protect\citeauthoryear{Ronneberger et~al.}{2015}]{ronneberger2015u}
\begin{bchapter}
\bauthor{\bsnm{Ronneberger}, \binits{O.}},
\bauthor{\bsnm{Fischer}, \binits{P.}},
\bauthor{\bsnm{Brox}, \binits{T.}}:
\bctitle{U-net: Convolutional networks for biomedical image segmentation}.
In: \bbtitle{Proceedings of the 18th International Conference on Medical Image Computing and Computer-Assisted Intervention (MICCAI)}.
\bsertitle{Lecture Notes in Computer Science},
vol. \bseriesno{9351},
pp. \bfpage{234}--\blpage{241}.
\bpublisher{Springer},
\blocation{Munich, Germany}
(\byear{2015}).
\doiurl{10.1007/978-3-319-24574-4_28} .
\bcomment{MICCAI Society}
\end{bchapter}
\endbibitem

\bibitem[\protect\citeauthoryear{Chen et~al.}{2018}]{chen2018fastp}
\begin{barticle}
\bauthor{\bsnm{Chen}, \binits{S.}},
\bauthor{\bsnm{Zhou}, \binits{Y.}},
\bauthor{\bsnm{Chen}, \binits{Y.}},
\bauthor{\bsnm{Gu}, \binits{J.}}:
\batitle{fastp: an ultra-fast all-in-one fastq preprocessor}.
\bjtitle{Bioinformatics}
\bvolume{34}(\bissue{17}),
\bfpage{884}--\blpage{890}
(\byear{2018})
\doiurl{10.1093/bioinformatics/bty560}
\end{barticle}
\endbibitem

\bibitem[\protect\citeauthoryear{Cheng et~al.}{2021}]{cheng2021haplotype}
\begin{barticle}
\bauthor{\bsnm{Cheng}, \binits{H.}},
\bauthor{\bsnm{Concepcion}, \binits{G.T.}},
\bauthor{\bsnm{Feng}, \binits{X.}},
\bauthor{\bsnm{Zhang}, \binits{H.}},
\bauthor{\bsnm{Li}, \binits{H.}}:
\batitle{Haplotype-resolved de novo assembly using phased assembly graphs with hifiasm}.
\bjtitle{Nat. Methods}
\bvolume{18}(\bissue{2}),
\bfpage{170}--\blpage{175}
(\byear{2021})
\doiurl{10.1038/s41592-020-01056-5}
\end{barticle}
\endbibitem

\bibitem[\protect\citeauthoryear{Durand et~al.}{2016}]{durand2016juicer}
\begin{barticle}
\bauthor{\bsnm{Durand}, \binits{N.C.}},
\bauthor{\bsnm{Shamim}, \binits{M.S.}},
\bauthor{\bsnm{Machol}, \binits{I.}},
\bauthor{\bsnm{Rao}, \binits{S.S.P.}},
\bauthor{\bsnm{Huntley}, \binits{M.H.}},
\bauthor{\bsnm{Lander}, \binits{E.S.}},
\bauthor{\bsnm{Aiden}, \binits{E.L.}}:
\batitle{Juicer provides a one-click system for analyzing loop-resolution hi-c experiments}.
\bjtitle{Cell Syst.}
\bvolume{3}(\bissue{1}),
\bfpage{95}--\blpage{98}
(\byear{2016})
\doiurl{10.1016/j.cels.2016.07.002}
\end{barticle}
\endbibitem

\bibitem[\protect\citeauthoryear{Zeng et~al.}{2024}]{zeng2024chromosome}
\begin{barticle}
\bauthor{\bsnm{Zeng}, \binits{X.}},
\bauthor{\bsnm{Yi}, \binits{Z.}},
\bauthor{\bsnm{Zhang}, \binits{X.}},
\bauthor{\bsnm{Du}, \binits{Y.}},
\bauthor{\bsnm{Li}, \binits{Y.}},
\bauthor{\bsnm{Zhou}, \binits{Z.}},
\bauthor{\bsnm{Chen}, \binits{S.}},
\bauthor{\bsnm{Zhao}, \binits{H.}},
\bauthor{\bsnm{Yang}, \binits{S.}},
\bauthor{\bsnm{Wang}, \binits{Y.}}, \betal:
\batitle{Chromosome-level scaffolding of haplotype-resolved assemblies using hi-c data without reference genomes}.
\bjtitle{Nat. Plants}
\bvolume{10}(\bissue{8}),
\bfpage{1184}--\blpage{1200}
(\byear{2024})
\doiurl{10.1038/s41477-024-01755-3}
\end{barticle}
\endbibitem

\bibitem[\protect\citeauthoryear{Durand et~al.}{2016}]{durand2016juicebox}
\begin{barticle}
\bauthor{\bsnm{Durand}, \binits{N.C.}},
\bauthor{\bsnm{Robinson}, \binits{J.T.}},
\bauthor{\bsnm{Shamim}, \binits{M.S.}},
\bauthor{\bsnm{Machol}, \binits{I.}},
\bauthor{\bsnm{Mesirov}, \binits{J.P.}},
\bauthor{\bsnm{Lander}, \binits{E.S.}},
\bauthor{\bsnm{Aiden}, \binits{E.L.}}:
\batitle{Juicebox provides a visualization system for hi-c contact maps with unlimited zoom}.
\bjtitle{Cell Syst.}
\bvolume{3}(\bissue{1}),
\bfpage{99}--\blpage{101}
(\byear{2016})
\doiurl{10.1016/j.cels.2015.07.012}
\end{barticle}
\endbibitem

\bibitem[\protect\citeauthoryear{Li}{2018}]{li2018minimap2}
\begin{barticle}
\bauthor{\bsnm{Li}, \binits{H.}}:
\batitle{Minimap2: pairwise alignment for nucleotide sequences}.
\bjtitle{Bioinformatics}
\bvolume{34}(\bissue{18}),
\bfpage{3094}--\blpage{3100}
(\byear{2018})
\doiurl{10.1093/bioinformatics/bty191}
\end{barticle}
\endbibitem

\bibitem[\protect\citeauthoryear{Vega et~al.}{2003}]{vega2003getting}
\begin{barticle}
\bauthor{\bsnm{Vega}, \binits{L.R.}},
\bauthor{\bsnm{Mateyak}, \binits{M.K.}},
\bauthor{\bsnm{Zakian}, \binits{V.A.}}:
\batitle{Getting to the end: telomerase access in yeast and humans}.
\bjtitle{Nat. Rev. Mol. Cell Biol.}
\bvolume{4}(\bissue{12}),
\bfpage{948}--\blpage{959}
(\byear{2003})
\doiurl{10.1038/nrm1256}
\end{barticle}
\endbibitem

\bibitem[\protect\citeauthoryear{Kurtz et~al.}{2004}]{kurtz2004versatile}
\begin{barticle}
\bauthor{\bsnm{Kurtz}, \binits{S.}},
\bauthor{\bsnm{Phillippy}, \binits{A.}},
\bauthor{\bsnm{Delcher}, \binits{A.L.}},
\bauthor{\bsnm{Smoot}, \binits{M.}},
\bauthor{\bsnm{Shumway}, \binits{M.}},
\bauthor{\bsnm{Antonescu}, \binits{C.}},
\bauthor{\bsnm{Salzberg}, \binits{S.L.}}:
\batitle{Versatile and open software for comparing large genomes}.
\bjtitle{Genome Biol.}
\bvolume{5},
\bfpage{1}--\blpage{9}
(\byear{2004})
\doiurl{10.1186/gb-2004-5-2-r12}
\end{barticle}
\endbibitem

\bibitem[\protect\citeauthoryear{Xu et~al.}{2020}]{xu2020tgs}
\begin{barticle}
\bauthor{\bsnm{Xu}, \binits{M.}},
\bauthor{\bsnm{Guo}, \binits{L.}},
\bauthor{\bsnm{Gu}, \binits{S.}},
\bauthor{\bsnm{Wang}, \binits{O.}},
\bauthor{\bsnm{Zhang}, \binits{R.}},
\bauthor{\bsnm{Peters}, \binits{B.A.}},
\bauthor{\bsnm{Fan}, \binits{G.}},
\bauthor{\bsnm{Liu}, \binits{X.}},
\bauthor{\bsnm{Xu}, \binits{X.}},
\bauthor{\bsnm{Deng}, \binits{L.}}, \betal:
\batitle{Tgs-gapcloser: a fast and accurate gap closer for large genomes with low coverage of error-prone long reads}.
\bjtitle{GigaScience}
\bvolume{9}(\bissue{9}),
\bfpage{094}
(\byear{2020})
\doiurl{10.1093/gigascience/giaa094}
\end{barticle}
\endbibitem

\bibitem[\protect\citeauthoryear{Walker et~al.}{2014}]{walker2014pilon}
\begin{barticle}
\bauthor{\bsnm{Walker}, \binits{B.J.}},
\bauthor{\bsnm{Abeel}, \binits{T.}},
\bauthor{\bsnm{Shea}, \binits{T.}},
\bauthor{\bsnm{Priest}, \binits{M.}},
\bauthor{\bsnm{Abouelliel}, \binits{A.}},
\bauthor{\bsnm{Sakthikumar}, \binits{S.}},
\bauthor{\bsnm{Cuomo}, \binits{C.A.}},
\bauthor{\bsnm{Zeng}, \binits{Q.}},
\bauthor{\bsnm{Wortman}, \binits{J.}},
\bauthor{\bsnm{Young}, \binits{S.K.}}, \betal:
\batitle{Pilon: an integrated tool for comprehensive microbial variant detection and genome assembly improvement}.
\bjtitle{PLoS One}
\bvolume{9}(\bissue{11}),
\bfpage{112963}
(\byear{2014})
\doiurl{10.1371/journal.pone.0112963}
\end{barticle}
\endbibitem

\bibitem[\protect\citeauthoryear{Li}{2023}]{li2023protein}
\begin{barticle}
\bauthor{\bsnm{Li}, \binits{H.}}:
\batitle{Protein-to-genome alignment with miniprot}.
\bjtitle{Bioinformatics}
\bvolume{39}(\bissue{1}),
\bfpage{014}
(\byear{2023})
\doiurl{10.1093/bioinformatics/btad014}
\end{barticle}
\endbibitem

\bibitem[\protect\citeauthoryear{Stanke et~al.}{2006}]{stanke2006augustus}
\begin{barticle}
\bauthor{\bsnm{Stanke}, \binits{M.}},
\bauthor{\bsnm{Keller}, \binits{O.}},
\bauthor{\bsnm{Gunduz}, \binits{I.}},
\bauthor{\bsnm{Hayes}, \binits{A.}},
\bauthor{\bsnm{Waack}, \binits{S.}},
\bauthor{\bsnm{Morgenstern}, \binits{B.}}:
\batitle{Augustus: ab initio prediction of alternative transcripts}.
\bjtitle{Nucleic Acids Res.}
\bvolume{34}(\bissue{suppl\_2}),
\bfpage{435}--\blpage{439}
(\byear{2006})
\doiurl{10.1093/nar/gkl200}
\end{barticle}
\endbibitem

\bibitem[\protect\citeauthoryear{Burge and Karlin}{1997}]{burge1997prediction}
\begin{barticle}
\bauthor{\bsnm{Burge}, \binits{C.}},
\bauthor{\bsnm{Karlin}, \binits{S.}}:
\batitle{Prediction of complete gene structures in human genomic dna}.
\bjtitle{J. Mol. Biol.}
\bvolume{268}(\bissue{1}),
\bfpage{78}--\blpage{94}
(\byear{1997})
\doiurl{10.1006/jmbi.1997.0951}
\end{barticle}
\endbibitem

\bibitem[\protect\citeauthoryear{Kim et~al.}{2019}]{kim2019graph}
\begin{barticle}
\bauthor{\bsnm{Kim}, \binits{D.}},
\bauthor{\bsnm{Paggi}, \binits{J.M.}},
\bauthor{\bsnm{Park}, \binits{C.}},
\bauthor{\bsnm{Bennett}, \binits{C.}},
\bauthor{\bsnm{Salzberg}, \binits{S.L.}}:
\batitle{Graph-based genome alignment and genotyping with hisat2 and hisat-genotype}.
\bjtitle{Nat. Biotechnol.}
\bvolume{37}(\bissue{8}),
\bfpage{907}--\blpage{915}
(\byear{2019})
\doiurl{10.1038/s41587-019-0201-4}
\end{barticle}
\endbibitem

\bibitem[\protect\citeauthoryear{Pertea et~al.}{2015}]{pertea2015stringtie}
\begin{barticle}
\bauthor{\bsnm{Pertea}, \binits{M.}},
\bauthor{\bsnm{Pertea}, \binits{G.M.}},
\bauthor{\bsnm{Antonescu}, \binits{C.M.}},
\bauthor{\bsnm{Chang}, \binits{T.-C.}},
\bauthor{\bsnm{Mendell}, \binits{J.T.}},
\bauthor{\bsnm{Salzberg}, \binits{S.L.}}:
\batitle{Stringtie enables improved reconstruction of a transcriptome from rna-seq reads}.
\bjtitle{Nat. Biotechnol.}
\bvolume{33}(\bissue{3}),
\bfpage{290}--\blpage{295}
(\byear{2015})
\doiurl{10.1038/nbt.3122}
\end{barticle}
\endbibitem

\bibitem[\protect\citeauthoryear{Holt and Yandell}{2011}]{holt2011maker2}
\begin{barticle}
\bauthor{\bsnm{Holt}, \binits{C.}},
\bauthor{\bsnm{Yandell}, \binits{M.}}:
\batitle{Maker2: an annotation pipeline and genome-database management tool for second-generation genome projects}.
\bjtitle{BMC Bioinformatics}
\bvolume{12},
\bfpage{1}--\blpage{14}
(\byear{2011})
\doiurl{10.1186/1471-2105-12-491}
\end{barticle}
\endbibitem

\bibitem[\protect\citeauthoryear{Sun et~al.}{2024}]{sun2024chromosomal}
\begin{barticle}
\bauthor{\bsnm{Sun}, \binits{S.}},
\bauthor{\bsnm{Han}, \binits{X.}},
\bauthor{\bsnm{Han}, \binits{Z.}},
\bauthor{\bsnm{Liu}, \binits{Q.}}:
\batitle{Chromosomal-scale genome assembly and annotation of the land slug (meghimatium bilineatum)}.
\bjtitle{Sci. Data}
\bvolume{11}(\bissue{1}),
\bfpage{35}
(\byear{2024})
\doiurl{10.1038/s41597-023-02893-7}
\end{barticle}
\endbibitem

\bibitem[\protect\citeauthoryear{Lowe and Eddy}{1997}]{lowe1997trnascan}
\begin{barticle}
\bauthor{\bsnm{Lowe}, \binits{T.M.}},
\bauthor{\bsnm{Eddy}, \binits{S.R.}}:
\batitle{trnascan-se: a program for improved detection of transfer rna genes in genomic sequence}.
\bjtitle{Nucleic Acids Res.}
\bvolume{25}(\bissue{5}),
\bfpage{955}--\blpage{964}
(\byear{1997})
\doiurl{10.1093/nar/25.5.955}
\end{barticle}
\endbibitem

\bibitem[\protect\citeauthoryear{Altschul et~al.}{1990}]{altschul1990basic}
\begin{barticle}
\bauthor{\bsnm{Altschul}, \binits{S.F.}},
\bauthor{\bsnm{Gish}, \binits{W.}},
\bauthor{\bsnm{Miller}, \binits{W.}},
\bauthor{\bsnm{Myers}, \binits{E.W.}},
\bauthor{\bsnm{Lipman}, \binits{D.J.}}:
\batitle{Basic local alignment search tool}.
\bjtitle{J. Mol. Biol.}
\bvolume{215}(\bissue{3}),
\bfpage{403}--\blpage{410}
(\byear{1990})
\doiurl{10.32388/rhq6vj}
\end{barticle}
\endbibitem

\bibitem[\protect\citeauthoryear{Griffiths-Jones et~al.}{2005}]{griffiths2005rfam}
\begin{barticle}
\bauthor{\bsnm{Griffiths-Jones}, \binits{S.}},
\bauthor{\bsnm{Moxon}, \binits{S.}},
\bauthor{\bsnm{Marshall}, \binits{M.}},
\bauthor{\bsnm{Khanna}, \binits{A.}},
\bauthor{\bsnm{Eddy}, \binits{S.R.}},
\bauthor{\bsnm{Bateman}, \binits{A.}}:
\batitle{Rfam: annotating non-coding rnas in complete genomes}.
\bjtitle{Nucleic Acids Res.}
\bvolume{33}(\bissue{suppl\_1}),
\bfpage{121}--\blpage{124}
(\byear{2005})
\doiurl{10.1093/nar/gki081}
\end{barticle}
\endbibitem

\bibitem[\protect\citeauthoryear{Bu et~al.}{2021}]{bu2021kobas}
\begin{barticle}
\bauthor{\bsnm{Bu}, \binits{D.}},
\bauthor{\bsnm{Luo}, \binits{H.}},
\bauthor{\bsnm{Huo}, \binits{P.}},
\bauthor{\bsnm{Wang}, \binits{Z.}},
\bauthor{\bsnm{Zhang}, \binits{S.}},
\bauthor{\bsnm{He}, \binits{Z.}},
\bauthor{\bsnm{Wu}, \binits{Y.}},
\bauthor{\bsnm{Zhao}, \binits{L.}},
\bauthor{\bsnm{Liu}, \binits{J.}},
\bauthor{\bsnm{Guo}, \binits{J.}}, \betal:
\batitle{Kobas-i: intelligent prioritization and exploratory visualization of biological functions for gene enrichment analysis}.
\bjtitle{Nucleic Acids Res.}
\bvolume{49}(\bissue{W1}),
\bfpage{317}--\blpage{325}
(\byear{2021})
\doiurl{10.1093/nar/gkab447}
\end{barticle}
\endbibitem

\bibitem[\protect\citeauthoryear{Jones et~al.}{2014}]{jones2014interproscan}
\begin{barticle}
\bauthor{\bsnm{Jones}, \binits{P.}},
\bauthor{\bsnm{Binns}, \binits{D.}},
\bauthor{\bsnm{Chang}, \binits{H.-Y.}},
\bauthor{\bsnm{Fraser}, \binits{M.}},
\bauthor{\bsnm{Li}, \binits{W.}},
\bauthor{\bsnm{McAnulla}, \binits{C.}},
\bauthor{\bsnm{McWilliam}, \binits{H.}},
\bauthor{\bsnm{Maslen}, \binits{J.}},
\bauthor{\bsnm{Mitchell}, \binits{A.}},
\bauthor{\bsnm{Nuka}, \binits{G.}}, \betal:
\batitle{Interproscan 5: genome-scale protein function classification}.
\bjtitle{Bioinformatics}
\bvolume{30}(\bissue{9}),
\bfpage{1236}--\blpage{1240}
(\byear{2014})
\doiurl{10.1093/bioinformatics/btu031}
\end{barticle}
\endbibitem

\bibitem[\protect\citeauthoryear{Mistry et~al.}{2013}]{mistry2013challenges}
\begin{barticle}
\bauthor{\bsnm{Mistry}, \binits{J.}},
\bauthor{\bsnm{Finn}, \binits{R.D.}},
\bauthor{\bsnm{Eddy}, \binits{S.R.}},
\bauthor{\bsnm{Bateman}, \binits{A.}},
\bauthor{\bsnm{Punta}, \binits{M.}}:
\batitle{Challenges in homology search: Hmmer3 and convergent evolution of coiled-coil regions}.
\bjtitle{Nucleic Acids Res.}
\bvolume{41}(\bissue{12}),
\bfpage{121}--\blpage{121}
(\byear{2013})
\doiurl{10.1093/nar/gkt263}
\end{barticle}
\endbibitem

\bibitem[\protect\citeauthoryear{Li and Durbin}{2009}]{li2009fast}
\begin{barticle}
\bauthor{\bsnm{Li}, \binits{H.}},
\bauthor{\bsnm{Durbin}, \binits{R.}}:
\batitle{Fast and accurate short read alignment with burrows--wheeler transform}.
\bjtitle{Bioinformatics}
\bvolume{25}(\bissue{14}),
\bfpage{1754}--\blpage{1760}
(\byear{2009})
\doiurl{10.1093/bioinformatics/btp324}
\end{barticle}
\endbibitem

\bibitem[\protect\citeauthoryear{Freed et~al.}{2017}]{freed2017sentieon}
\begin{botherref}
\oauthor{\bsnm{Freed}, \binits{D.}},
\oauthor{\bsnm{Aldana}, \binits{R.}},
\oauthor{\bsnm{Weber}, \binits{J.A.}},
\oauthor{\bsnm{Edwards}, \binits{J.S.}}:
The sentieon genomics tools--a fast and accurate solution to variant calling from next-generation sequence data.
BioRxiv,
115717
(2017)
\doiurl{10.1101/115717}
\end{botherref}
\endbibitem

\bibitem[\protect\citeauthoryear{DePristo et~al.}{2011}]{DePristo_2011}
\begin{barticle}
\bauthor{\bsnm{DePristo}, \binits{M.A.}},
\bauthor{\bsnm{Banks}, \binits{E.}},
\bauthor{\bsnm{Poplin}, \binits{R.}},
\bauthor{\bsnm{Garimella}, \binits{K.V.}},
\bauthor{\bsnm{Maguire}, \binits{J.R.}},
\bauthor{\bsnm{Hartl}, \binits{C.}},
\bauthor{\bsnm{Philippakis}, \binits{A.A.}},
\bauthor{\bsnm{Angel}, \binits{G.}},
\bauthor{\bsnm{Rivas}, \binits{M.A.}},
\bauthor{\bsnm{Hanna}, \binits{M.}},
\bauthor{\bsnm{McKenna}, \binits{A.}},
\bauthor{\bsnm{Fennell}, \binits{T.J.}},
\bauthor{\bsnm{Kernytsky}, \binits{A.M.}},
\bauthor{\bsnm{Sivachenko}, \binits{A.Y.}},
\bauthor{\bsnm{Cibulskis}, \binits{K.}},
\bauthor{\bsnm{Gabriel}, \binits{S.B.}},
\bauthor{\bsnm{Altshuler}, \binits{D.}},
\bauthor{\bsnm{Daly}, \binits{M.J.}}:
\batitle{A framework for variation discovery and genotyping using next-generation dna sequencing data}.
\bjtitle{Nature Genetics}
\bvolume{43}(\bissue{5}),
\bfpage{491}--\blpage{498}
(\byear{2011})
\doiurl{10.1038/ng.806}
\end{barticle}
\endbibitem

\bibitem[\protect\citeauthoryear{Yang et~al.}{2014}]{yang2014combining}
\begin{barticle}
\bauthor{\bsnm{Yang}, \binits{W.}},
\bauthor{\bsnm{Guo}, \binits{Z.}},
\bauthor{\bsnm{Huang}, \binits{C.}},
\bauthor{\bsnm{Duan}, \binits{L.}},
\bauthor{\bsnm{Chen}, \binits{G.}},
\bauthor{\bsnm{Jiang}, \binits{N.}},
\bauthor{\bsnm{Fang}, \binits{W.}},
\bauthor{\bsnm{Feng}, \binits{H.}},
\bauthor{\bsnm{Xie}, \binits{W.}},
\bauthor{\bsnm{Lian}, \binits{X.}}, \betal:
\batitle{Combining high-throughput phenotyping and genome-wide association studies to reveal natural genetic variation in rice}.
\bjtitle{Nat. Commun.}
\bvolume{5}(\bissue{1}),
\bfpage{5087}
(\byear{2014})
\doiurl{10.1038/ncomms6087}
\end{barticle}
\endbibitem

\bibitem[\protect\citeauthoryear{Isensee et~al.}{2021}]{isensee2021nnu}
\begin{barticle}
\bauthor{\bsnm{Isensee}, \binits{F.}},
\bauthor{\bsnm{Jaeger}, \binits{P.F.}},
\bauthor{\bsnm{Kohl}, \binits{S.A.A.}},
\bauthor{\bsnm{Petersen}, \binits{J.}},
\bauthor{\bsnm{Maier-Hein}, \binits{K.H.}}:
\batitle{nnu-net: a self-configuring method for deep learning-based biomedical image segmentation}.
\bjtitle{Nat. Methods}
\bvolume{18}(\bissue{2}),
\bfpage{203}--\blpage{211}
(\byear{2021})
\doiurl{10.1038/s41592-020-01008-z}
\end{barticle}
\endbibitem

\bibitem[\protect\citeauthoryear{Gjerlaug-Enger et~al.}{2012}]{gjerlaug2012genetic}
\begin{barticle}
\bauthor{\bsnm{Gjerlaug-Enger}, \binits{E.}},
\bauthor{\bsnm{Kongsro}, \binits{J.}},
\bauthor{\bsnm{{\O}deg{\aa}rd}, \binits{J.}},
\bauthor{\bsnm{Aass}, \binits{L.}},
\bauthor{\bsnm{Vangen}, \binits{O.}}:
\batitle{Genetic parameters between slaughter pig efficiency and growth rate of different body tissues estimated by computed tomography in live boars of landrace and duroc}.
\bjtitle{Anim.}
\bvolume{6}(\bissue{1}),
\bfpage{9}--\blpage{18}
(\byear{2012})
\doiurl{10.1017/s1751731111001455}
\end{barticle}
\endbibitem

\bibitem[\protect\citeauthoryear{Karras et~al.}{2019}]{karras2019style}
\begin{bchapter}
\bauthor{\bsnm{Karras}, \binits{T.}},
\bauthor{\bsnm{Laine}, \binits{S.}},
\bauthor{\bsnm{Aila}, \binits{T.}}:
\bctitle{A style-based generator architecture for generative adversarial networks}.
In: \bbtitle{2019 IEEE/CVF Conference on Computer Vision and Pattern Recognition (CVPR)},
pp. \bfpage{4396}--\blpage{4405}
(\byear{2019}).
\doiurl{10.1109/CVPR.2019.00453}
\end{bchapter}
\endbibitem

\bibitem[\protect\citeauthoryear{Karras et~al.}{2020}]{karras2020training}
\begin{bchapter}
\bauthor{\bsnm{Karras}, \binits{T.}},
\bauthor{\bsnm{Aittala}, \binits{M.}},
\bauthor{\bsnm{Hellsten}, \binits{J.}},
\bauthor{\bsnm{Laine}, \binits{S.}},
\bauthor{\bsnm{Lehtinen}, \binits{J.}},
\bauthor{\bsnm{Aila}, \binits{T.}}:
\bctitle{Training generative adversarial networks with limited data}.
In: \bbtitle{Proceedings of the 34th International Conference on Neural Information Processing Systems}.
\bsertitle{NIPS '20}.
\bpublisher{Curran Associates Inc.},
\blocation{Red Hook, NY, USA}
(\byear{2020}).
\doiurl{10.5555/3495724.3496739}
\end{bchapter}
\endbibitem

\bibitem[\protect\citeauthoryear{Kim et~al.}{2021}]{kim2021exploiting}
\begin{bchapter}
\bauthor{\bsnm{Kim}, \binits{H.}},
\bauthor{\bsnm{Choi}, \binits{Y.}},
\bauthor{\bsnm{Kim}, \binits{J.}},
\bauthor{\bsnm{Yoo}, \binits{S.}},
\bauthor{\bsnm{Uh}, \binits{Y.}}:
\bctitle{Exploiting spatial dimensions of latent in gan for real-time image editing}.
In: \bbtitle{Proceedings of the IEEE/CVF Conference on Computer Vision and Pattern Recognition (CVPR)},
pp. \bfpage{852}--\blpage{861}.
\bpublisher{IEEE Computer Society},
\blocation{Virtual}
(\byear{2021}).
\doiurl{10.1109/CVPR46437.2021.00091} .
\bcomment{IEEE/CVF}
\end{bchapter}
\endbibitem

\bibitem[\protect\citeauthoryear{Liu et~al.}{2021}]{liu2021swin}
\begin{bchapter}
\bauthor{\bsnm{Liu}, \binits{Z.}},
\bauthor{\bsnm{Lin}, \binits{Y.}},
\bauthor{\bsnm{Cao}, \binits{Y.}},
\bauthor{\bsnm{Hu}, \binits{H.}},
\bauthor{\bsnm{Wei}, \binits{Y.}},
\bauthor{\bsnm{Zhang}, \binits{Z.}},
\bauthor{\bsnm{Lin}, \binits{S.}},
\bauthor{\bsnm{Guo}, \binits{B.}}:
\bctitle{Swin transformer: Hierarchical vision transformer using shifted windows}.
In: \bbtitle{Proceedings of the IEEE/CVF International Conference on Computer Vision (ICCV)},
pp. \bfpage{10012}--\blpage{10022}.
\bpublisher{IEEE Computer Society},
\blocation{Virtual}
(\byear{2021}).
\doiurl{10.1109/ICCV48922.2021.00986} .
\bcomment{IEEE/CVF}
\end{bchapter}
\endbibitem

\bibitem[\protect\citeauthoryear{Li et~al.}{2022}]{li2022grounded}
\begin{bchapter}
\bauthor{\bsnm{Li}, \binits{L.H.}},
\bauthor{\bsnm{Zhang}, \binits{P.}},
\bauthor{\bsnm{Zhang}, \binits{H.}},
\bauthor{\bsnm{Yang}, \binits{J.}},
\bauthor{\bsnm{Li}, \binits{C.}},
\bauthor{\bsnm{Zhong}, \binits{Y.}},
\bauthor{\bsnm{Wang}, \binits{L.}},
\bauthor{\bsnm{Yuan}, \binits{L.}},
\bauthor{\bsnm{Zhang}, \binits{L.}},
\bauthor{\bsnm{Hwang}, \binits{J.-N.}}, \betal:
\bctitle{Grounded language-image pre-training}.
In: \bbtitle{Proceedings of the IEEE/CVF Conference on Computer Vision and Pattern Recognition (CVPR)},
pp. \bfpage{10965}--\blpage{10975}.
\bpublisher{IEEE Computer Society},
\blocation{New Orleans, LA, USA}
(\byear{2022}).
\doiurl{10.1109/CVPR52688.2022.01069} .
\bcomment{IEEE/CVF}
\end{bchapter}
\endbibitem

\bibitem[\protect\citeauthoryear{Russakovsky et~al.}{2015}]{russakovsky2015imagenet}
\begin{barticle}
\bauthor{\bsnm{Russakovsky}, \binits{O.}},
\bauthor{\bsnm{Deng}, \binits{J.}},
\bauthor{\bsnm{Su}, \binits{H.}},
\bauthor{\bsnm{Krause}, \binits{J.}},
\bauthor{\bsnm{Satheesh}, \binits{S.}},
\bauthor{\bsnm{Ma}, \binits{S.}},
\bauthor{\bsnm{Huang}, \binits{Z.}},
\bauthor{\bsnm{Karpathy}, \binits{A.}},
\bauthor{\bsnm{Khosla}, \binits{A.}},
\bauthor{\bsnm{Bernstein}, \binits{M.}}, \betal:
\batitle{Imagenet large scale visual recognition challenge}.
\bjtitle{Int. J. Comput. Vis.}
\bvolume{115}(\bissue{3}),
\bfpage{211}--\blpage{252}
(\byear{2015})
\doiurl{10.1007/s11263-015-0816-y}
\end{barticle}
\endbibitem

\bibitem[\protect\citeauthoryear{Zhang et~al.}{2018}]{zhang2018unreasonable}
\begin{bchapter}
\bauthor{\bsnm{Zhang}, \binits{R.}},
\bauthor{\bsnm{Isola}, \binits{P.}},
\bauthor{\bsnm{Efros}, \binits{A.A.}},
\bauthor{\bsnm{Shechtman}, \binits{E.}},
\bauthor{\bsnm{Wang}, \binits{O.}}:
\bctitle{The unreasonable effectiveness of deep features as a perceptual metric}.
In: \bbtitle{Proceedings of the IEEE Conference on Computer Vision and Pattern Recognition (CVPR)},
pp. \bfpage{586}--\blpage{595}.
\bpublisher{IEEE Computer Society},
\blocation{Salt Lake City, UT, USA}
(\byear{2018}).
\doiurl{10.1109/CVPR.2018.00068} .
\bcomment{IEEE}
\end{bchapter}
\endbibitem

\bibitem[\protect\citeauthoryear{Liu and Deng}{2015}]{simonyan2014very}
\begin{bchapter}
\bauthor{\bsnm{Liu}, \binits{S.}},
\bauthor{\bsnm{Deng}, \binits{W.}}:
\bctitle{Very deep convolutional neural network based image classification using small training sample size}.
In: \bbtitle{2015 3rd IAPR Asian Conference on Pattern Recognition (ACPR)},
pp. \bfpage{730}--\blpage{734}
(\byear{2015}).
\doiurl{10.1109/ACPR.2015.7486599}
\end{bchapter}
\endbibitem

\bibitem[\protect\citeauthoryear{Hyv{\"a}rinen and Oja}{2000}]{hyvarinen2000independent}
\begin{barticle}
\bauthor{\bsnm{Hyv{\"a}rinen}, \binits{A.}},
\bauthor{\bsnm{Oja}, \binits{E.}}:
\batitle{Independent component analysis: algorithms and applications}.
\bjtitle{Neural Netw.}
\bvolume{13}(\bissue{4-5}),
\bfpage{411}--\blpage{430}
(\byear{2000})
\doiurl{10.1016/s0893-6080(00)00026-5}
\end{barticle}
\endbibitem

\bibitem[\protect\citeauthoryear{Langlois et~al.}{2010}]{langlois2010introduction}
\begin{barticle}
\bauthor{\bsnm{Langlois}, \binits{D.}},
\bauthor{\bsnm{Chartier}, \binits{S.}},
\bauthor{\bsnm{Gosselin}, \binits{D.}}:
\batitle{An introduction to independent component analysis: Infomax and fastica algorithms}.
\bjtitle{Tutorials Quant. Methods Psychol.}
\bvolume{6}(\bissue{1}),
\bfpage{31}--\blpage{38}
(\byear{2010})
\doiurl{10.20982/tqmp.06.1.p031}
\end{barticle}
\endbibitem

\bibitem[\protect\citeauthoryear{Hao and Ho}{2019}]{hao2019machine}
\begin{barticle}
\bauthor{\bsnm{Hao}, \binits{J.}},
\bauthor{\bsnm{Ho}, \binits{T.K.}}:
\batitle{Machine learning made easy: a review of scikit-learn package in python programming language}.
\bjtitle{J. Educ. Behav. Stat.}
\bvolume{44}(\bissue{3}),
\bfpage{348}--\blpage{361}
(\byear{2019})
\doiurl{10.3102/1076998619832248}
\end{barticle}
\endbibitem

\bibitem[\protect\citeauthoryear{Shen and Zhou}{2021}]{shen2021closed}
\begin{bchapter}
\bauthor{\bsnm{Shen}, \binits{Y.}},
\bauthor{\bsnm{Zhou}, \binits{B.}}:
\bctitle{Closed-form factorization of latent semantics in gans}.
In: \bbtitle{Proceedings of the IEEE/CVF Conference on Computer Vision and Pattern Recognition (CVPR)},
pp. \bfpage{1532}--\blpage{1540}.
\bpublisher{IEEE Computer Society},
\blocation{Virtual}
(\byear{2021}).
\doiurl{10.1109/CVPR46437.2021.00158} .
\bcomment{IEEE/CVF}
\end{bchapter}
\endbibitem

\bibitem[\protect\citeauthoryear{Song et~al.}{2023}]{song2023householder}
\begin{bchapter}
\bauthor{\bsnm{Song}, \binits{Y.}},
\bauthor{\bsnm{Zhang}, \binits{J.}},
\bauthor{\bsnm{Sebe}, \binits{N.}},
\bauthor{\bsnm{Wang}, \binits{W.}}:
\bctitle{Householder projector for unsupervised latent semantics discovery}.
In: \bbtitle{Proceedings of the IEEE/CVF International Conference on Computer Vision (ICCV)},
pp. \bfpage{7712}--\blpage{7722}.
\bpublisher{IEEE Computer Society},
\blocation{Paris, France}
(\byear{2023}).
\doiurl{10.1109/ICCV51070.2023.00709} .
\bcomment{IEEE/CVF}
\end{bchapter}
\endbibitem

\bibitem[\protect\citeauthoryear{Yin et~al.}{2021}]{yin2021rmvp}
\begin{barticle}
\bauthor{\bsnm{Yin}, \binits{L.}},
\bauthor{\bsnm{Zhang}, \binits{H.}},
\bauthor{\bsnm{Tang}, \binits{Z.}},
\bauthor{\bsnm{Xu}, \binits{J.}},
\bauthor{\bsnm{Yin}, \binits{D.}},
\bauthor{\bsnm{Zhang}, \binits{Z.}},
\bauthor{\bsnm{Yuan}, \binits{X.}},
\bauthor{\bsnm{Zhu}, \binits{M.}},
\bauthor{\bsnm{Zhao}, \binits{S.}},
\bauthor{\bsnm{Li}, \binits{X.}}, \betal:
\batitle{rmvp: a memory-efficient, visualization-enhanced, and parallel-accelerated tool for genome-wide association study}.
\bjtitle{Genomics Proteomics Bioinformatics}
\bvolume{19}(\bissue{4}),
\bfpage{619}--\blpage{628}
(\byear{2021})
\doiurl{10.1016/j.gpb.2020.10.007}
\end{barticle}
\endbibitem

\bibitem[\protect\citeauthoryear{Yin et~al.}{2023}]{yin2023hiblup}
\begin{barticle}
\bauthor{\bsnm{Yin}, \binits{L.}},
\bauthor{\bsnm{Zhang}, \binits{H.}},
\bauthor{\bsnm{Tang}, \binits{Z.}},
\bauthor{\bsnm{Yin}, \binits{D.}},
\bauthor{\bsnm{Fu}, \binits{Y.}},
\bauthor{\bsnm{Yuan}, \binits{X.}},
\bauthor{\bsnm{Li}, \binits{X.}},
\bauthor{\bsnm{Liu}, \binits{X.}},
\bauthor{\bsnm{Zhao}, \binits{S.}}:
\batitle{Hiblup: an integration of statistical models on the blup framework for efficient genetic evaluation using big genomic data}.
\bjtitle{Nucleic Acids Res.}
\bvolume{51}(\bissue{8}),
\bfpage{3501}--\blpage{3512}
(\byear{2023})
\doiurl{10.1093/nar/gkad074}
\end{barticle}
\endbibitem

\bibitem[\protect\citeauthoryear{Dong et~al.}{2021}]{dong2021ldblockshow}
\begin{barticle}
\bauthor{\bsnm{Dong}, \binits{S.-S.}},
\bauthor{\bsnm{He}, \binits{W.-M.}},
\bauthor{\bsnm{Ji}, \binits{J.-J.}},
\bauthor{\bsnm{Zhang}, \binits{C.}},
\bauthor{\bsnm{Guo}, \binits{Y.}},
\bauthor{\bsnm{Yang}, \binits{T.-L.}}:
\batitle{Ldblockshow: a fast and convenient tool for visualizing linkage disequilibrium and haplotype blocks based on variant call format files}.
\bjtitle{Brief. Bioinform.}
\bvolume{22}(\bissue{4}),
\bfpage{227}
(\byear{2021})
\doiurl{10.1093/bib/bbaa227}
\end{barticle}
\endbibitem

\bibitem[\protect\citeauthoryear{Purcell et~al.}{2007}]{purcell2007plink}
\begin{barticle}
\bauthor{\bsnm{Purcell}, \binits{S.}},
\bauthor{\bsnm{Neale}, \binits{B.}},
\bauthor{\bsnm{Todd-Brown}, \binits{K.}},
\bauthor{\bsnm{Thomas}, \binits{L.}},
\bauthor{\bsnm{Ferreira}, \binits{M.A.R.}},
\bauthor{\bsnm{Bender}, \binits{D.}},
\bauthor{\bsnm{Maller}, \binits{J.}},
\bauthor{\bsnm{Sklar}, \binits{P.}},
\bauthor{\bsnm{De~Bakker}, \binits{P.I.W.}},
\bauthor{\bsnm{Daly}, \binits{M.J.}}, \betal:
\batitle{Plink: a tool set for whole-genome association and population-based linkage analyses}.
\bjtitle{Am. J. Hum. Genet.}
\bvolume{81}(\bissue{3}),
\bfpage{559}--\blpage{575}
(\byear{2007})
\doiurl{10.1086/519795}
\end{barticle}
\endbibitem

\bibitem[\protect\citeauthoryear{Yang et~al.}{2015}]{yang2015genome}
\begin{barticle}
\bauthor{\bsnm{Yang}, \binits{W.}},
\bauthor{\bsnm{Guo}, \binits{Z.}},
\bauthor{\bsnm{Huang}, \binits{C.}},
\bauthor{\bsnm{Wang}, \binits{K.}},
\bauthor{\bsnm{Jiang}, \binits{N.}},
\bauthor{\bsnm{Feng}, \binits{H.}},
\bauthor{\bsnm{Chen}, \binits{G.}},
\bauthor{\bsnm{Liu}, \binits{Q.}},
\bauthor{\bsnm{Xiong}, \binits{L.}}:
\batitle{Genome-wide association study of rice (oryza sativa l.) leaf traits with a high-throughput leaf scorer}.
\bjtitle{J. Exp. Bot.}
\bvolume{66}(\bissue{18}),
\bfpage{5605}--\blpage{5615}
(\byear{2015})
\doiurl{10.1093/jxb/erv100}
\end{barticle}
\endbibitem

\bibitem[\protect\citeauthoryear{Guo et~al.}{2018}]{guo2018genome}
\begin{barticle}
\bauthor{\bsnm{Guo}, \binits{Z.}},
\bauthor{\bsnm{Yang}, \binits{W.}},
\bauthor{\bsnm{Chang}, \binits{Y.}},
\bauthor{\bsnm{Ma}, \binits{X.}},
\bauthor{\bsnm{Tu}, \binits{H.}},
\bauthor{\bsnm{Xiong}, \binits{F.}},
\bauthor{\bsnm{Jiang}, \binits{N.}},
\bauthor{\bsnm{Feng}, \binits{H.}},
\bauthor{\bsnm{Huang}, \binits{C.}},
\bauthor{\bsnm{Yang}, \binits{P.}}, \betal:
\batitle{Genome-wide association studies of image traits reveal genetic architecture of drought resistance in rice}.
\bjtitle{Mol. Plant}
\bvolume{11}(\bissue{6}),
\bfpage{789}--\blpage{805}
(\byear{2018})
\doiurl{10.1016/j.molp.2018.03.018}
\end{barticle}
\endbibitem

\bibitem[\protect\citeauthoryear{Zhao et~al.}{2021}]{zhao2021inferred}
\begin{barticle}
\bauthor{\bsnm{Zhao}, \binits{H.}},
\bauthor{\bsnm{Li}, \binits{J.}},
\bauthor{\bsnm{Yang}, \binits{L.}},
\bauthor{\bsnm{Qin}, \binits{G.}},
\bauthor{\bsnm{Xia}, \binits{C.}},
\bauthor{\bsnm{Xu}, \binits{X.}},
\bauthor{\bsnm{Su}, \binits{Y.}},
\bauthor{\bsnm{Liu}, \binits{Y.}},
\bauthor{\bsnm{Ming}, \binits{L.}},
\bauthor{\bsnm{Chen}, \binits{L.-L.}}, \betal:
\batitle{An inferred functional impact map of genetic variants in rice}.
\bjtitle{Mol. Plant}
\bvolume{14}(\bissue{9}),
\bfpage{1584}--\blpage{1599}
(\byear{2021})
\doiurl{10.1016/j.molp.2021.06.025}
\end{barticle}
\endbibitem

\bibitem[\protect\citeauthoryear{Do et~al.}{2021}]{do2021pyrice}
\begin{barticle}
\bauthor{\bsnm{Do}, \binits{Q.}},
\bauthor{\bsnm{Bich~Hai}, \binits{H.}},
\bauthor{\bsnm{Larmande}, \binits{P.}}:
\batitle{Pyrice: a python package for querying oryza sativa databases}.
\bjtitle{Bioinformatics}
\bvolume{37}(\bissue{7}),
\bfpage{1037}--\blpage{1038}
(\byear{2021})
\doiurl{10.1093/bioinformatics/btaa694}
\end{barticle}
\endbibitem

\bibitem[\protect\citeauthoryear{Fu et~al.}{2023}]{fu2023ianimal}
\begin{barticle}
\bauthor{\bsnm{Fu}, \binits{Y.}},
\bauthor{\bsnm{Liu}, \binits{H.}},
\bauthor{\bsnm{Dou}, \binits{J.}},
\bauthor{\bsnm{Wang}, \binits{Y.}},
\bauthor{\bsnm{Liao}, \binits{Y.}},
\bauthor{\bsnm{Huang}, \binits{X.}},
\bauthor{\bsnm{Tang}, \binits{Z.}},
\bauthor{\bsnm{Xu}, \binits{J.}},
\bauthor{\bsnm{Yin}, \binits{D.}},
\bauthor{\bsnm{Zhu}, \binits{S.}}, \betal:
\batitle{Ianimal: a cross-species omics knowledgebase for animals}.
\bjtitle{Nucleic Acids Res.}
\bvolume{51}(\bissue{D1}),
\bfpage{1312}--\blpage{1324}
(\byear{2023})
\doiurl{10.1093/nar/gkac936}
\end{barticle}
\endbibitem

\end{thebibliography}

\begin{appendices}
\bmhead{Supplementary information}
\setcounter{figure}{0}
\renewcommand{\thefigure}{S\arabic{figure}}
\renewcommand{\theHfigure}{S\arabic{figure}}
\begin{figure}[p]
    \centering
    \includegraphics[width=1.0\textwidth]{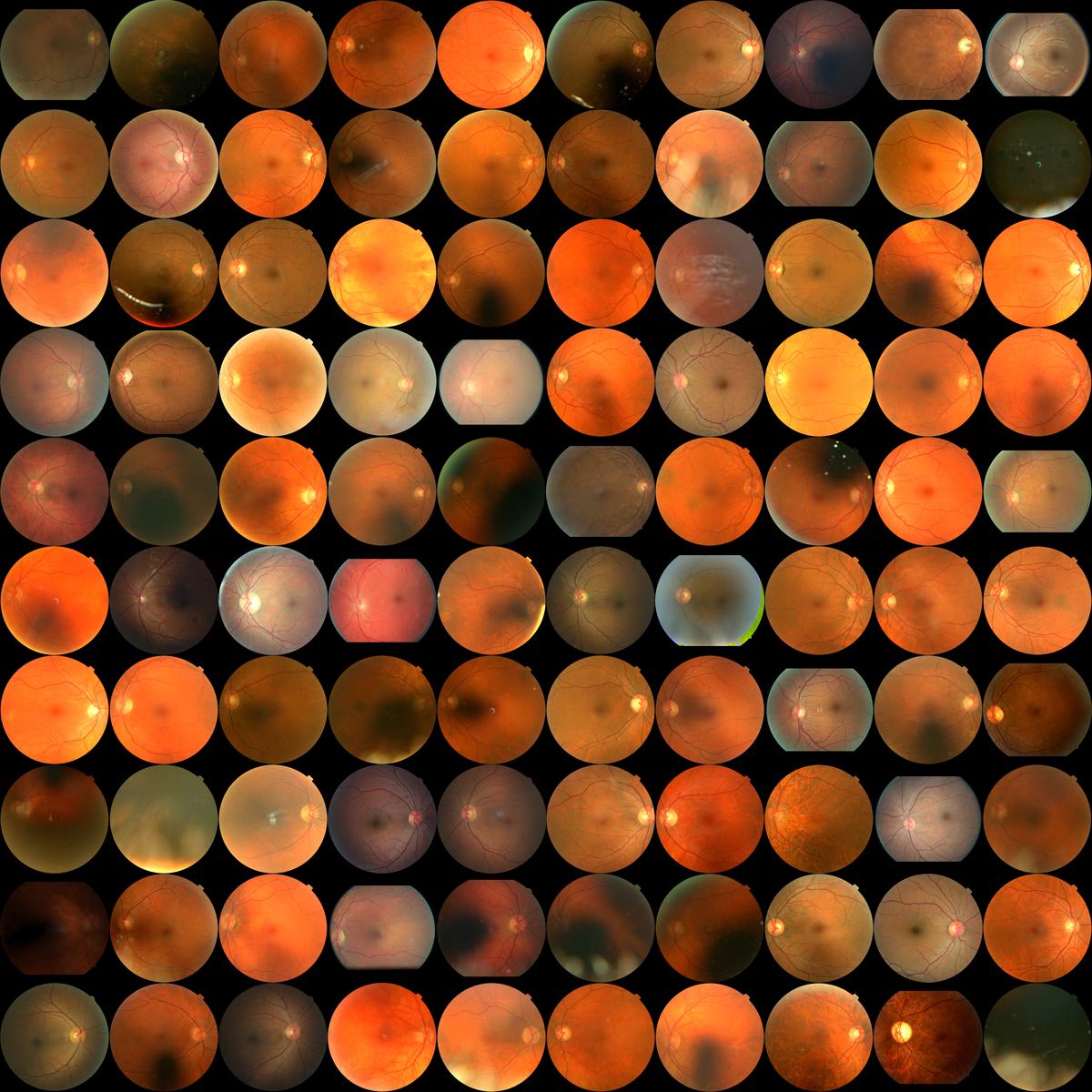}
    \caption{\textbf{Overview of the Human Retinal Fundus Dataset, related to Figure 1, Figure 2 and Methods "Data Collection and Preprocessing" section.} One hundred random images from dataset. 
    Retinal fundus images of left and right eyes from UKB and Kaggle datasets were integrated for model training.} \label{fig_S1}
\end{figure}
\clearpage

\begin{figure}[p]
    \centering
    \includegraphics[width=1.0\textwidth]{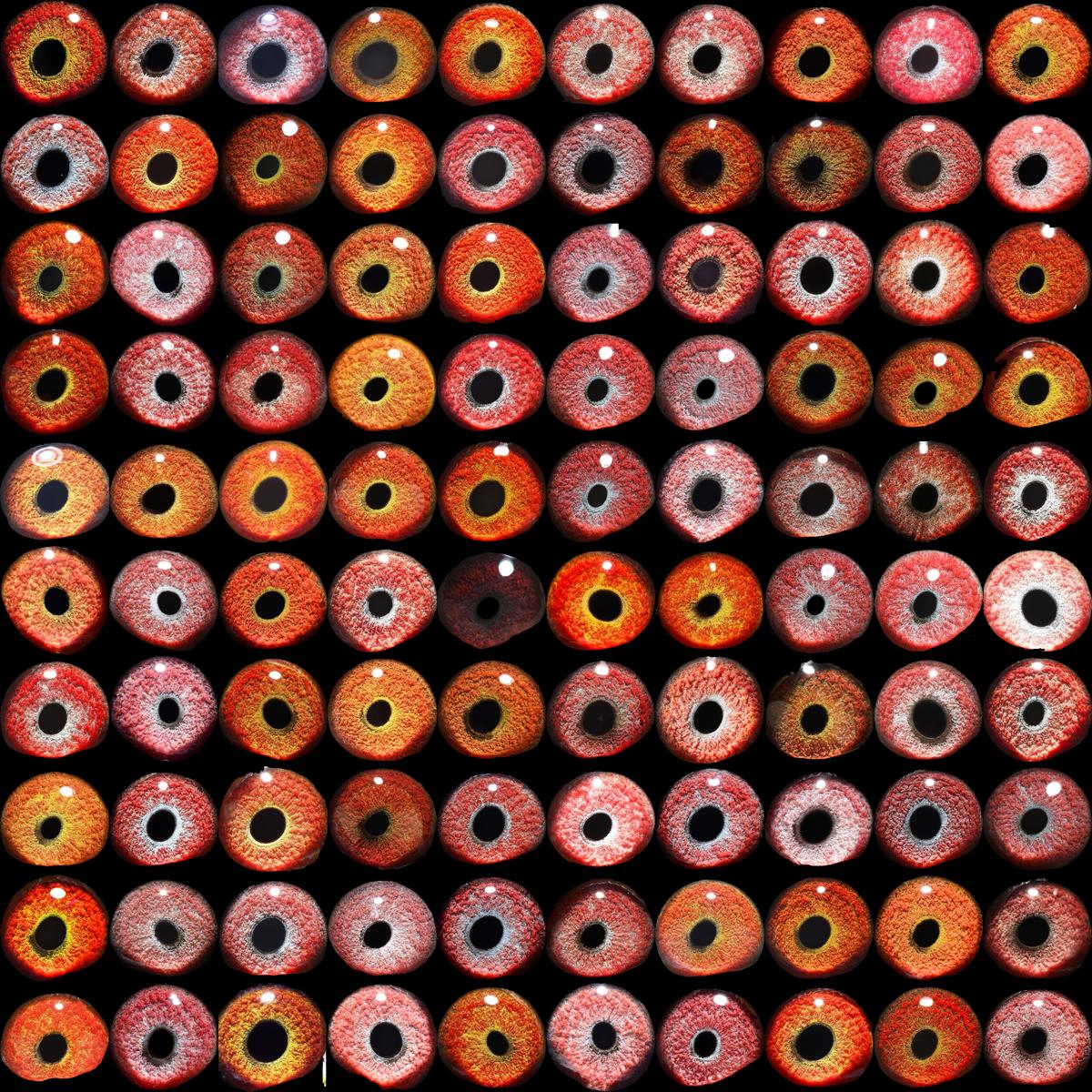}
    \caption{\textbf{Overview of the Domestic Pigeon Iris Dataset, related to Figure 1, Figure 2 and Methods "Data Collection and Preprocessing" section.} One hundred random images from dataset. Iris 
    segmented and scaled uniformly. The dataset is dominated by the gravel and pearl types; the bull type is less frequent.} \label{fig_S2}
\end{figure}
\clearpage

\begin{figure}[p]
    \centering
    \includegraphics[width=1.0\textwidth]{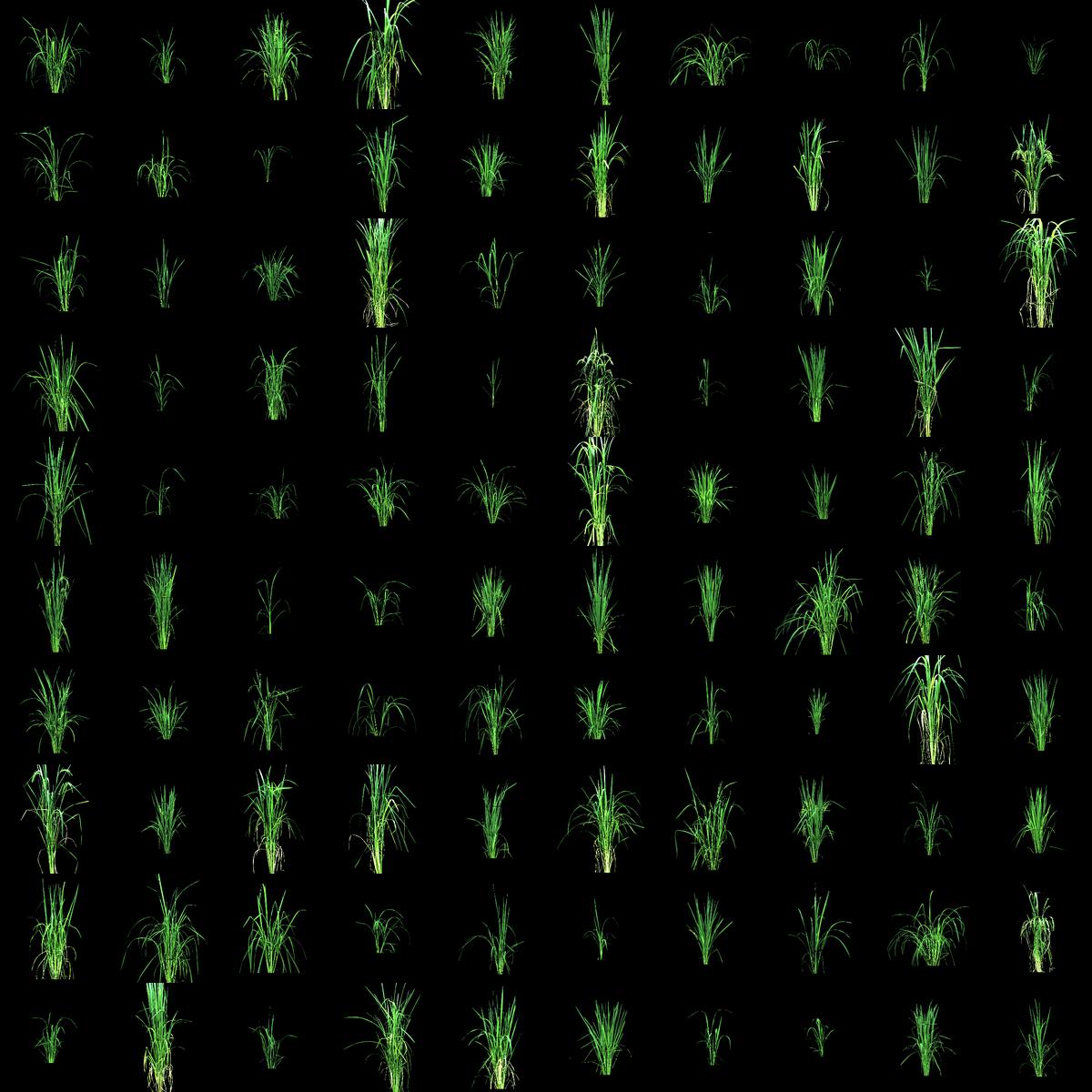}
    \caption{\textbf{Overview of the Potted Rice Dataset, related to Figure 1, Figure 2 and Methods "Data Collection and Preprocessing" section.} One hundred random images from dataset. 22 developmental 
    timepoints from seedling to maturity stages. Rice plants were segmented and centered in the middle of the image.} \label{fig_S3}
\end{figure}
\clearpage

\begin{figure}[p]
    \centering
    \includegraphics[width=1.0\textwidth]{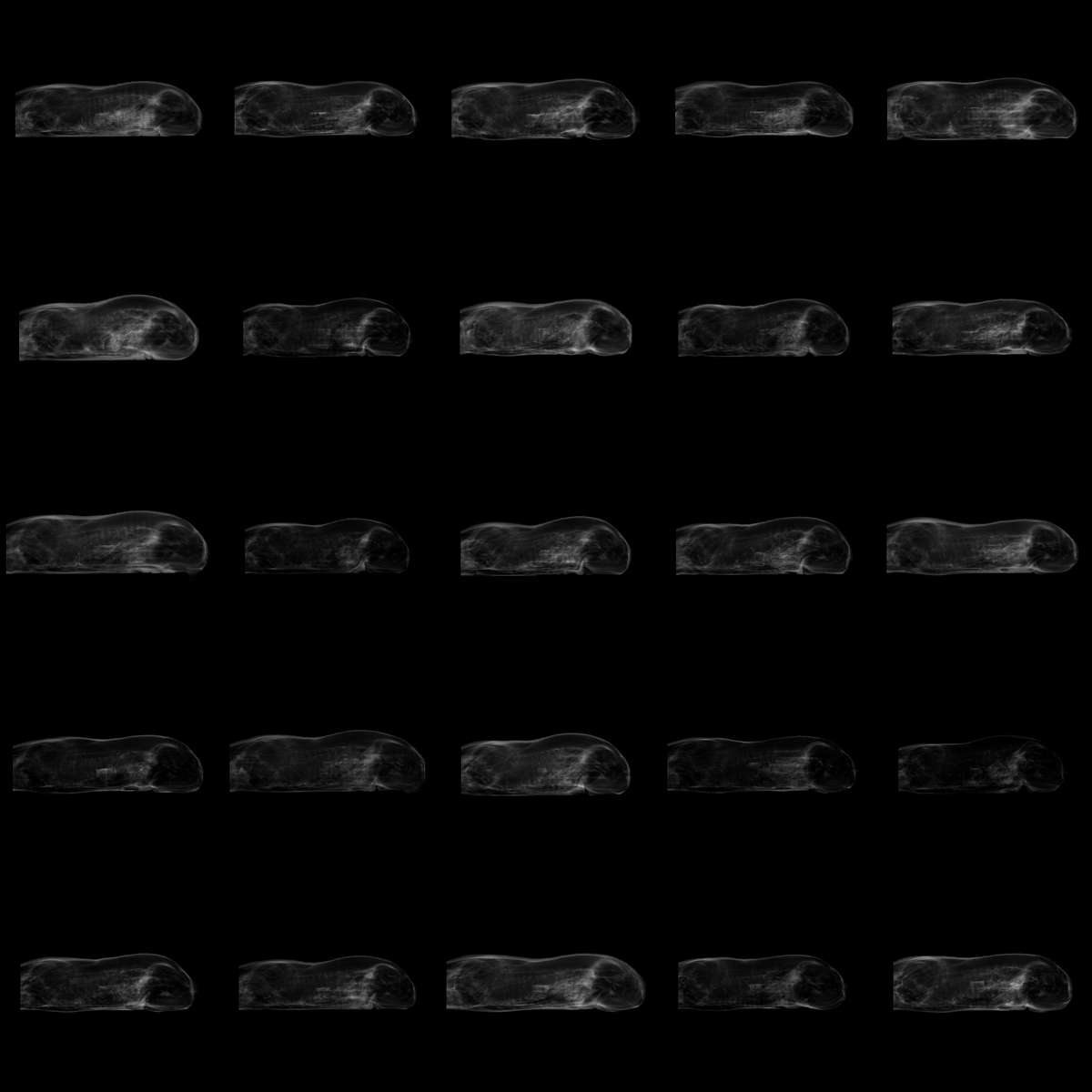}
    \caption{\textbf{Overview of the Swine CT Dataset, related to Figure 1, Figure 2 and Methods "Data Collection and Preprocessing" section.} One hundred random images from dataset. Pixel intensity in the 
    projection directly correlates with fat deposition density. The image highlights two main depots: backfat and abdominal (leaf) fat.} \label{fig_S4}
\end{figure}
\clearpage

\begin{sidewaysfigure}[p]
    \centering
    \includegraphics[width=1.0\textheight]{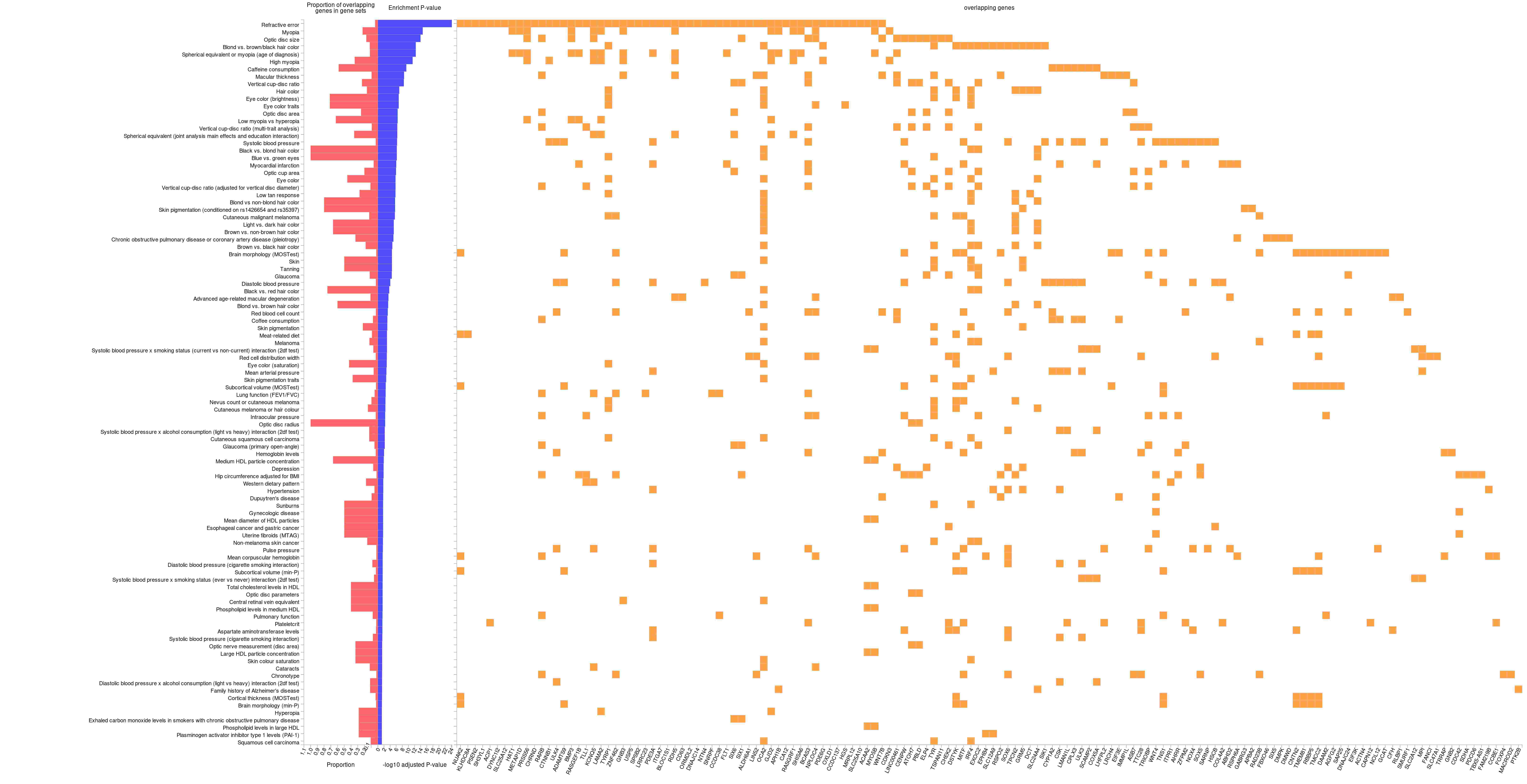}
    \caption{\textbf{GWAS Catalog enrichment analysis, related to Figure 2.} Enrichment analysis was performed with the GENE2FUNC module in FUMA.} \label{fig_S5}
\end{sidewaysfigure}
\clearpage

\begin{sidewaysfigure}[p]
    \centering
    \includegraphics[width=1.0\textheight]{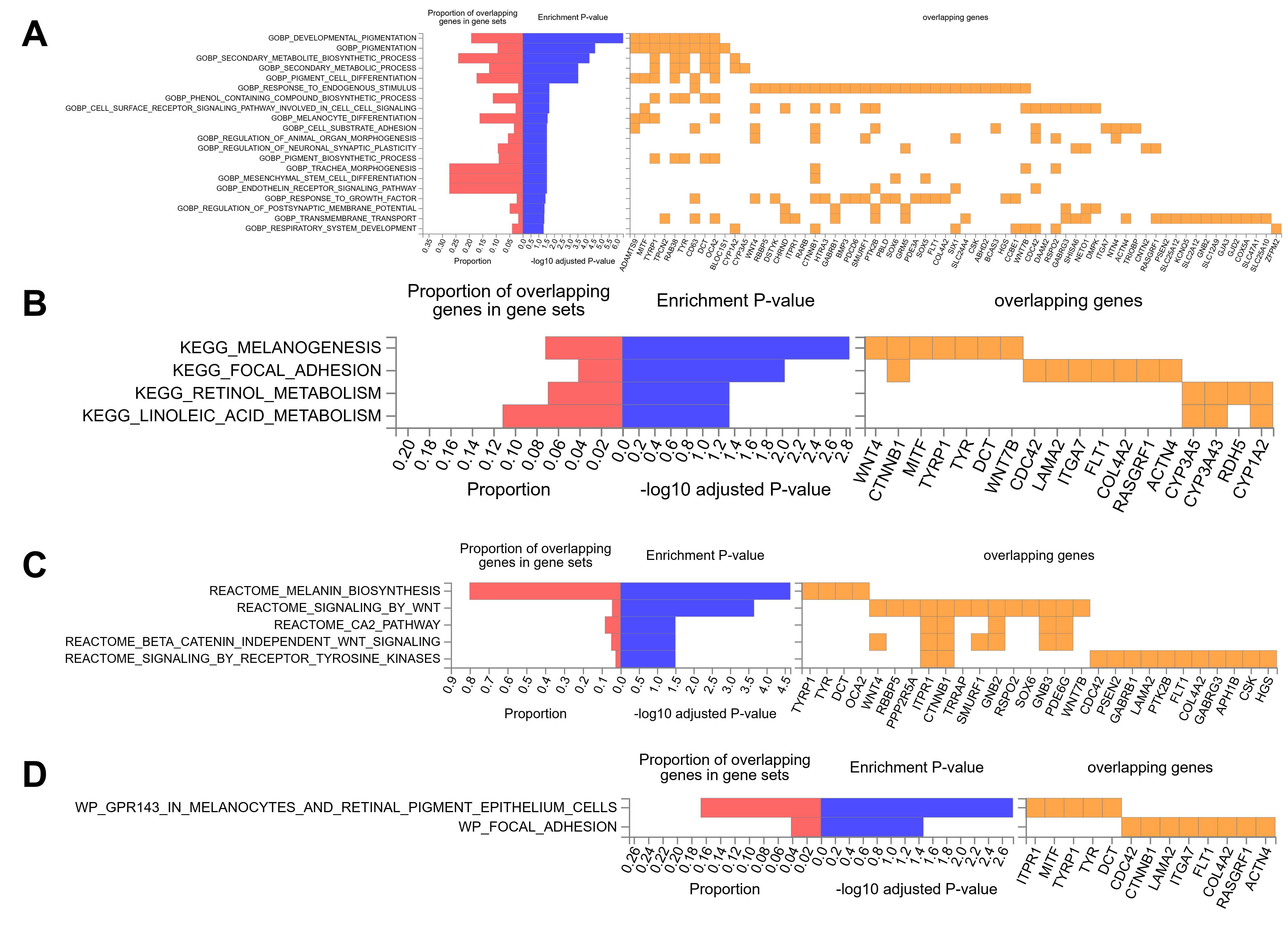}
    \caption{\textbf{Enrichment analyses, related to Figure 2.} In the enrichment analysis performed with the GENE2FUNC module in FUMA, the 
    top-ranked term based on \textit{p} value was related to pigmentation. \textbf{A,} Enrichment analyses of GO 
    Biological Process. \textbf{B,} Enrichment analyses of KEGG. \textbf{C,} Enrichment analyses of RECTOME. 
    \textbf{D,} Enrichment analyses of WikiPathways.} \label{fig_S6}
\end{sidewaysfigure}
\clearpage

\begin{figure}[p]
    \centering
    \includegraphics[width=1.0\textwidth]{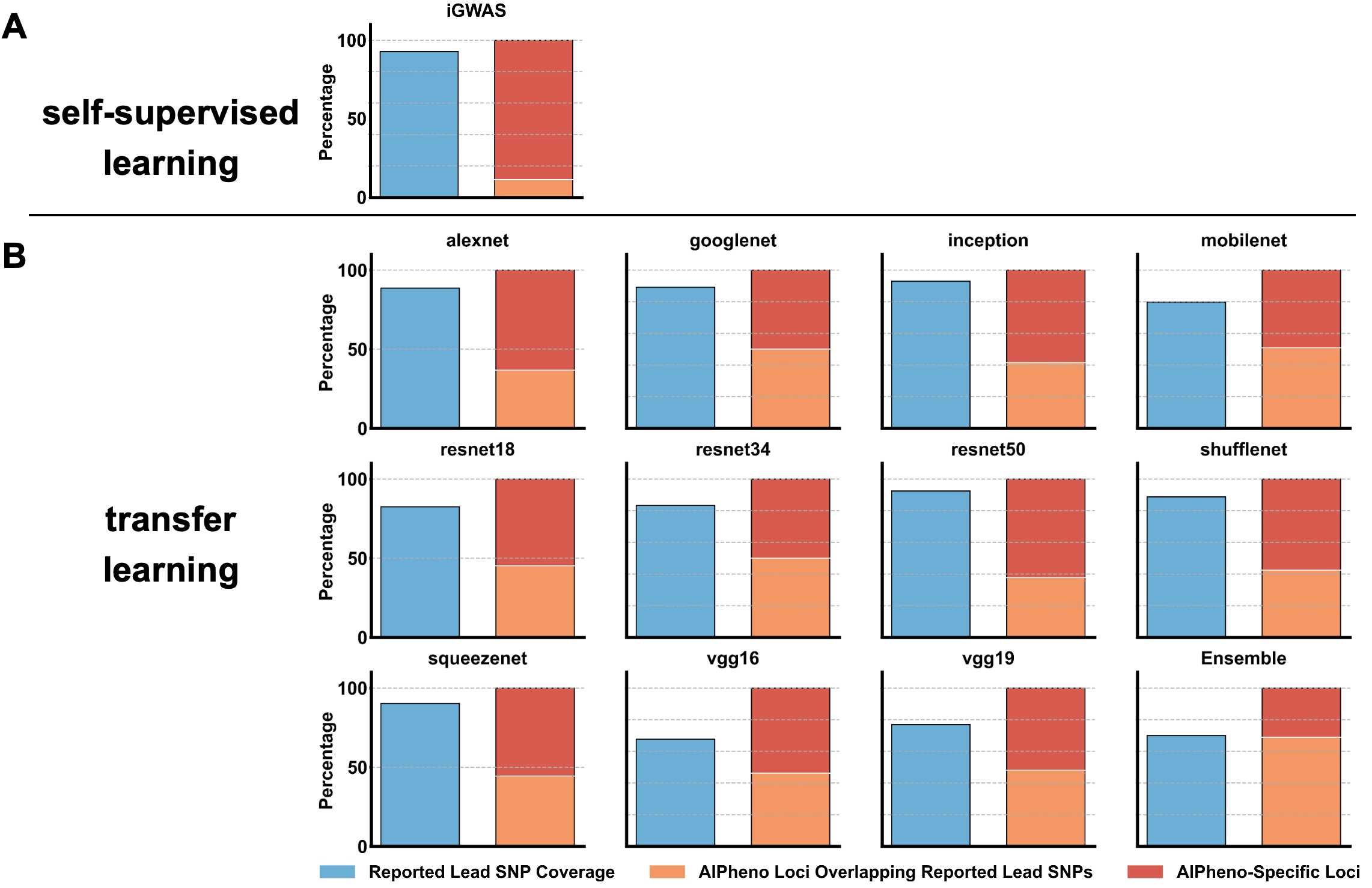}
    \caption{\textbf{Comparison of genetic discovery power between AIPheno and other deep learning-based phenotyping methods in \textit{Homo sapiens} (human), related to Figure 2.} \textbf{A,} a self-supervised learning model (iGWAS); and \textbf{B,} 11 ImageNet pre-trained transfer learning models (e.g., alexnet) and their Ensemble (the union of results from all 11 models). A lead SNP was considered to overlap with a locus if it fell within that locus. Bottom right bar: Percentage of AIPheno loci that overlapped with reported lead SNPs. Top right bar: Percentage of AIPheno-specific loci.} \label{fig_S7}
\end{figure}
\clearpage

\begin{figure}[p]
    \centering
    \includegraphics[width=1.0\textwidth]{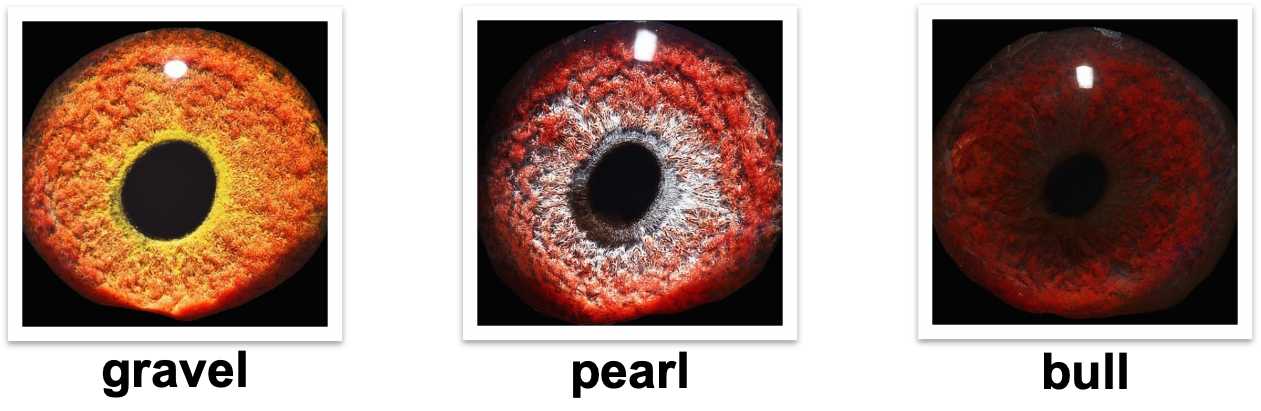}
    \caption{\textbf{Traditional classification of domestic pigeon irises, related to Figure 2.} In domestic pigeons, three principal types of iris coloration are observed: the wild-type "gravel," which is yellow to orange; the white "pearl"; and the black "bull" eye.} \label{fig_S8}
\end{figure}
\clearpage

\begin{figure}[p]
    \centering
    \includegraphics[width=1.0\textwidth]{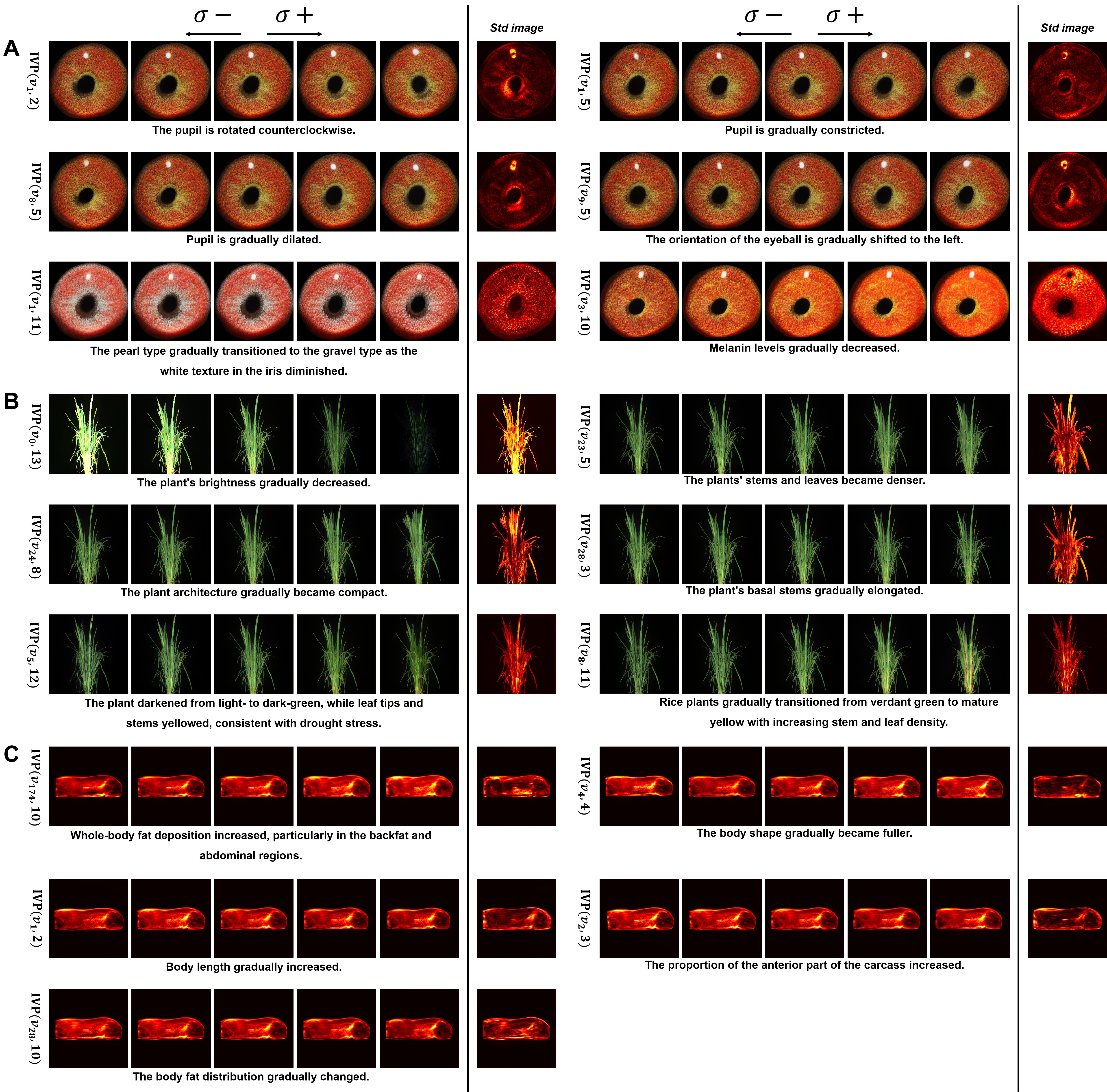}
    \caption{\textbf{Abundant image-variation phenotypes detected by AIPheno in domestic pigeon, rice, and swine, related to Figure 2.} \textbf{A,} For domestic pigeon, AIPheno detected image-variation phenotypes such as the orientation of the eyeball and pupil, pupil size, and iris color and texture. \textbf{B,} For rice, environmental variations, such as plant illumination, and biologically meaningful image variations, including plant architecture, stem and leaf morphology, color, and texture, detected by AIPheno. \textbf{C,} AIPheno detected not only changes in fat content and distribution but also alterations in body length and body shape.} \label{fig_S9}
\end{figure} 
\clearpage

\begin{figure}[p]
    \centering
    \includegraphics[width=1.0\textwidth]{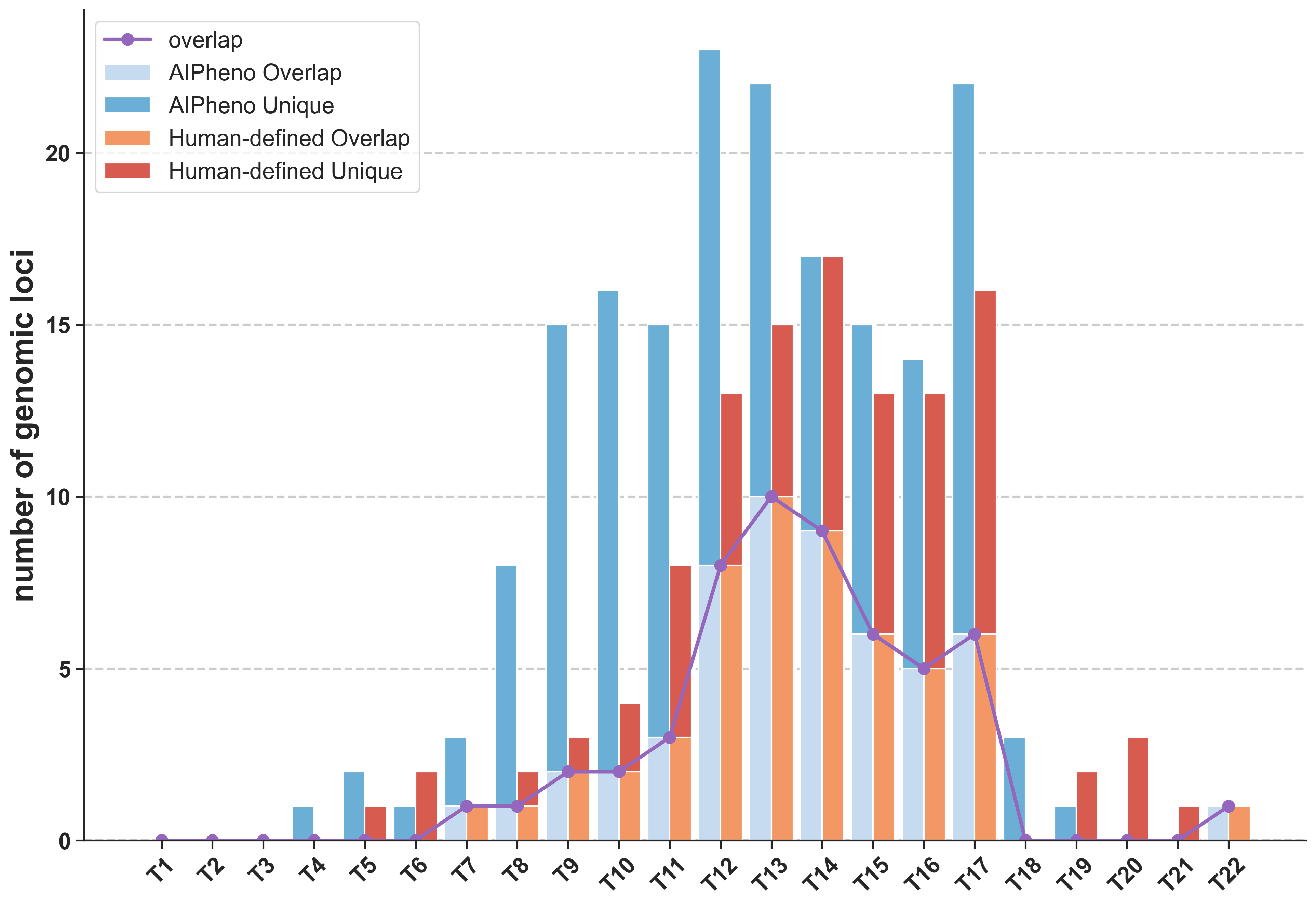}
    \caption{\textbf{Comparison of genetic discovery power between AIPheno and human-defined phenotypes at 22 time points, related to Figure 2.} At most time points, more genetic loci were discovered by AIPheno. AIPheno Overlap: Loci where lead SNPs from human-defined phenotypes lie within AIPheno loci. AIPheno Unique: Loci unique to AIPheno. Human-defined Overlap: Loci where lead SNPs from AIPheno lie within Human-defined phenotypes loci. Human-defined Unique: Loci unique to Human-defined phenotypes.} \label{fig_S10}
\end{figure}
\clearpage

\begin{figure}[p]
    \centering
    \includegraphics[width=1.0\textwidth]{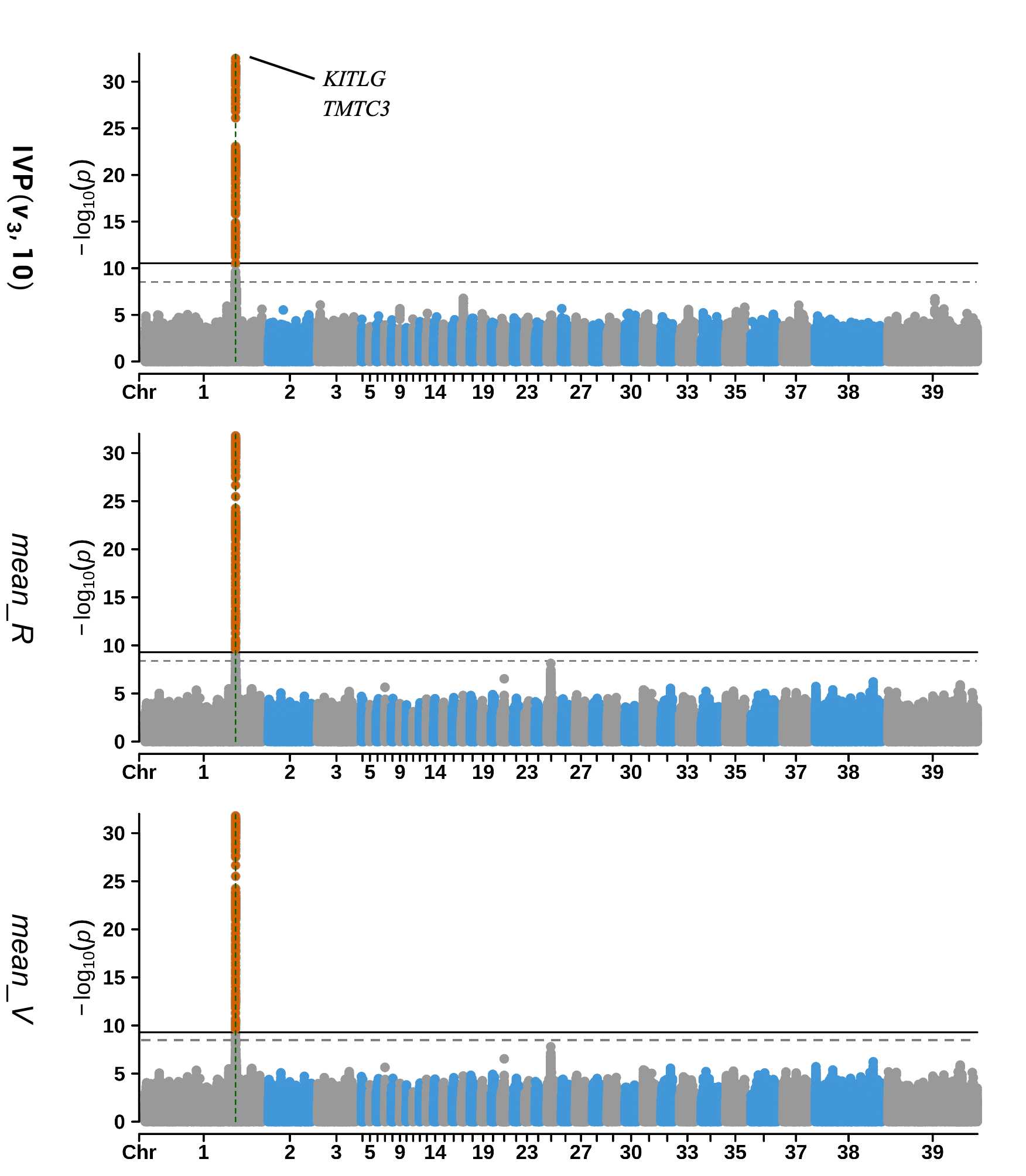}
    \caption{\textbf{Human-defined phenotypes also identified the \textit{KITLG}-\textit{TMTC3} locus, related to Figure 4.} This locus was also identified by two human-defined phenotypes: the mean R (redness) and V (brightness) channel values. However, it is difficult to recognize that it is related to melanin accumulation solely through the definitions of these phenotypes. The significance thresholds of AIPheno were adjusted using a Bonferroni correction for the number of phenotypes ($p_{\text{original}}=4.13\times10^{-9}$, $p_{\text{correction}}=2.95\times10^{-11}$ for image-variation phenotypes and $p_{\text{correction}}=5.16\times10^{-10}$ for human-defined phenotypes).} \label{fig_S11}
\end{figure}
\clearpage

\begin{figure}[p]
    \centering
    \includegraphics[width=0.55\textwidth]{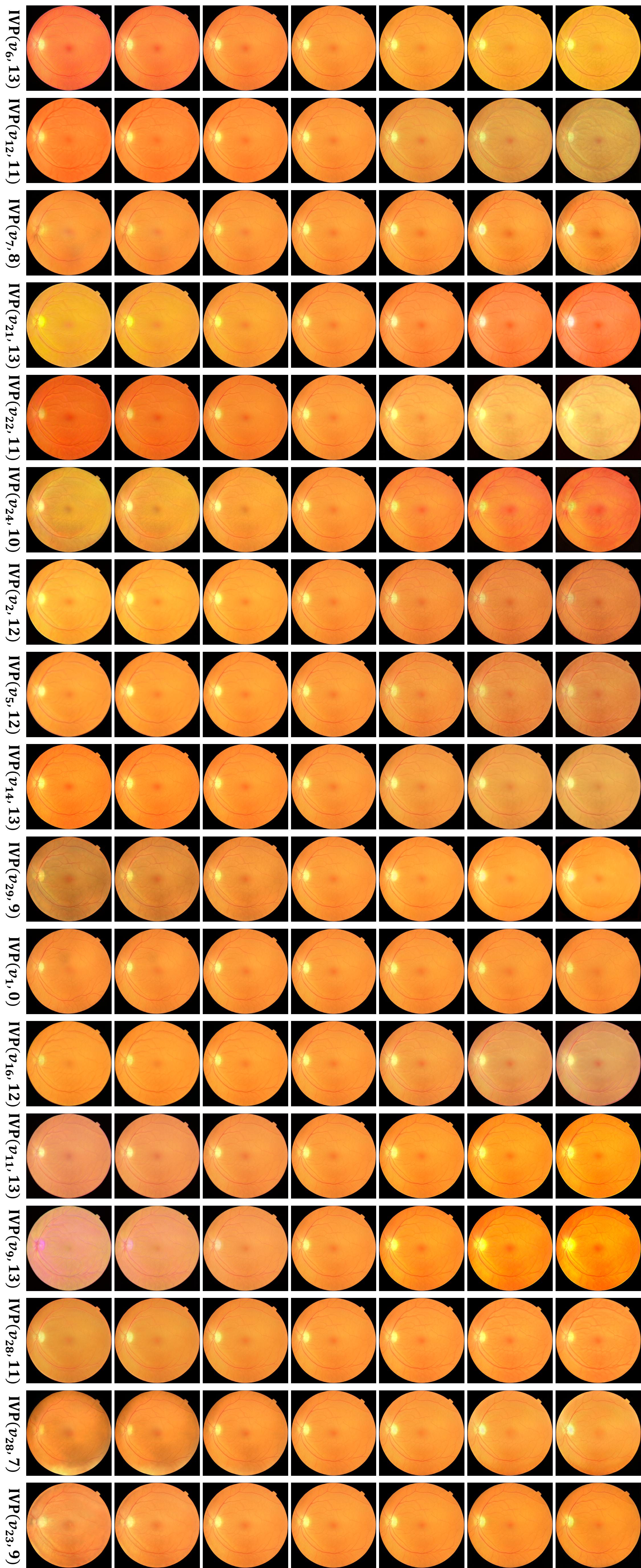}
    \caption{\textbf{Interpretability of image-variation phenotypes in Sankey diagram, related to Figure 4.} The image-variation phenotypes exhibiting the most significant association (i.e., the smallest \textit{p} value) with lead SNPs not previously reported in the GWAS Catalog database for associations with pigmentation and eye-related traits were selected.} \label{fig_S12}
\end{figure}
\clearpage

\begin{figure}[p]
    \centering
    \includegraphics[width=1.0\textwidth]{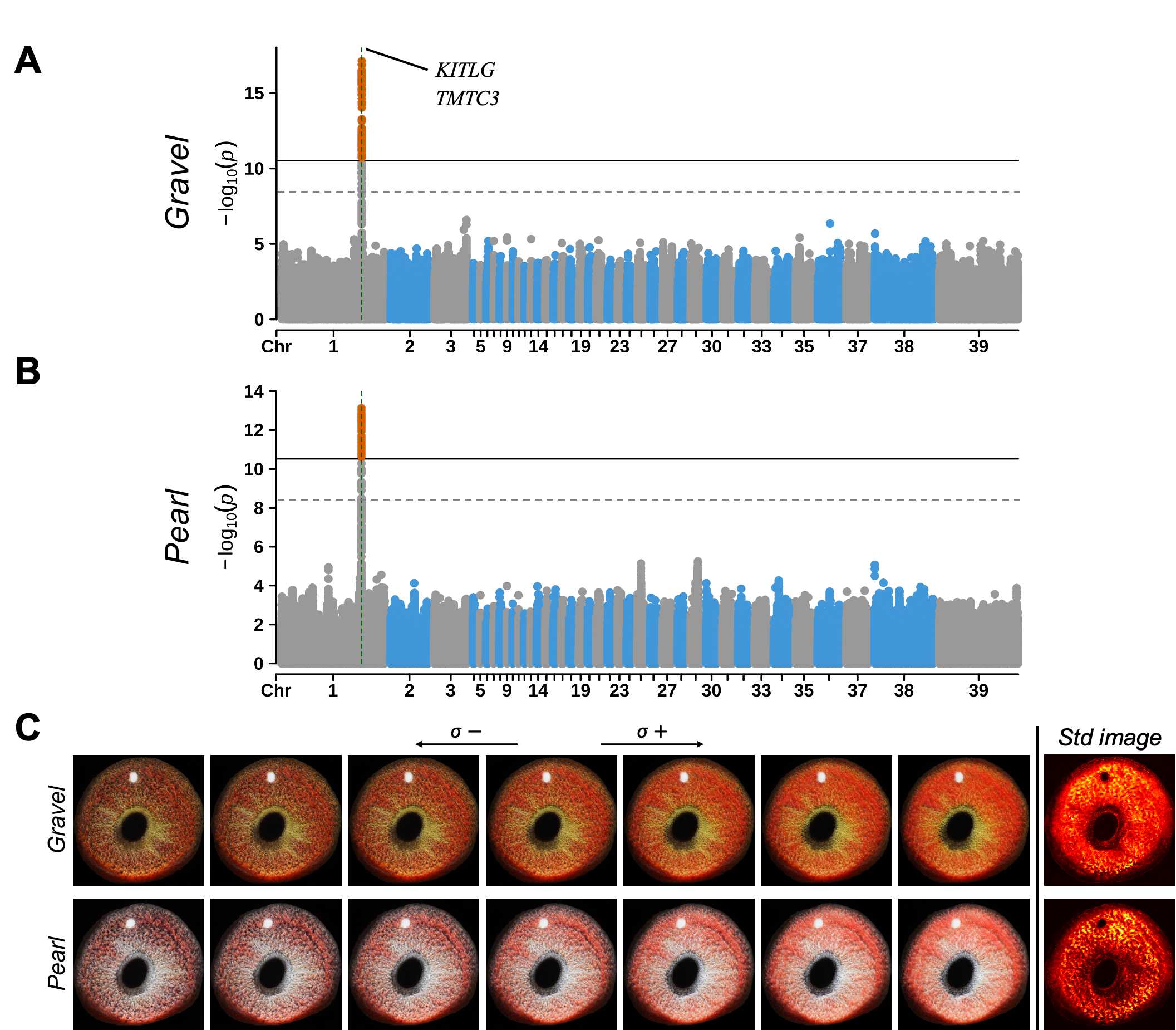}
    \caption{\textbf{Identification of \textit{KITLG}-\textit{TMTC3} locus in both gravel and pearl iris subpopulations with IVP$(\mathbf{v}_{3}, 10)$, related to Discussion.} \textbf{A,} Manhattan plot of the Gravel subpopulation ($n=328$). There are two threshold lines in the Manhattan plot ($p_{\text{original}} = 4.17\times10^{-9}$, $p_{\text{correction}} = 2.98\times10^{-11}$). \textbf{B,} Manhattan plot of the Pearl subpopulation ($n=311$). There are two threshold lines in the Manhattan plot ($p_{\text{original}} = 4.19\times10^{-9}$, $p_{\text{correction}} = 2.99\times10^{-11}$). \textbf{C,} Interpretability analysis of IVP$(\mathbf{v}_{3}, 10)$ within the gravel and pearl subgroups. It can be seen that differences in melanin levels exist in both the gravel and pearl groups. The std image further reveals that this variation occurs across the entire iris region, excluding the pupil and light spots.} 
    \label{sup_pigeon_two_class}
\end{figure}
\clearpage

\begin{figure}[p]
    \centering
    \includegraphics[width=1.0\textwidth]{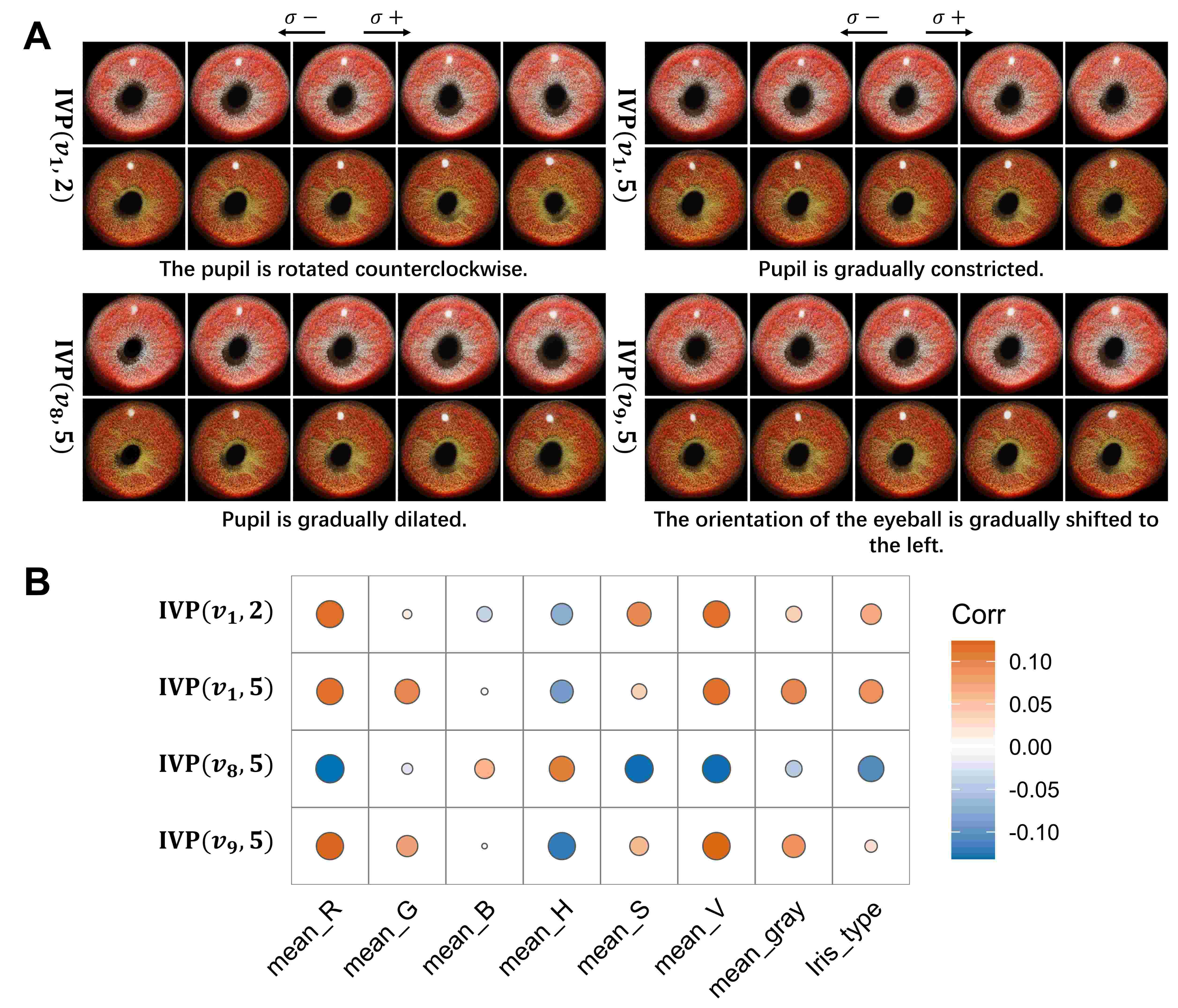}
    \caption{\textbf{Fine-grained image-variation phenotypes captured by AIPheno in the domestic pigeon, related to Discussion.} \textbf{A,} Interpretability Analysis of fine-grained image-variation phenotypes. \textbf{B,} Pearson correlation coefficients between fine-grained image-variation phenotypes and human-defined phenotypes. The image-variation phenotypes exhibit low correlation with the human-defined phenotypes, as they capture fine-grained features independent of color, whereas human-defined phenotypes primarily capture global color variations.}   
    \label{sup_pigeon_corr}
\end{figure}
\clearpage

\begin{figure}[p]
    \centering
    \includegraphics[width=1.0\textwidth]{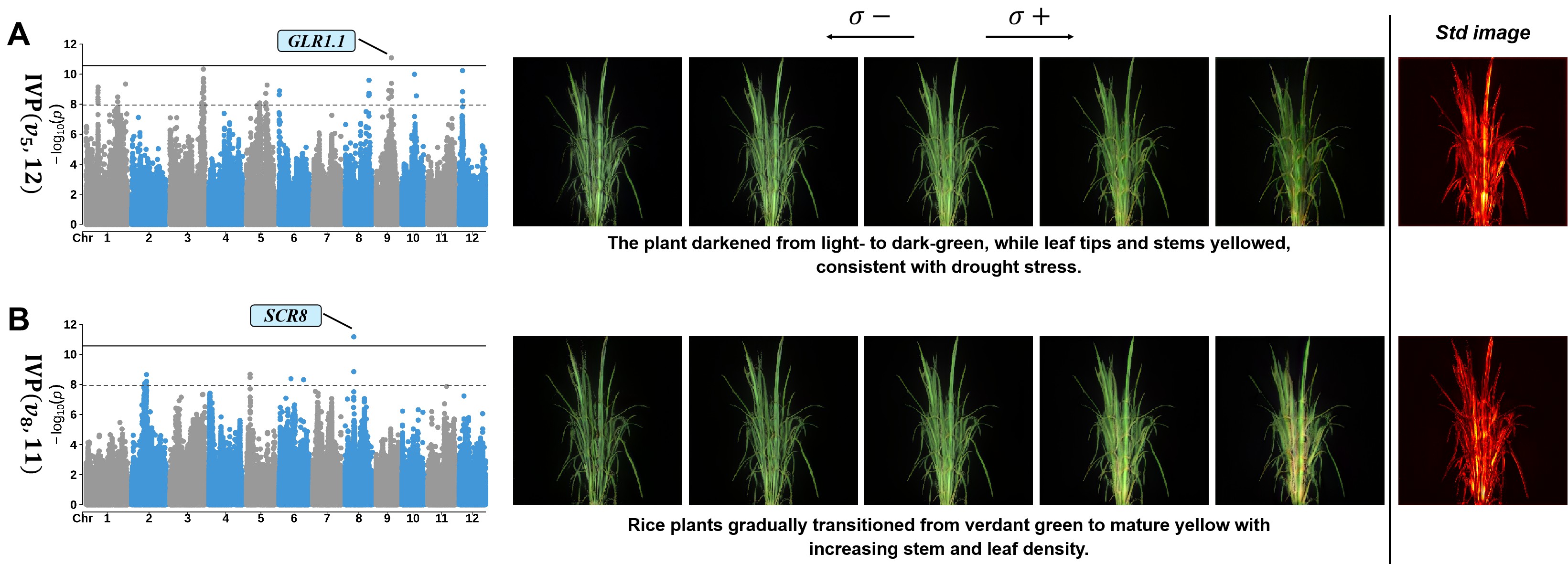}
    \caption{\textbf{Complex image variation quantified by AIPheno in rice, related to Discussion.}  \textbf{A,} Interpretability analysis of IVP$(\mathbf{v}_{5}, 12)$ in rice. The significance threshold was adjusted using a Bonferroni correction for the number of IVPs ($p_{\text{original}}=1.16\times10^{-8}$, $p_{\text{correction}}=2.75\times10^{-11}$). The plant darkened from light- to dark-green, while leaf tips and stems yellowed, which is consistent with drought stress. As shown in the std image, image variation is distributed throughout the entire plant. Also see a dynamic video included in the Data S18. \textbf{B,} Interpretability analysis of IVP$(\mathbf{v}_{8}, 11)$ in rice. The significance threshold was adjusted using a Bonferroni correction for the number of IVPs ($p_{\text{original}}=1.16\times10^{-8}$, $p_{\text{correction}}=2.75\times10^{-11}$). Rice plants transitioned gradually from verdant green to mature yellow with increasing stem and leaf density. As shown in the std image, image variation is distributed throughout the entire plant. Also see a dynamic video included in the Data S19.}\label{rice_discussion}
\end{figure}
\clearpage

\begin{figure}[p]
    \centering
    \includegraphics[width=1.0\textwidth]{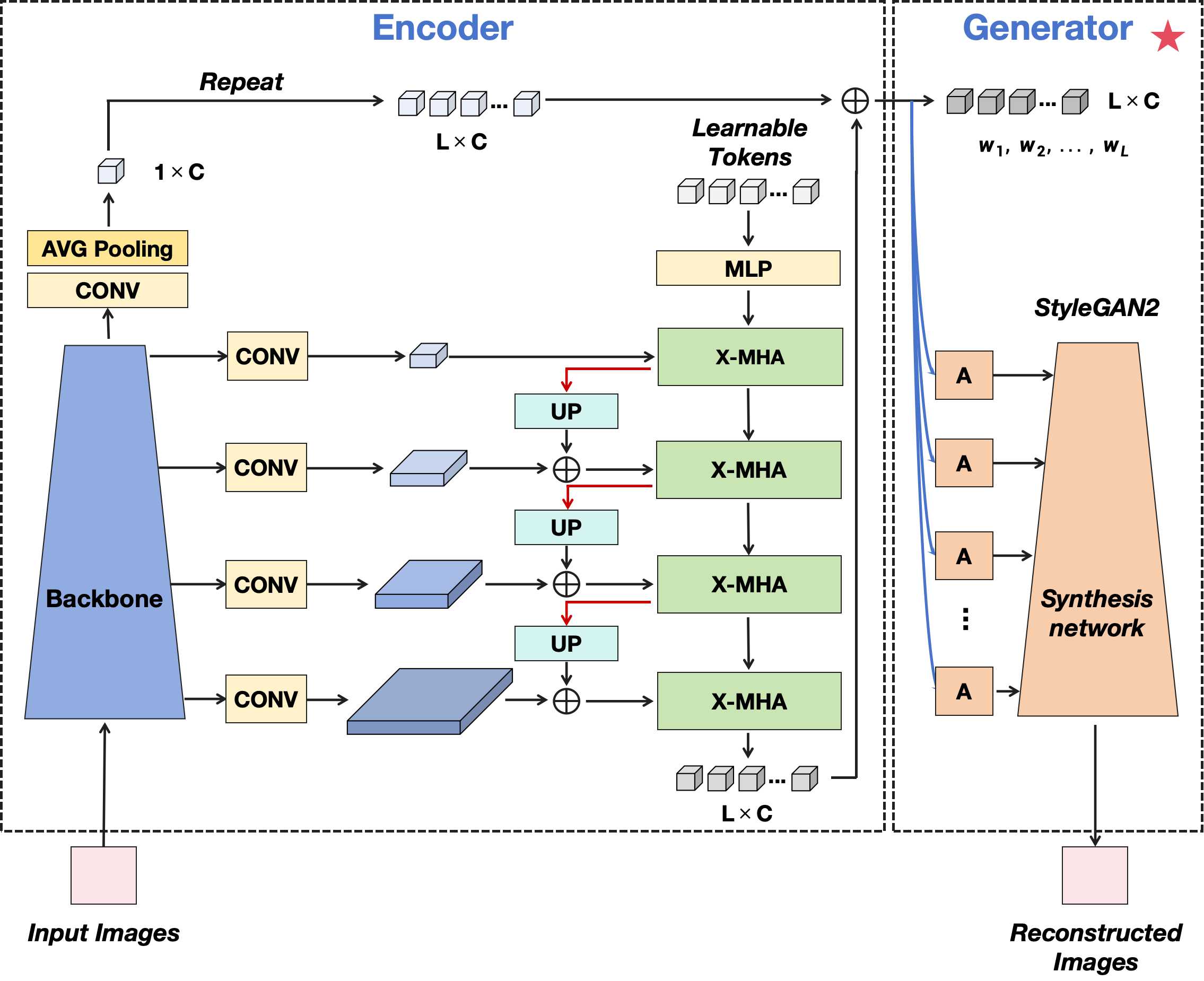}
    \caption{\textbf{The overall of the proposed AIPheno framework based on Encoder-Generator architecture, related to Figure 1 and Methods "The Encoder-Generator Framework of AIPheno" section.} The framework consists of two core modules: the Encoder and the Generator. Specifically, the overall structure of the Encoder can be divided into two components. The first component comprises global and local image representation pathways, while the second component is composed of $L$ learnable tokens ($\in \mathbb{R} ^{L\times C}$). Within the first component of the Encoder, the input image is encoded by a \emph{Backbone} (which can adopt CNN-based or Transformer-based network structure). The \emph{Backbone} generates multi-resolution local image features through different layers and a global image feature after passing through an \emph{AVG Pooling} layer, where \emph{CONV} denotes a convolution layer. The second component utilizes a Cross-Modality Multi-Head Attention (\emph{X-MHA}) mechanism to interact with the local image information (where the \emph{UP} represents an upsampling layer for aligning spatial resolutions), then fuses it with the global image representation to finally obtain the encoded representation in the $w+$ space ($\in \mathbb{R} ^{L\times C}$). Details within the \emph{X-MHA} block are depicted on the Fig. \ref{fig4}. For Generator, an unconditional StyleGAN2 style network is selected to construct the generator. Specifically, a series of affine layers $A_{i}$ (where $i=1,2,...,L$) and a \emph{synthesis network} is utilized to generate the reconstructed input image from $w+\in \mathbb{R} ^{L\times C}$.} \label{fig_S16}
\end{figure}
\clearpage
        
\begin{figure}[p]
    \centering
    \includegraphics[width=0.6\textwidth]{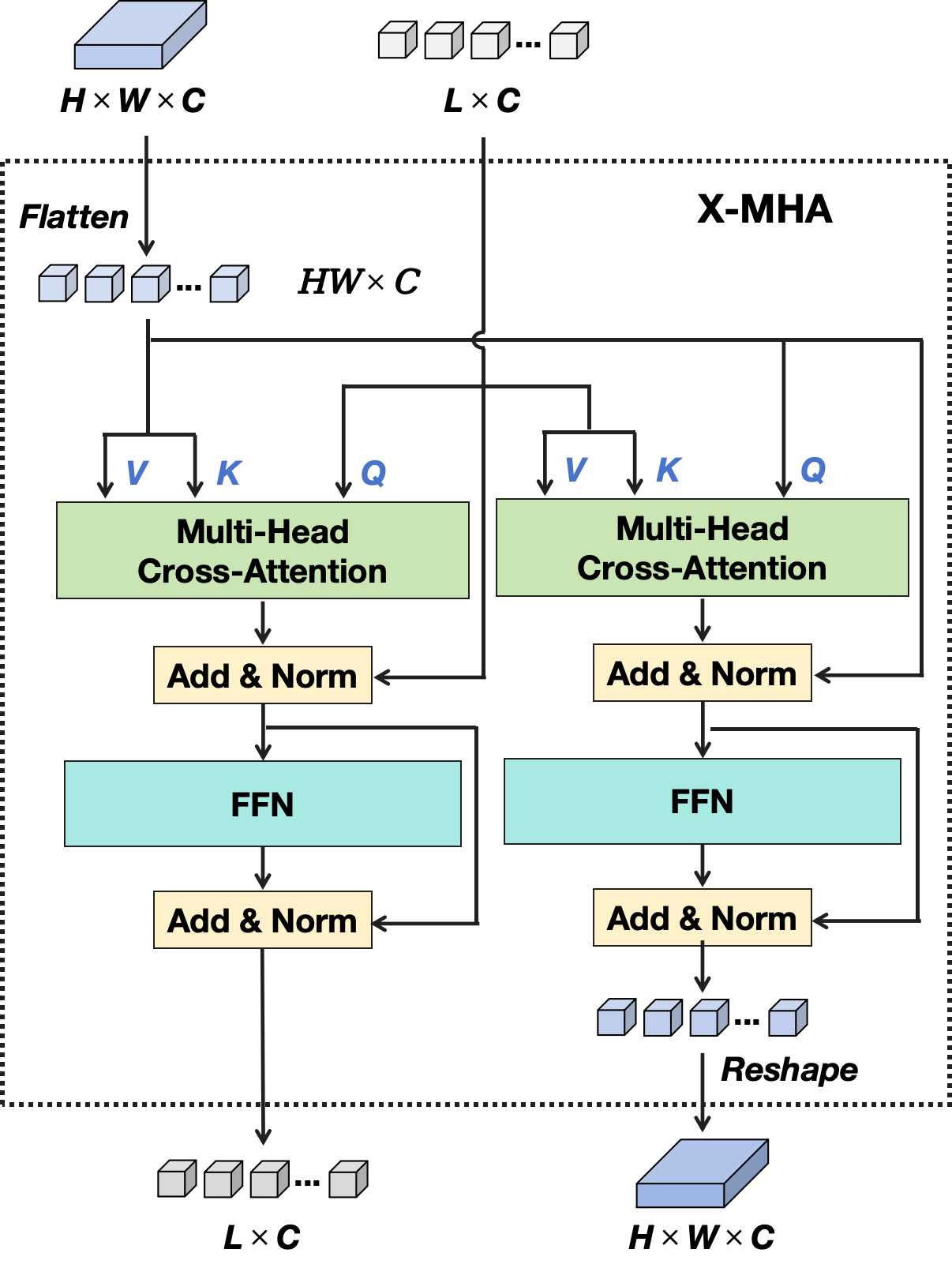}
    \caption{\textbf{The architecture of the Cross-Modality Multi-Head Attention (\emph{X-MHA}) block, related to Figure 1 and Methods "The Encoder-Generator Framework of AIPheno" section.} The \emph{X-MHA} block primarily serves to enable cross-modality information interaction between local image features ($\in \mathbb{R} ^{H\times W\times C}$) with $L$ learnable tokens ($\in \mathbb{R} ^{L\times C}$), facilitating mutual feature enhancement through bidirectional attention mechanisms. The main module in \emph{X-MHA} is two \emph{Multi-Head Cross-Attention} blocks, following the common routine in transformer model. Specifically, the flattened local feature $\in \mathbb{R} ^{HW\times C}$ of the image will function as Query and Key-Value respectively to participate in the cross-fusion process, each head computes the context vectors of one modality by attending to the other modality. Similarly, the $L$ learnable tokens will also serve as Query and Key-Value respectively to participate in the information interaction. Finally, the output of this block is the local fused features ($\in \mathbb{R} ^{H\times W\times C}$) and $L$ learnable tokens ($\in \mathbb{R} ^{L\times C}$) after interactive fusion, with their dimensionality remaining consistent with the input.} \label{fig_S17}
\end{figure}
\clearpage
        
\begin{figure}[p]
    \centering
    \includegraphics[width=1.0\textwidth]{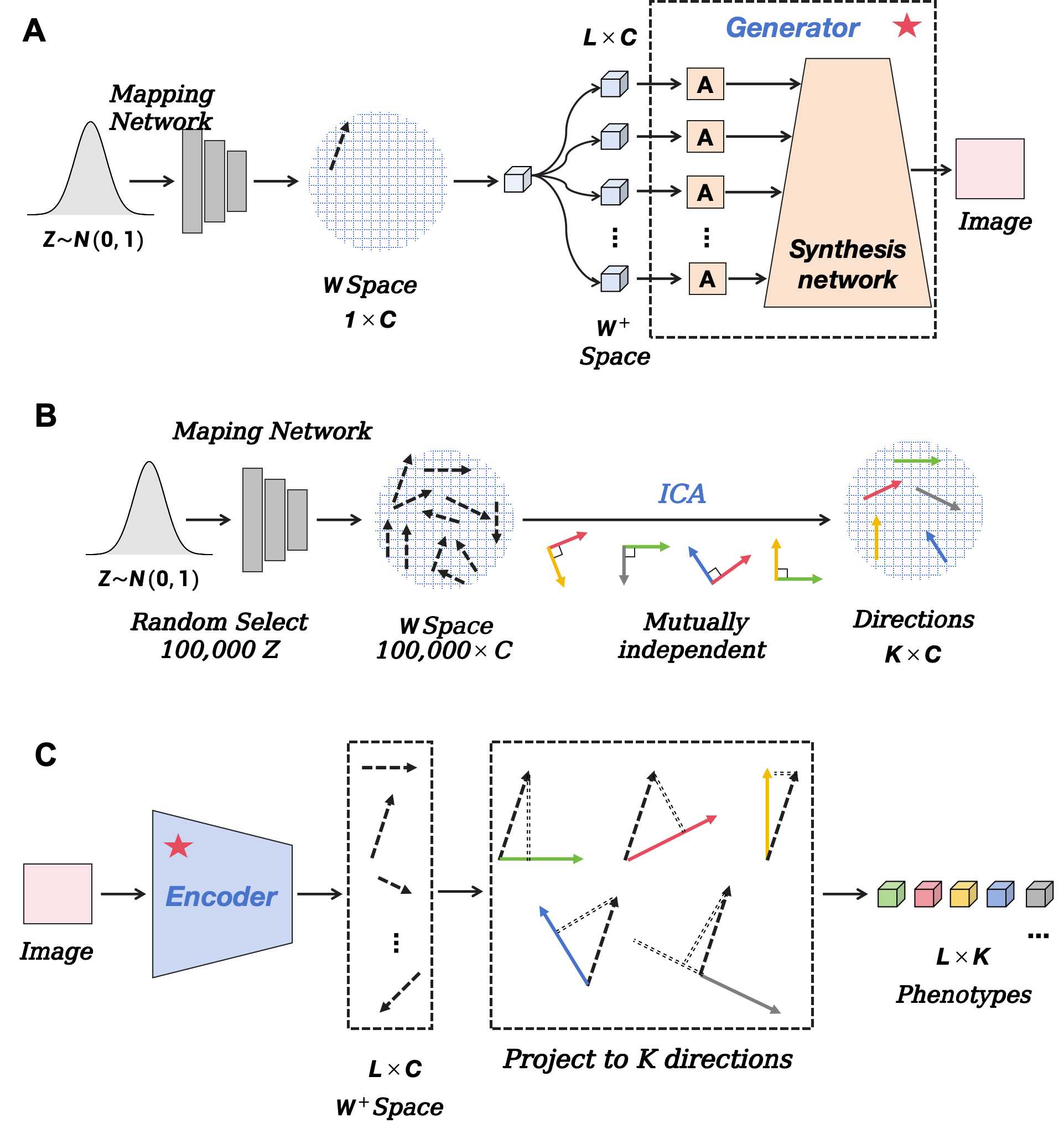}
    \caption{\textbf{The overall phenotype extraction process, related to Figure 1 and Methods "The Encoder-Generator Framework of AIPheno" section.} \textbf{A,} illustrates the training of an unconditional \emph{Synthesis Network} to fit the distribution of input images. Specifically, a random sample $z$ from the standard normal distribution $N(0,1)$ is mapped to $w\in \mathbb{R} ^{C}$ space via the \emph{Mapping Network}, corresponding to a generated image. \textbf{B,} involves sampling 100,000 times from $N(0,1)$ to simulate and generate a dataset of 100,000 images that conform to the original bio-image distribution. This generated set is then used to discover mutually independent directions ($\in \mathbb{R} ^{K\times C}$), where $K$ denotes the number of directions and $C$ represents their dimensionality (see methods). \textbf{C,} describes that for any image in the population subject to genetic analysis, a trained Encoder network is employed to obtain its representation in $W+\in \mathbb{R} ^{L\times C}$ space. The $L$-layer features in $W+$ space are then dot-multiplied with the $K$ directions to derive projection values, which serve as the final image phenotypes ($\in \mathbb{R} ^{L\times K}$).} \label{fig_S18}
\end{figure}
\clearpage

\begin{figure}[p]
    \centering
    \includegraphics[width=1.0\textwidth]{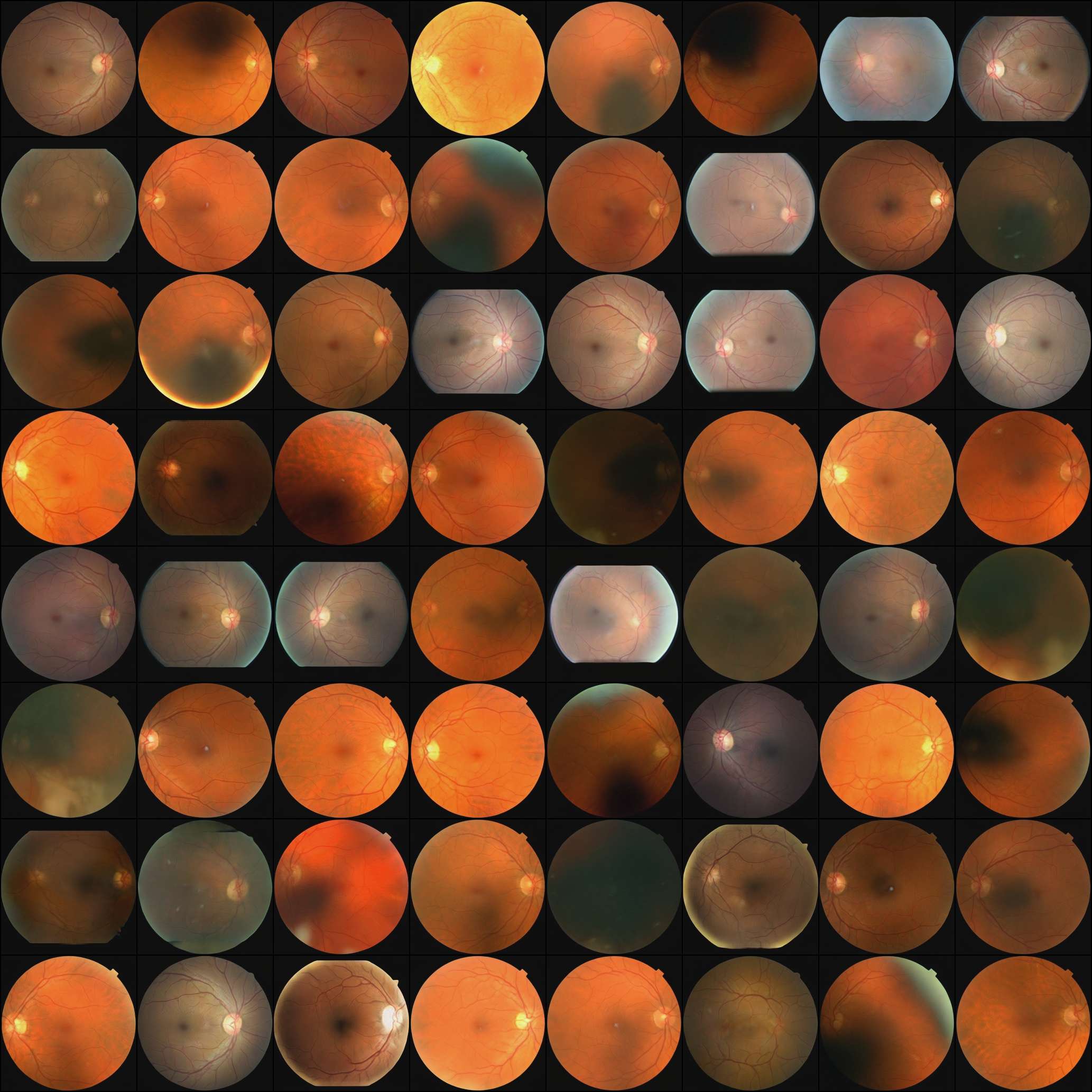}
    \caption{\textbf{Images generated unconditionally from retina fundus of human, related to Figure 1 and Methods "The Encoder-Generator Framework of AIPheno" section.} In the AIPheno framework, the Generator synthesizes high-quality human retinal fundus images unconditionally.}\label{sup_method_un_human}
\end{figure}
\clearpage

\begin{figure}[p]
    \centering
    \includegraphics[width=1.0\textwidth]{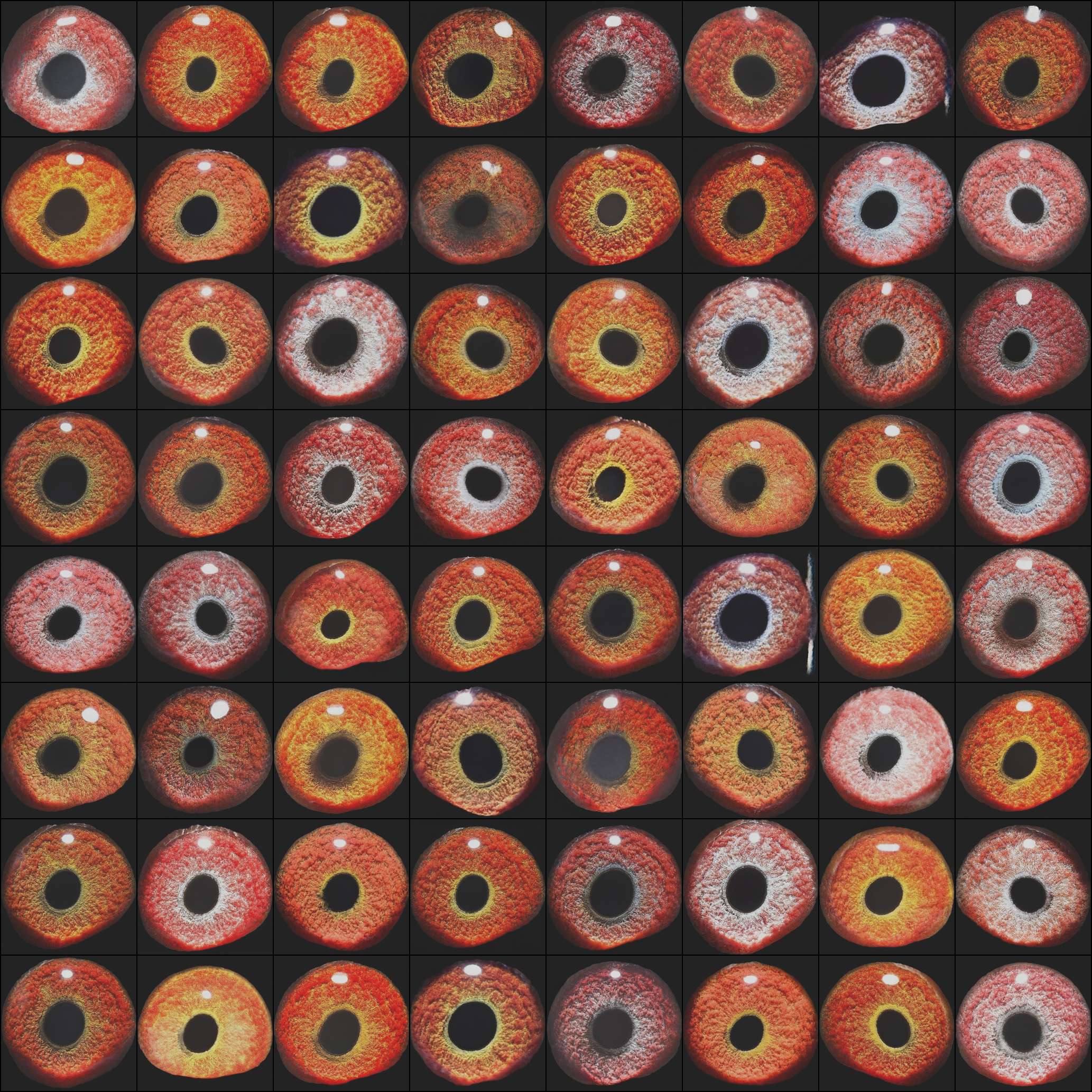}
    \caption{\textbf{Images generated unconditionally from iris of domestic pigeon, related to Figure 1 and Methods "The Encoder-Generator Framework of AIPheno" section.} In the AIPheno framework, the Generator synthesizes high-quality domestic pigeon iris images unconditionally.}\label{sup_method_un_pigeon}
\end{figure}
\clearpage

\begin{figure}[p]
    \centering
    \includegraphics[width=1.0\textwidth]{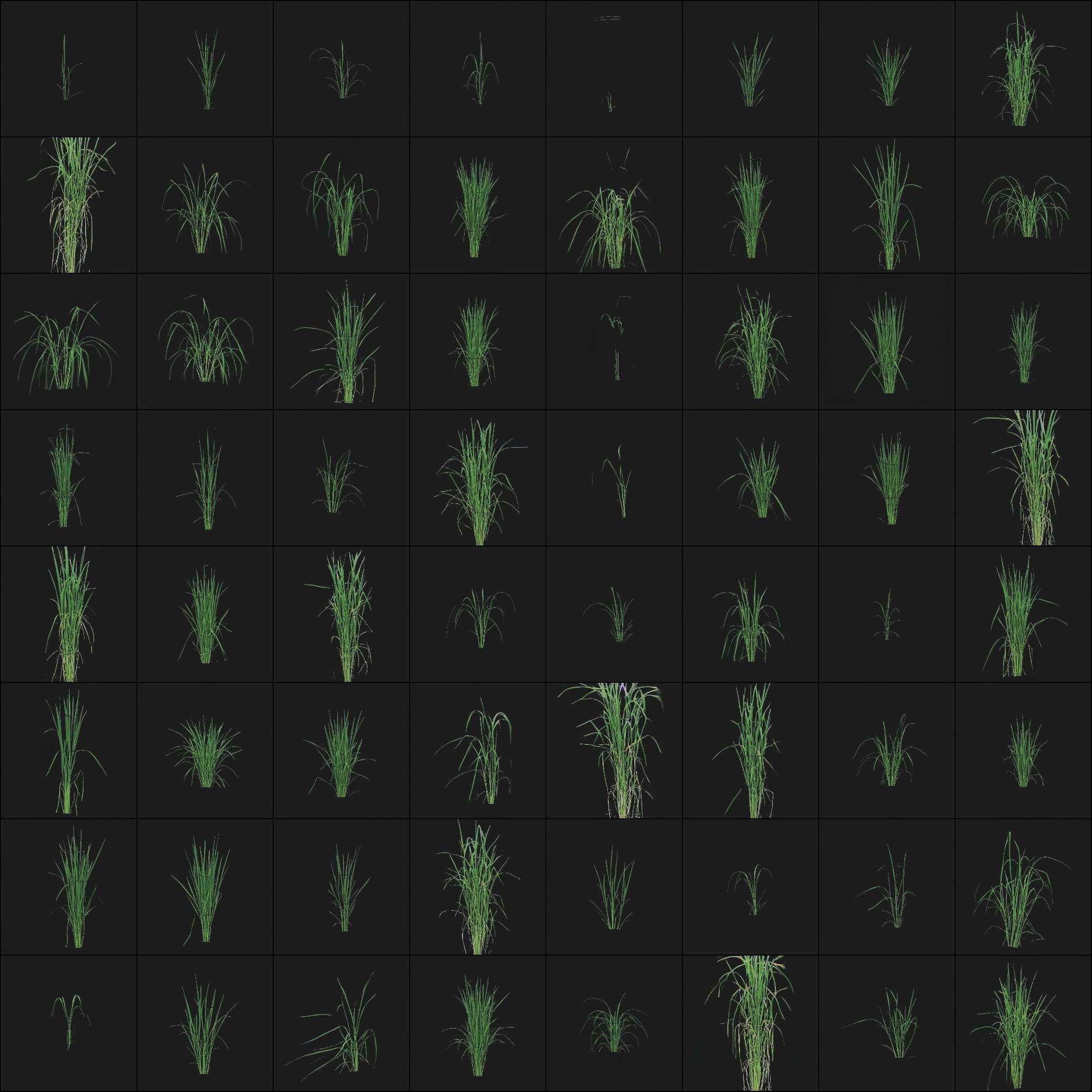}
    \caption{\textbf{Images generated unconditionally from potted rice, related to Figure 1 and Methods "The Encoder-Generator Framework of AIPheno" section.} In the AIPheno framework, the Generator synthesizes high-quality potted rice images unconditionally.}\label{sup_method_un_rice}
\end{figure}
\clearpage

\begin{figure}[p]
    \centering
    \includegraphics[width=1.0\textwidth]{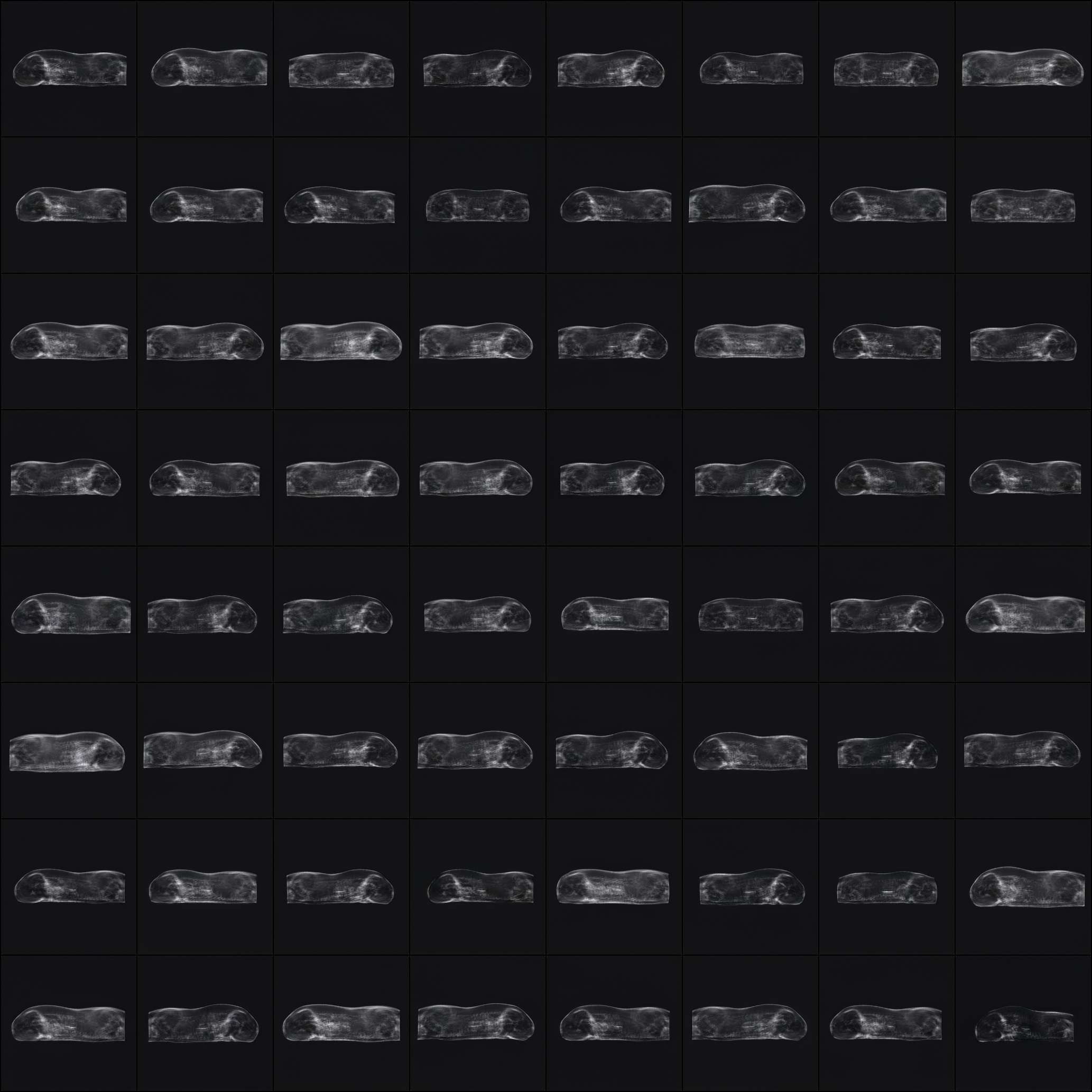}
    \caption{\textbf{Images generated unconditionally from swine CT, related to Figure 1 and Methods "The Encoder-Generator Framework of AIPheno" section.} In the AIPheno framework, the Generator synthesizes high-quality swine CT images unconditionally.}\label{sup_method_un_swine}
\end{figure}
\clearpage

\begin{figure}[p]
    \centering
    \includegraphics[width=1.0\textwidth]{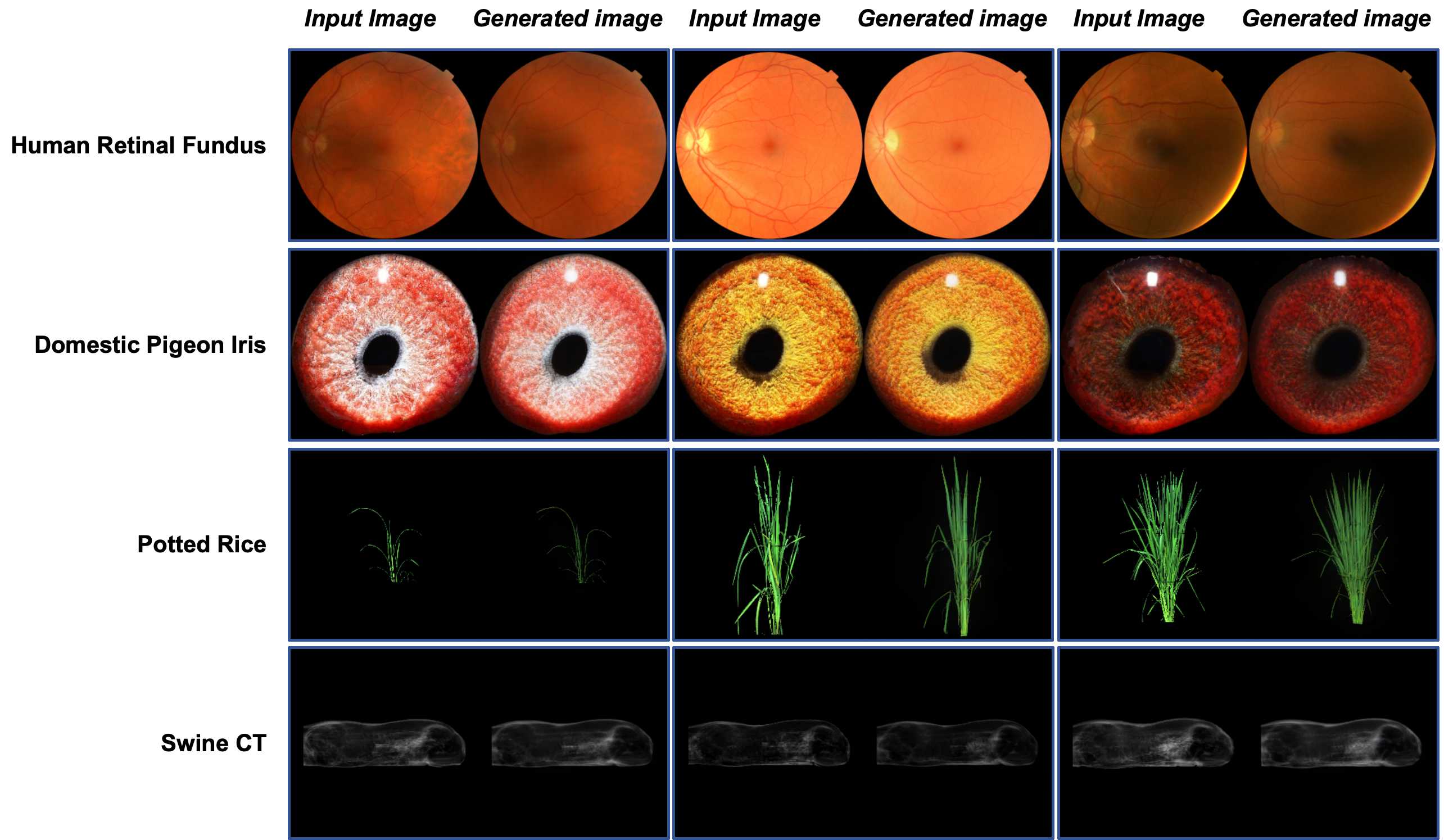}
    \caption{\textbf{High-quality reconstruction of the original image, related to Figure 1 and Methods "Model Training" section.} Generated images from four species show that the Encoder exhibits robust capability in restoring the overall image information, with consistent preservation of fine details.}
    \label{sup_framework_generation}
\end{figure}
\clearpage

\setcounter{table}{1}
\renewcommand{\thetable}{S\arabic{table}}
\newcommand{\cchead}[1]{\multicolumn{1}{c}{\textbf{#1}}}

\begin{sidewaystable}[p]
    \centering
    \caption{Lead SNPs and genomic loci in rice at T13, related to Figure 2} \label{table_s2}
\begin{tabular}{@{} l l l l l l l l r @{}}
        \toprule
        \textbf{\makecell{Lead SNP \\ (MSU6.1)}} & \textbf{\makecell{Lead SNP \\ (MSU7)}} & \textbf{Chr} &
        \textbf{\makecell{Position \\ (MSU6.1)}} & \textbf{\makecell{Position \\ (MSU7)}} &
        \textbf{\textit{p} value} & \textbf{Locus} &
        \textbf{\makecell{Start \\ (MSU7)}} & \textbf{\makecell{End \\ (MSU7)}} \\
        \midrule
        sf0113368830 & vg0113369856 & 1 & 13368830 & 13369856 & $2.41 \times 10^{-13}$ & 1 & 13069856 & 13669856 \\
        sf0114732944 & vg0114733970 & 1 & 14732944 & 14733970 & $1.94 \times 10^{-13}$ & 2 & 14433970 & 15033970 \\
        sf0115877533 & vg0115878559 & 1 & 15877533 & 15878559 & $1.72 \times 10^{-12}$ & 3 & 15578559 & 16178559 \\
        sf0118143691 & vg0118144737 & 1 & 18143691 & 18144737 & $1.02 \times 10^{-11}$ & 4 & 17844737 & 18444737 \\
        sf0136973473 & vg0136974516 & 1 & 36973473 & 36974516 & $1.10 \times 10^{-11}$ & 5 & 36674516 & 37274516 \\
        sf0137595036 & vg0137596079 & 1 & 37595036 & 37596079 & $1.89 \times 10^{-13}$ & 6 & 37296079 & 37896079 \\
        sf0323937770 & vg0323939596 & 3 & 23937770 & 23939596 & $1.75 \times 10^{-11}$ & 7 & 23639596 & 24239596 \\
\bottomrule
    \end{tabular}

    \begin{tablenotes}
        \item This table summarizes the lead SNPs identified in rice at T13, including their genomic positions, associated loci, and flanking regions. For each lead SNP, both the identifier and genomic coordinates are provided with respect to the MSU6.1 and MSU7 reference genomes. 
    \end{tablenotes}
\end{sidewaystable}
\clearpage

\begin{landscape}
    \setlength{\tabcolsep}{4pt}
    \begin{longtable}{@{}
        >{\itshape\raggedright\arraybackslash}p{0.15\linewidth}   
        >{\raggedright\arraybackslash}p{0.35\linewidth}           
        >{\itshape\raggedright\arraybackslash}p{0.16\linewidth}   
        >{\itshape\raggedright\arraybackslash}p{0.12\linewidth}   
        >{\itshape\raggedright\arraybackslash}p{0.22\linewidth}   
    @{}}
    \caption{Information about all potential candidate genes within 200 kb (100 kb upstream and downstream) of the lead SNP sf0113368830, related to Figure 2} \\
    \toprule
    \multicolumn{1}{l}{\textbf{\upshape RAP Locus ID}} &
    \multicolumn{1}{c}{\textbf{\upshape Description}} &
    \multicolumn{1}{l}{\textbf{\upshape RAP-DB}} &
    \multicolumn{1}{l}{\textbf{\upshape CGSNL}} &
    \multicolumn{1}{l}{\textbf{\upshape Oryzabase}} \\
    \midrule
    \endfirsthead
    
    \multicolumn{5}{c}{{\tablename} \thetable{} -- continued} \\
    \toprule
    \multicolumn{1}{l}{\textbf{\upshape RAP Locus ID}} &
    \multicolumn{1}{c}{\textbf{\upshape Description}} &
    \multicolumn{1}{l}{\textbf{\upshape RAP-DB}} &
    \multicolumn{1}{l}{\textbf{\upshape CGSNL}} &
    \multicolumn{1}{l}{\textbf{\upshape Oryzabase}} \\
    \midrule
    \endhead
    Os01g0337900 & Lipoamide dehydrogese 2, Paralog of OsLPD1 (lipoamide dehydrogese 1) & OsLPD2 & LPD2 & ptPDC-E3, OsptPDC-E3, OsLPD2 \\ 
    Os01g0338000 & Small GTPase, Rice storage proteins proglutelin and alpha-globulin export from the endoplasmic reticulum in endosperm & OsSar1a & SAR1A & OsARL1e, ARL1e, OsSar1b, Sar1b, OsSar1a, Sar1a \\
    Os01g0338150 & Hypothetical gene & {} & {} & {} \\
    Os01g0338100 & Similar to Transcription initiation factor IID, 18kD subunit family protein & {} & {} & {}\\
    Os01g0338200 & Mov34/MPN/PAD-1 family protein & {} & {} & {} \\
    Os01g0338401 & Hypothetical gene & {} & {} & {} \\
    Os01g0338600 & Similar to cysteine sulfite desulfise/cysteine desulfurase and related enzymes & {} & {} & {} \\
    Os01g0339001 & Non-protein coding transcript & {} & {} & {} \\
\bottomrule
    \end{longtable}
\end{landscape}

\begin{landscape}
    \setlength{\tabcolsep}{4pt}
    \begin{longtable}{@{}
        >{\itshape\raggedright\arraybackslash}p{0.15\linewidth}   
        >{\raggedright\arraybackslash}p{0.35\linewidth}           
        >{\itshape\raggedright\arraybackslash}p{0.16\linewidth}   
        >{\itshape\raggedright\arraybackslash}p{0.12\linewidth}   
        >{\itshape\raggedright\arraybackslash}p{0.22\linewidth}   
    @{}}
    \caption{Information about all potential candidate genes within 200 kb (100 kb upstream and downstream) of the lead SNP sf0114732944, related to Figure 2} \\
    \toprule
    \multicolumn{1}{l}{\textbf{\upshape RAP Locus ID}} &
    \multicolumn{1}{c}{\textbf{\upshape Description}} &
    \multicolumn{1}{l}{\textbf{\upshape RAP-DB}} &
    \multicolumn{1}{l}{\textbf{\upshape CGSNL}} &
    \multicolumn{1}{l}{\textbf{\upshape Oryzabase}} \\
    \midrule
    \endfirsthead
    
    \multicolumn{5}{c}{{\tablename} \thetable{} -- continued} \\
    \toprule
    \multicolumn{1}{l}{\textbf{\upshape RAP Locus ID}} &
    \multicolumn{1}{c}{\textbf{\upshape Description}} &
    \multicolumn{1}{l}{\textbf{\upshape RAP-DB}} &
    \multicolumn{1}{l}{\textbf{\upshape CGSNL}} &
    \multicolumn{1}{l}{\textbf{\upshape Oryzabase}} \\
    \midrule
    \endhead
    Os01g0360466 & {} & {} & {} & {} \\
    Os01g0360600 & Dephospho-CoA kise family protein & {} & {} & {} \\
    Os01g0361000 & Conserved hypothetical protein & {} & {} & {} \\
    Os01g0361500 & Phospholipid/glycerol acyltransferase domain containing protein & {} & PES1B & OsPES1b \\
    Os01g0361700 & Conserved hypothetical protein & {} & {} & {} \\
    Os01g0361800 & Hypothetical conserved gene & {} & {} & {} \\
    Os01g0362000 & Conserved hypothetical protein & {} & {} & {} \\
    Os01g0362100 & Esterase/lipase/thioesterase domain containing protein & {} & PES1A & OsPES1a \\
\bottomrule
    \end{longtable}
\end{landscape}

\begin{landscape}
    \setlength{\tabcolsep}{4pt}
    \begin{longtable}{@{}
        >{\itshape\raggedright\arraybackslash}p{0.15\linewidth}   
        >{\raggedright\arraybackslash}p{0.35\linewidth}           
        >{\itshape\raggedright\arraybackslash}p{0.16\linewidth}   
        >{\itshape\raggedright\arraybackslash}p{0.12\linewidth}   
        >{\itshape\raggedright\arraybackslash}p{0.22\linewidth}   
    @{}}
    \caption{Information about all potential candidate genes within 200 kb (100 kb upstream and downstream) of the lead SNP sf0115877533, related to Figure 2} \\
    \toprule
    \multicolumn{1}{l}{\textbf{\upshape RAP Locus ID}} &
    \multicolumn{1}{c}{\textbf{\upshape Description}} &
    \multicolumn{1}{l}{\textbf{\upshape RAP-DB}} &
    \multicolumn{1}{l}{\textbf{\upshape CGSNL}} &
    \multicolumn{1}{l}{\textbf{\upshape Oryzabase}} \\
    \midrule
    \endfirsthead
    
    \multicolumn{5}{c}{{\tablename} \thetable{} -- continued} \\
    \toprule
    \multicolumn{1}{l}{\textbf{\upshape RAP Locus ID}} &
    \multicolumn{1}{c}{\textbf{\upshape Description}} &
    \multicolumn{1}{l}{\textbf{\upshape RAP-DB}} &
    \multicolumn{1}{l}{\textbf{\upshape CGSNL}} &
    \multicolumn{1}{l}{\textbf{\upshape Oryzabase}} \\
    \midrule
    \endhead
    Os01g0379400 & Cyclin-like F-box domain containing protein & {} & {} & OsFbox010, OsFbox10, Os\_F0156, OsFBX6, OsFBX6, FBX6 \\
    Os01g0381400 & {} & {} & {} & {} \\
    Os01g0379601 & {} & {} & {} & {} \\
    Os01g0379800 & Similar to HAT family dimerisation domain containing protein & {} & {} & {} \\ 
    Os01g0380000 & {} & {} & {} & {} \\
    Os01g0380200 & {} & {} & ZOS1-07 & OsZOS1-07 \\
    Os01g0380800 & F-box domain, cyclin-like domain containing protein & {} & FBOX11 & OsFbox011, OsFbox11, Os\_F0196, OsFBX7, FBX7, OsFBX7, FBX7\\
    Os01g0381325 & Conserved hypothetical protein & {} & {} & {} \\
\bottomrule
    \end{longtable}
\end{landscape}

\begin{landscape}
    \setlength{\tabcolsep}{4pt}
    \begin{longtable}{@{}
      >{\itshape\raggedright\arraybackslash}p{0.15\linewidth}   
      >{\raggedright\arraybackslash}p{0.35\linewidth}           
      >{\itshape\raggedright\arraybackslash}p{0.16\linewidth}   
      >{\itshape\raggedright\arraybackslash}p{0.12\linewidth}   
      >{\itshape\raggedright\arraybackslash}p{0.22\linewidth}   
    @{}}
    \caption{Information about all potential candidate genes within 200 kb (100 kb upstream and downstream) of the lead SNP sf0118143691, related to Figure 2} \\
    \toprule
    \multicolumn{1}{l}{\textbf{\upshape RAP Locus ID}} &
    \multicolumn{1}{c}{\textbf{\upshape Description}} &
    \multicolumn{1}{l}{\textbf{\upshape RAP-DB}} &
    \multicolumn{1}{l}{\textbf{\upshape CGSNL}} &
    \multicolumn{1}{l}{\textbf{\upshape Oryzabase}} \\
    \midrule
    \endfirsthead
    
    \multicolumn{5}{c}{{\tablename} \thetable{} -- continued} \\
    \toprule
    \multicolumn{1}{l}{\textbf{\upshape RAP Locus ID}} &
    \multicolumn{1}{c}{\textbf{\upshape Description}} &
    \multicolumn{1}{l}{\textbf{\upshape RAP-DB}} &
    \multicolumn{1}{l}{\textbf{\upshape CGSNL}} &
    \multicolumn{1}{l}{\textbf{\upshape Oryzabase}} \\
    \midrule
    \endhead
        Os01g0512100 & Similar to Chaperone protein dJ 15 (Protein ALTERED RESPONSE TO GRAVITY) (AtARG1) (AtJ15) (AtDjB15) & {} & {} & OsDjC7 \\
        Os01g0512200 & Armadillo-type fold domain containing protein & {} & {} & {} \\ 
        Os01g0512400 & Conserved hypothetical protein & {} & GEBP2 & OsGeBP2, GeBP2 \\
        Os01g0512700 & Zinc finger, C2H2 domain containing protein & {} & ZOS1-08 & OsZOS1-08 \\
        Os01g0512800 & CS-like domain domain containing protein & {} & {} & {} \\
        Os01g0512950 & {} & {} & {} & {} \\
        Os01g0513100 & Similar to Protein phosphatase 2C ABI2 & {} & ABI2 & OsPP2C03, PP2C03, OsPP2C3, PP2C3, OsPP7, OsABI2 \\
        Os01g0513200 & Hypothetical conserved gene & {} & {} & {} \\
\bottomrule
    \end{longtable}
\end{landscape}
    
\begin{landscape}
    \setlength{\tabcolsep}{4pt}
    \begin{longtable}{@{}
      >{\itshape\raggedright\arraybackslash}p{0.15\linewidth}   
      >{\raggedright\arraybackslash}p{0.35\linewidth}           
      >{\itshape\raggedright\arraybackslash}p{0.16\linewidth}   
      >{\itshape\raggedright\arraybackslash}p{0.12\linewidth}   
      >{\itshape\raggedright\arraybackslash}p{0.22\linewidth}   
    @{}}
    \caption{Information about all potential candidate genes within 200 kb (100 kb upstream and downstream) of the lead SNP sf0136973473, related to Figure 2} \\
    \toprule
    \multicolumn{1}{l}{\textbf{\upshape RAP Locus ID}} &
    \multicolumn{1}{c}{\textbf{\upshape Description}} &
    \multicolumn{1}{l}{\textbf{\upshape RAP-DB}} &
    \multicolumn{1}{l}{\textbf{\upshape CGSNL}} &
    \multicolumn{1}{l}{\textbf{\upshape Oryzabase}} \\
    \midrule
    \endfirsthead
    
    \multicolumn{5}{c}{{\tablename} \thetable{} -- continued} \\
    \toprule
    \multicolumn{1}{l}{\textbf{\upshape RAP Locus ID}} &
    \multicolumn{1}{c}{\textbf{\upshape Description}} &
    \multicolumn{1}{l}{\textbf{\upshape RAP-DB}} &
    \multicolumn{1}{l}{\textbf{\upshape CGSNL}} &
    \multicolumn{1}{l}{\textbf{\upshape Oryzabase}} \\
    \midrule
    \endhead
        Os01g0855200 & Tetratricopeptide-like helical domain containing protein & {} & {} & {} \\
        Os01g0855301 & {} & {} & {} & {} \\
        Os01g0855400 & R2R3-MYB transcription factor 17 & Os2R\_MYB17 & DLN29 & OsDLN29, Os2R\_MYB17, 2R\_MYB17, MYB2-18, OsMYB2-18 \\
        Os01g0855600 & Similar to Hs1pro-1 protein & {} & {} & HS1 \\
        Os01g0855700 & Hypothetical protein {} & {} & {} & {} \\
        Os01g0855850 & {} & {} & {} & {} \\
        Os01g0855900 & Hypothetical gene & {} & {} & {} \\
        Os01g0856000 & Similar to CDC6 protein & {} & CDC6 & OsCDC6 \\
\bottomrule
    \end{longtable}
\end{landscape}

\begin{landscape}
    \setlength{\tabcolsep}{4pt}
    \begin{longtable}{@{}
      >{\itshape\raggedright\arraybackslash}p{0.15\linewidth}   
      >{\raggedright\arraybackslash}p{0.35\linewidth}           
      >{\itshape\raggedright\arraybackslash}p{0.16\linewidth}   
      >{\itshape\raggedright\arraybackslash}p{0.12\linewidth}   
      >{\itshape\raggedright\arraybackslash}p{0.22\linewidth}   
    @{}}
    \caption{Information about all potential candidate genes within 200 kb (100 kb upstream and downstream) of the lead SNP sf0137595036, related to Figure 2} \\
    \toprule
    \multicolumn{1}{l}{\textbf{\upshape RAP Locus ID}} &
    \multicolumn{1}{c}{\textbf{\upshape Description}} &
    \multicolumn{1}{l}{\textbf{\upshape RAP-DB}} &
    \multicolumn{1}{l}{\textbf{\upshape CGSNL}} &
    \multicolumn{1}{l}{\textbf{\upshape Oryzabase}} \\
    \midrule
    \endfirsthead
    
    \multicolumn{5}{c}{{\tablename} \thetable{} -- continued} \\
    \toprule
    \multicolumn{1}{l}{\textbf{\upshape RAP Locus ID}} &
    \multicolumn{1}{c}{\textbf{\upshape Description}} &
    \multicolumn{1}{l}{\textbf{\upshape RAP-DB}} &
    \multicolumn{1}{l}{\textbf{\upshape CGSNL}} &
    \multicolumn{1}{l}{\textbf{\upshape Oryzabase}} \\
    \midrule
    \endhead
        Os01g0866000 & Similar to E3 ubiquitin ligase EL5 (EC 6.3.2.-) & {} & RING189 & OsRING189 \\
        Os01g0865900 & Hypothetical protein & {} & {} & {} \\
        Os01g0866100 & Similar to Actin 4 & {} & AC7 & RAc7, OsAc7 \\
        Os01g0866133 & Hypothetical gene & {} & {} & {} \\
        Os01g0866200 & Similar to Histone H3 & OsHTR702 & {} & {} \\
        Os01g0866300 & VAMP-like protein YKT61 (AtYKT61) (Geranylgeranylated protein 1) (AtGP1) & {} & {} & {} \\
        Os01g0866400 & Cytosolic fructose-1, 6-bisphosphatase, Sucrose biosynthesis, Required for tiller bud outgrowth & OscFBP1, MOC2 & MOC2 & FR, OsFR, FBP1, OsFBP1, OsMOC2, OscFBP1, cFBP1, MOC2/FBP1 \\
        Os01g0866500 & Similar to Soluble inorganic pyrophosphatase (EC 3.6.1.1) (Pyrophosphate phospho- hydrolase) (PPase) & {} & {} & {} \\
\bottomrule
    \end{longtable}
\end{landscape}

\begin{landscape}
    \setlength{\tabcolsep}{4pt}
    \begin{longtable}{@{}
      >{\itshape\raggedright\arraybackslash}p{0.15\linewidth}   
      >{\raggedright\arraybackslash}p{0.35\linewidth}           
      >{\itshape\raggedright\arraybackslash}p{0.16\linewidth}   
      >{\itshape\raggedright\arraybackslash}p{0.12\linewidth}   
      >{\itshape\raggedright\arraybackslash}p{0.22\linewidth}   
    @{}}
    \caption{Information about all potential candidate genes within 200 kb (100 kb upstream and downstream) of the lead SNP sf0323937770, related to Figure 2} \\
    \toprule
    \multicolumn{1}{l}{\textbf{\upshape RAP Locus ID}} &
    \multicolumn{1}{c}{\textbf{\upshape Description}} &
    \multicolumn{1}{l}{\textbf{\upshape RAP-DB}} &
    \multicolumn{1}{l}{\textbf{\upshape CGSNL}} &
    \multicolumn{1}{l}{\textbf{\upshape Oryzabase}} \\
    \midrule
    \endfirsthead
    
    \multicolumn{5}{c}{{\tablename} \thetable{} -- continued} \\
    \toprule
    \multicolumn{1}{l}{\textbf{\upshape RAP Locus ID}} &
    \multicolumn{1}{c}{\textbf{\upshape Description}} &
    \multicolumn{1}{l}{\textbf{\upshape RAP-DB}} &
    \multicolumn{1}{l}{\textbf{\upshape CGSNL}} &
    \multicolumn{1}{l}{\textbf{\upshape Oryzabase}} \\
    \midrule
    \endhead
        Os03g0626200 & Conserved hypothetical protein & {} & {} & {} \\
        Os03g0626400 & Hypothetical conserved gene & {} & {} & {} \\
        Os03g0626500 & Similar to predicted protein & {} & {} & {} \\
        Os03g0626600 & Zinc finger, LIM-type domain containing protein & {} & HDR3.1 & OsHDR3.1 \\
        Os03g0626650 & Non-protein coding transcript & {} & {} & {} \\
        Os03g0626700 & Multi antimicrobial extrusion protein MatE family protein & {} & MATE16 & OsMATE16 \\
        Os03g0626750 & Hypothetical protein & {} & {} & {} \\
        Os03g0626800 & Calcineurin-B-like protein 3, Calcium-binding protein & OsCBL3 & CBL3 & OsCBL3 \\
\bottomrule
    \end{longtable}
\end{landscape}

\begin{landscape}
    \setlength{\tabcolsep}{4pt}
    \begin{longtable}{@{}
      >{\itshape\raggedright\arraybackslash}p{0.15\linewidth}   
      >{\raggedright\arraybackslash}p{0.35\linewidth}           
      >{\itshape\raggedright\arraybackslash}p{0.16\linewidth}   
      >{\itshape\raggedright\arraybackslash}p{0.12\linewidth}   
      >{\itshape\raggedright\arraybackslash}p{0.22\linewidth}   
    @{}}
    \caption{Information about all potential candidate genes within 200 kb (100 kb upstream and downstream) of the lead SNP sf0327583070, related to Figure 2} \\
    \toprule
    \multicolumn{1}{l}{\textbf{\upshape RAP Locus ID}} &
    \multicolumn{1}{c}{\textbf{\upshape Description}} &
    \multicolumn{1}{l}{\textbf{\upshape RAP-DB}} &
    \multicolumn{1}{l}{\textbf{\upshape CGSNL}} &
    \multicolumn{1}{l}{\textbf{\upshape Oryzabase}} \\
    \midrule
    \endfirsthead
    \multicolumn{5}{c}{{\tablename} \thetable{} -- continued} \\
    \toprule
    \multicolumn{1}{l}{\textbf{\upshape RAP Locus ID}} &
    \multicolumn{1}{c}{\textbf{\upshape Description}} &
    \multicolumn{1}{l}{\textbf{\upshape RAP-DB}} &
    \multicolumn{1}{l}{\textbf{\upshape CGSNL}} &
    \multicolumn{1}{l}{\textbf{\upshape Oryzabase}} \\
    \midrule
    \endhead
    Os03g0689100 & Histidine acid phosphatase family protein & {} & {} & {} \\
    Os03g0689300 & Plasma membrane H\textsuperscript{+}-ATPase 1, Regulation of ammonium (NH$_4^+$) uptake, Regulation of stomatal opening for C fixation, Control of photosynthesis & OSA1 & AHA1 & OSA1, OsA1, osAHA1, AHA1 \\
    Os03g0689350 & Hypothetical protein & {} & {} & {} \\
    Os03g0689400 & Similar to NB-ARC domain containing protein, expressed & {} & {} & {} \\
    Os03g0689544 & Hypothetical protein & {} & {} & {} \\
    Os03g0689688 & Hypothetical protein & {} & {} & {} \\
    Os03g0689900 & Similar to predicted protein & {} & {} & {} \\
    Os03g0690000 & Similar to Costars family protein & {} & {} & {} \\
\bottomrule
    \end{longtable}
\end{landscape}

\begin{landscape}
    \setlength{\tabcolsep}{4pt}
    \begin{longtable}{@{}
      >{\itshape\raggedright\arraybackslash}p{0.15\linewidth}   
      >{\raggedright\arraybackslash}p{0.35\linewidth}           
      >{\itshape\raggedright\arraybackslash}p{0.16\linewidth}   
      >{\itshape\raggedright\arraybackslash}p{0.12\linewidth}   
      >{\itshape\raggedright\arraybackslash}p{0.22\linewidth}   
    @{}}
    \caption{Information about all potential candidate genes within 200 kb (100 kb upstream and downstream) of the lead SNP sf0426935488, related to Figure 2} \\
    \toprule
    \multicolumn{1}{l}{\textbf{\upshape RAP Locus ID}} &
    \multicolumn{1}{c}{\textbf{\upshape Description}} &
    \multicolumn{1}{l}{\textbf{\upshape RAP-DB}} &
    \multicolumn{1}{l}{\textbf{\upshape CGSNL}} &
    \multicolumn{1}{l}{\textbf{\upshape Oryzabase}} \\
    \midrule
    \endfirsthead
    \multicolumn{5}{c}{{\tablename} \thetable{} -- continued} \\
    \toprule
    \multicolumn{1}{l}{\textbf{\upshape RAP Locus ID}} &
    \multicolumn{1}{c}{\textbf{\upshape Description}} &
    \multicolumn{1}{l}{\textbf{\upshape RAP-DB}} &
    \multicolumn{1}{l}{\textbf{\upshape CGSNL}} &
    \multicolumn{1}{l}{\textbf{\upshape Oryzabase}} \\
    \midrule
    \endhead
    Os04g0540000 & {} & {} & {} & {} \\
    Os04g0540200 & BBX transcription factor, Regulation of heading date & OsRIP1 & {} & OsBBX13, BBX13, OsRIP1, RIP1 \\
    Os04g0540300 & COBRA-like protein 7 & {} & PCS10 & OsPCS10 \\
    Os04g0540401 & Hypothetical protein & {} & {} & {} \\
    Os04g0540500 & Hypothetical protein & {} & {} & {} \\
    Os04g0540600 & Aldehyde dehydrogese domain containing protein & {} & ALDH3B1 & OsALDH3B1 \\
    Os04g0540750 & Hypothetical protein & {} & {} & {} \\
    Os04g0540900 & Receptor-like cytoplasmic kise, Salt tolerance, Oxidative stress tolerance & STRK1 & RLCK154 & OsRLCK154, STRK1, OsSTRK1 \\
\bottomrule
    \end{longtable}
\end{landscape}

\begin{landscape}
    \setlength{\tabcolsep}{4pt}
    \begin{longtable}{@{}
      >{\itshape\raggedright\arraybackslash}p{0.15\linewidth}   
      >{\raggedright\arraybackslash}p{0.35\linewidth}           
      >{\itshape\raggedright\arraybackslash}p{0.16\linewidth}   
      >{\itshape\raggedright\arraybackslash}p{0.12\linewidth}   
      >{\itshape\raggedright\arraybackslash}p{0.22\linewidth}   
    @{}}
    \caption{Information about all potential candidate genes within 200 kb (100 kb upstream and downstream) of the lead SNP sf0428557370, related to Figure 2} \\
    \toprule
    \multicolumn{1}{l}{\textbf{\upshape RAP Locus ID}} &
    \multicolumn{1}{c}{\textbf{\upshape Description}} &
    \multicolumn{1}{l}{\textbf{\upshape RAP-DB}} &
    \multicolumn{1}{l}{\textbf{\upshape CGSNL}} &
    \multicolumn{1}{l}{\textbf{\upshape Oryzabase}} \\
    \midrule
    \endfirsthead
    \multicolumn{5}{c}{{\tablename} \thetable{} -- continued} \\
    \toprule
    \multicolumn{1}{l}{\textbf{\upshape RAP Locus ID}} &
    \multicolumn{1}{c}{\textbf{\upshape Description}} &
    \multicolumn{1}{l}{\textbf{\upshape RAP-DB}} &
    \multicolumn{1}{l}{\textbf{\upshape CGSNL}} &
    \multicolumn{1}{l}{\textbf{\upshape Oryzabase}} \\
    \midrule
    \endhead
    Os04g0569300 & Similar to Membrane protein & {} & {} & OsRhmbd11, RHMBD11 \\
    Os04g0569500 & Hypothetical protein & {} & {} & {} \\
    Os04g0569400 & Similar to OSIGBa0139P06.9 protein & {} & {} & {} \\
    Os04g0569900 & IQ calmodulin-binding region domain containing protein & {} & {} & {} \\
    Os04g0570000 & Cytochrome P450 87A3, Auxin sigling in the regulation of coleoptile growth & CYP87A3 & CYP87A3 & OsCYP87A3 \\
    Os04g0570125 & Hypothetical protein & {} & {} & {} \\
    Os04g0570250 & Hypothetical gene & {} & {} & {} \\
    Os04g0570500 & Cytochrome P450 family protein & {} & {} & DWF1, DIM/DWF1, DIM \\
\bottomrule
    \end{longtable}
\end{landscape}

\begin{landscape}
    \setlength{\tabcolsep}{4pt}
    \begin{longtable}{@{}
      >{\itshape\raggedright\arraybackslash}p{0.15\linewidth}   
      >{\raggedright\arraybackslash}p{0.35\linewidth}           
      >{\itshape\raggedright\arraybackslash}p{0.16\linewidth}   
      >{\itshape\raggedright\arraybackslash}p{0.12\linewidth}   
      >{\itshape\raggedright\arraybackslash}p{0.22\linewidth}   
    @{}}
    \caption{Information about all potential candidate genes within 200 kb (100 kb upstream and downstream) of the lead SNP sf0428988813, related to Figure 2} \\
    \toprule
    \multicolumn{1}{l}{\textbf{\upshape RAP Locus ID}} &
    \multicolumn{1}{c}{\textbf{\upshape Description}} &
    \multicolumn{1}{l}{\textbf{\upshape RAP-DB}} &
    \multicolumn{1}{l}{\textbf{\upshape CGSNL}} &
    \multicolumn{1}{l}{\textbf{\upshape Oryzabase}} \\
    \midrule
    \endfirsthead
    \multicolumn{5}{c}{{\tablename} \thetable{} -- continued} \\
    \toprule
    \multicolumn{1}{l}{\textbf{\upshape RAP Locus ID}} &
    \multicolumn{1}{c}{\textbf{\upshape Description}} &
    \multicolumn{1}{l}{\textbf{\upshape RAP-DB}} &
    \multicolumn{1}{l}{\textbf{\upshape CGSNL}} &
    \multicolumn{1}{l}{\textbf{\upshape Oryzabase}} \\
    \midrule
    \endhead
    Os04g0576800 & Similar to predicted protein & {} & {} & {} \\
    Os04g0576850 & Non-protein coding transcript & {} & {} & {} \\
    Os04g0576900 & LRR receptor-like kise, Non-canonical, Control of organ size & XIAO & XIAO & OsXIAO, OsCLV2e, CLV2e \\
    Os04g0577000 & Similar to H0404F02.2 protein & {} & {} & {} \\
    Os04g0577200 & Similar to Rac GTPase activating protein 1 & {} & {} & {} \\
    Os04g0577300 & Similar to GDSL-motif lipase/hydrolase-like protein & {} & GELP58 & OsGELP58, OsGELP58a, OsGELP58b, OsGELP58c, OsGELP58d \\
    Os04g0577337 & Conserved hypothetical protein & {} & {} & {} \\
    Os04g0577375 & Hypothetical conserved gene & {} & {} & {} \\
\bottomrule
    \end{longtable}
\end{landscape}

\begin{landscape}
    \setlength{\tabcolsep}{4pt}
    \begin{longtable}{@{}
      >{\itshape\raggedright\arraybackslash}p{0.15\linewidth}   
      >{\raggedright\arraybackslash}p{0.35\linewidth}           
      >{\itshape\raggedright\arraybackslash}p{0.16\linewidth}   
      >{\itshape\raggedright\arraybackslash}p{0.12\linewidth}   
      >{\itshape\raggedright\arraybackslash}p{0.22\linewidth}   
    @{}}
    \caption{Information about all potential candidate genes within 200 kb (100 kb upstream and downstream) of the lead SNP sf0431202219, related to Figure 2} \\
    \toprule
    \multicolumn{1}{l}{\textbf{\upshape RAP Locus ID}} &
    \multicolumn{1}{c}{\textbf{\upshape Description}} &
    \multicolumn{1}{l}{\textbf{\upshape RAP-DB}} &
    \multicolumn{1}{l}{\textbf{\upshape CGSNL}} &
    \multicolumn{1}{l}{\textbf{\upshape Oryzabase}} \\
    \midrule
    \endfirsthead
    \multicolumn{5}{c}{{\tablename} \thetable{} -- continued} \\
    \toprule
    \multicolumn{1}{l}{\textbf{\upshape RAP Locus ID}} &
    \multicolumn{1}{c}{\textbf{\upshape Description}} &
    \multicolumn{1}{l}{\textbf{\upshape RAP-DB}} &
    \multicolumn{1}{l}{\textbf{\upshape CGSNL}} &
    \multicolumn{1}{l}{\textbf{\upshape Oryzabase}} \\
    \midrule
    \endhead
    Os04g0616400 & Similar to H0525G02.12 protein & {} & MRLK16 & OsMRLK16 \\
    Os04g0616500 & LRR-receptor-like kise (LRR-RLK) family protein & {} & MRLK17 & OsMRLK17 \\
    Os04g0616600 & Serine/threonine protein kise-related domain containing protein & {} & {} & {} \\
    Os04g0616700 & Hypothetical conserved gene & {} & MRLK18 & OsMRLK18 \\
    Os04g0616800 & Similar to H0525G02.13 protein & {} & {} & {} \\
    Os04g0616900 & Similar to H0525G02.14 protein & {} & {} & {} \\
    Os04g0617050 & Similar to H0313F03.1 protein & {} & SAUR21 & OsSAUR21 \\
    Os04g0617200 & Conserved hypothetical protein & {} & {} & {} \\
\bottomrule
    \end{longtable}
\end{landscape}
    
\begin{landscape}
    \setlength{\tabcolsep}{4pt}
    \begin{longtable}{@{}
      >{\itshape\raggedright\arraybackslash}p{0.15\linewidth}   
      >{\raggedright\arraybackslash}p{0.35\linewidth}           
      >{\itshape\raggedright\arraybackslash}p{0.16\linewidth}   
      >{\itshape\raggedright\arraybackslash}p{0.12\linewidth}   
      >{\itshape\raggedright\arraybackslash}p{0.22\linewidth}   
    @{}}
    \caption{Information about all potential candidate genes within 200 kb (100 kb upstream and downstream) of the lead SNP sf0501670937, related to Figure 2} \\
    \toprule
    \multicolumn{1}{l}{\textbf{\upshape RAP Locus ID}} &
    \multicolumn{1}{c}{\textbf{\upshape Description}} &
    \multicolumn{1}{l}{\textbf{\upshape RAP-DB}} &
    \multicolumn{1}{l}{\textbf{\upshape CGSNL}} &
    \multicolumn{1}{l}{\textbf{\upshape Oryzabase}} \\
    \midrule
    \endfirsthead
    \multicolumn{5}{c}{{\tablename} \thetable{} -- continued} \\
    \toprule
    \multicolumn{1}{l}{\textbf{\upshape RAP Locus ID}} &
    \multicolumn{1}{c}{\textbf{\upshape Description}} &
    \multicolumn{1}{l}{\textbf{\upshape RAP-DB}} &
    \multicolumn{1}{l}{\textbf{\upshape CGSNL}} &
    \multicolumn{1}{l}{\textbf{\upshape Oryzabase}} \\
    \midrule
    \endhead
    Os05g0128000 & Homeodomain-like containing protein & {} & MSL24 & OsMSL24 \\
    Os05g0128100 & Hypothetical conserved gene & {} & {} & {} \\
    Os05g0128200 & CCCH-tandem zinc finger protein, Drought resistance & OsTZF5, OsC3H33 & C3H33 & OsC3H33, OsTZF5, TZF5, OsCCCH-Zn-5, CCCH-Zn-5 \\
    Os05g0128300 & Conserved hypothetical protein & {} & {} & {} \\
    Os05g0128350 & Hypothetical protein & {} & {} & {} \\
    Os05g0128400 & Cation diffusion facilitator (CDF) family protein, Bivalent cation transporter, Translocation of Zn, Cd and other heavy metals, Ion homeostasis & OsMTP1 & MTP1 & OsMTP1 \\
    Os05g0128500 & {} & {} & {} & {} \\
    Os05g0128900 & Similar to Trehalose-phosphatase family protein & {} & {} & {} \\
\bottomrule
    \end{longtable}
\end{landscape}

\begin{landscape}
    \setlength{\tabcolsep}{4pt}
    \begin{longtable}{@{}
      >{\itshape\raggedright\arraybackslash}p{0.15\linewidth}   
      >{\raggedright\arraybackslash}p{0.35\linewidth}           
      >{\itshape\raggedright\arraybackslash}p{0.16\linewidth}   
      >{\itshape\raggedright\arraybackslash}p{0.12\linewidth}   
      >{\itshape\raggedright\arraybackslash}p{0.22\linewidth}   
    @{}}
    \caption{Information about all potential candidate genes within 200 kb (100 kb upstream and downstream) of the lead SNP sf0504419821, related to Figure 2} \\
    \toprule
    \multicolumn{1}{l}{\textbf{\upshape RAP Locus ID}} &
    \multicolumn{1}{c}{\textbf{\upshape Description}} &
    \multicolumn{1}{l}{\textbf{\upshape RAP-DB}} &
    \multicolumn{1}{l}{\textbf{\upshape CGSNL}} &
    \multicolumn{1}{l}{\textbf{\upshape Oryzabase}} \\
    \midrule
    \endfirsthead
    \multicolumn{5}{c}{{\tablename} \thetable{} -- continued} \\
    \toprule
    \multicolumn{1}{l}{\textbf{\upshape RAP Locus ID}} &
    \multicolumn{1}{c}{\textbf{\upshape Description}} &
    \multicolumn{1}{l}{\textbf{\upshape RAP-DB}} &
    \multicolumn{1}{l}{\textbf{\upshape CGSNL}} &
    \multicolumn{1}{l}{\textbf{\upshape Oryzabase}} \\
    \midrule
    \endhead
        Os05g0172400 & Conserved hypothetical protein & {} & {} & {} \\
        Os05g0172500 & Cyclin-like F-box domain containing protein & {} & {} & OsFbox250, Os\_F0011, OsFBX161, FBX161 \\
        Os05g0172800 & Conserved hypothetical protein & {} & {} & {} \\
        Os05g0173500 & Conserved hypothetical protein & {} & {} & {} \\
        Os05g0173600 & {} & {} & {} & {} \\
        Os05g0173700 & Similar to DNA replication licensing factor MCM3 homolog (Replication origin activator) (ROA protein) (Fragment) & {} & {} & {} \\
        Os05g0173800 & TPP-binding enzymes, conserved site domain containing protein & {} & {} & {} \\
        Os05g0173900 & {} & {} & {} & {} \\
\bottomrule
    \end{longtable}
\end{landscape}

\begin{landscape}
    \setlength{\tabcolsep}{4pt}
    \begin{longtable}{@{}
      >{\itshape\raggedright\arraybackslash}p{0.15\linewidth}   
      >{\raggedright\arraybackslash}p{0.35\linewidth}           
      >{\itshape\raggedright\arraybackslash}p{0.16\linewidth}   
      >{\itshape\raggedright\arraybackslash}p{0.12\linewidth}   
      >{\itshape\raggedright\arraybackslash}p{0.22\linewidth}   
    @{}}
    \caption{Information about all potential candidate genes within 200 kb (100 kb upstream and downstream) of the lead SNP sf0601367485, related to Figure 2} \\
    \toprule
    \multicolumn{1}{l}{\textbf{\upshape RAP Locus ID}} &
    \multicolumn{1}{c}{\textbf{\upshape Description}} &
    \multicolumn{1}{l}{\textbf{\upshape RAP-DB}} &
    \multicolumn{1}{l}{\textbf{\upshape CGSNL}} &
    \multicolumn{1}{l}{\textbf{\upshape Oryzabase}} \\
    \midrule
    \endfirsthead
    \multicolumn{5}{c}{{\tablename} \thetable{} -- continued} \\
    \toprule
    \multicolumn{1}{l}{\textbf{\upshape RAP Locus ID}} &
    \multicolumn{1}{c}{\textbf{\upshape Description}} &
    \multicolumn{1}{l}{\textbf{\upshape RAP-DB}} &
    \multicolumn{1}{l}{\textbf{\upshape CGSNL}} &
    \multicolumn{1}{l}{\textbf{\upshape Oryzabase}} \\
    \midrule
    \endhead
        Os06g0123750 & {} & {} & {} & {} \\
        Os06g0124300 & Protein of unknown function DUF594 family protein & {} & {} & {} \\
        Os06g0124400 & Conserved hypothetical protein & {} & {} & {} \\
        Os06g0124900 & Conserved hypothetical protein & {} & {} & {} \\
        Os06g0125000 & Nucleotide-binding LRR receptor (NLR) family protein & {} & {} & {} \\
        Os06g0125132 & Conserved hypothetical protein & {} & {} & {} \\
        Os06g0125200 & FCS-like zinc finger (FLZ) protein 18, DUF domain containing protein, Flooding tolerance during germition & OsFLZ18 & FLZ18 & OsFLZ18 \\
        Os06g0125300 & Pentatricopeptide repeat domain containing protein & {} & {} & {} \\
\bottomrule
    \end{longtable}
\end{landscape}

\begin{landscape}
    \setlength{\tabcolsep}{4pt}
    \begin{longtable}{@{}
      >{\itshape\raggedright\arraybackslash}p{0.15\linewidth}   
      >{\raggedright\arraybackslash}p{0.35\linewidth}           
      >{\itshape\raggedright\arraybackslash}p{0.16\linewidth}   
      >{\itshape\raggedright\arraybackslash}p{0.12\linewidth}   
      >{\itshape\raggedright\arraybackslash}p{0.22\linewidth}   
    @{}}
    \caption{Information about all potential candidate genes within 200 kb (100 kb upstream and downstream) of the lead SNP sf0719790827, related to Figure 2} \\
    \toprule
    \multicolumn{1}{l}{\textbf{\upshape RAP Locus ID}} &
    \multicolumn{1}{c}{\textbf{\upshape Description}} &
    \multicolumn{1}{l}{\textbf{\upshape RAP-DB}} &
    \multicolumn{1}{l}{\textbf{\upshape CGSNL}} &
    \multicolumn{1}{l}{\textbf{\upshape Oryzabase}} \\
    \midrule
    \endfirsthead
    \multicolumn{5}{c}{{\tablename} \thetable{} -- continued} \\
    \toprule
    \multicolumn{1}{l}{\textbf{\upshape RAP Locus ID}} &
    \multicolumn{1}{c}{\textbf{\upshape Description}} &
    \multicolumn{1}{l}{\textbf{\upshape RAP-DB}} &
    \multicolumn{1}{l}{\textbf{\upshape CGSNL}} &
    \multicolumn{1}{l}{\textbf{\upshape Oryzabase}} \\
    \midrule
    \endhead
        Os07g0513600 & Hypothetical conserved gene & {} & {} & {} \\
        Os07g0513700 & {} & {} & {} & {} \\
        Os07g0513900 & {} & {} & {} & {} \\
        Os07g0514000 & {} & {} & {} & {} \\
        Os07g0514101 & {} & {} & {} & {} \\
        Os07g0514200 & {} & {} & {} & {} \\
        Os07g0514300 & {} & {} & {} & {} \\
        Os07g0514400 & Domain of unknown function DUF629 domain containing protein & {} & {} & {} \\
\bottomrule
    \end{longtable}
\end{landscape}

\begin{landscape}
    \setlength{\tabcolsep}{4pt}
    \begin{longtable}{@{}
      >{\itshape\raggedright\arraybackslash}p{0.15\linewidth}   
      >{\raggedright\arraybackslash}p{0.35\linewidth}           
      >{\itshape\raggedright\arraybackslash}p{0.16\linewidth}   
      >{\itshape\raggedright\arraybackslash}p{0.12\linewidth}   
      >{\itshape\raggedright\arraybackslash}p{0.22\linewidth}   
    @{}}
    \caption{Information about all potential candidate genes within 200 kb (100 kb upstream and downstream) of the lead SNP sf0809027844, related to Figure 2} \\
    \toprule
    \multicolumn{1}{l}{\textbf{\upshape RAP Locus ID}} &
    \multicolumn{1}{c}{\textbf{\upshape Description}} &
    \multicolumn{1}{l}{\textbf{\upshape RAP-DB}} &
    \multicolumn{1}{l}{\textbf{\upshape CGSNL}} &
    \multicolumn{1}{l}{\textbf{\upshape Oryzabase}} \\
    \midrule
    \endfirsthead
    \multicolumn{5}{c}{{\tablename} \thetable{} -- continued} \\
    \toprule
    \multicolumn{1}{l}{\textbf{\upshape RAP Locus ID}} &
    \multicolumn{1}{c}{\textbf{\upshape Description}} &
    \multicolumn{1}{l}{\textbf{\upshape RAP-DB}} &
    \multicolumn{1}{l}{\textbf{\upshape CGSNL}} &
    \multicolumn{1}{l}{\textbf{\upshape Oryzabase}} \\
    \midrule
    \endhead
        Os08g0246200 & {} & {} & {} & {} \\
        Os08g0246300 & CC-NBS-LRR protein, Defense response, Control of crown root development & SCR8, RGly8 & SCR8 & OsSCR8 \\
        Os08g0246351 & Hypothetical protein & {} & {} & {} \\ 
        Os08g0246400 & Similar to Ubiquinol-cytochrome c reductase complex 14 kDa protein (EC 1.10.2.2) (CR14) & {} & {} & {} \\
        Os08g0246550 & Hypothetical gene & {} & {} & {} \\
        Os08g0246800 & Conserved hypothetical protein & {} & {} & {} \\
        Os08g0246950 & Hypothetical protein & {} & {} & {} \\
        Os08g0247600 & LRR-receptor-like kinase (LRR-RLK) family protein, Non-canonical & {} & {} & {} \\
\bottomrule
    \end{longtable}
\end{landscape}

\begin{landscape}
    \setlength{\tabcolsep}{4pt}
    \begin{longtable}{@{}
      >{\itshape\raggedright\arraybackslash}p{0.15\linewidth}   
      >{\raggedright\arraybackslash}p{0.35\linewidth}           
      >{\itshape\raggedright\arraybackslash}p{0.16\linewidth}   
      >{\itshape\raggedright\arraybackslash}p{0.12\linewidth}   
      >{\itshape\raggedright\arraybackslash}p{0.22\linewidth}   
    @{}}
    \caption{Information about all potential candidate genes within 200 kb (100 kb upstream and downstream) of the lead SNP sf0824675627, related to Figure 2} \\
    \toprule
    \multicolumn{1}{l}{\textbf{\upshape RAP Locus ID}} &
    \multicolumn{1}{c}{\textbf{\upshape Description}} &
    \multicolumn{1}{l}{\textbf{\upshape RAP-DB}} &
    \multicolumn{1}{l}{\textbf{\upshape CGSNL}} &
    \multicolumn{1}{l}{\textbf{\upshape Oryzabase}} \\
    \midrule
    \endfirsthead
    \multicolumn{5}{c}{{\tablename} \thetable{} -- continued} \\
    \toprule
    \multicolumn{1}{l}{\textbf{\upshape RAP Locus ID}} &
    \multicolumn{1}{c}{\textbf{\upshape Description}} &
    \multicolumn{1}{l}{\textbf{\upshape RAP-DB}} &
    \multicolumn{1}{l}{\textbf{\upshape CGSNL}} &
    \multicolumn{1}{l}{\textbf{\upshape Oryzabase}} \\
    \midrule
    \endhead
        Os08g0497600 & WD40 subfamily protein, Salt stress & SRWD1, SRWD1, OsWD40-162 & SRWD1 & SRWD1, OsWD40-162, OsDWD57, DWD57 \\
        Os08g0497750 & {} & {} & {} & {} \\
        Os08g0497900 & Conserved hypothetical protein & {} & {} & {} \\
        Os08g0498100 & O-methyltransferase & ROMT-15 & {} & OsOMT26, ROMT-15, ROMT15, COA20, OsCCoAOMT, CCoAOMT, OsCCoAOMT5, CCoAOMT5 \\
        Os08g0498400 & O-methyltransferase & ROMT-17 & {} & ROMT-17, OsROMT-17, ROMT17, Os ROMT17, CCoAOMT1, OsCCoAOMT1, OsCCoAOMT3, CCoAOMT3 \\
        Os08g0498600 & Similar to Caffeoyl-CoA 3-O-methyltransferase (Fragment) & {} & {} & ROMT-27, ROMT27 \\
        Os08g0499000 & Conserved hypothetical protein & {} & {} & {} \\
        Os08g0499100 & ATPase, AAA+ type, core domain containing protein & {} & {} & {} \\
\bottomrule
    \end{longtable}
\end{landscape}

\begin{landscape}
    \setlength{\tabcolsep}{4pt}
    \begin{longtable}{@{}
      >{\itshape\raggedright\arraybackslash}p{0.15\linewidth}   
      >{\raggedright\arraybackslash}p{0.35\linewidth}           
      >{\itshape\raggedright\arraybackslash}p{0.16\linewidth}   
      >{\itshape\raggedright\arraybackslash}p{0.12\linewidth}   
      >{\itshape\raggedright\arraybackslash}p{0.22\linewidth}   
    @{}}
    \caption{Information about all potential candidate genes within 200 kb (100 kb upstream and downstream) of the lead SNP sf0905061479, related to Figure 2} \\
    \toprule
    \multicolumn{1}{l}{\textbf{\upshape RAP Locus ID}} &
    \multicolumn{1}{c}{\textbf{\upshape Description}} &
    \multicolumn{1}{l}{\textbf{\upshape RAP-DB}} &
    \multicolumn{1}{l}{\textbf{\upshape CGSNL}} &
    \multicolumn{1}{l}{\textbf{\upshape Oryzabase}} \\
    \midrule
    \endfirsthead
    \multicolumn{5}{c}{{\tablename} \thetable{} -- continued} \\
    \toprule
    \multicolumn{1}{l}{\textbf{\upshape RAP Locus ID}} &
    \multicolumn{1}{c}{\textbf{\upshape Description}} &
    \multicolumn{1}{l}{\textbf{\upshape RAP-DB}} &
    \multicolumn{1}{l}{\textbf{\upshape CGSNL}} &
    \multicolumn{1}{l}{\textbf{\upshape Oryzabase}} \\
    \midrule
    \endhead
        Os09g0265800 & HGWP repeat domain containing protein & {} & {} & {} \\
        Os09g0266000 & Similar to predicted protein & {} & {} & {} \\
        Os09g0266100 & Conserved hypothetical protein & {} & {} & {} \\
        Os09g0266200 & Non-protein coding transcript & {} & {} & {} \\
        Os09g0266300 & Conserved hypothetical protein & {} & {} & {} \\
        Os09g0266400 & PGAP1-like family protein & {} & {} & {} \\
        Os09g0266600 & NPH3 domain containing protein & {} & {} & BTBN18, OsBTBN18 \\
        Os09g0266950 & Hypothetical protein & {} & {} & {} \\
\bottomrule
    \end{longtable}
\end{landscape}

\begin{landscape}
    \setlength{\tabcolsep}{4pt}
    \begin{longtable}{@{}
      >{\itshape\raggedright\arraybackslash}p{0.15\linewidth}   
      >{\raggedright\arraybackslash}p{0.35\linewidth}           
      >{\itshape\raggedright\arraybackslash}p{0.16\linewidth}   
      >{\itshape\raggedright\arraybackslash}p{0.12\linewidth}   
      >{\itshape\raggedright\arraybackslash}p{0.22\linewidth}   
    @{}}
    \caption{Information about all potential candidate genes within 200 kb (100 kb upstream and downstream) of the lead SNP sf0915724607, related to Figure 2} \\
    \toprule
    \multicolumn{1}{l}{\textbf{\upshape RAP Locus ID}} &
    \multicolumn{1}{c}{\textbf{\upshape Description}} &
    \multicolumn{1}{l}{\textbf{\upshape RAP-DB}} &
    \multicolumn{1}{l}{\textbf{\upshape CGSNL}} &
    \multicolumn{1}{l}{\textbf{\upshape Oryzabase}} \\
    \midrule
    \endfirsthead
    \multicolumn{5}{c}{{\tablename} \thetable{} -- continued} \\
    \toprule
    \multicolumn{1}{l}{\textbf{\upshape RAP Locus ID}} &
    \multicolumn{1}{c}{\textbf{\upshape Description}} &
    \multicolumn{1}{l}{\textbf{\upshape RAP-DB}} &
    \multicolumn{1}{l}{\textbf{\upshape CGSNL}} &
    \multicolumn{1}{l}{\textbf{\upshape Oryzabase}} \\
    \midrule
    \endhead
        Os09g0429500 & Similar to Glutamate receptor & {} & GLR2.4 & OsGLR2.4, GLR2.4 \\
        Os09g0429600 & Non-protein coding transcript & {} & {} & {} \\
        Os09g0429766 & Hypothetical protein & {} & {} & {} \\
        Os09g0429932 & {} & {} & {} & {} \\
        Os09g0430600 & Hypothetical conserved gene & {} & ZOS9-10 & OsZOS9-10 \\
        Os09g0430701 & Conserved hypothetical protein & {} & {} & {} \\
        Os09g0431001 & {} & {} & {} & {} \\
        Os09g0431050 & Hypothetical gene & {} & {} & {} \\
\bottomrule
    \end{longtable}
\end{landscape}

\begin{landscape}
    \setlength{\tabcolsep}{4pt}
    \begin{longtable}{@{}
      >{\itshape\raggedright\arraybackslash}p{0.15\linewidth}   
      >{\raggedright\arraybackslash}p{0.35\linewidth}           
      >{\itshape\raggedright\arraybackslash}p{0.16\linewidth}   
      >{\itshape\raggedright\arraybackslash}p{0.12\linewidth}   
      >{\itshape\raggedright\arraybackslash}p{0.22\linewidth}   
    @{}}
    \caption{Information about all potential candidate genes within 200 kb (100 kb upstream and downstream) of the lead SNP sf0922835847, related to Figure 2} \\
    \toprule
    \multicolumn{1}{l}{\textbf{\upshape RAP Locus ID}} &
    \multicolumn{1}{c}{\textbf{\upshape Description}} &
    \multicolumn{1}{l}{\textbf{\upshape RAP-DB}} &
    \multicolumn{1}{l}{\textbf{\upshape CGSNL}} &
    \multicolumn{1}{l}{\textbf{\upshape Oryzabase}} \\
    \midrule
    \endfirsthead
    \multicolumn{5}{c}{{\tablename} \thetable{} -- continued} \\
    \toprule
    \multicolumn{1}{l}{\textbf{\upshape RAP Locus ID}} &
    \multicolumn{1}{c}{\textbf{\upshape Description}} &
    \multicolumn{1}{l}{\textbf{\upshape RAP-DB}} &
    \multicolumn{1}{l}{\textbf{\upshape CGSNL}} &
    \multicolumn{1}{l}{\textbf{\upshape Oryzabase}} \\
    \midrule
    \endhead
        Os09g0569700 & Conserved hypothetical protein & {} & {} & {} \\
        Os09g0569800 & Serine/threonine protein kise-related domain containing protein & {} & RLCK280 & OsPUB52, PUB52, OsRLCK280, OsUSP33, USP33 \\
        Os09g0569900 & Hypothetical protein & {} & {} & {} \\
        Os09g0570000 & Serine/threonine protein kise-related domain containing protein & {} & RLCK281 & OsPUB56, PUB56, OsRLCK281, OsUSP34, USP34 \\
        Os09g0570100 & Protein kinase, catalytic domain domain containing protein & {} & RLCK282 & OsRLCK282 \\
        Os09g0570150 & Hypothetical conserved gene & {} & {} & {} \\
        Os09g0570200 & Zinc finger, C2H2-type domain containing protein & {} & ZOS9-21 & OsZOS9-21 \\
        Os09g0570300 & Similar to Short-chain dehydrogese Tic32 & {} & {} & {} \\
\bottomrule
    \end{longtable}
\end{landscape}

\begin{landscape}
    \setlength{\tabcolsep}{4pt}
    \begin{longtable}{@{}
      >{\itshape\raggedright\arraybackslash}p{0.15\linewidth}   
      >{\raggedright\arraybackslash}p{0.35\linewidth}           
      >{\itshape\raggedright\arraybackslash}p{0.16\linewidth}   
      >{\itshape\raggedright\arraybackslash}p{0.12\linewidth}   
      >{\itshape\raggedright\arraybackslash}p{0.22\linewidth}   
    @{}}
    \caption{Information about all potential candidate genes within 200 kb (100 kb upstream and downstream) of the lead SNP sf1010959303, related to Figure 2} \\
    \toprule
    \multicolumn{1}{l}{\textbf{\upshape RAP Locus ID}} &
    \multicolumn{1}{c}{\textbf{\upshape Description}} &
    \multicolumn{1}{l}{\textbf{\upshape RAP-DB}} &
    \multicolumn{1}{l}{\textbf{\upshape CGSNL}} &
    \multicolumn{1}{l}{\textbf{\upshape Oryzabase}} \\
    \midrule
    \endfirsthead
    \multicolumn{5}{c}{{\tablename} \thetable{} -- continued} \\
    \toprule
    \multicolumn{1}{l}{\textbf{\upshape RAP Locus ID}} &
    \multicolumn{1}{c}{\textbf{\upshape Description}} &
    \multicolumn{1}{l}{\textbf{\upshape RAP-DB}} &
    \multicolumn{1}{l}{\textbf{\upshape CGSNL}} &
    \multicolumn{1}{l}{\textbf{\upshape Oryzabase}} \\
    \midrule
    \endhead
        Os10g0357700 & Conserved hypothetical protein & {} & {} & {} \\
        Os10g0357750 & {} & {} & {} & {} \\
        Os10g0357800 & N2,N2-dimethylguanosine tRNA methyltransferase family protein & {} & {} & {} \\
        Os10g0357850 & Hypothetical gene & {} & {} & {} \\
        Os10g0357900 & {} & {} & {} & {} \\
        Os10g0357950 & {} & {} & {} & {} \\
        Os10g0358100 & Non-protein coding transcript & {} & {} & {} \\
        Os10g0358150 & Conserved hypothetical protein & {} & {} & {} \\
\bottomrule
    \end{longtable}
\end{landscape}

\begin{landscape}
    \setlength{\tabcolsep}{4pt}
    \begin{longtable}{@{}
      >{\itshape\raggedright\arraybackslash}p{0.15\linewidth}   
      >{\raggedright\arraybackslash}p{0.35\linewidth}           
      >{\itshape\raggedright\arraybackslash}p{0.16\linewidth}   
      >{\itshape\raggedright\arraybackslash}p{0.12\linewidth}   
      >{\itshape\raggedright\arraybackslash}p{0.22\linewidth}   
    @{}}
    \caption{Information about all potential candidate genes within 200 kb (100 kb upstream and downstream) of the lead SNP sf1014955970, related to Figure 2} \label{table_s25}\\
    \toprule
    \multicolumn{1}{l}{\textbf{\upshape RAP Locus ID}} &
    \multicolumn{1}{c}{\textbf{\upshape Description}} &
    \multicolumn{1}{l}{\textbf{\upshape RAP-DB}} &
    \multicolumn{1}{l}{\textbf{\upshape CGSNL}} &
    \multicolumn{1}{l}{\textbf{\upshape Oryzabase}} \\
    \midrule
    \endfirsthead
    \multicolumn{5}{c}{{\tablename} \thetable{} -- continued} \\
    \toprule
    \multicolumn{1}{l}{\textbf{\upshape RAP Locus ID}} &
    \multicolumn{1}{c}{\textbf{\upshape Description}} &
    \multicolumn{1}{l}{\textbf{\upshape RAP-DB}} &
    \multicolumn{1}{l}{\textbf{\upshape CGSNL}} &
    \multicolumn{1}{l}{\textbf{\upshape Oryzabase}} \\
    \midrule
    \endhead
        Os10g0422300 & Pentatricopeptide repeat domain containing protein & {} & {} & {} \\
        Os10g0422401 & Similar to T6D22.22 & {} & {} & {} \\
        Os10g0422500 & Endonuclease/exonuclease/phosphatase domain containing protein & {} & {} & {} \\
        Os10g0422583 & Hypothetical protein & {} & {} & {} \\
        Os10g0422700 & Conserved hypothetical protein & {} & {} & {} \\
        Os10g0422800 & Protein of unknown function DUF1677, plant domain containing protein & {} & {} & {} \\
        Os10g0423000 & Conserved hypothetical protein & {} & {} & {} \\
        Os10g0423300 & Similar to BTB/POZ domain containing protein & {} & {} & OsMB40, MB40, OsMBTB33, MBTB33 \\
\bottomrule
    \end{longtable}
\end{landscape}
\clearpage

\begin{table}[h]
    \centering
    \caption{Lead SNPs, genomic loci, and mapped genes of swine, related to Figure 2} \label{table_s26}

    \begin{tabularx}{\textwidth}{@{} l l l l l l l >{\raggedright\arraybackslash}X @{}}
        \toprule
        \textbf{Lead SNP} & \textbf{Chr} & \textbf{Position} & \textbf{\textit{p} value} & 
        \textbf{Locus} & \textbf{Start} & \textbf{End} & \textbf{Genes} \\
        \midrule
        1-7680166 & 1 & 7680166 & $2.00 \times 10^{-13}$ & 1 & 7180166 & 8180166 & \textit{ACAT2}, \textit{TCP1}, \textit{Mas1}, \textit{\textbf{IGF2R}}, \textit{MRPL18}, \textit{Pnldc1}, \textit{SLC22A1}, \textit{SLC22A2}, \textit{FNDC1}, \textbf{SOD2}, \textit{Wtap}, \textit{SLC22A3} \\
        6-44068807 & 6 & 44068807 & $1.12 \times 10^{-13}$ & 2 & 43568807 & 44568807 & \textit{LSM14A}, \textit{WTIP}, \textit{KIAA0355}, \textit{GPI}, \textit{ZNF181}, \textit{PDCD2L}, \textit{UBA2}, \textit{ZNF792}, \textit{ZNF543}, \textit{ZNF599} \\
        17-15860017 & 17 & 15860017 & $1.30 \times 10^{-13}$ & 3 & 15360017 & 16360017 & \textit{\textbf{BMP2}}, \textit{\textbf{FABP3}} \\
        \bottomrule
    \end{tabularx}

    \begin{tablenotes}
        \item This table summarizes the lead SNPs identified in swine, including their genomic positions, associated loci, flanking regions, and mapped genes. Genes highlighted in bold represent candidate genes inferred through AIPheno interpretability analysis and supported by previously reported studies.
    \end{tablenotes}

\end{table}

\end{appendices}

\end{document}